\documentclass[12pt]{article}
\usepackage{amsmath}
\usepackage{graphicx}
\usepackage{setspace}
\usepackage{url} 
\usepackage{algorithm}
\usepackage{algpseudocode}

\RequirePackage{amsthm,amsmath,amsfonts,amssymb}
\RequirePackage[authoryear]{natbib}
\newtheorem{theorem}{Theorem}[section]
\newtheorem{lemma}[theorem]{Lemma}
\newtheorem{corollary}{Corollary}
\usepackage{bbm}
\usepackage{enumerate}
\usepackage{mathtools}
\usepackage{indentfirst}
\usepackage[colorlinks=true,citecolor=blue,linkcolor=blue]{hyperref}
\newcommand{\vtt}[1]{%
  \text{\normalfont\ttfamily\detokenize{#1}}%
}
\graphicspath{{Fig/}}

\usepackage{algpseudocode}

\usepackage{bbm}
\usepackage{enumerate}
\usepackage{mathtools}
\usepackage{indentfirst}

\usepackage{mathrsfs}
\usepackage{caption}
\usepackage{bbm}
\usepackage{tabularx}      
\usepackage{multirow}
\usepackage[table]{xcolor}
\usepackage{mathrsfs}
\usepackage{caption}
\usepackage{tabularx}      
\usepackage{multirow}
\usepackage[table]{xcolor}

\RequirePackage{graphicx}
\newcommand{\blind}{0}

\usepackage{caption}
\usepackage{caption}
\begin{document}

\def\spacingset#1{\renewcommand{\baselinestretch}%
{#1}\small\normalsize} \spacingset{1}


\if0\blind
{
  \title{Bayesian Quantile Regression with Subset Selection: A Decision  Analysis Perspective}
  \author{Joseph Feldman \hspace{.2cm}\\
    Department of Statistical Science, Duke University\\
    and \\
    Daniel R. Kowal \\
    Department of Statistics and Data Science, Cornell University\\
    Department of Statistics, Rice University}
    \date{}
  \maketitle
} \fi

\if1\blind
{
  \bigskip
  \bigskip
  \bigskip
  \begin{center}
    {\LARGE\bf Title}
\end{center}
  \medskip
} \fi

\bigskip
\begin{abstract}
Quantile regression is a powerful tool for inferring how covariates affect specific percentiles of the response distribution. Existing methods either estimate conditional quantiles separately for each quantile of interest or estimate the entire conditional distribution using semi- or non-parametric models. The former often produce  inadequate models for real data and do not share information across quantiles, while the latter are characterized by complex and constrained models that can be difficult to interpret and computationally inefficient. Neither approach is well-suited for quantile-specific subset selection. Instead, we pose the fundamental problems of linear quantile estimation, uncertainty quantification, and subset selection from a Bayesian decision analysis perspective. For any Bayesian regression model---including, but not limited to existing Bayesian quantile regression models---we derive optimal point estimates, interpretable uncertainty quantification, and scalable subset selection techniques for all model-based conditional quantiles. Our approach introduces a quantile-focused squared error loss that enables efficient, closed-form computing and maintains a close relationship with Wasserstein-based density estimation. In an extensive simulation study, our methods demonstrate substantial gains in quantile estimation accuracy, inference, and variable selection over frequentist and Bayesian competitors. We use these tools to identify and quantify the heterogeneous impacts of multiple social stressors and environmental exposures on  educational outcomes across the full spectrum of low-, medium-, and high-achieving students in North Carolina.
\end{abstract}

\noindent%
{\it Variable selection,
interpretable machine learning,
Bayesian inference, robust regression}  
\vfill

\newpage
\spacingset{1.75} 
\section{Introduction}
Quantile regression estimates the functional relationship between covariates and specific percentiles of a response variable. In linear quantile regression, the $\tau$th conditional quantile for a random variable $Y$ is modeled  as a function of $p$-dimensional predictors $\boldsymbol x$ via
\begin{equation}\label{linearQR}
Q_{\tau}\{Y \mid \boldsymbol x, \boldsymbol \beta(\tau)\} = \boldsymbol{x}^{\intercal}\boldsymbol\beta(\tau)
\end{equation}
(a nonlinear version is in Section~\ref{sec-nl}). 
Estimated across $\tau$, the coefficients $\boldsymbol{\beta}(\tau)$ summarize how the covariates affect not only the location, but also the shape of the response distribution $Y \mid \boldsymbol{x}$. The unique insight from linear quantile regression comes from identifying predictors with \emph{heterogeneous} effects $\beta_{j}(\tau) \neq \beta_{j}(\tau')$  for some $\tau \neq \tau'$.  This capability is essential when the covariates affect higher order moments or other distributional features of $Y \mid \boldsymbol{x}$. 
For instance, when analyzing education data (see Section~\ref{realdat}), it is important not only to identify the factors that impact educational outcomes, but also to determine whether those effects are different for low-, medium, or high-achieving students.   
Such heterogeneous, quantile-specific effects can have far-reaching implications for policy interventions. In general, quantile regression provides 
a  robust and comprehensive view of the relationship between covariates and the response variable, leading to many important applications:  medicine \citep{kottas2001bayesian}, finance \citep{bassett2002portfolio}, and environmental studies \citep{pandey1999comparative}, among many others.

Broadly, there are several important components and considerations in quantile regression. First, quantile-specific linear coefficient \emph{estimates} are obtained to detect potentially heterogeneous effects, $\beta_{j}(\tau) \neq \beta_{j}(\tau')$  for some $\tau \neq \tau'$, including both magnitude and direction. Second,  quantile-specific  \emph{uncertainty quantification} provides important context for these coefficients and facilitates comparisons across both quantiles and variables. Third, when $p$ is moderate or large, quantile-specific \emph{subset selection} provides more parsimonious summaries and identifies the most impactful covariates across 
the distribution $Y \mid \boldsymbol{x}$. Each of these targets must simultaneously respect the fundamental \emph{smoothness} across quantiles: estimates, uncertainties, and selections should be similar for adjacent quantiles. Finally, the algorithms that deliver these results should be scalable in both the number of observations $n$ and the number of covariates $p$. 

The confluence of these demands creates a challenging environment for general statistical procedures. Existing approaches diverge into separate paradigms, each with their own limitations (see Sections~\ref{sec-sep}~and~\ref{simult}). With this in mind, we seek to 
\begin{enumerate}
    \item Develop a decision analysis framework for quantile regression that is foundational, coherent, and compatible with {any} Bayesian regression model.
\end{enumerate}
Our decision analysis integrates quantile regression into a data-centric Bayesian workflow that prioritizes suitable modeling of observed data, rather than adherence to certain quantile modeling requirements (see Sections~\ref{sec-sep}~and~\ref{simult}). This broadens the utility of quantile regression, since it can be incorporated alongside traditional posterior summaries such as posterior means and credible intervals for model parameters. At the same time,  the foundational and general nature of the decision analysis provides a new path to 
\begin{enumerate}
    \item[{2.}] Rigorously unify divergent approaches for Bayesian quantile regression.
\end{enumerate}
From this perspective, the decision analysis provides essential context for comparing and contrasting the large array of existing Bayesian quantile regression models. More specifically, the inferential goal of the decision analysis is to
\begin{enumerate}
    \item[{3.}] Deliver optimal (in a decision theory sense) point estimates and uncertainty quantification for conditional quantile functions and quantile-specific linear coefficients.
\end{enumerate}
This goal is shared by many quantile regression approaches, and thus presents an opportunity for competitive evaluations (Section~\ref{sim}). More uniquely,  via decision analysis we 
\begin{enumerate}
    \item[{4.}] Provide efficient algorithms for quantile-specific subset selection.
\end{enumerate}
Subset selection is a daunting task---especially for quantile regression. Our decision analysis leverages state-of-the-art search algorithms and subset selection strategies for mean regression, and neatly adapts them to quantile regression. 

Before expanding upon these goals and contributions, we first review existing Bayesian and frequentist methods for quantile regression---both to showcase the successes in this area and to highlight the need for methodological advances.

\subsection{Separate Quantile Regressions}\label{sec-sep}
Separate quantile regression techniques estimate independent models for any set of quantiles, providing targeted estimation for each $\tau$.  Given paired  data $\{(\boldsymbol x_i, y_i)\}_{i=1}^{n}$, \cite{koenker1978regression} introduced this approach from a frequentist perspective, obtaining quantile-specific coefficient estimates by minimizing the check loss  
\begin{equation}\label{CL}
      \hat{\boldsymbol \beta}(\tau) =
\mbox{arg min}_{\boldsymbol \beta} \sum_{i=1}^{n}\rho_{\tau}\{y_{i},\boldsymbol x_{i}^{\intercal}\boldsymbol{\beta}(\tau)\}
\end{equation}
where $\rho_{\tau}(a,b) = \{a -b(\tau - \mathbbm{1}_{a-b<0})\}$.  
Since no closed-form solutions exist, \eqref{CL} is usually solved by linear programming. However, the solutions are computed separately for each $\tau$ with no  mechanism for information-sharing between $\hat{\boldsymbol \beta}(\tau)$ and $\hat{\boldsymbol \beta}(\tau')$ at nearby quantiles $\tau,\tau'$. As a result, the estimates can be erratic and non-smooth across $\tau$, especially for extreme quantiles near zero or one.  Confidence intervals are obtained through bootstrapping, which is computationally intensive, or asymptotic approximations, which can be inaccurate for small to moderate $n$ \citep{koenker2017handbook}. Similar to point estimation, there is no information-sharing across quantiles for these interval estimates. 



The Bayesian analog to \eqref{CL} uses separate, quantile-specific asymmetric Laplace (AL) likelihoods for $\boldsymbol y = \{y_i\}_{i=1}^{n}$ given $\boldsymbol X = \{\boldsymbol x_i\}_{i=1}^{n}$ with centrality parameters $\{ \boldsymbol{x}_i^{\intercal}\boldsymbol{\beta}(\tau)\}_{i=1}^{n}$ \citep{yu2001bayesian}:
\begin{equation}\label{AL}
      p_\tau\{\boldsymbol{y} \mid \boldsymbol X, \boldsymbol{\beta}(\tau)\}=
    \prod_{i=1}^{n}\tau(1-\tau) \mbox{exp}[-\rho_{\tau}\{y_{i},\boldsymbol x_{i}^{\intercal}\boldsymbol{\beta}(\tau)\}]. 
\end{equation}
Bayesian inference proceeds by placing priors on  $\boldsymbol \beta(\tau)$ and inferring posterior distributions separately for each $\tau$. Under a flat prior on $\boldsymbol \beta(\tau)$ and the likelihood \eqref{AL}, the maximum \textit{a posteriori} estimator yields the same solution as  \eqref{CL}. Broader theoretical justification  is provided by \cite{ALconsist}, who detail sufficient conditions under which the posterior for $\boldsymbol \beta(\tau)$ is strongly consistent. Convenient parameter expansions have been developed to facilitate posterior sampling under \eqref{AL} \citep{kozumi2011gibbs, fasiolo2021fast}, which has led to widespread implementation with open source software \citep{benoit2017bayesqr, alhamzawi2020brq}. 

In general, Bayesian quantile regression with the AL likelihood faces significant limitations. First, there is no information-sharing among nearby quantiles, which results in excessively large posterior uncertainties and underpowered inference for $\boldsymbol \beta(\tau)$, especially for extreme quantiles near zero or one (see Sections~\ref{sim}-\ref{realdat}). Second, an AL likelihood must be specified for each  $\tau$, which induces distinct Bayesian models for the same data, typically with no attempt to reconcile or combine them. Finally, the AL likelihood often produces a substantially inadequate model for real data across many, if not all $\tau$, which undermines the interpretability of the resulting inferences.  \cite{kowal2024monte} proposed a transformation-based generalization of \eqref{AL} to improve model adequacy,  but did not address the previous two concerns.

\subsection{Simultaneous Quantile Regression Methods}\label{simult}
The lack of information-sharing across quantiles for  separate (frequentist or Bayesian) quantile regression estimators commonly results in probabilistically incoherent quantile estimates. Specifically, separate quantile estimates often exhibit undesirable \emph{quantile crossing}, which occurs when $\boldsymbol x^{\intercal}\boldsymbol \beta(\tau) > \boldsymbol x^{\intercal}\boldsymbol\beta(\tau')$ for $\tau < \tau'$, violating basic probability properties for the implied conditional distribution of $Y \mid \boldsymbol x$. In response, a variety of Bayesian and frequentist methodologies have been developed to provide quantile regressions that ensure both smoothness and non-crossing of the coefficients across quantiles $\tau$.  In contrast to separate quantile regression methods, these \emph{simultaneous} quantile regression methods fit a singular model to the data that  estimates all quantiles of $Y \mid \boldsymbol{x}$.

\cite{bondell2010noncrossing} proposed to estimate linear quantiles \eqref{linearQR} jointly across $\tau$ and subject to  constraints that enforce quantile non-crossing. \cite{Tokdar} introduced a suitable Bayesian version. Alternative likelihoods to the AL \eqref{AL} have included  empirical likelihoods \citep{emplike} and substitution likelihoods \citep{dunson2005approximate}, which allow simultaneous inference for multiple quantiles.  Other approaches  estimate \eqref{linearQR} by specifying semi- or non-parametric distributions for the errors $\{y_i - \boldsymbol x_{i}^{\intercal}\boldsymbol{\beta}(\tau)\}$ that satisfy certain quantile restrictions  \citep{kottas2001bayesian, kottas2009bayesian,reichBNPqr,reich2013bayesian}. \cite {taddy2010bayesian} specified a joint model for $(Y, \boldsymbol{x})$ and then inferred the conditional quantiles of $Y\mid \boldsymbol{x}$, which enforces quantile non-crossing but sacrifices a convenient form for $Q_{\tau}(Y \mid \boldsymbol x)$ such as linearity \eqref{linearQR}.

A principal criticism of simultaneous quantile regression methods is their complexity, both for modelling and  computing. For many semi- or non-parametric simultaneous methods, the  functional form of the conditional quantiles deviates from the linear parameterization \eqref{linearQR}. Consequently, inference and comparisons among quantile-specific covariate effects are difficult to interpret and detection of heterogeneous covariate effects is more challenging. Further, models constrained to prevent quantile crossing require sophisticated optimization techniques or sampling algorithms, which present significant burdens for even moderately-sized  data. Related, open source software for these methods is lacking. Finally, the modeling and computational complexity of these methods inhibits \emph{quantile-specific subset selection}, which we discuss below. 

\subsection{Subset Selection in Quantile Regression}
The goal of subset selection is to identify parsimonious representations of the regression function without sacrificing predictive power. This is particularly useful when $p$ is moderate or large: with fewer active (or selected) covariates, it is easier to interpret the effects of each variable.  Further, subset selection reduces storage requirements and can lower the variability of the estimated effects. For quantile regression, there is the added complexity that subset selection must be quantile-specific. This further aids in detecting covariate heterogeneity, as the subsets are allowed to vary between quantiles.

Among separate quantile regression methods, modifications to \eqref{CL} have been developed for quantile-specific variable selection. Sparse estimates are obtained by appending a penalty term to   \eqref{CL}, such as an $\ell_1$-penalty \citep{wu2009variable, 10.1214/10-AOS827,wang2012quantile, lee2014model}. The sparsity among penalized quantile regression estimates is controlled by a tuning parameter, typically selected via cross-validation, which can be computationally intensive for the (penalized) objective \eqref{CL}. These methods do not provide uncertainty quantification and are limited in that common sparsity penalties i) introduce overshrinkage of nonzero effects and ii) severely restrict the subset search path. Critically, these methods focus on selecting a single ``best" subset. However, even for  moderate $p$ with correlated covariates or weak signals, there are often many subsets that offer similar predictive accuracy. Thus, it can be misleading to report only a single ``best" subset; consideration of a broader collection of ``near-optimal" subsets can be more comprehensive and informative, but requires a robust subset search \citep{DKJMLR}. 

Although frequentist subset search and selection is well-studied for mean regression \citep{furnival2000regressions,hofmann2007efficient,Bertsimas2006}, extensions to quantile regression are so far unavailable. 




For Bayesian variable selection with separate AL likelihoods \eqref{AL}, it is common to use sparsity or shrinkage priors for $\boldsymbol \beta(\tau)$ \citep{li20081, alhamzawi2012bayesian,chen2013bayesian, keil2020quantile, dao2022bayesian}. However, these priors do not resolve the fundamental inadequacies of the AL model. Further, each of these methods selects variables marginally, either via posterior inclusion probabilities or  credible intervals that exclude zero. Under the AL likelihood, marginal posteriors for $\ \beta_j(\tau)$ are often characterized by large uncertainty, which results in severely underpowered variable selection (see Section~\ref{sim}).

For simultaneous quantile regression methods, no obvious path toward quantile-specific subset selection emerges. These methods, already burdened by complex constraints or semi- or non-parametric specifications, are not well-suited to incorporate cardinality constraints for each quantile. Thus, we consider alternative approaches.


\subsection{Overview of the Proposed Approach}
To address the gaps in quantile regression methodology, we develop a Bayesian decision analysis  for estimation, uncertainty quantification, and subset selection for  quantile regression. 
Our methodology centers around first building a Bayesian model to capture salient features in the data, and then extracting targeted summaries of the \emph{model-based} conditional quantiles. The general procedure is outlined in Algorithm~\ref{alg1}.
\begin{algorithm}[h]
\caption{Bayesian decision analysis for quantile regression}\label{alg1}
      \begin{enumerate}
          \item  Fit  a Bayesian regression model $\mathcal{M}$ (e.g., \eqref{LLS})
          \begin{quote}
              {\bf Inputs:} paired data $\{(\boldsymbol{x}_{i}, y_{i})\}_{i=1}^{n}$ 

              {\bf Output:} posterior distribution (e.g., posterior samples) of the model-based conditional quantiles $\{Q_\tau(Y \mid \boldsymbol x_i, \boldsymbol \theta)\}_{i=1}^n$ (e.g., \eqref{QF}) for each quantile $\tau$ 
          \end{quote}
          

        \item Conduct Bayesian decision analysis for quantile regression: 
        \begin{enumerate}
            \item Compute quantile-specific point estimates (Sections~\ref{LinActions}~and~\ref{sec-nl})
                \begin{quote}
                {\bf Output:} \emph{optimal actions}  (e.g., \eqref{OLS_sol}) that minimize the quantile-focused loss \eqref{relax-wass} for each quantile $\tau$ (nonlinear case: Section~\ref{sec-nl})
            \end{quote}

            \item Provide quantile-specific posterior uncertainty quantification (Section~\ref{projecting})
            \begin{quote}
                {\bf Output:} distribution for the \emph{posterior action}   \eqref{proj}   for each quantile $\tau$
            \end{quote}

            \item Search, filter, and select quantile-specific subsets (Section~\ref{subset search})

              \begin{quote}
                {\bf Output:}  quantile-specific \emph{acceptable family} \eqref{accept} that collects near-optimal subsets; quantile-specific \emph{smallest acceptable subset} \eqref{smallest}; and quantile-specific \emph{variable importance} metrics \eqref{VI}. 

                 
                
                
            \end{quote}
        \end{enumerate}

      \end{enumerate}
\end{algorithm}

Our formulation introduces several overarching advances to Bayesian quantile regression. First, the machinery is compatible with \emph{any} Bayesian regression model $\mathcal{M}$ for $Y\in\mathbb{R}$, including non-linear and non-parametric models \citep{pratola2020heteroscedastic}, Bayesian model averaging \citep{raftery1997bayesian}, and Bayesian model stacking \citep{stacking}, among many others. Our framework remains compatible with separate Bayesian AL models and Bayesian simultaneous quantile regression models, thus offering a unified framework for these divergent approaches. However, the enhanced generality of our approach allows the Bayesian modeler to build $\mathcal{M}$ to be suitable for the observed data, rather than to satisfy certain quantile-specific modeling requirements, such as AL likelihoods or unwieldy constraints. 

Next, we design a decision analysis that extracts point estimates of quantile-specific coefficients from the model $\mathcal{M}$ conditional quantile functions. Like any decision analysis, we must select a loss function; our approach features a \emph{quantile-focused squared error loss}. This choice is motivated through its close connection with the Wasserstein distance between measures \citep{frechet1948elements} and enables efficient, closed-form computation of optimal (in a decision analysis sense)   coefficients for any quantile and any subset of predictors. Crucially, the model $\mathcal{M}$ conveys regularization, both in the traditional sense (i.e., shrinkage of extraneous coefficients to zero) and via smoothness and information-sharing across nearby quantiles. Thus, while $\mathcal{M}$ need not feature linear quantiles \eqref{linearQR}, the conditional quantiles under a singular model $\mathcal{M}$ are probabilistically coherent and typically smooth across $\tau$. 

The underlying Bayesian regression model $\mathcal{M}$ also delivers uncertainty quantification for the quantile-specific linear coefficients. In particular, the solution to each quantile-specific decision analysis is a posterior functional, which provides posterior inference for the linear quantile coefficients. We emphasize that this procedure derives from a single model $\mathcal{M}$ and does not lead to data re-use or  model re-fitting for multiple quantiles. 

Finally, we leverage the decision analysis framework to achieve quantile-specific subset search and selection. Our approach yields closed-form linear coefficient estimates for any quantile and any subset, and unlocks decision analysis strategies for variable selection previously deployed only for mean regression  \citep{HahnCarvalho, woody, kowal2021fast, DKJMLR}. 
We extend these approaches for quantile regression and design a subset search procedure to accumulate an \emph{acceptable family}  of subsets that provide strong predictive accuracy for each quantile. From this acceptable family, we recommend one subset based on the parsimony principle (i.e., the smallest subset in this family), but also construct variable importance metrics to avoid the overreliance on any single ``best" subset. This is especially important when multiple subsets offer similar predictive accuracy. Our variable importance metrics seek to provide additional context in this common scenario.

The remainder of the paper is organized as follows. In Section \ref{sec1}, we introduce posterior decision analysis for  quantile regression. Section \ref{subset search} describes  our quantile-specific subset search and selection procedure. We provide a  simulation study in Section \ref{sim} to evaluate the proposed methodology for prediction, inference, selection,  and quantile crossing. In Section \ref{realdat} we conduct a quantile regression analysis using data on end-of-grade test scores for children in North Carolina. We conclude in Section~\ref{conclude}. Supplementary materials include proofs to all results, Bayesian model specifications, additional simulation results, and an \vtt{R} package\footnote{Codes implementing the proposed subset search and selection methodology are available at \href{https://github.com/jfeldman396/QRSubsets}{https://github.com/jfeldman396/QRSubsets}} implementing the proposed methodology.

\section{Bayesian Decision Analysis for Quantile Regression}\label{sec1}

Our approach to quantile regression first constructs a Bayesian regression model to capture salient features of $Y\mid \boldsymbol{x}$. Then, we use Bayesian decision analysis to provide linear summaries of the model-based quantiles. Crucially, the user chooses the underlying Bayesian model, which is unhindered by rigid structures imposed by models that require quantile-specific likelihoods (Section~\ref{sec-sep}) or constraints (Section~\ref{simult}).  

Given paired data $\{(\boldsymbol x_i, y_i)\}_{i=1}^{n}$, consider a Bayesian regression model $\mathcal{M}$ for  response variable $Y \in \mathbb{R}$ parameterized by $\boldsymbol{\theta}$. Consistent with a data-centric Bayesian workflow,  suppose $\mathcal{M}$ has been curated to best capture the conditional distribution of $Y \mid \boldsymbol x$, which may include nonlinearity, skewness, and  heteroscedasticity, among many other features. 

Importantly, $\mathcal{M}$  implicitly models the quantiles  of $Y \mid \boldsymbol{x}$. To see this, consider the class of additive location-scale models: 
\begin{equation}\label{LLS}
    y_{i} = f(\boldsymbol x_{i}) + s(\boldsymbol x_{i})\epsilon_{i}, \quad \epsilon_i \overset{iid}{\sim} F
\end{equation}
where $F$ is a cumulative distribution function (CDF) such that $\epsilon_i$ has mean zero and variance one. Under \eqref{LLS}, the conditional quantile function for any $\tau$ and covariate value $\boldsymbol x$ is a function of model parameters $\boldsymbol \theta = (f, s)$:
\begin{equation} \label{QF}
    Q_{\tau}(Y\mid \boldsymbol x_i, \boldsymbol{\theta}) = f(\boldsymbol x_i) + s(\boldsymbol x_i)F^{-1}(\tau).
\end{equation}
Given a prior $p(\boldsymbol{\theta})$, Bayesian inference for \eqref{LLS} targets the posterior distribution $p(\boldsymbol{\theta} \mid \boldsymbol y)$. Then,  for any $\tau$, the model $\mathcal{M}$ posterior propagates uncertainty to the conditional quantiles  via \eqref{QF}. Accordingly, we view $Q_{\tau}(Y \mid \boldsymbol x, \boldsymbol{\theta})$ as a \emph{posterior functional}. When the error quantile function $F^{-1}$ is smooth in $\tau$, then the implied model-based quantiles \eqref{QF} inherit this smoothness. Assuming $\mathcal{M}$ is a valid probability model for $Y \mid \boldsymbol{x}$,  $Q_{\tau}(Y\mid \boldsymbol x_i, \boldsymbol{\theta})$ is guaranteed to  avoid quantile crossing. 

 Leveraging the Bayesian model $\mathcal{M}$, our goal is to provide linear quantile estimates, uncertainty quantification, and subset selection. We design a \emph{decision analysis} for each task. Central to this approach is a loss function $\mathcal{L}\{Q_{\tau}(Y \mid \boldsymbol{x}, \boldsymbol \theta), \boldsymbol x^{\intercal} \boldsymbol \delta_S (\tau)\}$ that evaluates the accuracy of a quantile-specific \emph{linear action}, $\boldsymbol x^{\intercal} \boldsymbol \delta_S (\tau)$, with active (i.e., nonzero) coefficients for a given subset $S \subseteq \{1,\ldots,p\}$ of covariates. Nonlinear actions such as quantile-specific trees or additive functions are also compatible within this framework (see Section~\ref{sec-nl}). 
 
 Since the model-based quantiles $Q_{\tau}(Y \mid \boldsymbol{x}, \boldsymbol \theta)$ are unknown but inherit a posterior distribution under $\mathcal{M}$ through $p(\boldsymbol \theta \mid \boldsymbol y)$, the optimal coefficients are obtained by minimizing the posterior expected loss: 
\begin{equation}\label{post-dec}
    \hat{\boldsymbol \delta}_S(\tau) = \mbox{arg min}_{\boldsymbol \delta_S} E_{\boldsymbol \theta \mid \boldsymbol y} \mathcal{L}\{Q_{\tau}(Y \mid \boldsymbol{x}, \boldsymbol \theta), \boldsymbol x^{\intercal}\boldsymbol \delta_S(\tau)\}.
\end{equation}
Thus, \eqref{post-dec} provides linear quantile estimation, specific to each quantile $\tau$ and each subset $S$, under the model $\mathcal{M}$ and the loss $\mathcal{L}$.

These steps---fitting a Bayesian model $\mathcal{M}$ and then minimizing a posterior expected loss---are the foundational and uncontroversial components of a Bayesian decision analysis. 
The principal  task is to specify the loss function $\mathcal{L}$. We propose a \emph{quantile-focused squared error loss}, aggregated over the covariate values $\{\boldsymbol x_i\}_{i=1}^{n}$: 
  \begin{equation}\label{relax-wass}
      \mathcal{L}\{Q_{\tau}(Y \mid \boldsymbol{x}, \boldsymbol \theta), \boldsymbol x^{\intercal}\boldsymbol \delta_S(\tau)\}= 
        \sum_{i=1}^{n}\big\lVert Q_{\tau}(Y_i \mid \boldsymbol{x}_i, \boldsymbol \theta) - \boldsymbol x_i^{\intercal}\boldsymbol \delta_S(\tau)\big\rVert_{2}^{2}.
  \end{equation} 
    The loss function  \eqref{relax-wass} is exceptionally convenient for point estimation (Section~\ref{LinActions}), uncertainty quantification (Section~\ref{projecting}), and subset selection (Section~\ref{subset search}), with closed-form estimation and efficient subset search strategies. More formally, we establish deeper theoretical justification for \eqref{relax-wass} via connections to Wasserstein-based density regression (Section~\ref{wass}). 

\subsection{Point Estimation for Linear Quantiles}\label{LinActions}
A significant advantage of using the quantile-focused squared error loss \eqref{relax-wass} is in the availability of a closed-form solution for \eqref{post-dec}. For any quantile $\tau$ and subset of covariates $S$, we derive the optimal linear action:
\begin{lemma}\label{OLS} Suppose $E_{\boldsymbol \theta \mid \boldsymbol y}\lVert Q_{\tau}(Y_i \mid \boldsymbol x_i, \boldsymbol \theta)\rVert_{2}^{2}< \infty$ for $i = 1, \ldots, n$. For any quantile $\tau \in (0,1)$ and any subset of predictors $S \subseteq \{1,\dots,p\}$, the optimal action~\eqref{post-dec} under the quantile-focused squared error loss \eqref{relax-wass} is 
\begin{align}
\hat{\boldsymbol \delta}_S(\tau) =({\boldsymbol{X}}^{\intercal}_{S} {\boldsymbol{X}}_{S})^{-1}{\boldsymbol{X}}^{\intercal}_{S} \boldsymbol{\hat{Q}}_{\tau}({\boldsymbol{X}})\label{OLS_sol}\end{align}
with zeros for indices $j \notin S$, where ${\boldsymbol{X}}_{S}$ is the $n \times \lvert S\rvert $  matrix of active covariates for subset $S$, $\boldsymbol{\hat{Q}}_{\tau}(\boldsymbol{X}) =\{\hat{Q}_{\tau}(\boldsymbol x_1),\dots,\hat{Q}_{\tau}(\boldsymbol x_n)\}^\intercal$,  and 
 $\hat{Q}_{\tau}(\boldsymbol x_i) = E_{\boldsymbol \theta \mid \boldsymbol y}\{Q_{\tau}(Y_i \mid \boldsymbol x_i, \boldsymbol \theta)\}$.
\end{lemma}

The optimal linear action under the quantile-focused squared error loss  is the least squares solution with response vector $\boldsymbol{\hat{Q}}_{\tau}({\boldsymbol{X}})$ and covariate submatrix ${\boldsymbol{X}}_{S}$. The result is a linear point estimate of the $\tau$th conditional quantile of $Y \mid \boldsymbol{x}$, which provides the magnitude and direction of the relationship between each covariate and a specific percentile of the response. These coefficients may also be used for quantile-specific prediction. 


In general, the posterior expected conditional quantiles $\hat{Q}_{\tau}(\boldsymbol x_i) = E_{\boldsymbol \theta \mid \boldsymbol y}\{Q_{\tau}(Y_i \mid \boldsymbol x_i, \boldsymbol \theta)\}$ under $\mathcal{M}$ are not available in closed-form.  Monte Carlo approximations are easily obtained via $E_{\boldsymbol \theta \mid \boldsymbol y}[Q_{\tau}(Y_i \mid \boldsymbol x_i, \boldsymbol \theta)] \approx M^{-1}\sum_{m=1}^{M}Q_{\tau}(Y_i \mid \boldsymbol x_i, \boldsymbol \theta^{m})$ where $\{\boldsymbol{\theta}^m\}_{m=1}^M \sim p(\boldsymbol{\theta} \mid \boldsymbol{y})$.


It is apparent from Lemma \ref{OLS}  that different models will provide different optimal linear actions \eqref{post-dec}, since  $\boldsymbol{\hat{Q}}_{\tau}({\boldsymbol{X}})$ will vary from model to model. An important special case emerges for homoscedastic linear regression:
\begin{corollary} \label{cor1}
For the location-scale model \eqref{LLS} with linearity $f(\boldsymbol x_i) = \boldsymbol{x}_i^{\intercal}\boldsymbol \beta(\tau)$,  homoscedasticity $s(\boldsymbol x_i) = \sigma$, and an intercept $x_{i1} = 1$, 
the optimal action \eqref{post-dec} under \eqref{relax-wass} for the full set of covariates is $\boldsymbol {\hat{\boldsymbol \delta}}_{\{1,\dots,p\}}(\tau)= [\hat{\beta}_{1}(\tau), \{\hat{\beta}_{j}\}_{j = 2}^{p}]$ for any $\tau$, where $\hat{\beta}_{1}(\tau) = E_{\boldsymbol \theta \mid \boldsymbol{y}} [\beta_{1} + \sigma F^{-1}(\tau)]$ and $\hat{\beta}_{j} = E_{\boldsymbol \theta \mid \boldsymbol{y}} \beta_{j}$ for $j=2,\ldots,p$.
\end{corollary}

For homoscedastic linear  regression,  the quantile-specific linear action for the full set of covariates $S = \{1,\ldots,p\}$ is precisely the posterior expectation of the linear model coefficients, with a quantile-specific shift for the intercept. This result also emphasizes the inability of homoscedastic linear  regression to detect heterogeneous covariate effects; excluding the intercept, the quantile-specific linear coefficients are constant across $\tau$. 

Alternatively, consider any Bayesian model $\mathcal{M}$ with linear quantiles \eqref{linearQR}. Prominent examples include separate Bayesian AL regressions  along with several simultaneous quantile methods (Section~\ref{simult}). In this case, the optimal action \eqref{post-dec} under \eqref{relax-wass}  with all covariates $S = \{1,\ldots,p\}$ is the posterior expectation of the quantile-specific linear regression  coefficients:
\begin{corollary} \label{cor2}
For any Bayesian model with with linear quantiles $Q_{\tau}(Y_i \mid \boldsymbol x_i, \boldsymbol \theta)  = \boldsymbol{x}^{\intercal}\boldsymbol\beta(\tau)$,  the optimal action \eqref{post-dec} under \eqref{relax-wass} for the full set of covariates is $\boldsymbol {\hat{\boldsymbol \delta}}_{\{1,\dots,p\}}(\tau)= \hat{\boldsymbol{\beta}}(\tau)$, where $\hat{\boldsymbol{\beta}}(\tau) = E_{\boldsymbol \theta \mid \boldsymbol{y}} \boldsymbol{\beta}(\tau)$.
\end{corollary}


Corollaries~\ref{cor1}~and~\ref{cor2} show that, under common (mean and quantile) linear regression models, the optimal actions \eqref{post-dec} are directly related to the posterior expectations of the regression coefficients.  Thus, the point estimates \eqref{post-dec} inherit regularization (shrinkage, sparsity, smoothness, etc.) from $\mathcal{M}$, which improves estimation, especially for large $p$. 

While Corollaries~\ref{cor1}~and~\ref{cor2} confirm reasonable behavior for  the optimal action
\eqref{post-dec} under linear regression models, we emphasize the utility of Lemma \ref{OLS} for quantile regression estimation under \emph{any} Bayesian regression model. In particular, the analyst can prioritize curation of $\mathcal{M}$ to capture complex features of the entire conditional distribution $Y \mid \boldsymbol{x}$, including nonlinearity, skewness, and heteroscedasticity, while \eqref{OLS_sol} provides the optimal linear approximation of the model-based quantiles under the quantile-focused loss  \eqref{relax-wass}.


\subsection{Posterior Uncertainty Quantification} \label{projecting}
To enable posterior uncertainty quantification for the quantile-specific linear coefficients, we revisit \eqref{post-dec}, but \emph{without} the posterior expectation. Using the quantile-focused squared error loss \eqref{relax-wass},  
we define the quantile-specific \emph{posterior action} as the solution to
\begin{align}
\boldsymbol {{\boldsymbol \delta}}_{S}(\boldsymbol \theta; \tau) &= \mbox{arg min}_{  \boldsymbol{\delta}_{S}(\tau)} \sum_{i=1}^{ n}\lVert  Q_{\tau}(Y_i \mid \boldsymbol x_i, \boldsymbol \theta) - \boldsymbol x_i^{\intercal}   \boldsymbol{\delta}_{S}(\tau)\rVert_{2}^{2} \nonumber \\
 &= ({\boldsymbol X}^{\intercal}_{S} {\boldsymbol X}_{S})^{-1}\boldsymbol {X}^{\intercal}_{S} \boldsymbol{Q}_{\tau}(\boldsymbol{Y} \mid {\boldsymbol X}, \boldsymbol \theta)\label{proj}
\end{align}
for any subset of covariates $S$ and any quantile $\tau$, where $\boldsymbol{Q}_{\tau}(\boldsymbol{Y} \mid \boldsymbol X, \boldsymbol \theta) = \{Q_{\tau}(y_{1} \mid \boldsymbol{x}_1, \boldsymbol \theta),\dots, Q_{\tau}(y_{n} \mid \boldsymbol{x}_n, \boldsymbol \theta)\}^{\intercal}$. By design, \eqref{proj} projects the posterior functional $\boldsymbol{Q}_{\tau}(\boldsymbol Y \mid \boldsymbol X, \boldsymbol \theta)$ onto the corresponding sub-matrix of covariates, so $\boldsymbol {{\boldsymbol \delta}}_{S}(\boldsymbol \theta; \tau)$ inherits a posterior distribution under $\mathcal{M}$. 


Similar to the optimal action \eqref{post-dec}, the posterior action \eqref{proj} can be linked directly to the model $\mathcal{M}$ parameters by considering linear (mean or quantile) regression models. For homoscedastic linear regression, the posterior action  returns the posterior distribution of the linear regression coefficients, with a quantile-specific shift for the intercept:
\begin{corollary}\label{cor3}
    For the location-scale model \eqref{LLS} with linearity $f(\boldsymbol x_i) = \boldsymbol{x}_i^{\intercal}\boldsymbol \beta(\tau)$, homoscedasticity $s(\boldsymbol x_i) = \sigma$, and an intercept $x_{i1} = 1$,
  the  posterior action \eqref{proj} for the full set of covariates $S = \{1,\ldots,p\}$ satisfies
    \[
    \boldsymbol \delta_{\{1,\dots, p\}}(\boldsymbol \theta; \tau) \sim p(\boldsymbol \theta^* \mid \boldsymbol y)
    \]
    where  $\boldsymbol \theta^{*} = [\beta_{1} + \sigma F^{-1}(\tau), \{\beta_{j}\}_{j=2}^{p}]$.
\end{corollary}
As in Corollary~\ref{cor2}, a similar result is available when $\mathcal{M}$ features linear quantiles \eqref{linearQR}:  the posterior action for the full set of covariates is distributed according to the posterior distribution of the quantile-specific regression coefficients $\boldsymbol \beta (\tau)$. 

Once again, the advantage of the posterior action \eqref{proj} is that it delivers quantile- and subset-specific uncertainty quantification under  \emph{any}  Bayesian model $\mathcal{M}$. Crucially, the posterior action does not require Bayesian model re-fitting for each choice of quantile $\tau$ or subset $S$: all uncertainty derives from the single model $\mathcal{M}$ posterior. The distribution of \eqref{proj} is easily accessed given posterior samples $\{\boldsymbol \theta^{m}\}_{m=1}^{M}$: we simply  compute $Q_{\tau}(Y \mid \boldsymbol X, \boldsymbol\theta^{m})$ and plug the result into \eqref{proj} for any identified subset $S$. 


\subsection{Connection with the Wasserstein Geometry}\label{wass}
The quantile-focused squared error loss  \eqref{relax-wass} is  motivated through its connection with the Wasserstein geometry on the space of probability measures. Consider a posterior decision analysis for point-estimation of the probability density function (PDF) of $Y \mid \boldsymbol x$ under $\mathcal{M}$.  
Density regression can be accomplished under Wasserstein geometry, which defines valid metrics over the space of random probability measures \citep{PetersenAlexander2021W}. Formally, let $\mathcal{D}$ be the space of  univariate PDFs with finite second moments and consider the random PDF $g \in \mathcal{D}$ for $Y \mid \boldsymbol{x}$ with associated (conditional) CDF $G$ and quantile function $G^{-1}$. Regression of $g$ on covariates $\boldsymbol x$  finds $h \in \mathcal{D}$ (with  CDF $H$ and quantile function $H^{-1}$) that minimizes the expected squared-Wasserstein distance
\begin{equation}\label{sqwass}
    h^{*}(\boldsymbol{x}) = \mbox{argmin}_{h \in \mathcal{D}}E_{g \mid \boldsymbol{x}} d^{2}_{W}(g,h).
\end{equation}
Equivalently, \eqref{sqwass} minimizes the expected $L^{2}$ distance between conditional (on $\boldsymbol{x}$) quantiles $G^{-1}$ and $H^{-1}$ integrated over all $\tau$:
\begin{equation}\label{wass_int}
    h^{*}(\boldsymbol{x}) =\mbox{argmin}_{h \in \mathcal{D}}E_{g \mid \boldsymbol{x}} \int_{0}^{1}\big\lVert G^{-1}_{\tau} - H^{-1}_{\tau}\big\rVert_{2}^{2}\ d\tau.
\end{equation}
Clasically, data-driven estimation of \eqref{wass_int} requires multiple realizations from $Y_i \mid \boldsymbol x_i$ for each covariate value $\boldsymbol{x}_i$ in order to compute  empirical quantiles $\hat{G}^{-1}_{\tau}(Y_i \mid \boldsymbol x_i)$. By computing these empirical quantiles   over a fine grid $\tau \in\{0 < \tau_{1},\dots, \tau_{\ell}<1\}$,  \cite{PetersenAlexander2021W} proposed to estimate \eqref{wass_int} by minimizing
\begin{equation}\label{emp-wass}
    \sum_{\tau = \tau_{1}}^{\tau_{\ell}}\sum_{i=1}^{n}\lVert \hat{G}^{-1}_{\tau}(Y_i \mid \boldsymbol x_i)-     H^{-1}_{\tau}(Y_i \mid \boldsymbol{x}_i)\rVert_{2}^{2}.
\end{equation}
over densities $h \in \mathcal{D}$, and showed that the solution is a consistent estimator of $h^*$. 

For Bayesian decision analysis, the analogous approach is to replace the empirical quantiles $\hat{G}^{-1}_{\tau}(Y_i \mid \boldsymbol x_i)$  by the model $\mathcal{M}$ quantiles ${Q}_{\tau}(Y_i \mid \boldsymbol{x}_i,\boldsymbol \theta)$. Unlike the empirical quantiles, the model-based quantiles do \emph{not} require multiple realizations of $Y_i \mid \boldsymbol{x}_i$ at each $\boldsymbol{x}_i$. Then, like in \eqref{post-dec}, we minimize the posterior expected loss
\begin{equation}\label{post-dec-density}
    E_{\boldsymbol \theta \mid \boldsymbol{y}}\sum_{\tau = \tau_{1}}^{\tau_{\ell}}\sum_{i=1}^{n}\lVert {Q}_{\tau}(Y_i \mid \boldsymbol{x}_i,\boldsymbol \theta) - \boldsymbol \delta(\boldsymbol x_i; \tau)\rVert_{2}^{2}
\end{equation}
over densities with corresponding quantile functions $\boldsymbol \delta(\boldsymbol x_i; \tau)$. Thus, the minimizer of \eqref{post-dec-density} is a model-based point estimate of the PDF (or quantile function) of $Y \mid \boldsymbol{x}$ under squared-Wasserstein loss. 
 

To establish a connection with the proposed decision analysis for quantile regression in \eqref{post-dec}-\eqref{relax-wass}, we emphasize two critical choices for \eqref{post-dec-density}. First, we require linearity of the quantile function, $\boldsymbol \delta_{S}(\boldsymbol x;\tau) = \boldsymbol x^{\intercal}\boldsymbol \delta_{S} (\tau)$, possibly with a subset of active covariates $S \subseteq\{1,\ldots,p\}$. This requirement aligns with our goals of linear quantile regression \eqref{linearQR} and quantile-specific subset search and selection (Section~\ref{subset search}). Second, the decision analysis in  \eqref{post-dec}-\eqref{relax-wass} elects \emph{not} to impose the requirement that the estimated quantiles $\boldsymbol x^{\intercal}\boldsymbol \delta_{S} (\tau)$, taken jointly across $\tau$, yield a valid density function for each $\boldsymbol{x}$.  The primary motivation is tractability---including for point estimation, uncertainty quantification, and subset selection. Then, the collection of quantile-specific estimates $\hat{\boldsymbol \delta}_S(\tau)$ from \eqref{post-dec}-\eqref{OLS_sol}, taken across $\tau \in \{\tau_{1},\dots, \tau_{\ell}\}$, minimizes a modified version of the posterior expected squared-Wasserstein loss \eqref{post-dec-density}:
\begin{lemma}\label{form_relax_wass}
    Let $\boldsymbol \delta(\boldsymbol x;\tau) = \boldsymbol x^{\intercal}\boldsymbol \delta_{S} (\tau)$ for any $\boldsymbol{x}$ and any subset $S$ of covariates. Then the minimizer of \eqref{post-dec-density}, \emph{without} a density restriction, is given by the quantile estimators $\hat{\boldsymbol \delta}(\boldsymbol x;\tau_m) = \boldsymbol{x}^{\intercal}\hat{\boldsymbol \delta}_{S}(\tau_m)$ with $\hat{\boldsymbol \delta}_{S}(\tau_m)$ computed from \eqref{OLS_sol} separately for each $m=1,\ldots,\ell$. 
  
\end{lemma}

   

This result provides additional motivation for the quantile-focused squared error loss \eqref{relax-wass}, which is linked to Bayesian decision analysis for density estimation of $Y \mid \boldsymbol{x}$ under squared-Wasserstein loss.

We emphasize that the relaxation of the density requirement maintains alignment with our primary goals: estimation, uncertainty quantification, and selection of quantile-specific regression coefficients. We do not claim to provide valid density estimates. We do, however, observe that the estimated quantiles tend to preserve probabilistic coherence, such as quantile non-crossing (see Section E.5 of the supplementary material). Furthermore, this relaxation does not affect the validity of the proposed decision analysis: we specify a carefully-chosen loss function and minimize the posterior expected loss under the model $\mathcal{M}$. Indeed,  quantile regression often focuses on a few select quantiles $\tau$, such as upper or lower extremes, quartiles, and medians. Consideration of a fine grid $\tau \in \{\tau_{1},\dots, \tau_{\ell}\}$, while feasible with the proposed approach,  may offer little additional information. 


%



\subsection{Extensions for Nonlinear Quantiles}\label{sec-nl}
The proposed decision analysis framework for quantile estimation and uncertainty quantification is readily modified for nonlinear quantiles. First, recall that the Bayesian model $\mathcal{M}$ is generic, and may or may not specify linear quantiles as in \eqref{linearQR}. For example, the additive location-scale model \eqref{LLS} may incorporate a nonlinear mean function $f(\boldsymbol{x})$ or scale function $s(\boldsymbol{x})$, each of which induces nonlinear quantiles via \eqref{QF}. In this case, the linear summaries from Sections~\ref{LinActions}-\ref{projecting} may not be suitable. Instead, the  quantile-specific action $\boldsymbol \delta_{S}(\boldsymbol x; \tau)$ (e.g., as in \eqref{post-dec-density}) may be parametrized using decision trees or additive functions of $\boldsymbol x$, akin to  \cite{woody} for mean regression. Crucially, this does not completely sacrifice the computational convenience of our decision analysis: extending  Lemma~\ref{OLS}, the optimal action (similar to \eqref{post-dec}) is still obtained by minimizing a least squares objective with response vector $\hat{\boldsymbol Q}_{\tau}(\boldsymbol X)$. Similar extensions are available for the posterior action (akin to \eqref{proj}) to provide uncertainty quantification. However, nonlinear quantiles are not without drawbacks: they are typically more difficult to interpret, less amenable to subset selection, and require a choice for the parametrization of $\boldsymbol \delta_{S}(\boldsymbol x; \tau)$. Thus, we focus on linear quantile regression.

\section{Quantile-Specific Subset Search and Selection}\label{subset search}

A primary benefit of the quantile-focused squared error loss  is that for any $\tau$ and subset $S$ of covariates, the optimal linear action is given by the least squares solution with covariate matrix $\boldsymbol{X}_{S}$ and pseudo-response vector $\hat{\boldsymbol Q}_{\tau}(\boldsymbol X)$. We leverage this result (Lemma~\ref{OLS}) to provide new and efficient strategies for quantile-specific subset search and selection. 

\subsection{Subset Search with Decision Analysis}
Subset search requires i) evaluation criteria to compare among subsets and ii) algorithms to efficiently explore the (typically massive) space of all $2^p$ subsets.  For any quantile $\tau$ and subset $S$, we evaluate the optimal action $\hat{\boldsymbol{\delta}}_{S}(\tau)$ from \eqref{OLS_sol} using two quantities. First, 
\begin{equation}\label{exp-loss}
L_{S}(\tau) = E_{\boldsymbol \theta \mid \boldsymbol y} \sum_{i=1}^{ n}\lVert  Q_{\tau}(Y_i \mid \boldsymbol x_i, \boldsymbol \theta) - \boldsymbol x_i^{\intercal}   \hat{\boldsymbol{\delta}}_{S}(\tau)\rVert_{2}^{2}
\end{equation}
is the minimum achieved for each subset $S$ and quantile $\tau$ under the quantile-focused squared error loss \eqref{relax-wass}. Thus, \eqref{exp-loss} is a single number summary that compares the quantile estimates under the optimal action to the model-based conditional quantiles. Second, we compute  
\begin{equation} \label{post-loss}
L_{S}(\boldsymbol{\theta}; \tau)= \sum_{i=1}^{ n}\lVert  Q_{\tau}(Y_i \mid \boldsymbol x_i, \boldsymbol \theta) - \boldsymbol x_i^{\intercal}   \hat{\boldsymbol{\delta}}_{S}(\tau)\rVert_{2}^{2},
\end{equation}
which resembles \eqref{exp-loss} but instead 
inherits a posterior distribution via  $p(\boldsymbol{\theta}\mid \boldsymbol{y})$ under model $\mathcal{M}$. Thus, \eqref{post-loss} enables uncertainty quantification for the performance of each subset. The evaluation criteria in \eqref{exp-loss} and \eqref{post-loss} are used in conjunction to search and subsequently filter to the "near-optimal" or \emph{acceptable} family of subsets \citep{kowal2021fast}, specific to each quantile $\tau$. 

A key observation is that $L_{S_1}(\tau)  - L_{S_2}(\tau) = \hat L_{S_1}(\tau)  - \hat L_{S_2}(\tau)$, where $\hat L_S(\tau) = \sum_{i=1}^{ n}\lVert  \hat{Q}_{\tau}(\boldsymbol x_i) - \boldsymbol x_i^{\intercal}   \hat{\boldsymbol{\delta}}_S(\tau)\rVert_{2}^{2}$, so we need only consider the residual sum-of-squares (RSS) $\hat L_S(\tau)$ from a linear predictor with response $\hat{Q}_{\tau}(\boldsymbol x_i) = E_{\boldsymbol \theta \mid \boldsymbol y}\{Q_{\tau}(Y_i \mid \boldsymbol x_i, \boldsymbol \theta)\}$. Notably, classical and state-of-the-art subset search algorithms for mean regression rely on RSS comparisons among subsets \citep{furnival2000regressions,hofmann2007efficient,Bertsimas2006}. Thus, our deployment of the quantile-focused squared error loss \eqref{exp-loss} enables adaptation of these strategies for quantile-specific subset search and selection. 

When the complete enumeration of all possible $2^p$ subsets is not feasible, we apply the branch-and-bound (BBA) search algorithm \citep{furnival2000regressions} to efficiently eliminate non-competitive subsets. BBA uses a tree-based enumeration of all possible subsets to extract the most promising $m_{k} \leq {p \choose k}$ subsets for each size $k \in \{1,\dots,p\}$ according to RSS, which in our setting is equivalent to \eqref{exp-loss}. A key benefit of BBA is that it provides a large number of candidate subsets $m_k$ of each size $k$. Thus, we apply BBA as a pre-screening procedure to obtain \emph{candidate subsets} $\mathbbm{S}({\tau})$ for each quantile $\tau$. 

The inputs for the BBA algorithm are i) the model-based fitted quantiles $\{\hat{Q}_{\tau}(\boldsymbol x_i)\}$, ii) the covariates $\{\boldsymbol x_i\}$, and iii) the maximum number of subsets $m_{k}$ to return for each subset size $k$, which should be chosen as large as computationally feasible. In our simulation and application, we set $m_{k} = 50$.  An efficient implementation of the BBA algorithm is available in the \vtt{leaps} package in \vtt{R}, which provides fast filtering for $p \leq 35$; see the Section D of the supplement  for details on our model-assisted pre-screening approach when $p > 35$.

\subsection{Subset Filtration}\label{filt}
Although it is tempting to consider subset \emph{selection} based on $L_{S}(\tau)$, the well-known ordering properties of RSS apply in the current setting: for nested subsets $S_1 \subseteq S_2$, it is guaranteed that $L_{S_1}(\tau) \ge L_{S_2}(\tau)$. Thus, selection based on \eqref{exp-loss} alone will invariably and trivially select the full set of covariates, $S=\{1,\ldots,p\}$. This motivates our consideration of $L_S(\boldsymbol{\theta}; \tau)$ in \eqref{post-loss}, which provides an alternative path for subset selection.

We compare the posterior loss \eqref{post-loss} between any given subset, $L_S(\boldsymbol{\theta}; \tau)$, and the  model-based fitted quantiles, $L_{\hat Q}(\boldsymbol{\theta}; \tau) = \sum_{i=1}^{ n}\lVert  Q_{\tau}(Y_i \mid \boldsymbol x_i, \boldsymbol \theta) - \hat Q_\tau(\boldsymbol x_i)\rVert_{2}^{2}$, which provide a useful and competitive benchmark. Specifically, we compute the percent increase in loss between the optimal linear action and the model-based fitted quantiles:
 \begin{equation}\label{post-dif}
     D_S(\tau) = 100 \times \{L_S(\boldsymbol{\theta}; \tau)  -L_{\hat Q}(\boldsymbol{\theta}; \tau) \}/L_{\hat Q}(\boldsymbol{\theta}; \tau).
 \end{equation}
Crucially, $D_S(\tau)$ inherits a posterior distribution under $\mathcal{M}$, and is easily computed for each candidate subset $S \in \mathbbm{S}({\tau})$ using posterior samples of $\boldsymbol{\theta}$. Then, from \eqref{post-dif} we  collect the subsets   for which the optimal linear action has nonnegligble probability $\varepsilon > 0$ (under $\mathcal{M}$) of matching the performance of the fitted quantiles:
 \begin{equation}\label{accept}
    \mathbbm{A}_{\varepsilon}(\tau) = \{S \in \mathbbm{S}^{\tau}: p\{D_S(\tau)  \leq 0 \mid \boldsymbol{y}\} \geq \varepsilon\}. 
\end{equation}
We refer to $\mathbbm{A}_{\varepsilon}(\tau)$ as the  \emph{quantile-specific acceptable family}, which extends the acceptable family from mean regression to quantile regression \citep{kowal2021fast,DKJMLR}. Equivalently, $S \in \mathbbm{A}_{\varepsilon}(\tau)$ if and only if there is a lower $(1-\varepsilon)$ credible interval for $D_S(\tau)$ that includes zero \citep{kowal2021fast}. Given this correspondence, we set $\varepsilon = 0.05$ by default. 

Within the acceptable family, we select a subset based on the parsimony principle: 
\begin{equation}\label{smallest}
    \mbox{S}_{small}(\tau) = \mbox{argmin}_{S \in \mathbbm{A}_\varepsilon(\tau)} \lvert S\rvert
\end{equation}
which is the quantile-specific \emph{smallest acceptable subset}. By construction, $\mbox{S}_{small}(\tau)$ reports the simplest (most parsimonious) explanation that maintains competitive estimation for the $\tau$th quantile. The selected subset is quantile-specific, and thus it is informative to compare $\mbox{S}_{small}(\tau)$ across quantiles $\tau$.

When \eqref{smallest} is nonunique, we select the subset that minimizes \eqref{exp-loss}. In some cases, $\mathbbm{A}_\varepsilon(\tau)$ may be empty and $\mbox{S}_{small}(\tau)$ will be undefined, such as when the quantiles are highly nonlinear in $\boldsymbol{x}$. This outcome is informative: it suggests that the linear actions \eqref{post-dec} are inadequate for the $\tau$th quantile and must be replaced by alternative quantile summaries, such as trees or additive functions.

\subsection{Quantifying Variable Importance}\label{Vimps}
A principal advantage of curating the quantile-specific acceptable family \eqref{accept} is that it contains more information than any single ("best") subset. In the common applied setting with moderate $p$, correlated covariates, or weak signals, there are typically many subsets that offer similar accuracy. Thus, we de-emphasize  the selection of a single "best" subset and instead provide more comprehensive summaries of the quantile-specific acceptable family. 

For each covariate $j$, we introduce a quantile-specific \emph{variable importance} metric: 
\begin{equation}\label{VI}
   \mbox{VI}_{j}(\tau) = \vert \mathbbm{A}_{\varepsilon}(\tau)\vert^{-1} \sum_{S \in  \mathbbm{A}_{\varepsilon}(\tau)}\mathbbm{I}\{j\in S\}
\end{equation}
which reports the proportion of acceptable subsets that include variable $j$. Informally, \eqref{VI} measures how essential the $j$th covariate is for linearly predicting the $\tau$th quantile. We are particularly interested in \emph{keystone covariates} that appear in nearly all acceptable subsets, $\mbox{VI}_{j}(\tau) > q$ for some large cutoff $q \in [0,1]$. Alternatively, while a variable $j$ may be omitted from the "best" subset or $\mbox{S}_{small}(\tau)$, observing $\mbox{VI}_{j}(\tau) > 0$ implies that variable $j$ belongs to at least one subset with near-optimal quantile prediction. Naturally, $\mbox{VI}_{j}(\tau)$  may vary with $\tau$, which implies that the importance of covariate $j$ is heterogeneous across the percentiles of $Y\mid\boldsymbol{x}$.


\section{Simulation Study}\label{sim}
\subsection{Simulation Design}
We design a simulation study to compare the proposed methodology with competing methods for quantile regression. Response variables are simulated from the linear location-scale model
\begin{equation}\label{DGP}
    y_i =  \boldsymbol x_i^{\intercal}\boldsymbol \xi^{*} + (\boldsymbol{x}_i^{\intercal}\boldsymbol \gamma^{*})\epsilon_i, \quad \epsilon_i \overset{iid}{\sim} N(0,1).
\end{equation}
By design, \eqref{DGP} yields linear conditional quantile functions  $Q^{*}_\tau(\boldsymbol x) =  \boldsymbol x^{\intercal} \boldsymbol \beta^{*}(\tau)$, where $\beta^{*}_{1}(\tau) = 2 + \Phi^{-1}(\tau)$ and $ \beta^{*}_{j}(\tau) = \xi^{*}_{j} + \Phi^{-1}(\tau) \gamma^{*}_{j}, j = 2, \ldots, p$, with $\Phi$ the standard normal CDF. The covariate values $\boldsymbol{x}_{i}$ are simulated independently for $i=1,\ldots,n$ from a Gaussian copula with uniform marginals and copula correlation $\rho_{\ell j} = 0.5^{\lvert \ell -j\rvert}$, which provides varying degrees of correlation among the predictors. Additional simulations with independent covariates ($\rho_{\ell j} = 0$) are available in Section E.2 of the supplement.

We enforce both sparsity and heterogeneous effects in the response distribution through $\boldsymbol \xi^{*}$ and $\boldsymbol \gamma^{*}$. First, we incorporate homogeneous quantile effects by setting $\xi^{*}_{j} = 2$ and $ \gamma^{*}_{j} = 0$  for a pre-specified set of indices $j \in hom \subseteq \{2,\dots, p\}$, so the linear quantile coefficients for these variables  satisfy $\beta^{*}_{j}(\tau) = 2$ for all $\tau$. The intercepts are $\xi_{1}^{*} = 2$ and $\gamma_{1}^{*} = 1$.  Next, for one covariate $het 
 \in \{2,\dots, p\} \setminus hom$, we set $\gamma^{*}_{het} = h$ and $\xi^{*}_{het} = 0$, where, $h$ controls the strength of the heteroscedasticity in \eqref{DGP} and thus the  magnitude of $\beta^{*}_{het}(\tau)  = \Phi^{-1}(\tau)h$.  We determine $h$ by selecting a heterogeneity ratio,  $\mbox{HetRatio} = var(\boldsymbol x_i^{\intercal}\boldsymbol \xi^*)/\gamma^{*}_{het}$, from $\{0.5,1\}$ for weak or strong heterogeneity, respectively. The remaining coefficients are fixed at zero.

  To determine $hom$ and $het$ for $p$-dimensional (correlated) covariates, we isolate every fourth index using the sequence $\mbox{I}= \{2 + 4j, j \in (0,\dots,\lfloor p/4 \rfloor)\}$ and let $hom = \{\mbox{I} \setminus \mbox{median}(\mbox{I})\}$ and $het = \lfloor \mbox{median}(\mbox{I})\rfloor$. Given the dependence among the covariates, this ensures that the covariate with heterogeneous effects is at least moderately correlated with the remaining covariates, which is challenging for selection. A representation of this scheme is given below for $p = 10$, $\mbox{I} = \{2,6,10\}, \ het = 6, \ hom = \{2,10\}$:
\begin{align}
    \boldsymbol \gamma^{*} &= (\gamma^{*}_{1} = 1,\underbrace{\gamma^{*}_{2}}_{hom} =0,\dots,\underbrace{ \gamma^{*}_{6}}_{het} = h, \dots, \underbrace{\gamma^{*}_{10}}_{hom} = 0)\nonumber\\
     \boldsymbol \xi^{*} &= (\xi^{*}_{1} = 2,\underbrace{\xi^{*}_{2}}_{hom} = 2,\dots,\underbrace{ \xi^{*}_{6}}_{het} = 0, \dots, \underbrace{\xi^{*}_{10}}_{hom} = 2).\nonumber
\end{align}
We carry out our simulation design for $(n,p) \in \{(500,20),(200,50), (100,100)\}$ and $\mbox{HetRatio} \in \{0.5,1\}$ across 50 independent repetitions of the data generating process in each setting.  For brevity and because they more closely mirror our application, we present and discuss results for the $(n,p) \in \{(500,20),(200,50)\}$ settings. However, the $n = 100, p = 100$ results were qualitatively similar, and Section E of the supplement reports the results for all outstanding settings with nearly identical conclusions.

For the Bayesian model $\mathcal M$, we fit a  linear location-log scale (LL-LS) regression model to our simulated data:
\begin{equation}\label{M}
    y_i \mid \boldsymbol \theta \stackrel{indep}{\sim} \mbox{Normal}[\boldsymbol x_i^{\intercal} \boldsymbol{\xi}, \{\sigma\exp(\boldsymbol{x}_i^{\intercal}\boldsymbol{\gamma})\}^2]
\end{equation}
with model parameters    $\boldsymbol \theta = (\sigma, \boldsymbol \xi, \boldsymbol \gamma)$. By design, both the mean and variance functions are misspecified under $\mathcal{M}$: both include all covariates, even those with true null effects, while the variance function is nonlinear. The full Bayesian specification of $\mathcal{M}$ is available in Section C of the supplement, but we note here that regularizing priors are specified on both $\boldsymbol \xi$ and $\boldsymbol \gamma$ and the model is estimated using  \verb|STAN| \citep{carpenter2017stan}. Posterior samples of the $\tau$th conditional quantile function under $\mathcal{M}$ are easily computed under \eqref{M}: for each posterior sample $\boldsymbol \theta^{s}$ and any covariate value $\boldsymbol x$, we simply compute $Q_{\tau}(Y \mid \boldsymbol{x}, \boldsymbol \theta^{s}) =  \boldsymbol x^{\intercal}\boldsymbol{\xi}^{s} +\sigma^s\exp(\boldsymbol{x}^{\intercal}\boldsymbol{\gamma}^{s})\Phi^{-1}(\tau)$.



Under each $(n,p,\mbox{HetRatio})$ combination, we curate acceptable families for $\tau \in \{0.01,0.05,\\0.25, 0.5,0.75,0.95, 0.99\}$ using the proposed methodology in Sections~\ref{sec1}-\ref{subset search} under default settings ($m_{k} = 50, \varepsilon = 0.05$). Among the subsets in each $\mathbbm{A}_{0.05}(\tau)$, we evaluate quantile prediction, inference, and variable selection for  $\mbox{S}_{small}(\tau)$. 

For a frequentist competitor, we fit  adaptive LASSO quantile regressions separately for each $\tau$, which use an $\ell_1$-penalized check loss for sparse estimation \citep{wu2009variable}. The (sparsity) tuning parameter is selected by 
5-fold cross-validation with the one-standard-error rule  $(\mbox{aLASSO})$, and an efficient implementation is available in the \vtt{R} package \vtt{rqPen} \citep{sherwood2017rqpen}. 

For a Bayesian competitor, we fit Bayesian linear quantile regressions with the AL likelihood  and  adaptive LASSO priors   ($\mbox{AL}_{Bayes}$; \cite{alhamzawi2012bayesian}) using default settings in the \vtt{bayesQR} package in \vtt{R} \citep{benoit2017bayesqr}. Point estimates are computed using posterior means of the coefficients. Variable selection with $\mbox{AL}_{Bayes}$ identifies those covariates for which the 95\% highest posterior density intervals for the coefficients do not include zero. 

Finally, we compare to the quantile estimates from the model-based fitted quantiles  $\boldsymbol{\hat{Q}}(\boldsymbol{X})$ under  \eqref{M} ($\mbox{Q}_{hat}$). This competitor does not provide \emph{linear} coefficient estimates, uncertainty quantification, or selection. Nonetheless, it is a useful benchmark to evaluate whether our linear quantile estimators sacrifice any predictive performance relative to an unrestricted estimator under  $\mathcal{M}$. 
 Posterior inference for both $\mathcal{M}$ and $\mbox{AL}_{Bayes}$ is based on 2500 posterior samples accumulated after a burn-in of 2500.

\subsection{Quantile Prediction}\label{pred}
We first summarise the predictive performance of the competing methods on hold-out datasets $(\boldsymbol{X}_{test},\boldsymbol{y}_{test})$ with $n_{test} = 1000$, which are generated independently and identically distributed to the training data. We compute out-of-sample quantile predictions, say $\tilde{Q}_{\tau}(\boldsymbol{X}_{test})$, for each competitor.  
For each $\tau$, we evaluate the performance using three metrics: i) the mean-squared error (MSE) between the ground-truth $Q_{\tau}^*(\boldsymbol X_{test})$ and predictions $\tilde{Q}_{\tau}(\boldsymbol{X}_{test})$, ii) the average check loss between between $\boldsymbol{y}_{test}$ and $\tilde{Q}_{\tau}(\boldsymbol{X}_{test})$, and iii) the calibration of $\tilde{Q}(\boldsymbol X_{test})$ computed via $n_{test}^{-1}\sum_{i=1}^{n_{test}} \mathbbm{1}\{y_{test_i}\leq \tilde Q_{\tau}(\boldsymbol x_{test_i})\}$, which should be near $\tau$ for a well-calibrated quantile estimator (Section E.5 of the supplementary material). The results for MSE and check loss for $\tau \in \{0.01,0.5,0.99\}$ and $n =200, p = 50, \mbox{HetRatio} = 1$ across simulations are presented in Figure \ref{Fig2}; results for other simulation settings are qualitatively similar, see Section E of the supplement. 

\begin{figure*}[ht]
    \centering

     \includegraphics[width = .49\textwidth,keepaspectratio]
    {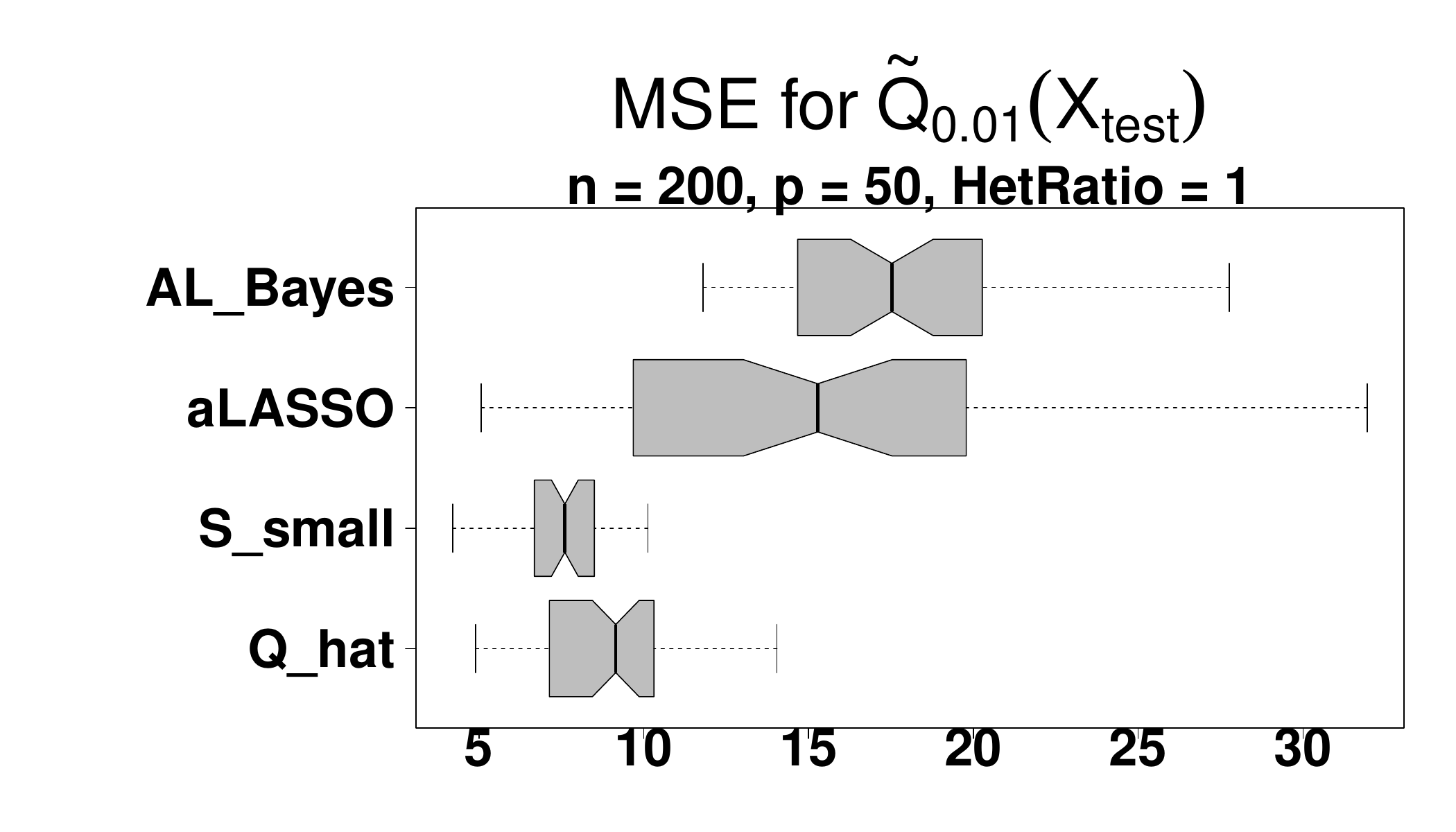}
    \includegraphics[width = .49\textwidth,keepaspectratio]
    {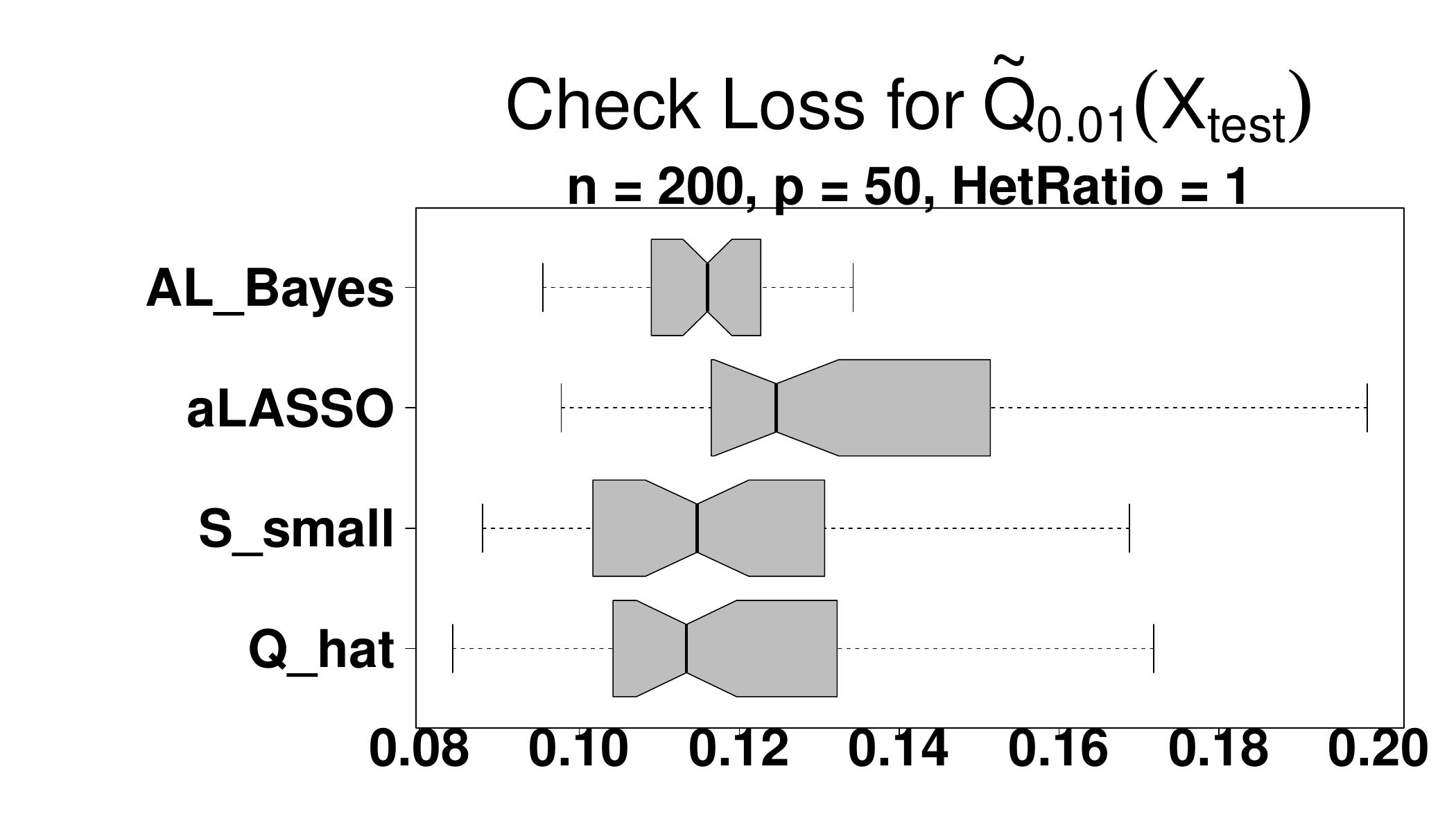}
    \includegraphics[width = .49\textwidth,keepaspectratio]
    {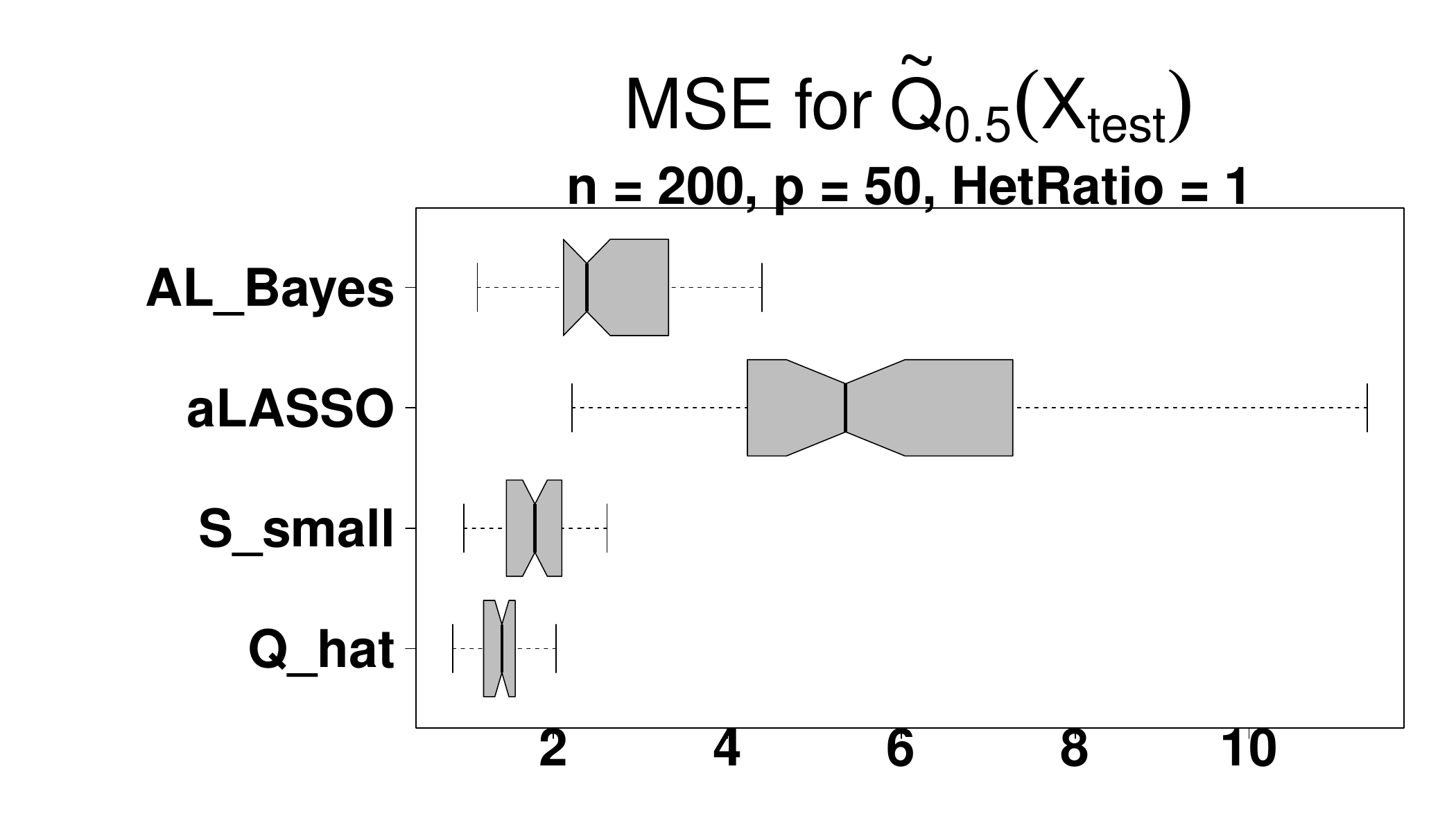}
        \includegraphics[width = .49\textwidth,keepaspectratio]
    {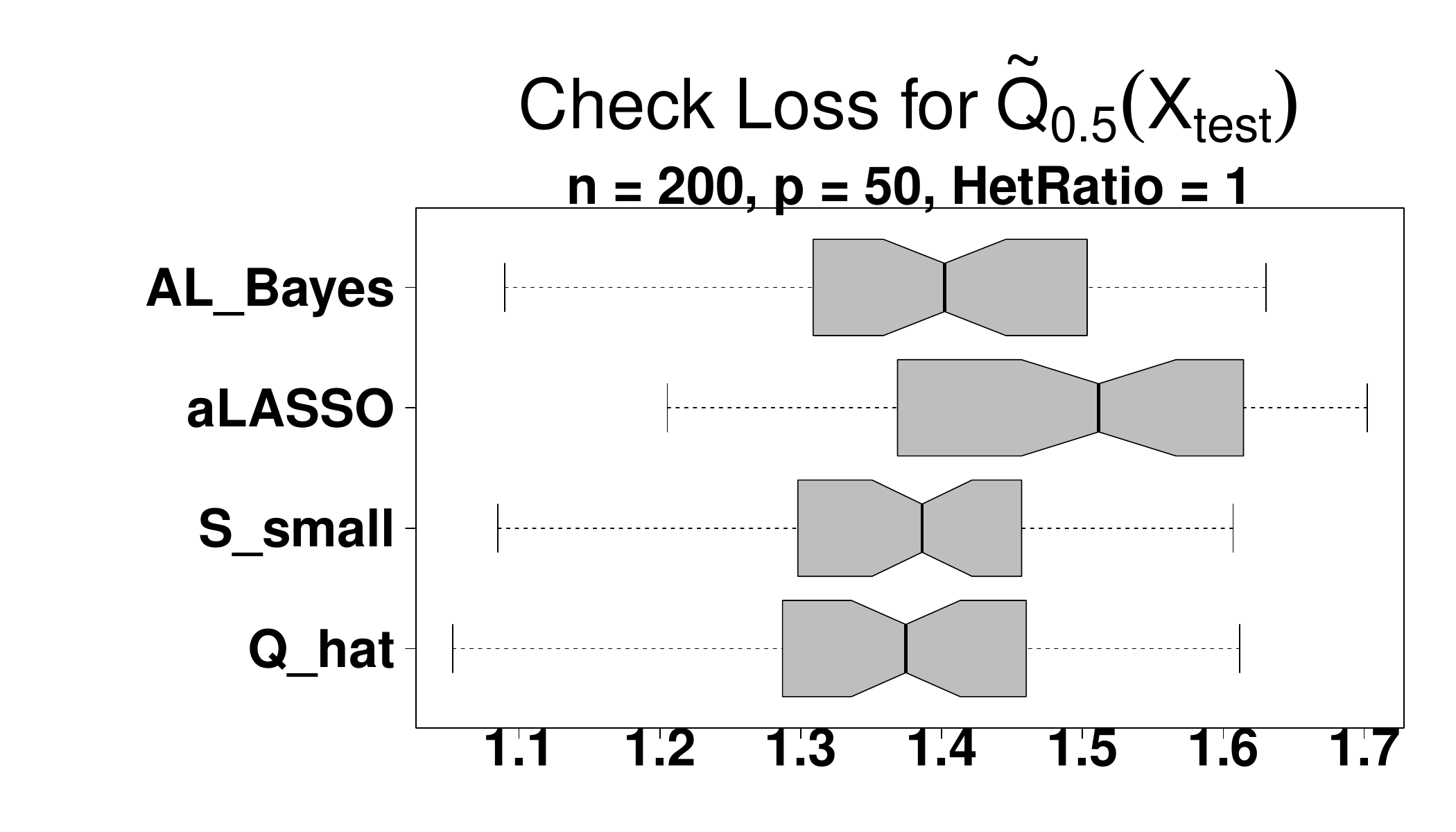}
    \includegraphics[width = .49\textwidth,keepaspectratio]
    {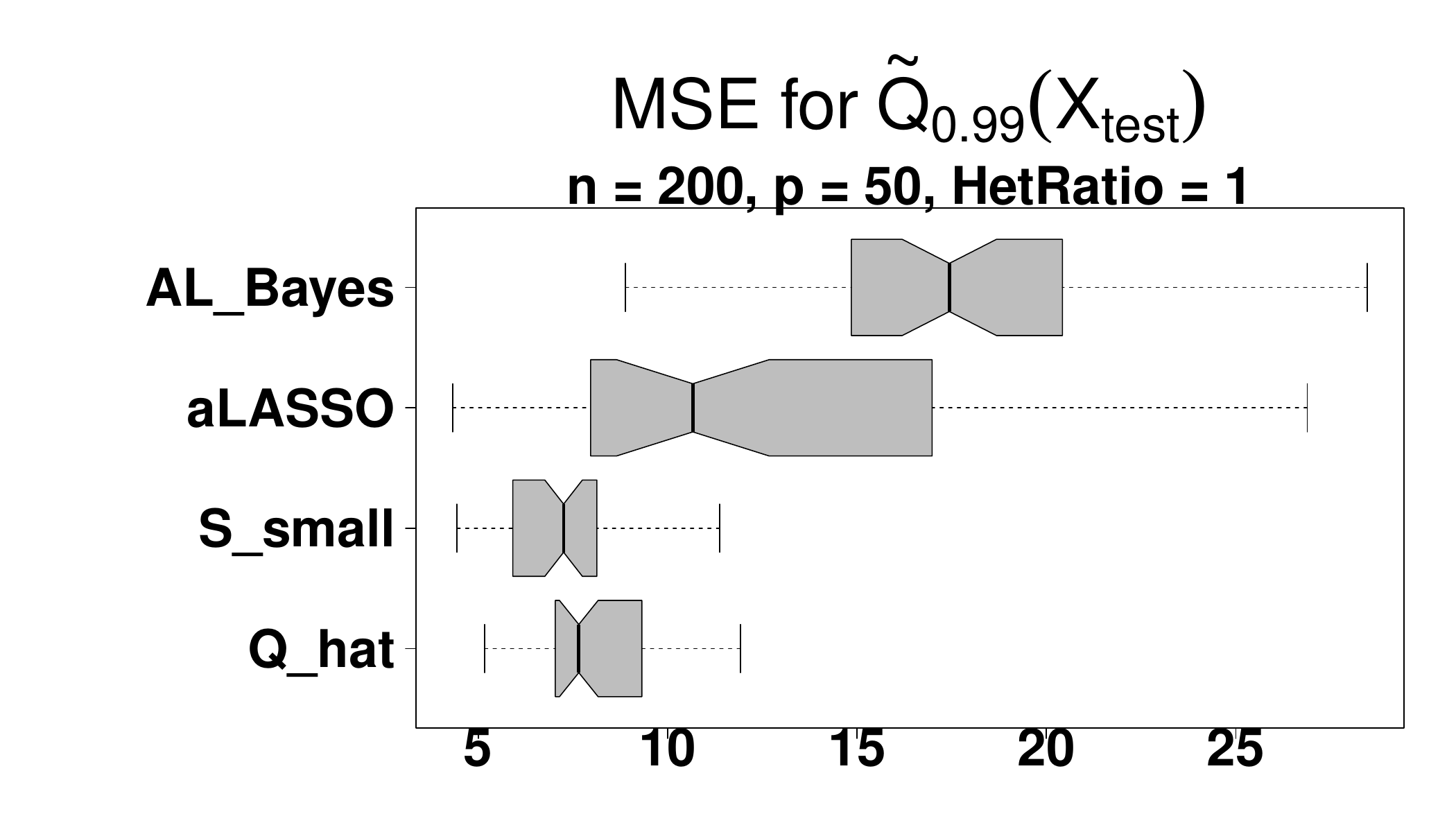}
    \includegraphics[width = .49\textwidth,keepaspectratio]
    {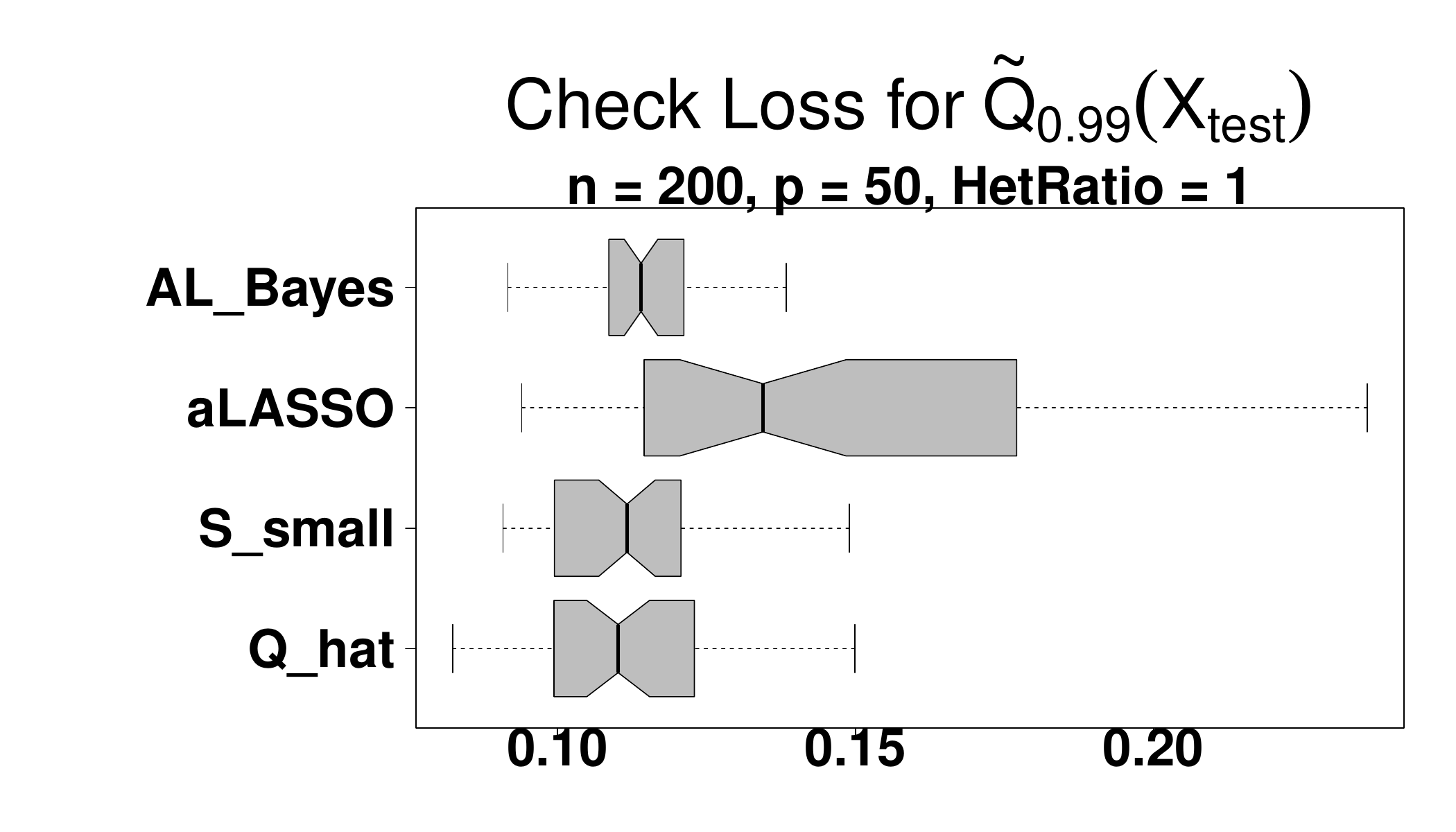}
    \caption{ \small Quantile mean squared error (MSE; left) and check loss (right) among competing quantile regression methods; nonoverlapping notches indicate significant differences between medians, and the dashed vertical line (right) denotes $\tau$. The proposed approach ($\mbox{S}_{small}(\tau)$)  offers substantial improvements in quantile prediction relative to both frequentist and Bayesian competitors, and matches or improves upon the model-based fitted quantiles ($\mbox{Q}_{hat}$).}
    \label{Fig2}
\end{figure*}


    

Across all quantiles, the proposed approach demonstrates exceptional quantile prediction. The quantile MSEs for $\mbox{S}_{small}(\tau)$  are significantly lower than those for competing frequentist and Bayesian methods, especially for extreme quantiles near zero or one.  $\mbox{S}_{small}(\tau)$ decisively outperforms aLASSO in check loss, even though aLASSO---unlike $\mbox{S}_{small}(\tau)$---directly optimizes (penalized) check loss via \eqref{CL}. Finally, the linear quantile predictions from $\mbox{S}_{small}(\tau)$ match or improve upon model-based fitted quantiles ($\mbox{Q}_{hat}$), despite the nonlinearity of the quantile functions in the model  \eqref{M}.



\subsection{Uncertainty Quantification}\label{uq}
Next, we assess uncertainty quantification for the posterior action \eqref{proj} via 95\% posterior interval estimates. Specifically, we construct intervals from the posterior draws of \eqref{proj} for  two subsets: the smallest acceptable subset $\mbox{S}_{small}(\tau)$ and the full set of covariates $\mbox{S}_{full} = \{1,\ldots,p\}$. By design, $\mbox{S}_{small}(\tau)$ is sparse, so the interval estimates under \eqref{proj} for any covariates omitted from $\mbox{S}_{small}(\tau)$  are null. Since $\mbox{S}_{full}$ includes all covariates, it circumvents this issue. We compare these interval estimates to the 95\% posterior credible intervals from $\mbox{AL}_{Bayes}$ for each $\tau$.

The intervals are evaluated for \emph{calibration}, measured by the empirical coverage, and \emph{sharpness}, measured by the average interval widths, in Table \ref{covrates}. Narrow intervals that achieve the 95\% nominal coverage are preferred. Most notably, the posterior actions for both $\mbox{S}_{full}$ and $\mbox{S}_{small}(\tau)$ yield \emph{significantly} more narrow intervals than $\mbox{AL}_{Bayes}$, often by a factor of 2-5. These effects are especially pronounced for extreme quantiles near zero or one, where $\mbox{AL}_{Bayes}$ is excessively conservative and underpowered. Importantly, both $\mbox{S}_{full}$ and $\mbox{S}_{small}(\tau)$ maintain the nominal coverage in nearly all settings. As expected, the sparsity of $\mbox{S}_{small}(\tau)$ produces the most narrow intervals, but also sacrifices empirical coverage for any active variables excluded from $\mbox{S}_{small}(\tau)$. The results for the other simulation settings are similar, and can be found in the supplementary material (Section E).


 \begin{table*}[h]
 \centering
    $\boldsymbol{n = 200, p = 50, \textbf{\mbox{HetRatio}} =1}$ 

    \begin{tabular}{c r| c  c c c c c c}    

      & $\tau$  & 0.01&0.05 &0.25&0.5&0.75&0.95&0.99  \\
       \hline
       \multirow{ 3}{*}{Coverage Rate}& $\mbox{S}_{small}(\tau)$  & 0.83 & 0.86 & 0.88 & 0.90 & 0.88 &  0.85 & 0.83\\
        & $\mbox{S}_{full}$ & 0.98 & 0.97 &  0.95 & 0.94 & 0.95 &  0.97 & 0.97\\
        & $\mbox{AL}_{Bayes}$ & 0.98 & 0.97 &  0.94 & 0.94 & 0.95 &  0.97 & 0.98 \\\hline
        \multirow{ 3}{*}{Avg. 95\% CI Width}& $\mbox{S}_{small}(\tau)$  & 1.23 & 1.13 &  0.94 & 0.81 & 0.94 &  1.17 & 1.27\\
        & $\mbox{S}_{full}$ & 3.56 & 2.87 &  2.02 & 1.76 & 2.03 &  2.90 & 3.60\\
        & $\mbox{AL}_ {Bayes}$ & 8.64 & 4.95 &  3.10 & 2.73 & 3.04&  4.99 & 9.19\\\hline\\
    \end{tabular}

     $\boldsymbol{n = 500, p = 20, \textbf{\mbox{HetRatio}} =1}$

   \begin{tabular}{c r| c c c c c c c c c}    

      & $\tau$  & 0.01&0.05 &0.25&0.5&0.75&0.95&0.99  \\
       \hline
       \multirow{ 3}{*}{Coverage Rate}& $\mbox{S}_{small}(\tau)$  & 0.92 & 0.91 &  0.90 & 0.89 & 0.89 &  0.94 & 0.95\\
        & $\mbox{S}_{full}$ & 0.93 & 0.92 & 0.91 & 0.91 & 0.91 & 0.96 & 0.97\\
        & $\mbox{AL}_{Bayes}$ & 0.94 & 0.94 &  0.93 & 0.94 & 0.95 &  0.99 & 0.99 \\\hline
        \multirow{ 3}{*}{Avg. 95\% CI Width}& $\mbox{S}_{small}(\tau)$  & 0.63 & 0.51 & 0.36 & 0.31 & 0.37 &  0.52 & 0.65\\
        & $\mbox{S}_{full}$ & 1.41 & 1.14 & 0.83 & 0.75 & 0.84 &  1.15 & 1.43 \\
        & $\mbox{AL}_ {Bayes}$ & 4.01 & 2.40 &  1.43 & 1.25 & 1.41 &  2.42 & 4.32\\ \hline\\
    \end{tabular}
    \caption{\small Coverage rates and average widths of 95\% credible intervals for the posterior action \eqref{proj} with $\mbox{S}_{small}(\tau)$ or $\mbox{S}_{full}$ as well as $\mbox{AL}_{Bayes}$. The intervals for $\mbox{S}_{small}(\tau)$ and $\mbox{S}_{full}$ are significantly more narrow than those for $\mbox{AL}_{Bayes}$, especially for  quantiles near zero or one, and typically maintain  nominal coverage. $\mbox{S}_{small}(\tau)$ sacrifices some coverage  in favor of sparsity, and thus provides the most narrow intervals. \label{covrates}
    }

\end{table*}

\subsection{Selection}\label{select}
We evaluate quantile-specific subset selection by computing true positive rates (TPR) and true negative rates (TNR)  for the variables selected by $\mbox{S}_{small}(\tau)$, $\mbox{aLASSO}$ and $\mbox{AL}_{Bayes}$. Here, $\mathcal{M}$ is not a competitor: it does not specify linear quantiles or any mechanism for quantile-specific variable selection. The results are presented in Table~\ref{Table1}. 

 Across all settings, $\mbox{S}_{small}(\tau)$ provides a superior balance between TPR and TNR. $\mbox{AL}_{Bayes}$ is underpowered and overconservative due to the excessively wide posterior credible intervals that are used for selection (see Table~\ref{covrates}). The frequentist competitor $\mbox{aLASSO}$ often produces similar TPRs but lower TNRs compared to $\mbox{S}_{small}(\tau)$, and thus selects too many variables. 

Improvements in TPRs for $\mbox{S}_{small}(\tau)$ over $\mbox{aLASSO}$ are most notable for extreme quantiles, though $\mbox{S}_{small}(\tau)$ appears to offer relatively low power in the larger $p$ settings. Considering this exception more carefully, the coefficients $\beta_{het}^*(.01)$ and $\beta_{het}^*(.99)$ are about four times larger in magnitude than  $\boldsymbol{\beta}_{hom}^*(\tau)$, so that $\boldsymbol X_{het}\beta_{het}^*(\tau)$ explains the vast majority of the variability in the quantiles for $\tau \in\{0.01, 0.99\}$. This effect is even more pronounced for the $(n,p) = (100,100)$ setting, where the heterogeneous coefficient is often five to ten times larger for each quantile. For these quantiles, $\mbox{S}_{small}(\tau)$ always includes the heterogeneous predictor, but leaves out the homogeneous predictors, which are less important for predicting those quantiles. This is consistent with the definition of $\mbox{S}_{small}(\tau)$, which seeks to find the \emph{smallest} subset that nearly matches the model-based quantile estimation from $\mathcal{M}$, and indeed achieves this latter objective (Figure \ref{Fig2}).

\begin{table*}[h]
\centering
       $\boldsymbol{n = 200, p = 50, \textbf{\mbox{HetRatio}} =1}$

    \begin{tabular}{c r| c  c c c c c c}    
    
      & $\tau$  & 0.01&0.05 &0.25&0.5&0.75&0.95&0.99  \\
       \hline
       \multirow{ 3}{*}{TPR}& $\mbox{S}_{small}(\tau)$  & 0.43& 0.54& 0.69 & 0.79&0.63 &0.57 &0.46\\
        & $\mbox{aLASSO}$ & 0.21 & 0.76 & 0.89 &0.95& 0.89  & 0.71 & 0.27\\
        & $\mbox{AL}_{Bayes}$ & 0.00 &0.02 &0.22 &0.35 & 0.23 & 0.03& 0.00 \\\hline
        \multirow{ 3}{*}{TNR}& $\mbox{S}_{small}(\tau)$  & 0.88& 0.85 & 0.82 & 0.84&0.83 & 0.85 &0.87\\
        & $\mbox{aLASSO}$ & 0.88 & 0.55 & 0.29 &0.27& 0.28  & 0.54 & 0.83\\
        & $\mbox{AL}_ {Bayes}$ & 0.97 & 0.97 & 0.97 & 0.96 & 0.97 &  0.97 & 0.93\\\hline\\

    \end{tabular}





   $\boldsymbol{n = 500, p = 20, \textbf{\mbox{HetRatio}} =1}$

    \begin{tabular}{c r| c c c c c c c c c} 

      & $\tau$  & 0.01&0.05 &0.25&0.5&0.75&0.95&0.99  \\
       \hline
        \multirow{ 3}{*}{TPR}& $\mbox{S}_{small}(\tau)$& 0.96 & 0.98 &  0.95 &  1.00 & 0.92 &  0.99 & 0.96\\
        & $\mbox{aLASSO}$ & 0.65 & 0.99 &  0.99 & 1.00 & 0.98 & 0.99 & 0.72\\
        & $\mbox{AL}_ {Bayes}$ & 0.00 & 0.31 & 0.83 & 0.99 & 0.86 & 0.36 & 0.00\\\hline
        \multirow{ 3}{*}{TNR}& $\mbox{S}_{small}(\tau)$ & 0.94 & 0.93 &  0.93 & 0.91 & 0.92 & 0.93& 0.94 \\
        & $\mbox{aLASSO}$ & 0.87 & 0.71 & 0.58 & 0.52 & 0.55 &  0.72 & 0.82\\
        & $\mbox{AL}_ {Bayes}$ & 1.00 & 1.00 &  0.99 & 0.99 & 0.99 &  1.00 & 1.00\\\hline\\

    \end{tabular}




    \caption{\small True positive rates (TPR) and true negative rates (TNR) for variable selection averaged across simulations. The proposed approach ($\mbox{S}_{small}(\tau)$) neatly balances TPR and TNR. The Bayesian competitor ($\mbox{AL}_ {Bayes}$) is significantly underpowered due to excessively wide posterior credible intervals, while the frequentist competitor (aLASSO) often overselects variables (low TNRs). The results for the $(n,p) = (100,100)$ are qualitatively similar, but with lower TPR due to the disproportionately large effect of the heterogeneous covariate on the response}.\label{Table1}

\end{table*}

\section{Social Stressors, Environmental Exposures, and Childhood Educational Outcomes}\label{realdat}
Childhood educational outcomes are affected by  social stressors (e.g., poverty, structural racism, high unemployment) and environmental exposures (e.g., poor air quality, lead exposure), which often cumulate in the same communities    \citep{miranda2007relationship,bravo2024spatial}. Thus, there is urgency to quantify and characterize these effects and initiate well-informed policy interventions, especially given the strong links between educational attainment and adult health. However, far less is known about the differential impacts of social stressors and environmental exposures on low-, medium-, and high-achieving students. Such information is critical for understanding the factors that shape educational outcomes and for recommending policy interventions, especially to support at-risk students. Thus, quantile regression is an informative tool---especially with the ability to quantify effect directions and magnitudes, measure and report uncertainty, and provide quantile-specific subset selection.

To analyze these relationships, we construct a large cohort ($n = 23,232$) of North Carolina (NC) students by linking three administrative datasets \citep{CEHI}: \textbf{NC Detailed Birth Records}, which  provides maternal and infant characteristics for all documented live births in NC; \textbf{NC Blood Lead Surveillance}, which includes blood lead level (\vtt{Blood_lead}) measurements for each child; and \textbf{NC Standardized Testing Data}, which  contains standardized (by the year of test, 2010--2012) fourth end-of-grade (EoG) reading scores for each student $(\vtt{Reading_Score})$. Because these datasets also include residential information, we are able to calculate indices of neighborhood deprivation (\vtt{NDI}) and racial residential isolation (\vtt{RI}), which measure students' exposure to poverty and structural racism, respectively. 
The full collection of variables is summarized in Table~\ref{tabdata}.  

\begin{table*}[h]
\centering 
\label{tab:variable}
\begin{tabular}{|l|l|}
\hline 
\multicolumn{2}{|l|}{\cellcolor[HTML]{C0C0C0}\textbf{Birth information}}                                                                                                                                                                                     \\ \hline
\texttt{mEdu}         & \begin{tabular}[c]{@{}l@{}}Mother's education group at the time of birth \\(No high school diploma, High school diploma, \\College diploma) \end{tabular} \\ \hline
\texttt{mRace}        & \begin{tabular}[c]{@{}l@{}}Mother's race/ethnicity group (Non-Hispanic (NH) White, \\ NH Black)\end{tabular}                                                                                         \\ \hline
\texttt{BWTpct}        &  Birthweight percentile  \\ \hline
\texttt{mAge}         & Mother's age at the time of birth                                                                                                                                                                                                    \\ \hline
\texttt{Male}         & Male infant? (1 = Yes)                                                                                                                                                                                                               \\ \hline
\texttt{Smoker}       & Mother smoked? (1 = Yes)                                                                                                                                                                                                         \\ \hline
\texttt{NotMarried} & Not married at time of birth (1 = Yes) 

\\ \hline

\texttt{NOPNC} & \begin{tabular}[c]{@{}l@{}}Mother received pre-natal care before birth?\\ (1 = No prenatal care)\end{tabular}
\\ \hline
\texttt{Weeks\_Gestation} & Gestational period (in weeks)
\\ \hline
\multicolumn{2}{|l|}{\cellcolor[HTML]{C0C0C0}\textbf{Education/End-of-grade (EoG) test information}}                                                                                                                                                                       \\ \hline

\texttt{Reading\_Score} & \begin{tabular}[c]{@{}l@{}} Standardized score for the (chronologically first)  4th EoG\\  reading test  \end{tabular} \\ \hline
                                         
\multicolumn{2}{|l|}{\cellcolor[HTML]{C0C0C0}\textbf{Blood lead surveillance}}                                                                                                                                                                                     \\ \hline
\texttt{Blood\_lead}  & Blood lead level (micrograms per deciliter) \\ \hline

\multicolumn{2}{|l|}{\cellcolor[HTML]{C0C0C0}\textbf{Social/Economic status}}                                                                                                                                                                                     \\ \hline

\texttt{EconDisadvantage}         & \begin{tabular}[c]{@{}l@{}} Participation in Child Nutrition Lunch Program? (1 = Yes)\end{tabular} \\ \hline
\texttt{NDI} & Neighborhood Deprivation Index, at time of EoG test\\ \hline
\texttt{RI} & Racial residential isolation, at time of EoG test\\ \hline

\end{tabular}
\vspace{1em}
\caption{\small Variables in the North Carolina dataset. Data are restricted to children with  NH Black or NH White mothers \citep{bravo2022racial}, 30-42 weeks of gestation, 0-104 weeks of age-within-cohort, mother's age 15-44, \texttt{Blood\_Lead} $\leq$ 10, birth order $\leq$ 4, no status as an English language learner, and residence in NC at the time of birth and the time of 4th EoG test. Numeric covariates are centered and scaled to mean zero and standard deviation 0.5.}
\label{tabdata}
\end{table*}

We augment the covariates in Table~\ref{tabdata} with interactions between mother's race (\vtt{mRace}) and each of blood lead level (\vtt{blood_lead}), neighborhood deprivation (\vtt{NDI}), and racial residential isolation (\vtt{RI}). Crucially, these interactions allow us to assess whether the (possibly heterogeneous) effects of environmental exposures and social stressors on educational outcomes also differ between race groups.  

We fit the Bayesian LL-LS model \eqref{M} to this dataset of $n =$ 23,232 students with $p =$ 18 covariates. Posterior  and posterior predictive diagnostics \citep{gelman1996posterior}  demonstrate that the model is well-calibrated to the data, and notably provide key evidence for heteroscedasticity of $Y \mid \boldsymbol{x}$  (Section F.2 of the supplement). Thus, we anticipate that quantile regression may detect
heterogeneous covariate effects.

 Quantile-specific acceptable families $\mathbbm{A}_{\varepsilon}(\tau)$ are constructed for $\tau \in \{0.01,0.05,0.25,0.5,\\0.75,0.95,0.99\}$ under $\mathcal{M}$ using the subset search and selection techniques from Sections~\ref{sec1}-\ref{subset search}.  Many of the demographic and socioeconomic covariates (Table~\ref{tabdata}) are strongly correlated (see Section F.1 of the supplement for the correlations between predictors). As a result, there are likely many subsets that perform similarly---which is captured by the acceptable family $\mathbbm{A}_{\varepsilon}(\tau)$, but \emph{not} any single ``best" subset. In particular, we identify several hundred acceptable subsets for each $\tau$. We summarize the acceptable family $\mathbbm{A}_{\varepsilon}(\tau)$ using quantile-specific coefficient estimations and intervals for $\mbox{S}_{small}(\tau)$ along with the quantile-specific variable importance \eqref{VI}. For comparison, we include  point estimates from $\mbox{aLASSO}$ and posterior means and 95\% credible intervals from $\mbox{AL}_{Bayes}$ for the quantile-specific linear coefficients.



\begin{figure*}[h]
    \centering
    \includegraphics[width = \textwidth, keepaspectratio]{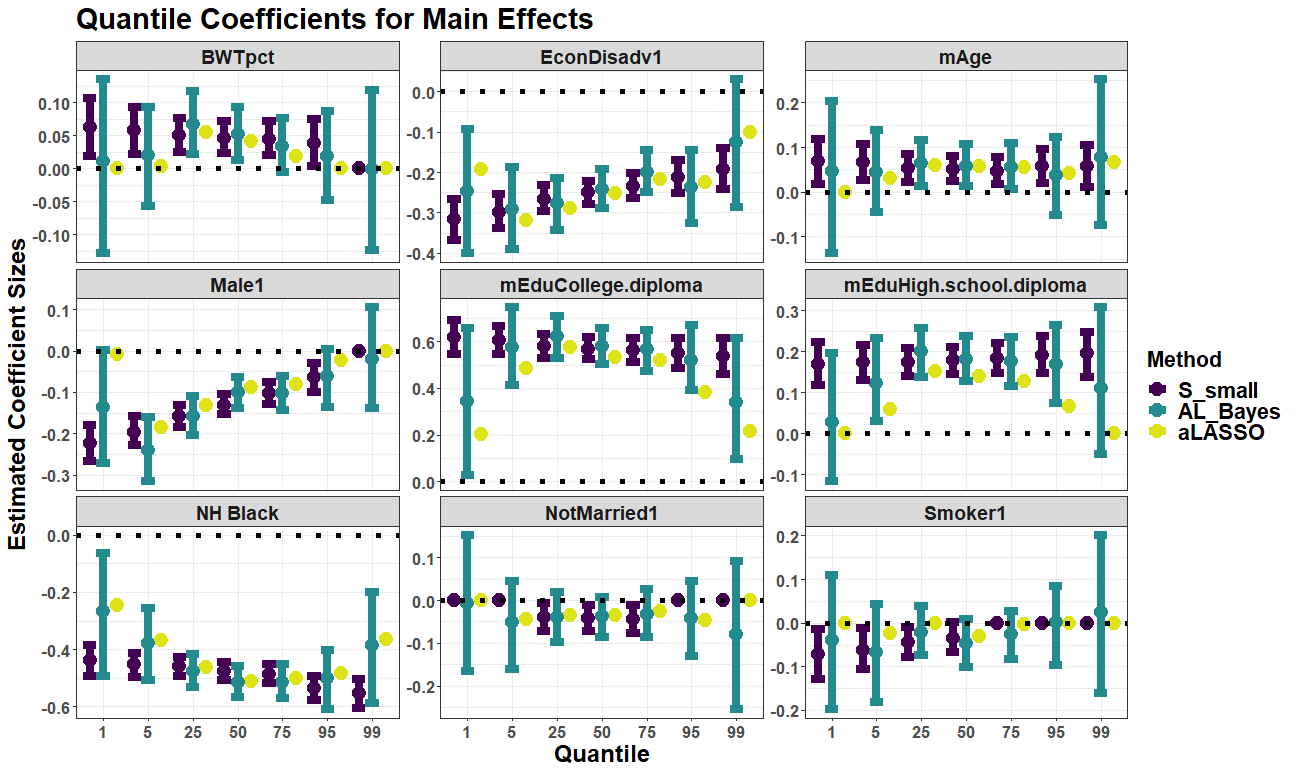}
    \caption{\small Quantile-specific point estimates and 95\% credible intervals for main effects (not interacted with mother's race) under $\mbox{S}_{small}$ (purple), $\mbox{AL}_{Bayes}$ (green) and $\mbox{aLASSO}$ (yellow); the horizontal line denotes zero. Under the proposed approach ($\mbox{S}_{small}$), we identify several heterogeneous effects on reading scores, including gender and economically disadvantaged students.  Other effects, such birthweight percentile and mother's age, are relatively homogeneous.  By contrast, $\mbox{AL}_{Bayes}$ exhibits excessively wide credible intervals, especially for extreme quantiles, while the $\mbox{aLASSO}$ estimates are highly variable and nonsmooth across $\tau$.}
    \label{main}
\end{figure*}

 
 In Figure~\ref{main} we report the estimates and uncertainty quantification among the competing methods for the main effects that do not include interactions with race. Some of the covariates are not selected by  $\mbox{S}_{small}(\tau)$ or $\mbox{aLASSO}$ for certain quantiles, so the resulting point estimates (and intervals for $\mbox{S}_{small}(\tau)$) are fixed at zero. The variables \vtt{Weeks_Gestation} and \vtt{NOPNC} do not belong to $\mbox{S}_{small}(\tau)$  for any $\tau$, and thus are omitted.

The magnitude and direction of the coefficient estimates from  $\mbox{S}_{small}$ reveal numerous interesting patterns. First, the point estimates for \vtt{EconDisadv} and \vtt{Male} are negative across all quantiles, but the effects on reading scores are more pronounced for \emph{lower} quantiles. This suggests that the discrepancies between male and female students, and between students who are economically disadvantaged and students who are not, are greater for lower-scoring students. Other covariates have little variability over $\tau$; the coefficients for \vtt{BWTpct} and \vtt{mAge} are relatively flat, demonstrating that birthweight percentile and mother's age at the time of birth have significant, yet homogeneous impacts across the distribution of $Y \mid \boldsymbol{x}$. 

The 95\% credible intervals for the coefficients in $\mbox{S}_{small}(\tau)$ are substantially more narrow than $\mbox{AL}_{Bayes}$, which is consistent with the simulation results  (Section~\ref{select}). This advantage is most pronounced in the extreme quantiles near zero or one, and helps to uncover clear patterns in covariate heterogeneity. The $\mbox{aLASSO}$ estimates vary erratically across $\tau$, often with large jumps for extreme quantiles. This effect is notable for the \vtt{Male} coefficients: the estimated coefficients are increasingly negative for smaller $\tau$, yet the estimate at $\tau=0.01$ is zero. Similar patterns persist for \vtt{mEdu}, \vtt{EconDisadv} and \vtt{mRace}, and undermine the interpretability of the quantile-specific coefficients under $\mbox{aLASSO}$.
 \begin{figure*}[h]
    \centering \includegraphics[width = 1\textwidth, keepaspectratio]{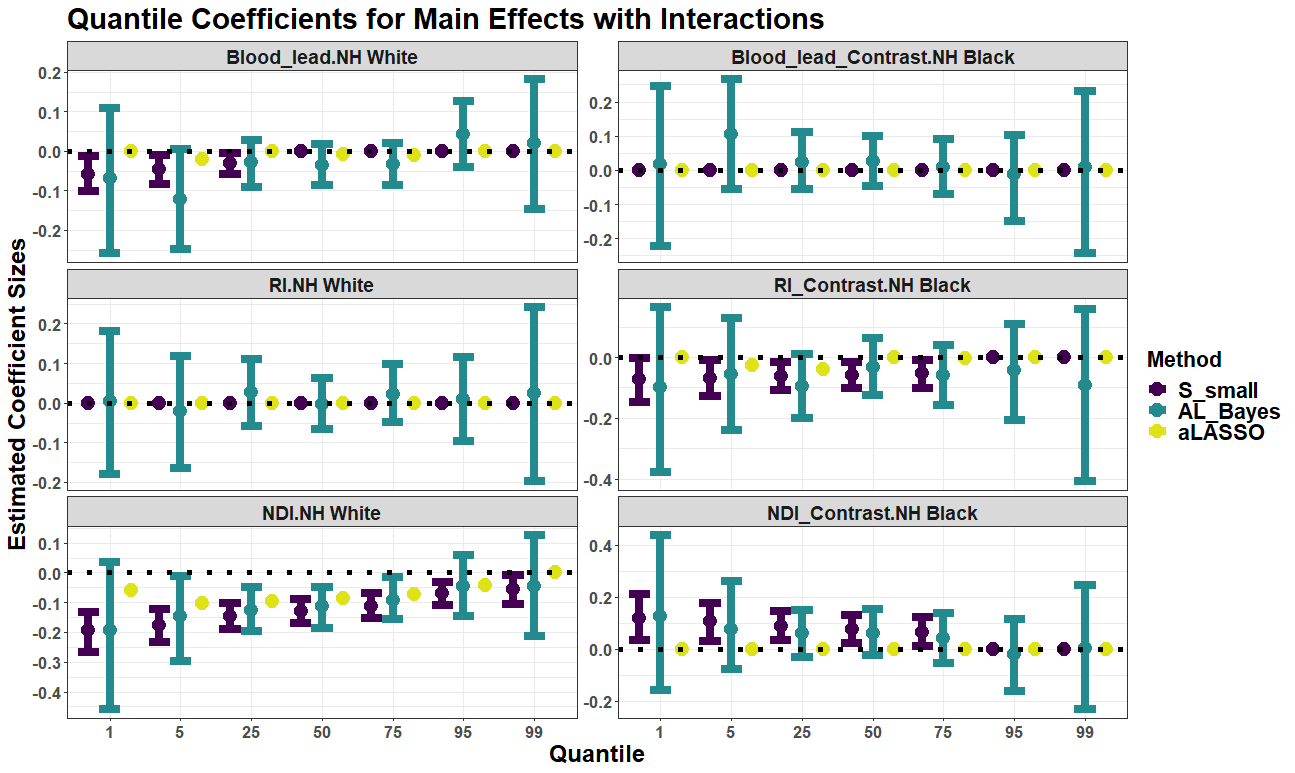}
    \caption{\small Quantile-specific point estimates and 95\% credible intervals for social and environmental exposures that are interacted with mother's race under $\mbox{S}_{small}$ (purple), $\mbox{AL}_{Bayes}$ (green) and $\mbox{aLASSO}$ (yellow); the horizontal line denotes zero. The main effects (left) refer specifically to the NH White group,  while the interaction effects (right) refer to the differences between these effects for NH Black students and NH White students. Under $\mbox{S}_{small}(\tau)$, the  \vtt{blood_lead} effects are increasingly negative for $\tau < 0.5$, with no clear differences between NH Black and NH White students. \vtt{RI} similarly exhibits an increasingly negative effect for lower quantiles, but only for NH Black students. Finally, the \vtt{NDI} effects are heterogeneous and increasingly negative for lower quantiles among NH White students, but for NH Black students, the effects are negative yet homogeneous across $\tau$.  Again, $\mbox{AL}_{Bayes}$ produces excessively wide credible intervals, while aLASSO only identifies nonzero effects for \vtt{NDI} among NH White students.}
    \label{intrctt}
\end{figure*}

The main and race-interaction effects for the social stressors and environmental exposures are presented in Figure~\ref{intrctt}. Here, the main effects refer specifically to the NH White group,  while the interaction effects refer to the differences between these effects for NH Black students and NH White students.  First, lead exposure is  especially detrimental for lower-scoring students, with no estimated differences between the race groups. Second, the estimated RI effect is similarly detrimental for lower-scoring students, but only for NH Black students. Finally, NDI exhibits a heterogeneous effect for NH White students, with increasingly negative effects for lower-scoring students. The positive and heterogenous effects for the NDI contrast term must be interpreted carefully: in conjunction with the main effect estimates, these estimates indicate that the effect of NDI is still negative for NH Black students, but now homogeneous across $\tau$. 

Once again, $\mbox{AL}_{Bayes}$ produces excessively wide credible intervals, which obscures important heterogeneity patterns across $\tau$. Similarly, aLASSO only identifies nonzero effects for NDI among NH White students; yet even these  estimated effects fail to satisfy monotonicity across $\tau$.



\begin{figure*}[h]
    \centering
    \includegraphics[width = .9\textwidth, keepaspectratio]{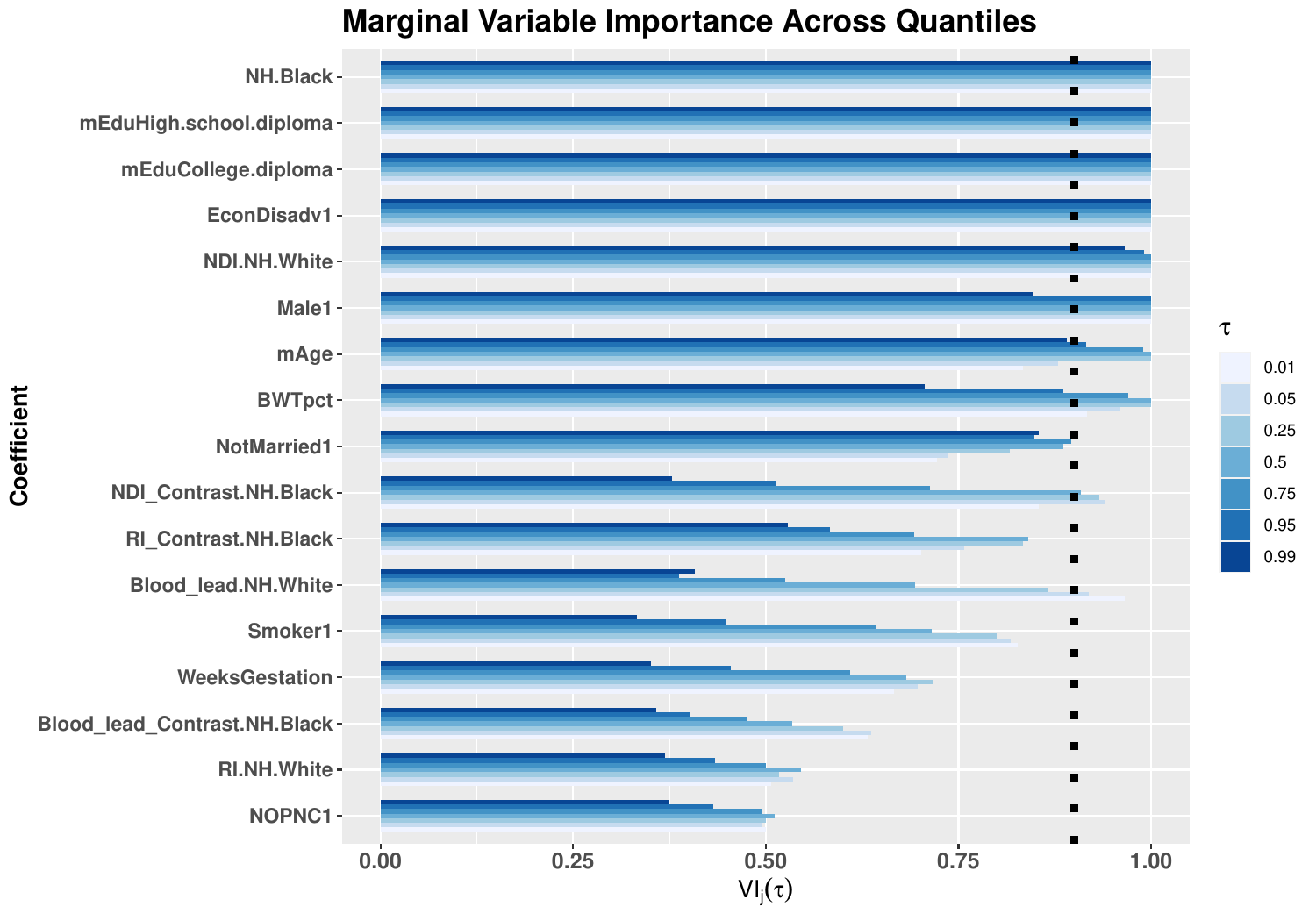}
    \caption{\small Variable importance $\mbox{VI}_{j}(\tau)$ from \eqref{VI}, colored by quantile; the dashed line indicates 0.90. Large values indicate that the covariate appears in many of the acceptable subsets. \vtt{mEdu}, \vtt{mRace} and \vtt{EconDisadvantage} appear in all acceptable subsets for all quantiles, while several of the environmental and social factors exhibit heterogeneous variable importance, often with increasing $\mbox{VI}_{j}(\tau)$ for smaller quantiles (See Section E.}
    \label{vimps}
\end{figure*}

Finally, we present the variable importance 
$\mbox{VI}_{j}(\tau)$ for each covariate $j$ and quantile $\tau$ in Figure \ref{vimps}. $\mbox{VI}_{j}(\tau)$ seeks to summarize the acceptable family $\mathbbm{A}_{\varepsilon}(\tau)$ of near-optimal subsets by tallying the proportion of acceptable subsets in which each variable belongs. In particular, we identify \emph{keystone covariates} that satisfy  $\mbox{VI}_{j}(\tau)>0.9$ for any $\tau$, and thus are valuable covariates that belong to at least 90\% of acceptable subsets.

Most notably, \vtt{mEdu}, \vtt{mRace} and \vtt{EconDisadvantage} appear in all acceptable subsets across quantiles. These keystone covariates are essential for near-optimal prediction across the entire distribution of reading scores $Y \mid \boldsymbol{x}$. Heterogeneous and large $\mbox{VI}_{j}(\tau)$ effects are apparent for \vtt{blood_lead} main effects (i.e., for NH White students) and \vtt{NDI} and \vtt{RI} contrasts (i.e., differences between NH Black and NH White students), with increasing variable importance for lower quantiles. Similarly heterogeneous and large variable importance are observed for \vtt{mAge}, \vtt{BWTpct}, and \vtt{NotMarried}.

In aggregate, our analysis uncovers substantial heterogeneity in the effects of environmental exposures, social stressors, and other key factors on reading scores. Notably, both the coefficient (point and interval) estimates (Figures~\ref{main}-\ref{intrctt}) and the variable importance (Figure~\ref{vimps}) highlight numerous increasingly adverse effects for lower-scoring students.  These effects include lead exposure, economic disadvantagement, racial residential isolation, and neighborhood deprivation, among others, with some differential effects by race. We also note that this discovery was consistent when applying the subset search and selection techniques to the $\text{AL}_{Bayes}$ model, see Section F.3 of the supplement for complete results. These new findings contribute to our understanding of the disparities in childhood development and educational outcomes.

\section{Conclusion}\label{conclude}
We  proposed a novel approach to Bayesian linear quantile regression with subset selection. The procedure features two stages, but operates within a single, coherent, Bayesian modeling and decision analysis framework. First, the analyst curates a Bayesian regression model to best represent the conditional distribution  $Y\mid\boldsymbol{x}$. Then, based on the conditional quantiles from the Bayesian model, we apply a decision analysis to extract linear and quantile-specific coefficient estimates, uncertainty quantification, and subset selection. This approach uses a quantile-focused squared error loss function, which maintains a close, theoretical connection to density regression on the Wasserstein geometry. Crucially, this loss function enables closed-form computation of optimal linear coefficients and uncertainty quantification for any subset of predictors. We leverage these computational results to unlock state-of-the-art subset search and selection algorithms that were previously available only for mean regression. Our strategy prioritizes accumulation of many, highly predictive subsets to form a quantile-specific acceptable family, which we summarize by reporting the smallest acceptable subset (in accordance with the parsimony principle) and  measures of variable importance.

There are several advantages of the proposed approach relative to existing frequentist and Bayesian quantile regression methods. First, the framework is valid under \emph{any} Bayesian regression model. Thus, the analyst can prioritize calibrated modeling of the observed data without the need to accommodate quantile-specific modeling requirements, such as  inadequate likelihoods or unwieldy constraints. Second, the decision analysis conveys regularization, uncertainty quantification, and smoothness across quantiles from the underlying Bayesian regression model. This occurs despite the fact that the decision analysis is applied separately for each quantile, yielding straightforward and efficient implementations. Finally, our approach delivers  subset search and selection for quantile regression, which has remained elusive among frequentist and Bayesian methods. 

These benefits translate to significant empirical improvements. In an extensive simulation study, we found that the proposed approach produces more accurate quantile predictions, more precise (yet calibrated) uncertainty quantification, and more powerful variable selection  for quantile regression coefficients.

We applied our methods to analyze the effects of social stressors, environmental exposures, and other key factors on 4th end-of-grade reading test scores for a large cohort of children in North Carolina. Our analysis revealed several important and unique insights into educational inequities. Most notably, we found that lead exposure, economic disadvantagement, racial residential isolation, and neighborhood deprivation more adversely impact reading test scores for lower-scoring students. These effects exhibited heterogeneity not only across quantiles, but also across race groups.  Furthermore, we found that these covariates tended to be more important for prediction of the lower quantiles of educational outcomes. These alarming results have important implications for childhood development and educational attainment, and may inform policy to develop and target intervention strategies. 



With such encouraging results, there are numerous promising directions for future research. Although we have focused on linear quantile regression, our decision analysis framework can be readily extended to nonlinear quantile-specific summaries, such as trees or additive models. This enhanced flexibility may be useful for quantile estimation under nonlinear Bayesian models, such as heteroscedastic Bayesian additive regression trees \citep{pratola2020heteroscedastic}. Further, our approach is not limited to Bayesian models with Gaussian errors. A useful extension would be to consider Bayesian models for extreme events \citep{fagnant2020characterizing}, with a customized decision analysis for estimation, uncertainty quantification, and selection. Finally, the proposed (linear) quantile predictions are quick to compute with minimal storage requirements, and thus may be used to provide fast, model-based prediction intervals, especially when the underlying Bayesian model does not admit efficient posterior predictive sampling.

\section{Acknowledgments}

The findings and conclusions in this presentation or publication are those of the authors and do not necessarily represent the views of the North Carolina Department of Health and Human Services, Division of Public Health, the National Institutes of Health, the Army Research Office, or the U.S. Government. The U.S. Government is authorized to reproduce and distribute reprints for Government purposes notwithstanding any copyright notation herein.

Research reported in this publication was supported by the National Institute of Environmental Health Sciences of the National Institutes of Health under award number R01ES028819, National Science Foundation under award number SES-2214726, and the Army Research Office under award number W911NF-20-1-0184.

\bibliographystyle{jasa3.bst}
\bibliography{reference}

\end{document}


\title{Supplement to \\ ``Bayesian Quantile Regression with Subset Selection: A Decision Analysis Perspective''}
\author{}
\date{}
\maketitle
\large 

\vspace{-10mm}
\appendix
\section{Proofs}
We present proofs for the main theoretical results presented in the main paper.

\begin{lemma}\label{OLS2} Suppose $E_{\boldsymbol \theta \mid \boldsymbol y}\lVert Q_{\tau}(Y_i \mid \boldsymbol x_i, \boldsymbol \theta)\rVert_{2}^{2}< \infty$ for $i = 1, \ldots, n$. For any quantile $\tau \in (0,1)$ and any subset of predictors $S \subseteq \{1,\dots,p\}$, the optimal action~\eqref{post-dec} under the quantile-focused squared error loss \eqref{relax-wass} is 
\begin{align}
\hat{\boldsymbol \delta}_S(\tau) =({\boldsymbol{X}}^{\intercal}_{S} {\boldsymbol{X}}_{S})^{-1}{\boldsymbol{X}}^{\intercal}_{S} \boldsymbol{\hat{Q}}_{\tau}({\boldsymbol{X}})\label{OLS_sol2}\end{align}
with zeros for indices $j \notin S$, where $\hat{Q}_{\tau}(\boldsymbol x_i) = E_{\boldsymbol \theta \mid \boldsymbol y}\{Q_{\tau}(Y_i \mid \boldsymbol x_i, \boldsymbol \theta)\}$, $\boldsymbol{\hat{Q}}_{\tau}(\boldsymbol{X}) =\{\hat{Q}_{\tau}(\boldsymbol x_1),\dots\\,\hat{Q}_{\tau}(\boldsymbol x_n)\}^\intercal$,  and ${\boldsymbol{X}}_{S}$ the $n \times \lvert S\rvert $  matrix of active covariates for subset $S$.
\end{lemma}
\begin{proof}
It suffices to observe that $E_{\boldsymbol \theta \mid \boldsymbol y}\lVert Q_{\tau}(Y_i \mid \boldsymbol x_i, \boldsymbol \theta) - \boldsymbol{ x}_i^{\intercal}   \boldsymbol{\delta}_{S}(\tau)\rVert_{2}^{2} = E_{\boldsymbol \theta \mid \boldsymbol y}\lVert \{Q_{\tau}(Y_i \mid \boldsymbol x_i, \boldsymbol \theta) -\hat{Q}_{\tau}({\boldsymbol x}_i) \} + \{\hat{Q}_{\tau}({\boldsymbol x}_i) -\boldsymbol{ x}_i^{\intercal}   \boldsymbol{\delta}_{S}(\tau)\}\rVert_{2}^{2} = E_{\boldsymbol \theta \mid \boldsymbol y}\lVert Q_{\tau}(Y_i \mid \boldsymbol x_i, \boldsymbol \theta) -  \hat{Q}_{\tau}({\boldsymbol x}_i)\rVert_{2}^{2} + \\ E_{\boldsymbol \theta \mid \boldsymbol y}\lVert \hat{Q}_{\tau}({\boldsymbol x}_i) -  \boldsymbol{ x}_i^{\intercal}   \boldsymbol{\delta}_{S}(\tau)\rvert_{2}^{2}$ where the first term is finite and does not depend on $  \boldsymbol{\delta}_{S}(\tau)$. The remaining steps constitute an ordinary least squares solution.
\end{proof}

\begin{corollary} \label{cor1}
For the location-scale model \eqref{LLS} with linearity $f(\boldsymbol x_i) = \boldsymbol{x}_i^{\intercal}\boldsymbol \beta(\tau)$,  homoscedasticity $s(\boldsymbol x_i) = \sigma$, and an intercept $x_{i1} = 1$, 
the optimal action \eqref{post-dec} under \eqref{relax-wass} for the full set of covariates is $\boldsymbol {\hat{\boldsymbol \delta}}_{\{1,\dots,p\}}(\tau)= [\hat{\beta}_{1}(\tau), \{\hat{\beta}_{j}\}_{j = 2}^{p}]$ for any $\tau$, where $\hat{\beta}_{1}(\tau) = E_{\boldsymbol \theta \mid \boldsymbol{y}} [\beta_{1} + \sigma F^{-1}(\tau)]$ and $\hat{\beta}_{j} = E_{\boldsymbol \theta \mid \boldsymbol{y}} \beta_{j}$ for $j=2,\ldots,p$.
\end{corollary}
\begin{proof}
    With $\hat{s}(\boldsymbol x_i) = E_{\boldsymbol \theta \mid \boldsymbol{y}} \sigma$ for any $\boldsymbol x_i$, it suffices to observe that under \eqref{LLS},  ${\hat{Q}}_{\tau}({\boldsymbol{x}_i}) = \hat{\beta}_{1}(\tau) + \sum_{j=2}^{p} {x_{ij}} \hat{\beta}_j$. In addition, for the action using the full set of covariates, the covariate matrix is ${\boldsymbol{X}}$. Plugging this into \eqref{OLS_sol} yields the result.
\end{proof}

\begin{corollary} 
For any Bayesian model with with linear quantiles $Q_{\tau}(Y_i \mid \boldsymbol x_i, \boldsymbol \theta)  = \boldsymbol{x}^{\intercal}\boldsymbol\beta(\tau)$,  the optimal action \eqref{post-dec} under \eqref{relax-wass} for the full set of covariates is $\boldsymbol {\hat{\boldsymbol \delta}}_{\{1,\dots,p\}}(\tau)= \hat{\boldsymbol{\beta}}(\tau)$, where $\hat{\boldsymbol{\beta}}(\tau) = E_{\boldsymbol \theta \mid \boldsymbol{y}} \boldsymbol{\beta}(\tau)$.
\end{corollary}
\begin{proof} 
    The result is immediate using the argument in Corollary \ref{cor1} and observing that $\boldsymbol{\hat{Q}}_{\tau}({\boldsymbol{X}}) =  {\boldsymbol{X}} \hat{\boldsymbol{\beta}}(\tau)$.
\end{proof}

\begin{corollary}
    For the location-scale model \eqref{LLS} with linearity $f(\boldsymbol x_i) = \boldsymbol{x}_i^{\intercal}\boldsymbol \beta(\tau)$, homoscedasticity $s(\boldsymbol x_i) = \sigma$, and an intercept $x_{i1} = 1$,
  the  posterior action \eqref{proj} for the full set of covariates $S = \{1,\ldots,p\}$ satisfies
    \[
    \boldsymbol \delta_{\{1,\dots, p\}}(\boldsymbol \theta; \tau) \sim p(\boldsymbol \theta^* \mid \boldsymbol y)
    \]
    where  $\boldsymbol \theta^{*} = [\beta_{1} + \sigma F^{-1}(\tau), \{\beta_{j}\}_{j=2}^{p}]$.
\end{corollary}
\begin{proof}
    For the homoscedastic linear regression, $Q_{\tau}(\boldsymbol Y \mid \boldsymbol X, \boldsymbol \theta) = [\{\beta_{1} + \sigma F^{-1}(\tau)\} + \sum_{j=2}^{p} x_{ij} \beta_{j}]_{i=1}^{n}$. Therefore, for the full set of covariates and any quantile, \eqref{proj} is equal in distribution to the regression coefficients under the model $\mathcal{M}$ posterior, with only the intercept term varying as a function of the quantile. .
\end{proof}

\begin{lemma}\label{form_relax_wass}
    Let $\boldsymbol \delta(\boldsymbol x;\tau) = \boldsymbol x^{\intercal}\boldsymbol \delta_{S} (\tau)$ for any $\boldsymbol{x}$ and any subset $S$ of covariates. Then the minimizer of \eqref{post-dec-density}, \emph{without} a density restriction, is given by the quantile estimators $\hat{\boldsymbol \delta}(\boldsymbol x;\tau_m) = \boldsymbol{x}^{\intercal}\hat{\boldsymbol \delta}_{S}(\tau_m)$ with $\hat{\boldsymbol \delta}_{S}(\tau_m)$ computed from \eqref{OLS_sol} separately for each $m=1,\ldots,\ell$. 
  
\end{lemma}

\begin{proof}
    It suffices to observe that objective in \eqref{post-dec-density} may be expanded as 
    $\sum_{\tau = \tau_{1}}^{\tau_{\ell}} E_{\boldsymbol \theta \mid \boldsymbol{y}}\sum_{i=1}^{n}\lVert {Q}_{\tau}(Y_i \mid \boldsymbol{x}_i,\boldsymbol \theta) - \boldsymbol{x}_{i}^{\intercal}\boldsymbol \delta_{S}(\tau)\rVert_{2}^{2}$, and since the optimization is unconstrained for $\boldsymbol \delta_{S}(\tau)$ across $\tau$, the summands may be optimized separately. Applying Lemma \ref{OLS} separately for each $\tau_m$ yields the result. 
   
\end{proof}

\section{Detailed Algorithm 1}
We present  specific computations required for each step in Algorithm \ref{alg1}:

\begin{algorithm}[h]
\caption{Bayesian quantile regression, inference and subset selection using posterior summarization}\label{D-compute}
\begin{algorithmic}\label{alg1}

    \STATE 1. Fit  a Bayesian regression model $\mathcal{M}$
        \begin{itemize}
            \item \textbf{Description}: Fit a Bayesian  model to $\{\boldsymbol{x}_{i}, y_{i}\}_{i=1}^{n}$ and extract posterior samples of model-based conditional quantile functions at each  $\boldsymbol x_i$ and any quantile $\tau$ of interest; \item \textbf{Inputs}: Observed data $\{y_{i},\boldsymbol x_i\}$
            \item \textbf{Outputs}: $M$ posterior samples of model-based conditional quantile functions $\{Q_{\tau}(Y_i \mid \boldsymbol x_i, \boldsymbol \theta^{m})\}_{i=1}^{M}$
            \begin{itemize}
                \item In the case of the LL-LS model \eqref{M}, these samples are easily computed with each posterior sample of $\boldsymbol \theta$, e.g. $Q_{\tau}(Y \mid \boldsymbol{x}, \boldsymbol \theta^{m}) =  \boldsymbol x^{\intercal}\boldsymbol{\xi}^{m} +\sigma^m\exp(\boldsymbol{x}^{\intercal}\boldsymbol{\gamma}^{m})\Phi^{-1}(\tau)$
            \end{itemize}
        \end{itemize}
            
    \STATE 2a. Quantile estimation and uncertainty quantification
    \begin{itemize}
            \item \textbf{Description}: For any $\tau$ and subset $S \subseteq \{1,\dots,p\}$ of predictors, apply decision analysis to obtain quantile-specific linear coefficient estimates and uncertainty quantification.
             \item \textbf{Inputs}: Posterior samples $\{Q_{\tau}(Y_i \mid \boldsymbol x_i, \boldsymbol \theta^{m})\}_{m=1}^{M}$, covariate submatrix $\boldsymbol X_{S}$
            \item \textbf{Outputs}: The optimal action \eqref{OLS_sol} with accompanying uncertainty provided by the posterior action \eqref{proj}

            \begin{itemize}
                \item For any subset $S$ of predictors, the optimal action is simply given by $({\boldsymbol{X}}^{\intercal}_{S} {\boldsymbol{X}}_{S})^{-1}{\boldsymbol{X}}^{\intercal}_{S} \boldsymbol{\hat{Q}}_{\tau}({\boldsymbol{X}})$ as in Lemma \ref{OLS}. The posterior action is obtained by projecting each draw of $\{Q_{\tau}(Y_i \mid \boldsymbol x_i, \boldsymbol \theta^{m})\}_{i=1}^{M}$ onto $\boldsymbol X_{S}$. 
            \end{itemize}
    \end{itemize}
    \STATE 2b.  Subset search, filtration, and selection
        \begin{itemize}
            \item \textbf{Description}:  For any $\tau$, conduct a quantile-specific  subset search, accumulate a family of subsets with strong predictive power, and summarize this family via i) a single subset that balances parsimony and predictive power and ii) measures of variable importance across all subsets in the family.
            \item \textbf{Inputs}: Posterior samples $\{Q_{\tau}(Y_i \mid \boldsymbol x_i, \boldsymbol \theta^{m})\}_{i=1}^{M}$, covariates $X$, $m_{k}$ for the BBA filtration, and $\epsilon$ for the acceptable family criteria.
            \item \textbf{Outputs}: Posterior samples of \eqref{post-dif} for each $S \in \mathbbm{S}({\tau})$ obtained from the BBA search, which are used to determine acceptable subsets $\mathbbm{A}_{0.05}(\tau)$ based on the criteria outlined by \eqref{accept}. The subset with the smallest cardinality is $\mbox{S}_{small}(\tau)$
            \begin{itemize}
                \item For the BBA algorithm, the key inputs are simply the point-wise posterior mean of the quantile function $\boldsymbol{\hat{Q}}_{\tau}({\boldsymbol{X}}) $, the covariates $\boldsymbol X$ and $m_{k}$. The output is $\mathbbm{S}(\tau)$, which is an $L \times p$ matrix of indicators, with each row corresponding to a subset. The indicators determine the member active predictors in each subset.
                \item For each subset in $\mathbbm{S}({\tau})$, posterior samples of \eqref{post-dif} are obtained by extracting the  optimal action for that subset, forming point predictions using that action, and evaluating the posterior distribution of the aggregated squared error loss using samples $\{Q_{\tau}(Y_i \mid \boldsymbol x_i, \boldsymbol \theta^{m})\}_{i=1}^{M}$. The same process is repeated using the anchor action $\boldsymbol{\hat{Q}}_{\tau}({\boldsymbol{X}})$. These samples are combined to measure \eqref{post-dif}. 
                \item Acceptable subsets are those whose corresponding posterior distribution of \eqref{post-dif} meets the criteria outlined by \eqref{accept}. Variable importance \eqref{VI} for each variable $j$ is computed by measuring the proportion of subsets which include variable $j$
            \end{itemize}
            \end{itemize}
\end{algorithmic}
\end{algorithm}

\section{Hierarchical Specification of the LL-LS Model}
We specify the following priors for estimation of the LL-LS model \eqref{M} for the simulation study and real data analysis in Sections \ref{sim}-\ref{realdat}.

\begin{align}
    y_i \sim \mbox{Normal}(&\boldsymbol{x}_i^{\intercal}\boldsymbol{\xi}, \{\boldsymbol\sigma \mbox{exp}(\boldsymbol x_i^{\intercal}\boldsymbol \gamma)\}^{2}]\\
    \xi_{j} \overset{indep}{\sim} \mbox{Normal}(0,\lambda_{\xi_j}), j \in\{2,\dots,p\}, \quad &    \gamma_{j} \overset{indep}{\sim} \mbox{Normal}(0,\lambda_{\gamma_j}), j \in\{ 2,\dots,p\}  \\
    \lambda_{\gamma_j}, \lambda_{\xi_j} &\sim \mbox{Cauchy}^{+}(0,5), j \in\{2,\dots,p\}\\
  \sigma^{2} &\sim \mbox{Inverse-Gamma}(1/2,1/2)
\end{align}
Here, $\mbox{Cauchy}^{+}$ is the half cauchy distribution.  In addition, we specify flat priors for the intercept terms, i.e. $\beta_{1}, \gamma_{1} \propto 1$. The model is estimated in the \vtt{STAN} programming language in \verb|R|.
\section{Prescreening for the BBA Algorithm}
For the simulation setting involving $p = 50$ covariates in Section \ref{sim}, we adopt a prescreening strategy for the LL-LS model $\mathcal{M}$ to narrow the class of candidate subsets. In general, the BBA algorithm is efficient for $p \leq 35$, and acceptable subsets are more interpretable with greater levels of sparsity.

The prescreening strategy proceeds as follows: We first compute the posterior mean for each $\boldsymbol \xi_j$ and $\boldsymbol \gamma_j$, denoted $\hat{\boldsymbol \xi}_j$ and $\hat{\boldsymbol{\gamma}}_j$, respectively. Then, we identify the top $35$ covariates for which $\lvert\hat{\boldsymbol \gamma}_{j}\rvert + \lvert \hat{\boldsymbol\xi}_{j}\rvert$ is greatest. These covariates, and the ensuing submatrix is passed into the BBA algorithm to provide candidate subsets for curation of the acceptable family, as described in Section \ref{filt}.

\section{Further Simulation Results}

For the $(n,p) = (100,100)$ setting, we utilized the \vtt{Brq} package in \vtt{R} under default settings for the adaptive LASSO prior \citep{alhamzawi2020brq}. Default settings for the adaptive LASSO quantile regressions in \vtt{BayesQR} repeatedly caused our local machine to crash whenever $p \geq n$. Upon correspondence with the authors of \vtt{BayesQR}, the code malfunction could be attributed to regression coefficient initialization at  OLS estimates, which are not identified in $p \geq n$ regimes. Fortunately, the MCMC algorithms are quite similar between software packages and the results were minimally affected and consistent with what is presented in the paper.

\subsection{Uncertainty Quantification}
We complete the information presented in Table \ref{Table1} with the nominal coverage rates and interval widths in the other simulation settings in Table \ref{uncert}. The results are consistent with what is presented in the main paper - $S_{small}$ and $S_{full}$ are significantly more sharp and calibrated than the Bayesian and frequentist alternatives.

 \begin{table*}[h]
\centering


    \captionsetup{labelformat = empty}
    \caption*{$\boldsymbol{n = 200, p = 50, \textbf{\mbox{HetRatio}} =0.5}$}
    \begin{tabular}{c r| c c c c c c c c c}   

      & $\tau$  & 0.01&0.05 &0.25&0.5&0.75&0.95&0.99  \\
       \hline
       \multirow{ 3}{*}{Coverage Rate}& $\mbox{S}_{small}(\tau)$  & 0.92 & 0.94 & 0.92 & 0.91 & 0.92 &  0.93 & 0.91\\
        & $\mbox{S}_{full}$ & 0.97 & 0.97 &  0.95 & 0.95 & 0.95 &  0.96 & 0.97\\
        & $\mbox{AL}_{Bayes}$ & 0.99 & 0.98 &  0.97 & 0.98 & 0.97 & 0.98 & 0.98 \\\hline
        \multirow{ 3}{*}{Avg. 95\% CI Width}& $\mbox{S}_{small}(\tau)$  & 1.02 & 0.95 &  0.81 & 0.77 & 0.80 &  0.94 & 1.02\\
        & $\mbox{S}_{full}$ & 3.56 & 2.96 & 2.33 & 2.18 & 2.34 & 2.96 & 3.57 \\
        & $\mbox{AL}_ {Bayes}$ & 7.87 & 4.30 & 2.62 & 2.33 & 2.61 & 4.34 & 8.30\\ \hline\\
    \end{tabular}
    \captionsetup{labelformat = empty}
      \caption{$\boldsymbol{n = 500, p = 20, \textbf{\mbox{HetRatio}} =1}$}

   \begin{tabular}{c r| c c c c c c c c c}    

      & $\tau$  & 0.01&0.05 &0.25&0.5&0.75&0.95&0.99  \\
       \hline
       \multirow{ 3}{*}{Coverage Rate}& $\mbox{S}_{small}(\tau)$  & 0.92 & 0.91 &  0.90 & 0.89 & 0.89 &  0.94 & 0.95\\
        & $\mbox{S}_{full}$ & 0.93 & 0.92 &  0.91 & 0.91 & 0.91 &  0.96 & 0.97\\
        & $\mbox{AL}_{Bayes}$ & 0.94 & 0.94 &  0.93 & 0.94 & 0.95 &  0.99 & 0.99 \\\hline
        \multirow{ 3}{*}{Avg. 95\% CI Width}& $\mbox{S}_{small}(\tau)$  & 0.63 & 0.51 &  0.36 & 0.31 & 0.37 &  0.52 & 0.65\\
        & $\mbox{S}_{full}$ & 1.41 & 1.14 &  0.83 & 0.75 & 0.84 & 1.15 & 1.43 \\
        & $\mbox{AL}_ {Bayes}$ & 4.01 & 2.40 &  1.43 & 1.25 & 1.41 &  2.42 & 4.32\\ \hline\\
    \end{tabular}
    \captionsetup{labelformat=empty}
    \caption{$\boldsymbol{n = 500, p = 20, \textbf{\mbox{HetRatio}} =0.5}$}

    \begin{tabular}{c r| c c c c c c c c c}    

      & $\tau$  & 0.01&0.05 &0.25&0.5&0.75&0.95&0.99  \\
       \hline
       \multirow{ 3}{*}{Coverage Rate}& $\mbox{S}_{small}(\tau)$  & 0.92 & 0.91 &  0.89 & 0.89 & 0.89 & 0.91 & 0.92\\
        & $\mbox{S}_{full}$ & 0.93 & 0.92 & 0.91 & 0.90 & 0.91 & 0.92 & 0.94\\
        & $\mbox{AL}_{Bayes}$ & 0.95 & 0.95 &  0.94 & 0.94 &0.95 & 0.96 & 0.98 \\\hline
        \multirow{ 3}{*}{Avg. 95\% CI Width}& $\mbox{S}_{small}(\tau)$  & 0.39 & 0.34 &  0.28 & 0.26 & 0.29 &  0.34 & 0.38\\
        & $\mbox{S}_{full}$ & 1.54 & 1.29 &  1.05 & 0.99 & 1.05 &  1.30 &1.54 \\
        & $\mbox{AL}_ {Bayes}$ & 3.71 & 2.15 &  1.29 & 1.13 & 1.26 &  2.19 &4.00\\ \hline\\
    \end{tabular}
 \caption{$\boldsymbol{n = 100, p = 100, \textbf{\mbox{HetRatio}} =1}$}

    \begin{tabular}{c r| c c c c c c c c c}    

      & $\tau$  & 0.01&0.05 &0.25&0.5&0.75&0.95&0.99  \\
       \hline
       \multirow{ 3}{*}{Coverage Rate}& $\mbox{S}_{small}(\tau)$  & 0.73 &0.73& 0.73 & 0.75& 0.73 & 0.74 & 0.73\\
        & $\mbox{S}_{full}$ & 0.82 & 0.82 & 0.82 & 0.84 & 0.81 & 082 & 0.81\\
        & $\mbox{AL}_{Bayes}$ & 0.92 & 0.94 & 0.96 & 0.98 & 0.96 & 0.93 & 0.91 \\\hline
        \multirow{ 3}{*}{Avg. 95\% CI Width}& $\mbox{S}_{small}(\tau)$  & 0.44 & 0.38 & 0.30 & 0.25 & 0.34 & 0.48 & 0.51\\
        & $\mbox{S}_{full}$ & 2.04 & 1.69 & 1.15 & 1.06 & 1.18 & 1.82 & 2.30 \\
        & $\mbox{AL}_ {Bayes}$ & 6.05 & 6.9 & 7.56 & 7.62 & 7.50 & 6.80 & 5.96\\ \hline\\
    \end{tabular}
   
     \captionsetup{labelformat=empty}
    \caption{$\boldsymbol{n = 100, p = 100, \textbf{\mbox{HetRatio}} =0.5}$}

    \begin{tabular}{c r| c c c c c c c c c}    

      & $\tau$  & 0.01&0.05 &0.25&0.5&0.75&0.95&0.99  \\
       \hline
       \multirow{ 3}{*}{Coverage Rate}& $\mbox{S}_{small}(\tau)$  & 0.75 & 0.76 &  0.77 & 0.79 & 0.77 & 0.76 & 0.75\\
        & $\mbox{S}_{full}$ & 0.94 & 0.95 & 0.95 & 0.91 & 0.95 & 0.96 & 0.95\\
        & $\mbox{AL}_{Bayes}$ & 0.90 & 0.92 &  0.94 & 0.96 &0.94 & 0.91 & 0.90 \\\hline
        \multirow{ 3}{*}{Avg. 95\% CI Width}& $\mbox{S}_{small}(\tau)$  & 0.57 & 0.60 &  0.71 & 0.71 & 0.68 &  0.74 & 0.66\\
        & $\mbox{S}_{full}$ & 2.43 & 2.05 &  1.56 & 1.45 & 1.58 &  2.14 &2.51 \\
        & $\mbox{AL}_ {Bayes}$ & 3.99 & 4.74 &  4.88 & 4.96 & 4.86 &  4.42 &3.90\\ \hline\\
    \end{tabular}

\label{uncert}
    \caption*{Coverage rates and average widths of 95\% credible intervals for the posterior action \eqref{proj} with $\mbox{S}_{small}(\tau)$ or $\mbox{S}_{full}$ as well as $\mbox{AL}_{Bayes}$. The intervals for $\mbox{S}_{small}(\tau)$ and $\mbox{S}_{full}$ are significantly more narrow than those for $\mbox{AL}_{Bayes}$, especially for  quantiles near zero or one, and typically maintain  nominal coverage. $\mbox{S}_{small}(\tau)$ sacrifices some coverage  in favor of sparsity, and thus provides the most narrow intervals. 
    }

\end{table*}
\subsection{Selection with Independent Covariates}
 We compute average TPRs and TNRs for each $(n,p,\mbox{HetRatio})$ setting where the covariates $x_{ij}$ are simulated independently from a $\mbox{uniform}(0,1)$ in Table \ref{Tableindep}.  Once again, the proposed approach balances TPR and TNR well, with the frequentist competitor once again overly dense (low TNRs), and $\mbox{AL}_{Bayes}$ underpowered (low TPRs). 

\begin{table*}[h]
     \caption{Independent Covariates: $\boldsymbol{n = 200, p = 50, \textbf{\mbox{HetRatio}} =1}$}    \label{Tableindep}
     \centering
    \begin{tabular}{c r| c c c c c c c c c}    

      & $\tau$  & 0.01&0.05 &0.25&0.5&0.75&0.95&0.99  \\
       \hline
       \multirow{ 3}{*}{TPR}& $\mbox{S}_{small}(\tau)$  & 0.34 & 0.53 &  0.75 & 0.82 & 0.78 & 0.62 & 0.50\\
        & $\mbox{aLASSO}$ & 0.14 & 0.76 &0.93 & 0.95 & 0.96 & 0.83 & 0.25\\
        & $\mbox{AL}_{Bayes}$ &0.00& 0.01 &  0.32 & 0.49 & 0.36 &  0.01 & 0.00 \\\hline
        \multirow{ 3}{*}{TNR}& $\mbox{S}_{small}(\tau)$  & 0.84 & 0.80 &  0.76 & 0.75 & 0.76 &  0.80 & 0.84\\
        & $\mbox{aLASSO}$ & 0.87 & 0.48 &  0.24 & 0.23 &  0.34 & 0.47 & 0.86\\
        & $\mbox{AL}_ {Bayes}$ &1.00 & 0.99 & 0.93 & 0.90 & 0.93 & 0.99 & 1.00\\\hline\\

    \end{tabular}

    \captionsetup{labelformat = empty}
    \caption*{Independent Covariates: $\boldsymbol{n = 200, p = 50, \textbf{\mbox{HetRatio}} =0.5}$}

    \begin{tabular}{c r| c c c c c c c c c}  

      & $\tau$  & 0.01&0.05 &0.25&0.5&0.75&0.95&0.99  \\
       \hline
        \multirow{ 3}{*}{TPR}& $\mbox{S}_{small}(\tau)$ & 0.87 & 0.95 &0.99 & 0.98 & 0.96 &  0.92 & 0.89 \\
        & $\mbox{aLASSO}$ & 0.20 & 0.91 & 0.99 & 1.00 & 0.99 & 0.95 & 0.35 \\
        & $\mbox{AL}_ {Bayes}$ & 0.00 & 0.00 & 0.28 & 0.43 & 0.31 & 0.01& 0.00\\\hline
        \multirow{ 3}{*}{TNR}& $\mbox{S}_{small}(\tau)$& 0.78 & 0.75 &  0.69 & 0.67 & 0.69 &  0.74 & 0.78 \\
        & $\mbox{aLASSO}$  & 0.89 & 0.49 & 0.26 & 0.27 & 0.29 & 0.49 & 0.89\\
        & $\mbox{AL}_{Bayes}$ & 1.00 & 0.99 &  0.91 & 0.87 & 0.91 & 0.99 & 1.00\\\hline\\

    \end{tabular}
    \captionsetup{labelformat = empty}
      \caption*{Independent Covariates: $\boldsymbol{n = 500, p = 20, \textbf{\mbox{HetRatio}} =1}$}

    \begin{tabular}{c r| c c c c c c c c c} 

      & $\tau$  & 0.01&0.05 &0.25&0.5&0.75&0.95&0.99  \\
       \hline
        \multirow{ 3}{*}{TPR}&$\mbox{S}_{small}(\tau)$& 0.98 & 0.99 &  0.95 & 1.00 & 0.94 &  0.97 & 0.97\\
        & $\mbox{aLASSO}$ &0.77 & 1.00 &  0.99 & 1.00 & 0.99 & 1.00 & 1.00  \\
        & $\mbox{AL}_ {Bayes}$ & 0.00 & 0.58 & 0.88 & 1.00 & 0.88  & 0.60 & 0.01\\\hline
        \multirow{ 3}{*}{TNR}& $\mbox{S}_{small}(\tau)$& 0.95 & 0.94 &  0.93 & 0.92 & 0.93 &  0.95 & 0.96\\
        & $\mbox{aLASSO}$ & 0.87 & 0.71 &  0.57 & 0.53 & 0.56 & 0.72 & 0.85\\
        & $\mbox{AL}_ {Bayes}$ & 1.00 & 1.00 &  0.99 & 0.99 & 0.99 &  1.00 & 1.00\\\hline\\

    \end{tabular}
    \captionsetup{labelformat=empty}
    \caption*{Independent Covariates: $\boldsymbol{n = 500, p = 20, \textbf{\mbox{HetRatio}} =0.5}$}

    \begin{tabular}{c r| c c c c c c c c c}

      &$\tau$  & 0.01&0.05 &0.25&0.5&0.75&0.95&0.99  \\
       \hline
        \multirow{ 3}{*}{TPR}& $\mbox{S}_{small}(\tau)$  & 0.86 & 0.86 &  0.84 & 1.00 & 0.84 & 0.86 & 0.86\\
        & $\mbox{aLASSO}$ & 0.86 & 0.99 & 0.95 & 1.00 & 0.95 & 0.98 & 0.83\\
        & $\mbox{AL}_ {Bayes}$  & 0.00 & 0.59 &  0.80 & 1.00 & 0.82 &  0.66 & 0.00\\\hline
        \multirow{ 3}{*}{TNR}& $\mbox{S}_{small}(\tau)$ &0.97 & 0.95 &  0.93 & 0.91 & 0.92 & 0.95 & 0.98 \\
        & $\mbox{aLASSO}$ &0.87 & 0.76 &  0.61 & 0.56 & 0.59 & 0.77 & 0.86\\
        & $\mbox{AL}_ {Bayes}$ & 1.00 & 1.00 & 0.99 & 0.99 & 0.99 &  1.00 & 1.00\\\hline\\
    \end{tabular}
\end{table*}

 \begin{table*}[h]
\centering

    \captionsetup{labelformat = empty}
      \caption*{Independent Covariates: $\boldsymbol{n = 100, p = 100, \textbf{\mbox{HetRatio}} =1}$}

    \begin{tabular}{c r| c c c c c c c c c} 

      & $\tau$  & 0.01&0.05 &0.25&0.5&0.75&0.95&0.99  \\
       \hline
        \multirow{ 3}{*}{TPR}&$\mbox{S}_{small}(\tau)$& 0.03 & 0.04 &  0.04 & 0.03 & 0.04 &  0.03 & 0.03\\
        & $\mbox{aLASSO}$ &0.05&0.50&0.84 & 0.88&0.83& 0.47 & 0.43 \\
        & $\mbox{AL}_ {Bayes}$ & 0.05 & 0.04 & 0.05 & 0.04 & 0.05 & 0.05 & 0.05\\\hline
        \multirow{ 3}{*}{TNR}& $\mbox{S}_{small}(\tau)$& 0.99& 0.99 & 0.98 & 0.99 & 0.99 & 0.99 & 0.99\\
        & $\mbox{aLASSO}$ & 0.97 & 0.66 & 0.20 & 0.17 & 022 & 0.68 & 0.98\\
        & $\mbox{AL}_ {Bayes}$ & 1.00 & 1.00 &  0.99 & 0.99 & 0.99 &  1.00 & 1.00\\\hline\\

    \end{tabular}
    \captionsetup{labelformat=empty}
    \caption*{Independent Covariates: $\boldsymbol{n = 100, p = 100, \textbf{\mbox{HetRatio}} =0.5}$}

    \begin{tabular}{c r| c c c c c c c c c}

      &$\tau$  & 0.01&0.05 &0.25&0.5&0.75&0.95&0.99  \\
       \hline
        \multirow{ 3}{*}{TPR}& $\mbox{S}_{small}(\tau)$  & 0.14 &0.18 & 0.22 & 0.22 & 0.23& 0.16 & 0.13\\
        & $\mbox{aLASSO}$ & 0.05 & 0.55 & 0.85 & 0.88 & 0.84 & 0.55 & 0.04\\
        & $\mbox{AL}_ {Bayes}$  & 0.09 & 0.09 & 0.13 & 0.13 & 0.14 & 0.11 & 0.10\\\hline
        \multirow{ 3}{*}{TNR}& $\mbox{S}_{small}(\tau)$ &0.97 & 0.95 & 0.94 & 0.95& 0.95 & 0.97 & 0.97 \\
        & $\mbox{aLASSO}$ &0.98 & 0.70 & 0.25 & 0.20 & 0.26 & 0.71 & 0.98\\
        & $\mbox{AL}_ {Bayes}$ & 0.99 & 0.99 & 0.99 & 0.98 & 0.98 & 0.99 & 0.98\\\hline\\
    \end{tabular}    

    \caption*{True positive rates (TPR) and true negative rates (TNR) for variable selection averaged across simulations with independent covariates. Once again, the proposed approach ($\mbox{S}_{small}(\tau)$) neatly balances TPR and TNR. Low TPRs for the $n = 100, p = 100$ settings are due to the overwhelmingly strong impact of the covariate with heterogeneous effects, which are five to ten times stronger than those corresponding to homogeneous covariates across quantiles. }\label{indep}

\end{table*}

\subsection{Predictive Evaluations}
We complete the results presented in Figure \ref{Fig2} for the remaining simulation settings, metrics and quantiles in Figures \ref{Fig1_append}-
\ref{Fig11_append}. The results are consistent with what is presented in the paper: $S_{small}(\tau)$ demonstrates better predictive capabilities and is more calibrated than the frequentist and Bayesian competitors, with pronounced improvement at extreme quantiles. We note that the results presented here and in the main paper are qualitatively similar to simulations carried out under the same generating model for the response and $(n,p, \mbox{HetRatio})$ settings, but with independent covariates. 

\begin{figure}[H]
    \centering
        \caption{\textbf{MSE}: $\boldsymbol{n = 500, p= 20, \mbox{\textbf{HetRatio} }= 0.5}$} \label{Fig1_append}
  \includegraphics[width = .32\textwidth,keepaspectratio]{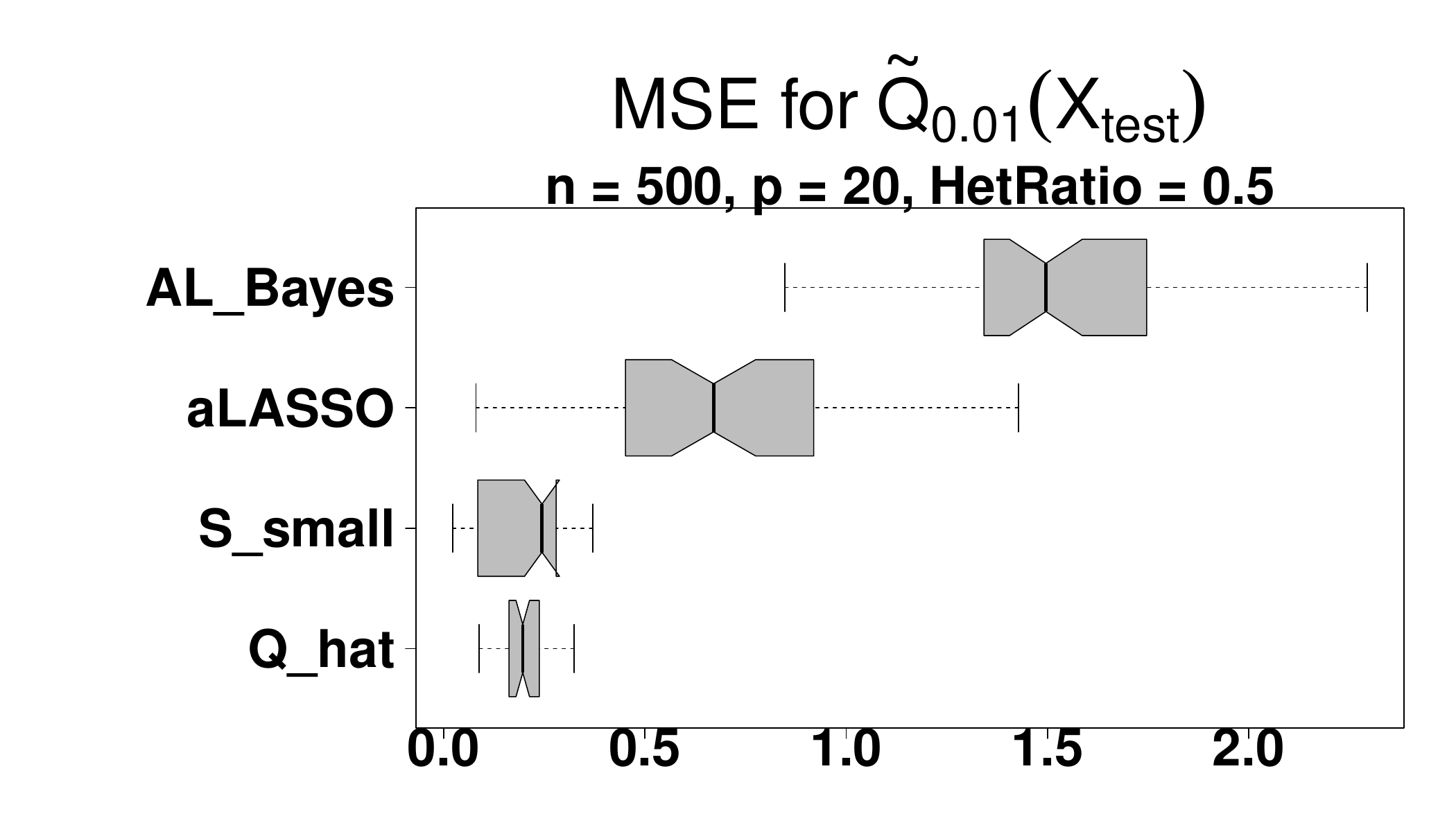}
    \includegraphics[width = .32\textwidth,keepaspectratio]{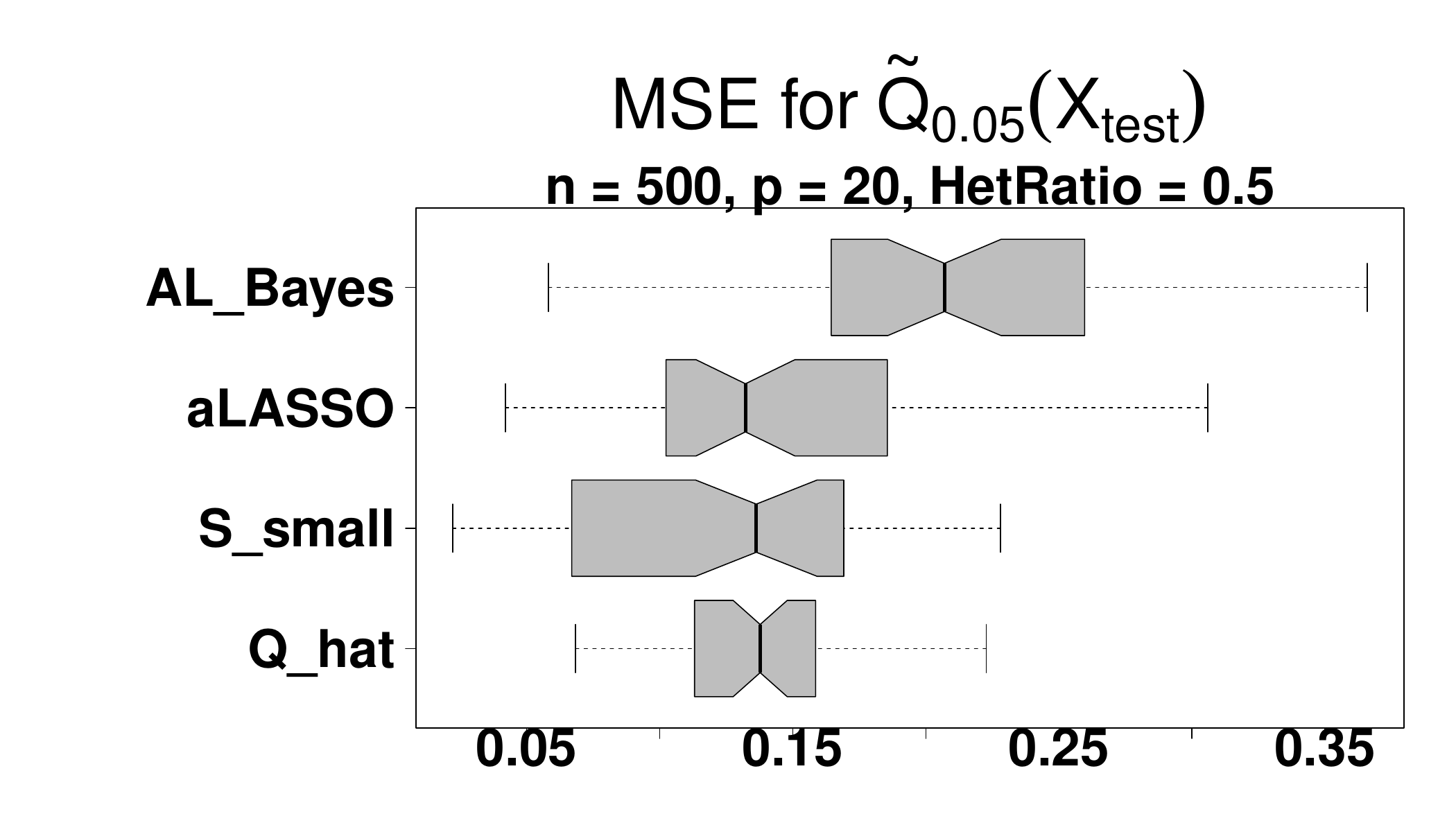}
    \includegraphics[width = .32\textwidth,keepaspectratio]{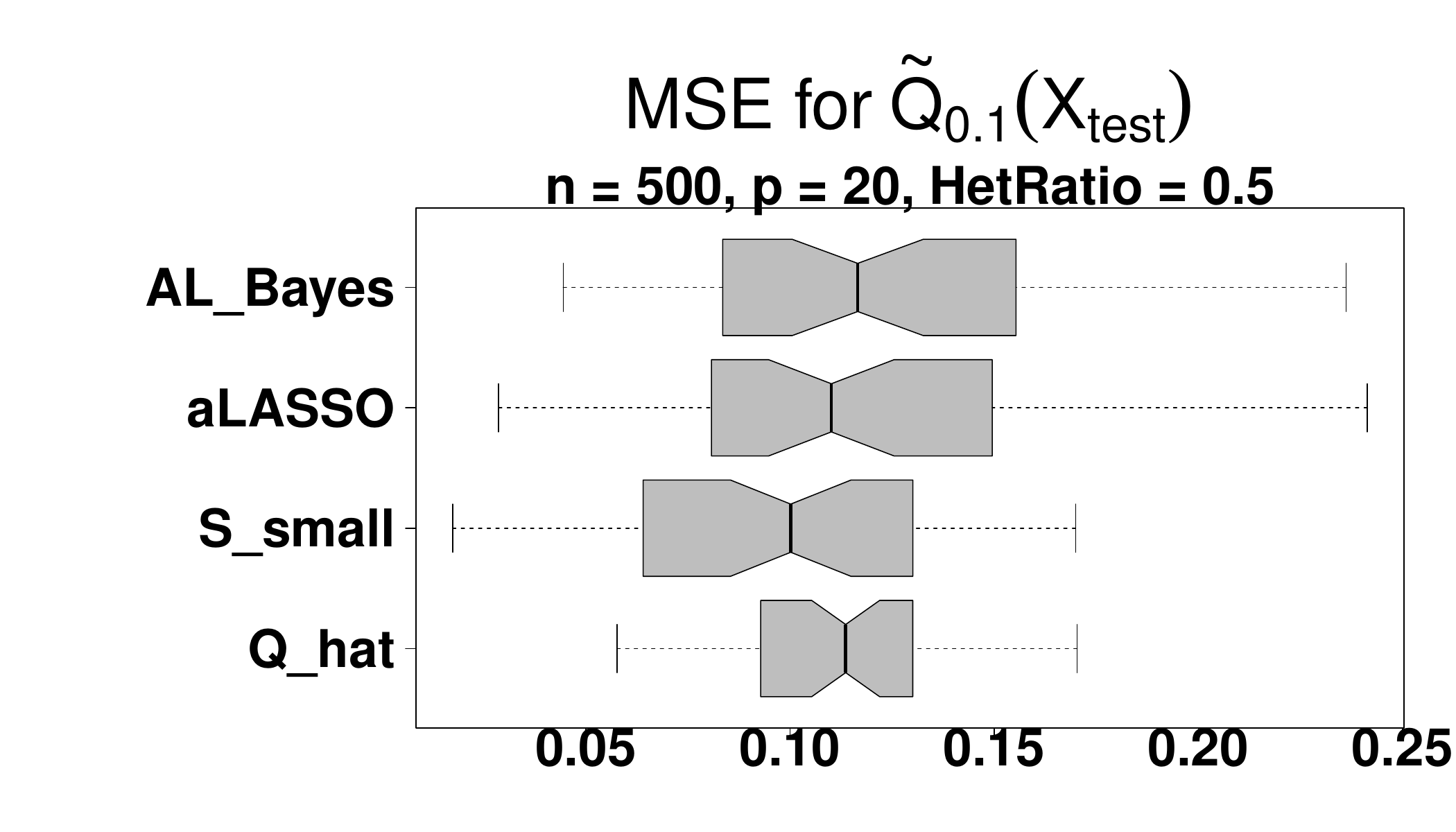}
    \includegraphics[width = .32\textwidth,keepaspectratio]{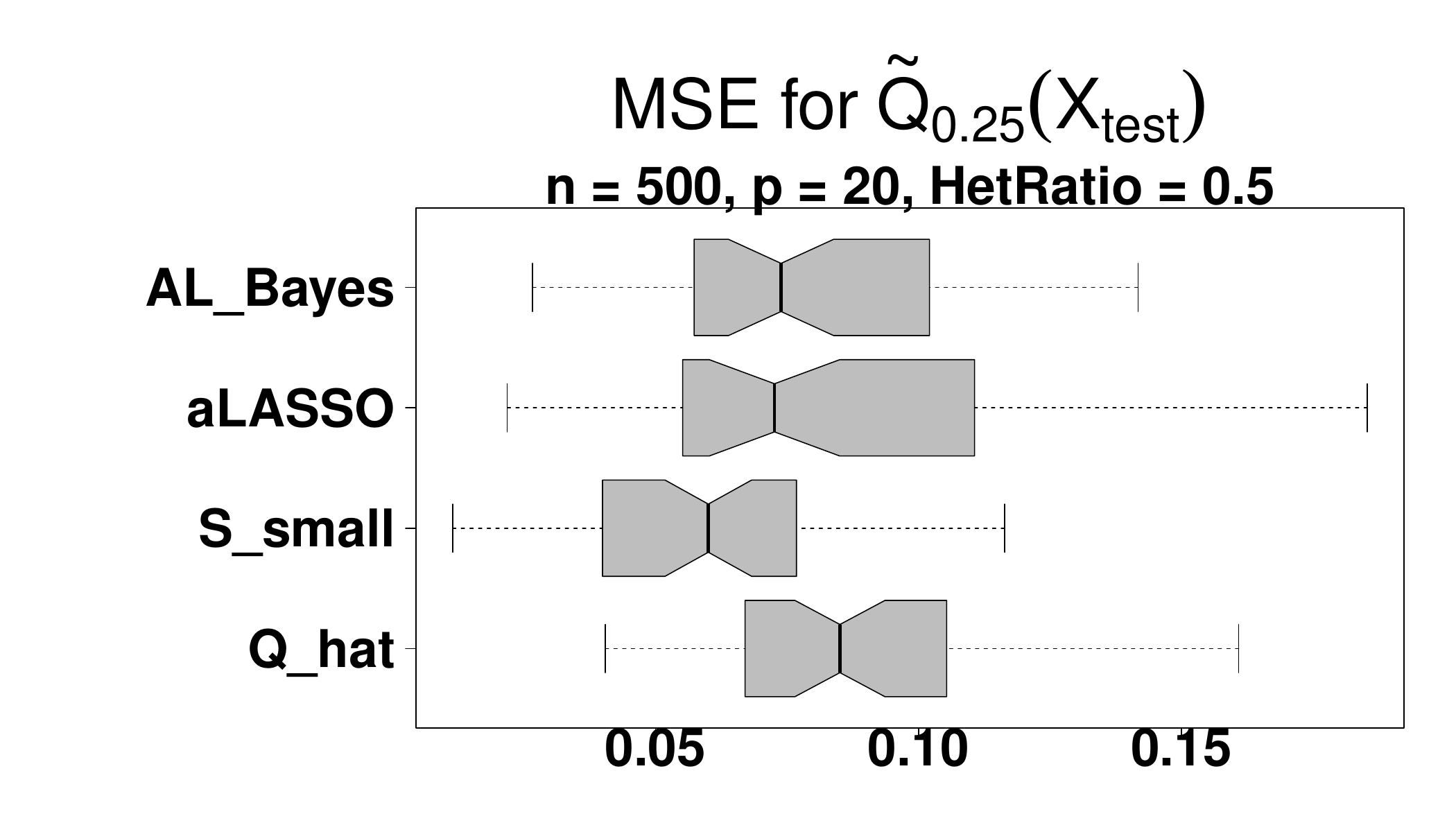}
    \includegraphics[width = .32\textwidth,keepaspectratio]{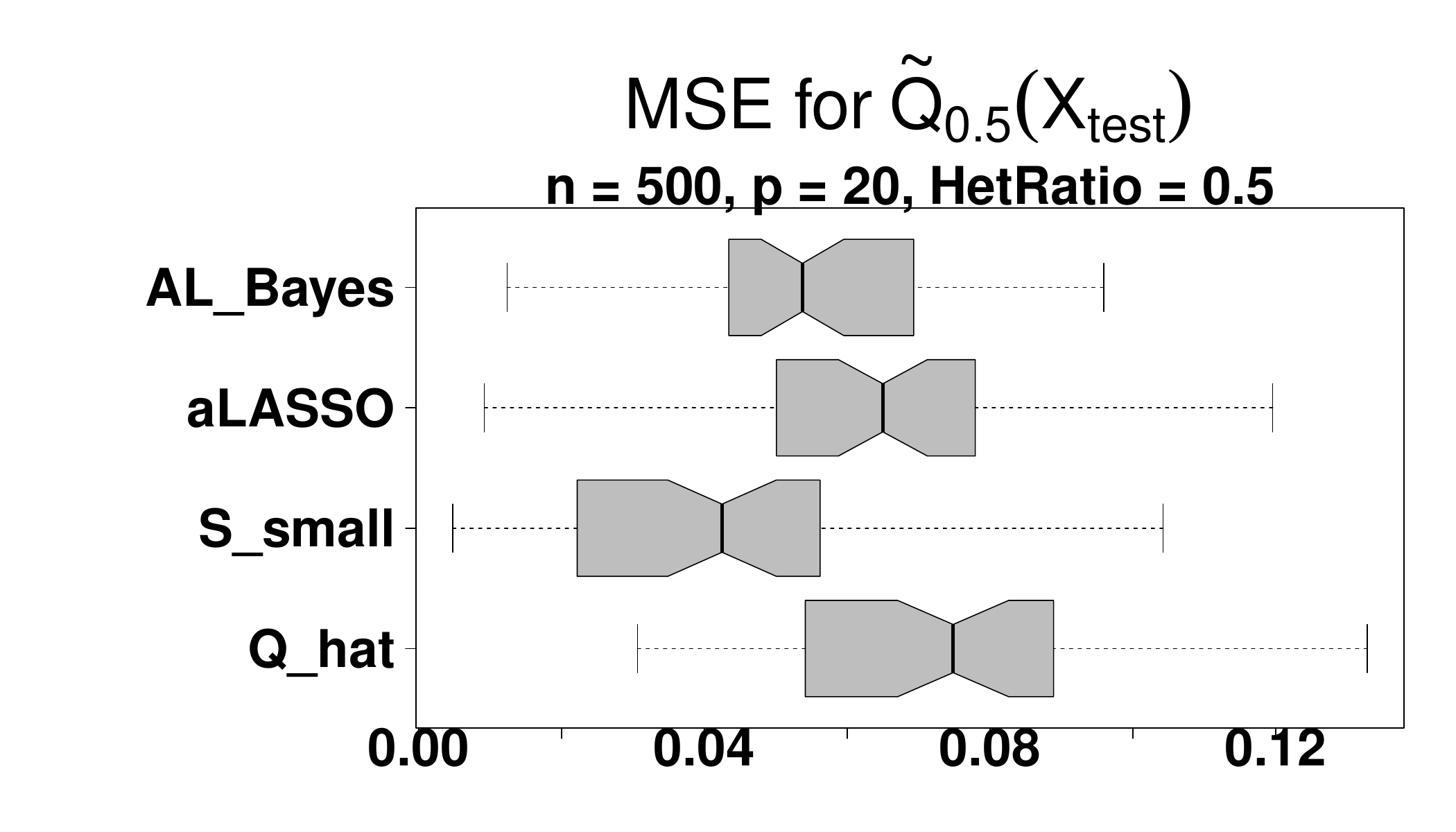}
    \includegraphics[width = .32\textwidth,keepaspectratio]{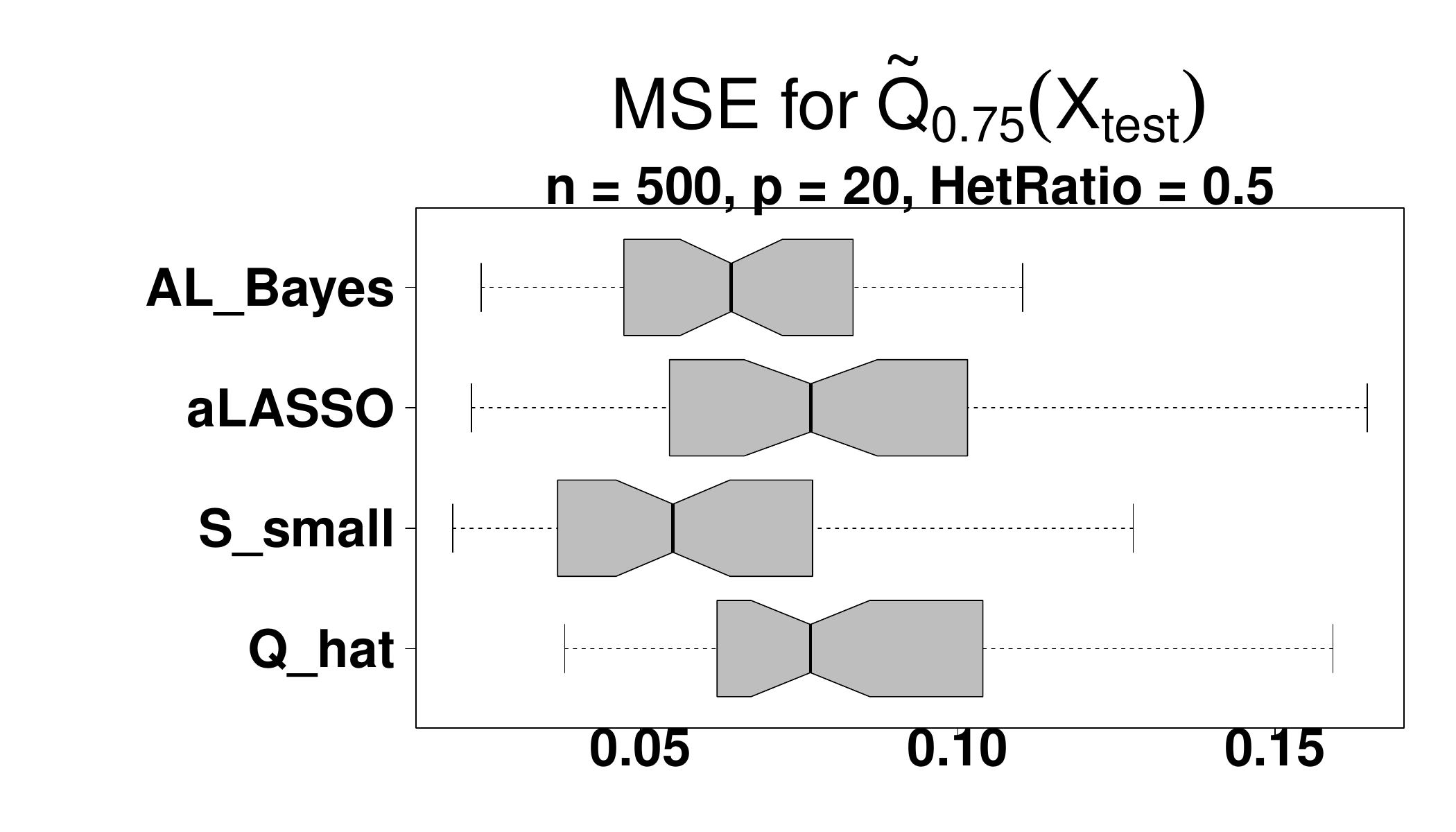}
   \includegraphics[width = .32\textwidth,keepaspectratio]{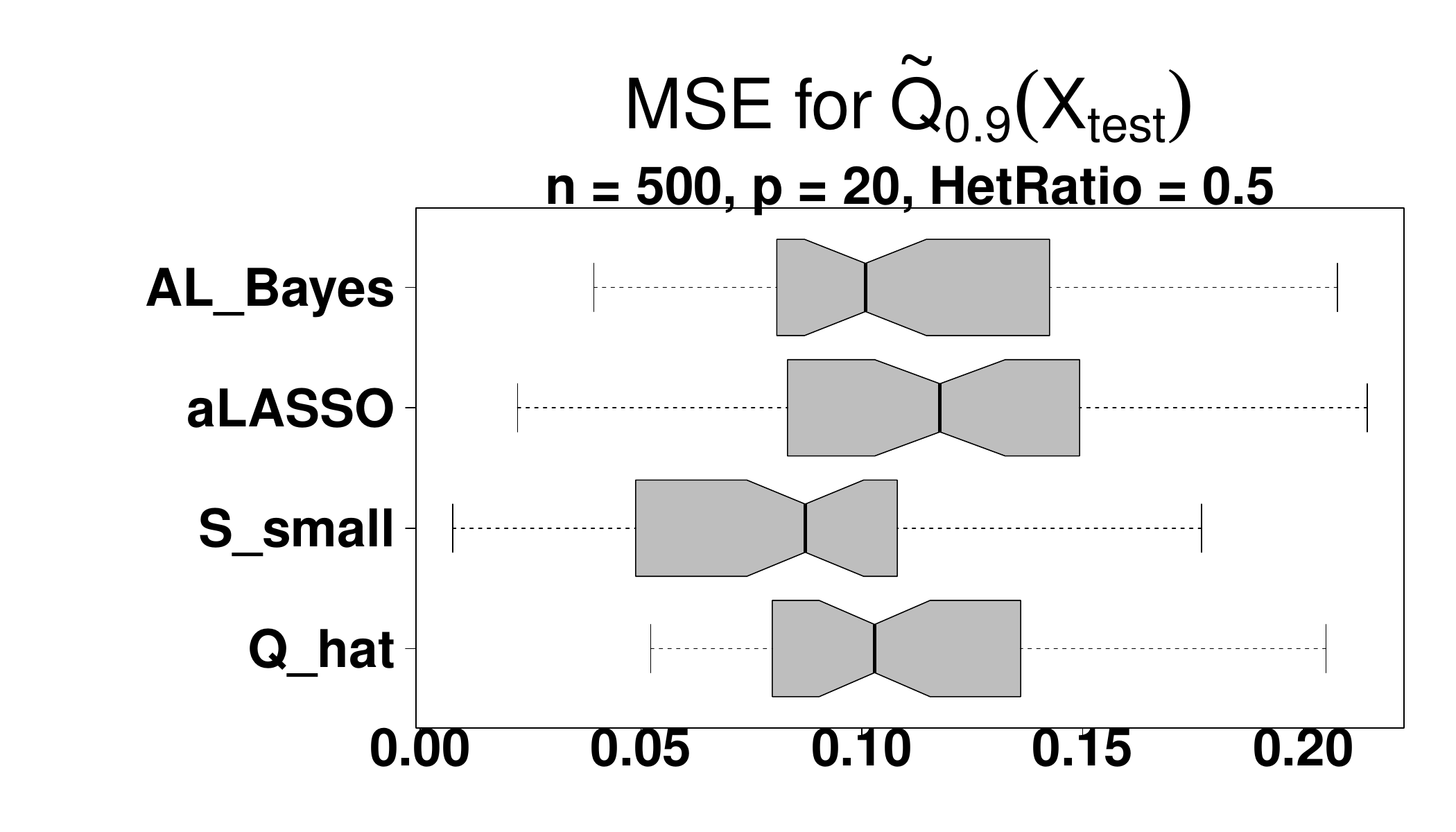}
    \includegraphics[width = .32\textwidth,keepaspectratio]{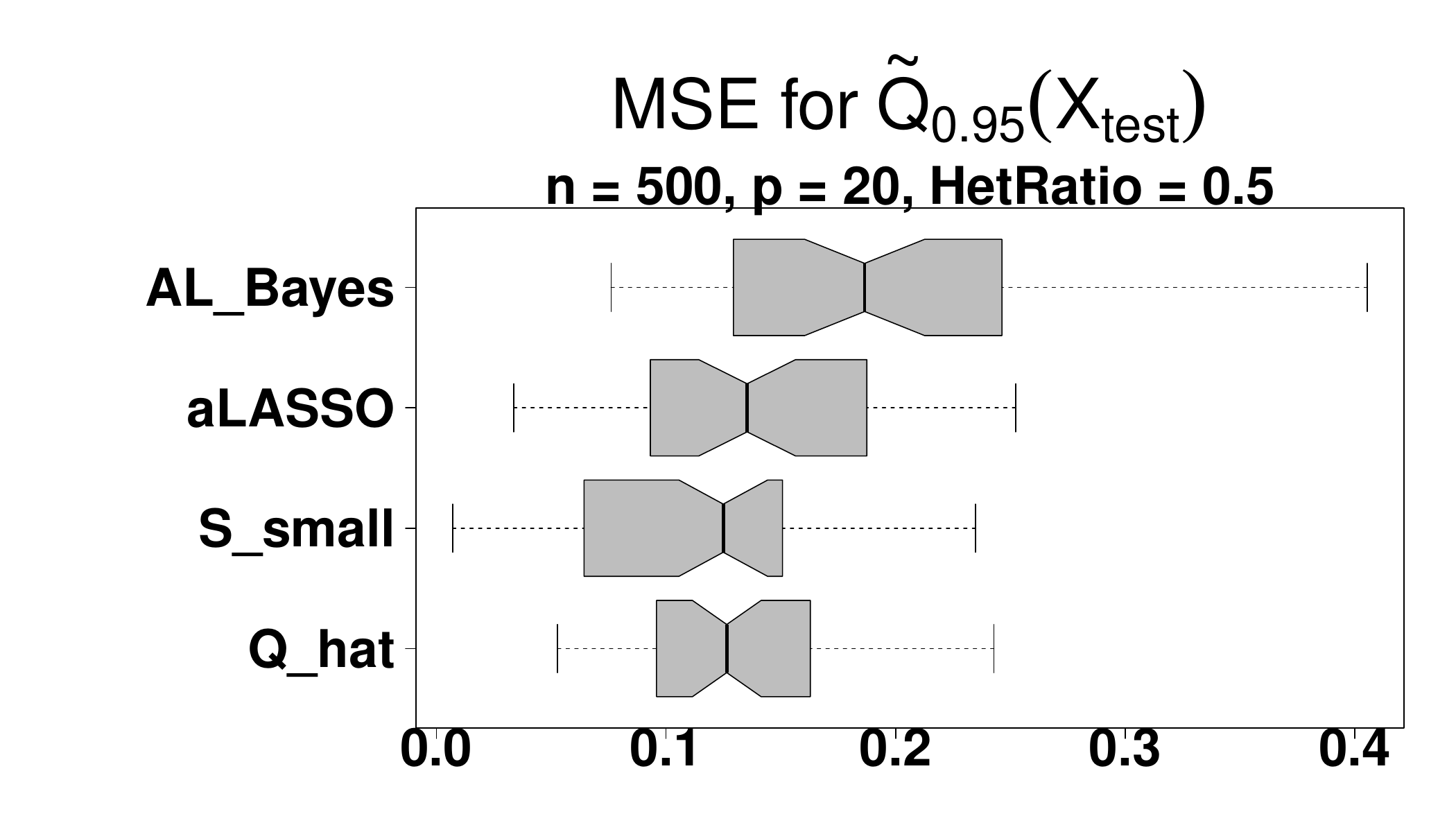}
   \includegraphics[width = .32\textwidth,keepaspectratio]{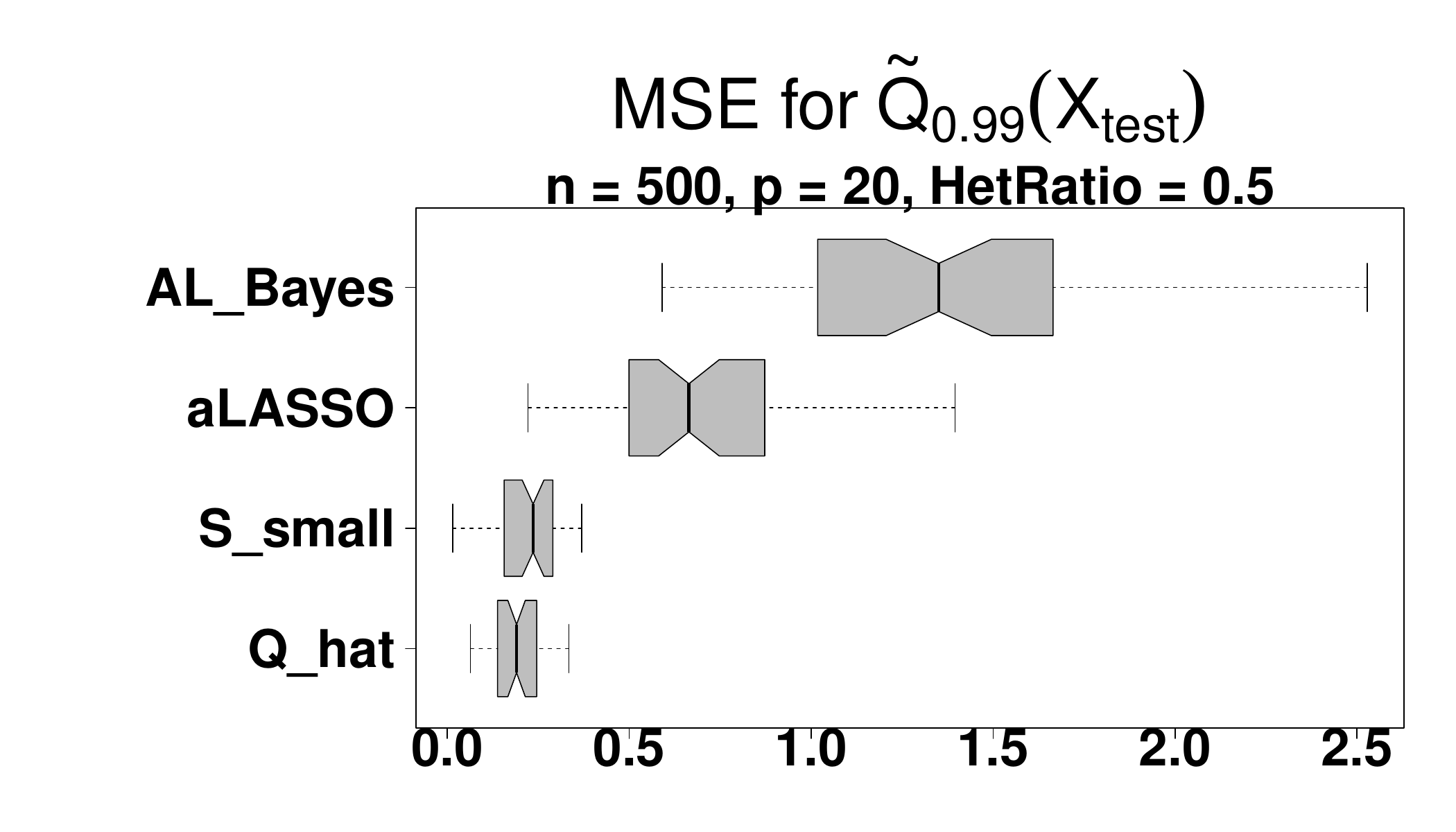}

\end{figure}

\begin{figure}[H]
    \centering
        \caption{\textbf{MSE}: $\boldsymbol{n = 500, p= 20, \mbox{\textbf{HetRatio} }= 1}$}
    \includegraphics[width = .32\textwidth,keepaspectratio]{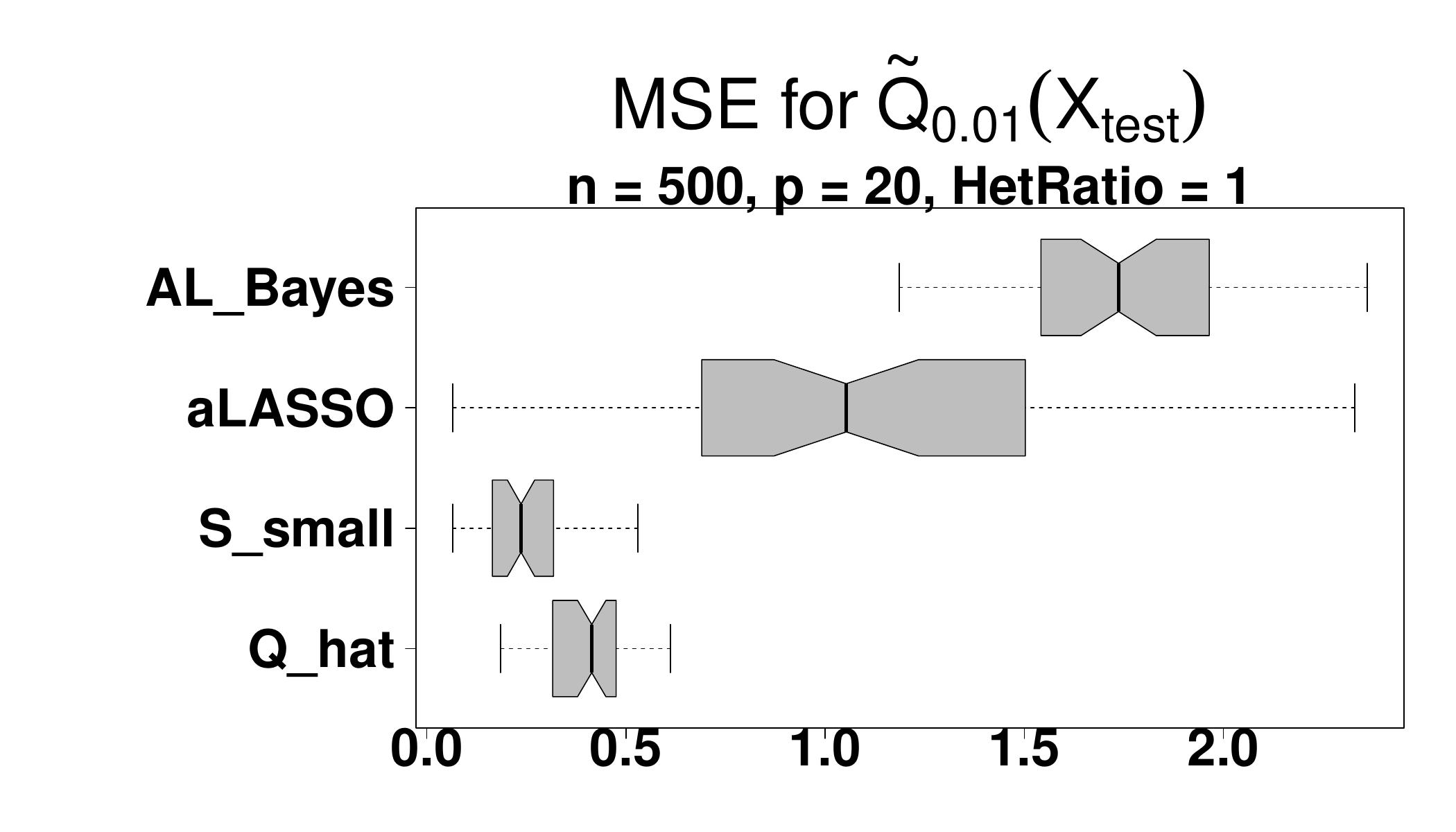}
    \includegraphics[width = .32\textwidth,keepaspectratio]{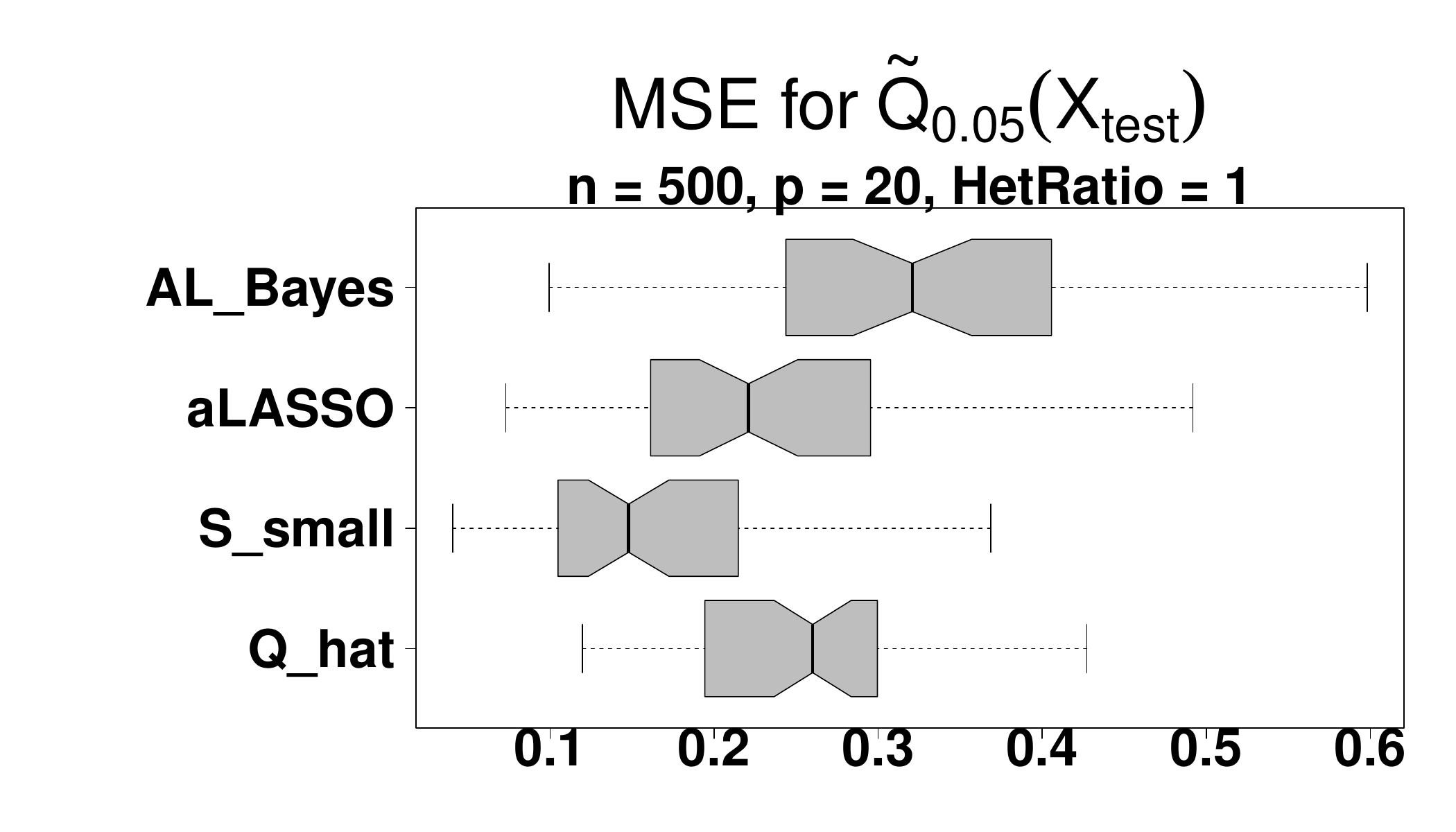}
    \includegraphics[width = .32\textwidth,keepaspectratio]{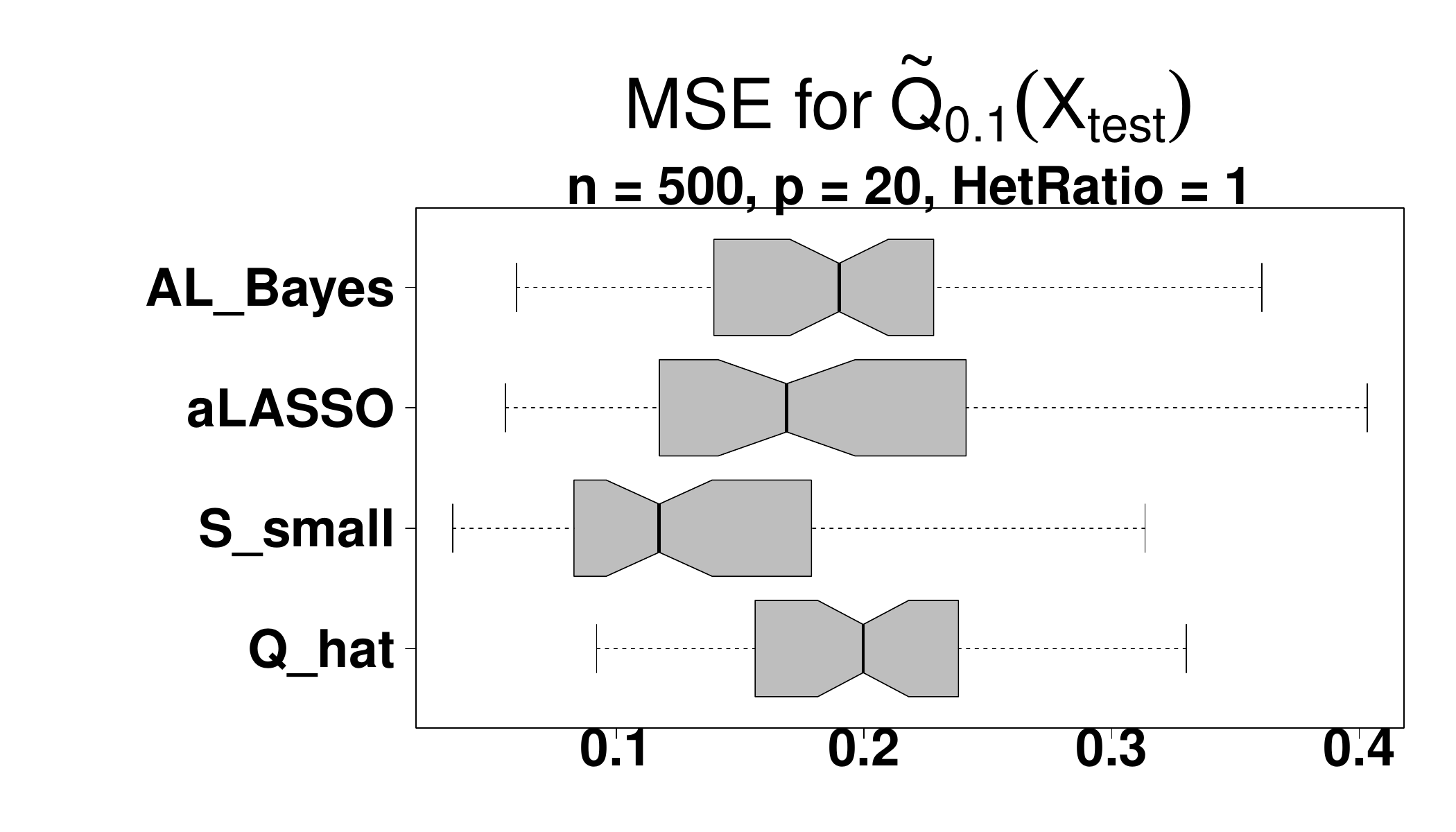}
    \includegraphics[width = .32\textwidth,keepaspectratio]{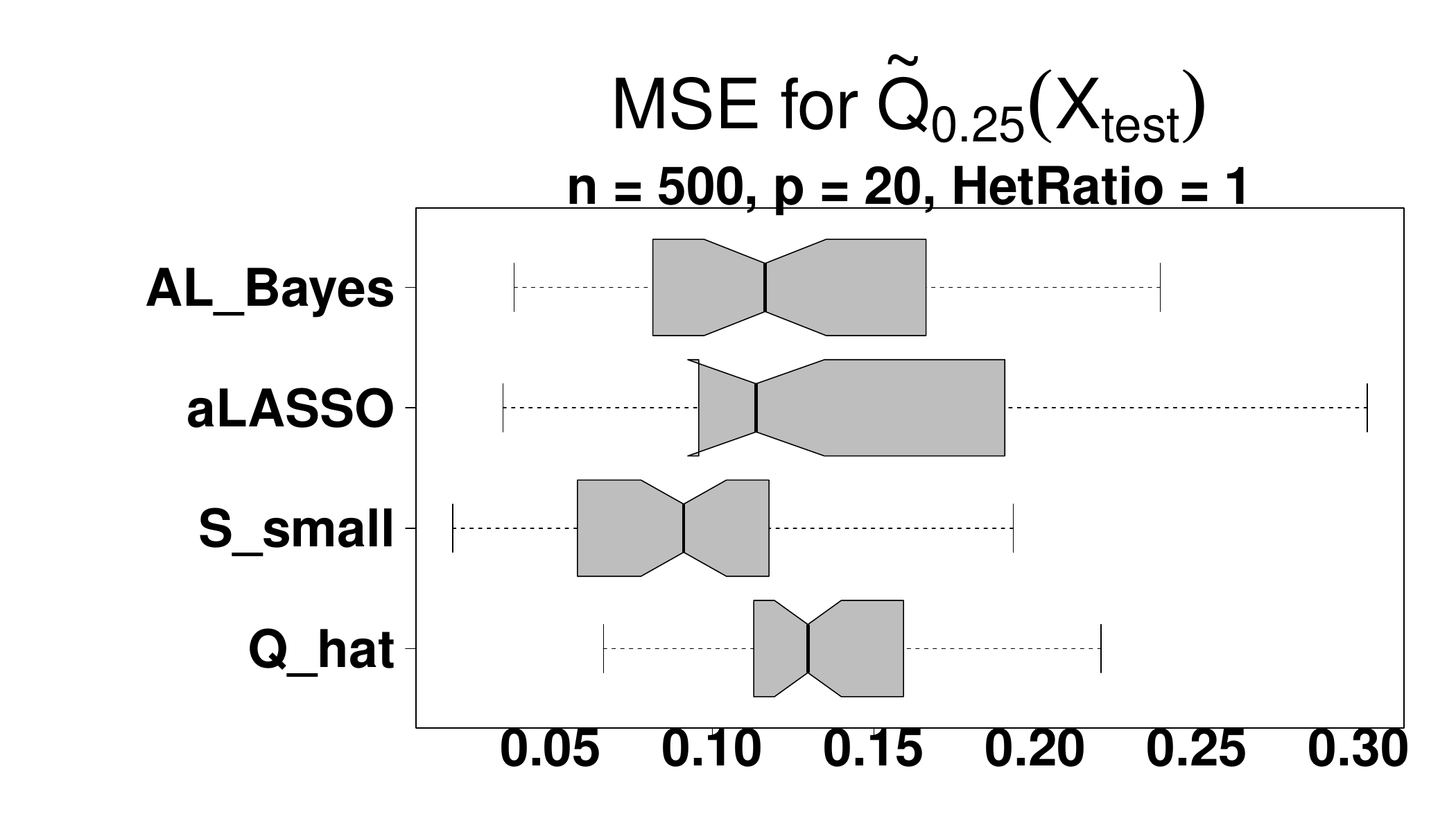}
    \includegraphics[width = .32\textwidth,keepaspectratio]{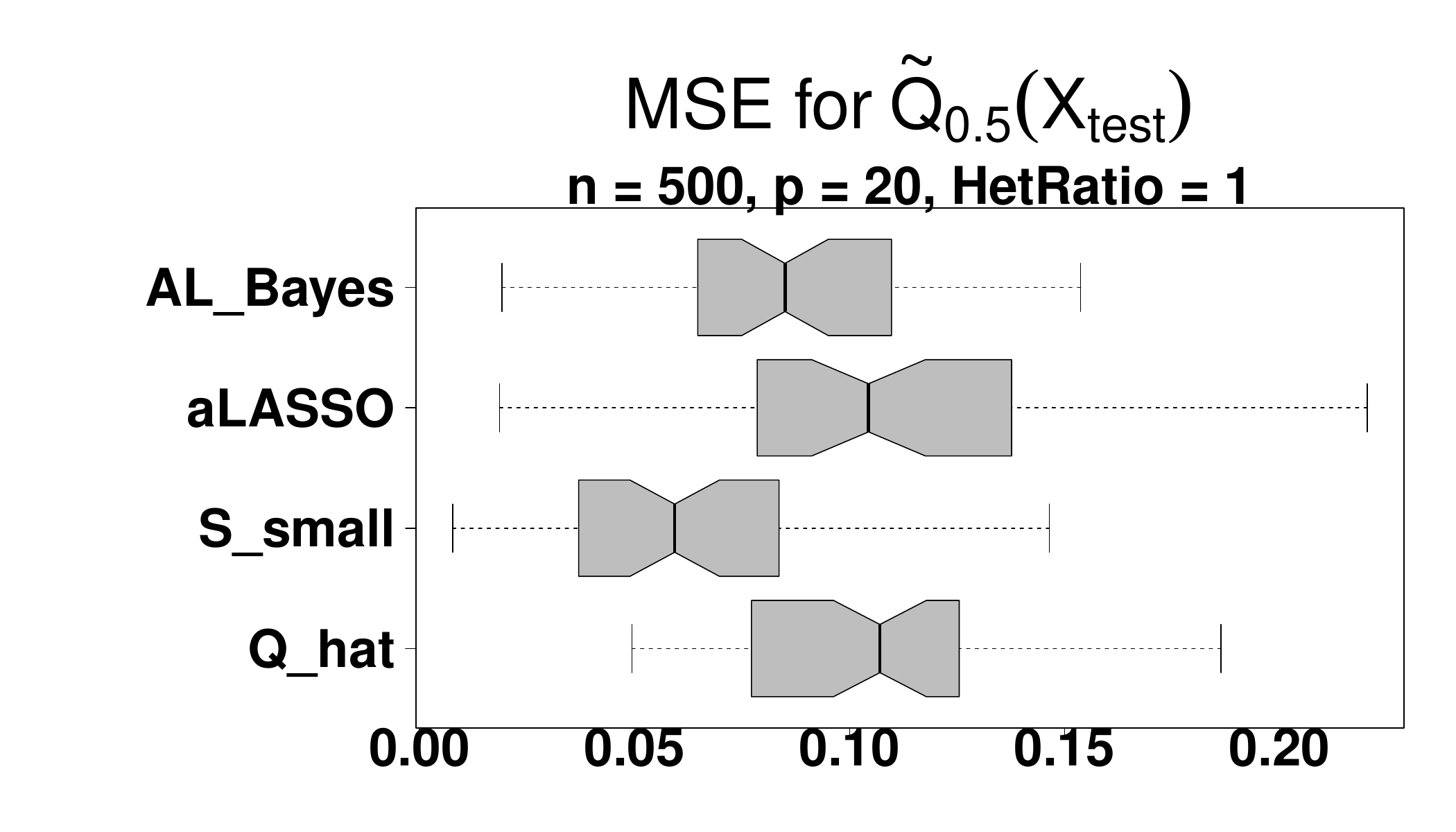}
    \includegraphics[width = .32\textwidth,keepaspectratio]{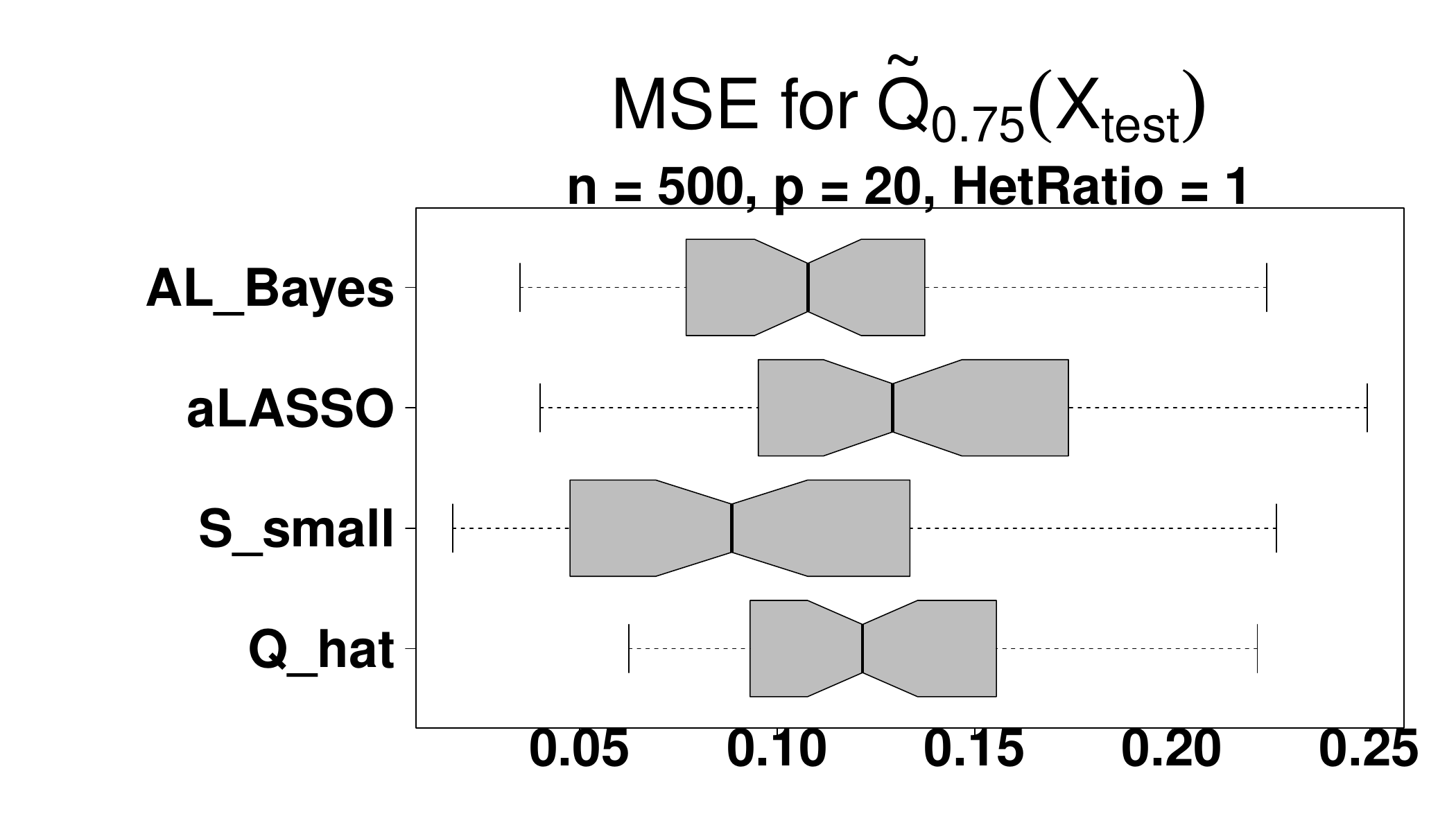}
   \includegraphics[width = .32\textwidth,keepaspectratio]{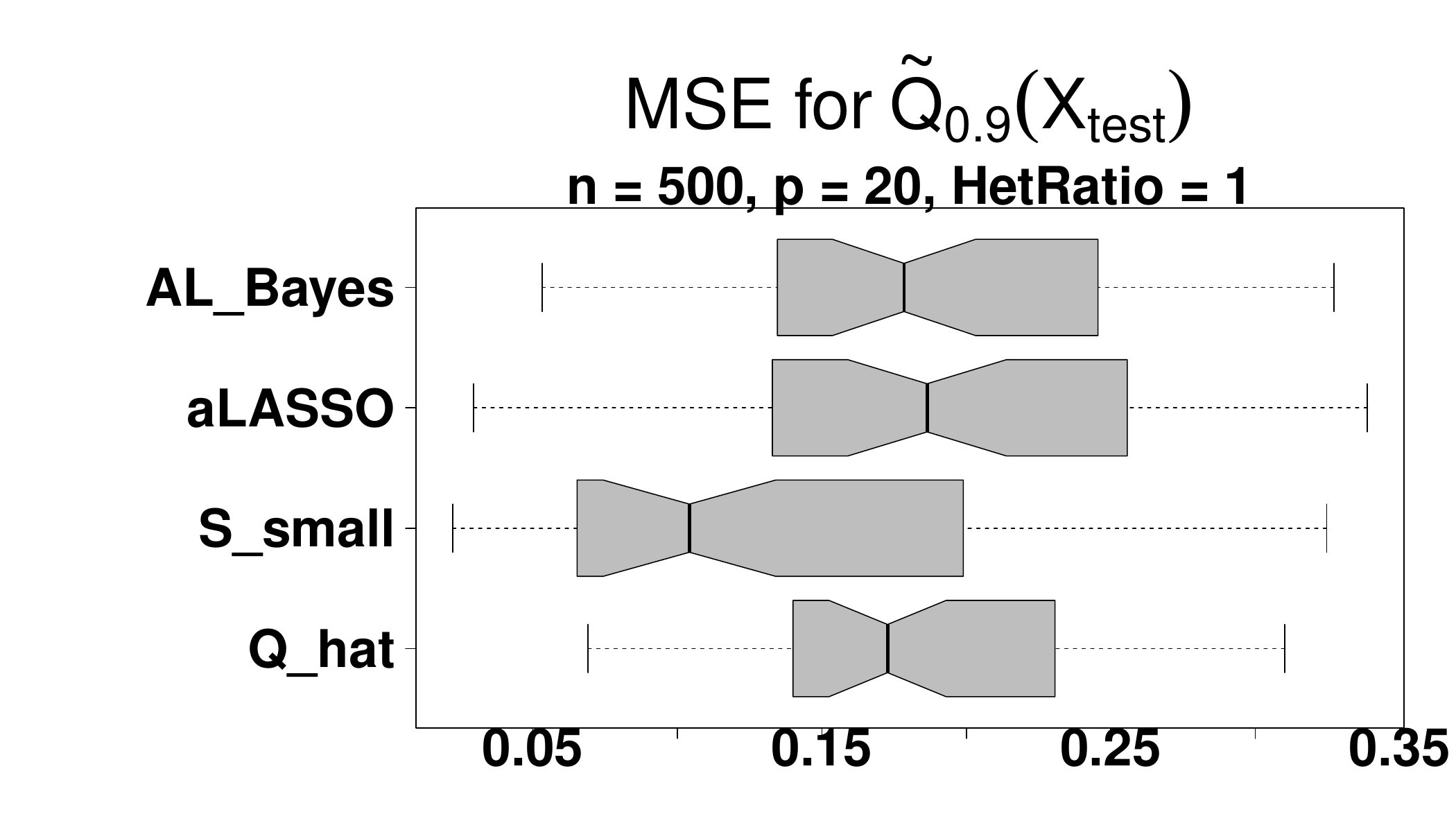}
    \includegraphics[width = .32\textwidth,keepaspectratio]{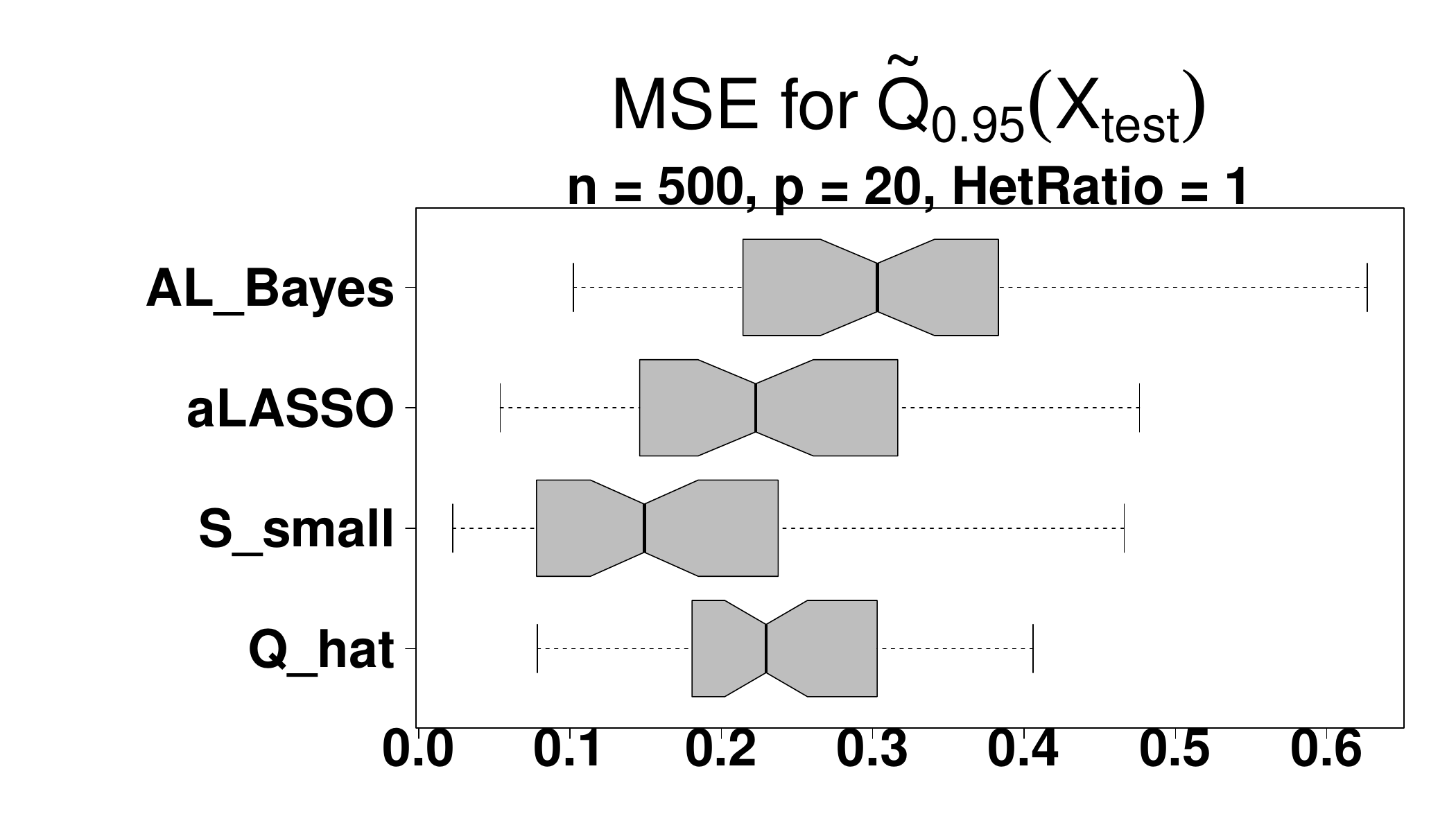}
   \includegraphics[width = .32\textwidth,keepaspectratio]{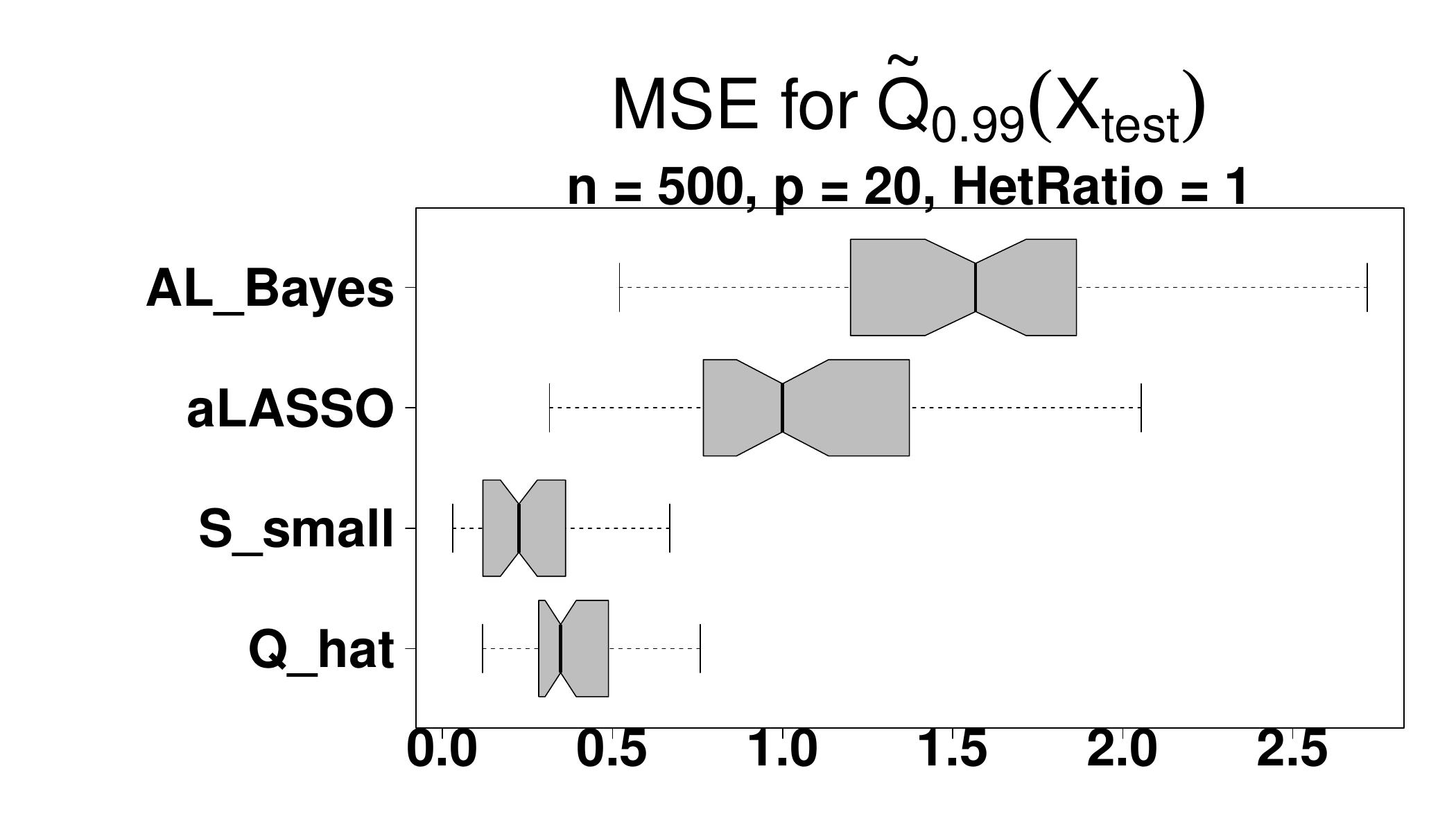}

    \label{fig:enter-label}
\end{figure}
\begin{figure}[H]
    \centering
        \caption{\textbf{MSE}: $\boldsymbol{n = 100, p= 100, \mbox{\textbf{HetRatio} }= 0.5}$} \label{Fig1_append}
  \includegraphics[width = .32\textwidth,keepaspectratio]{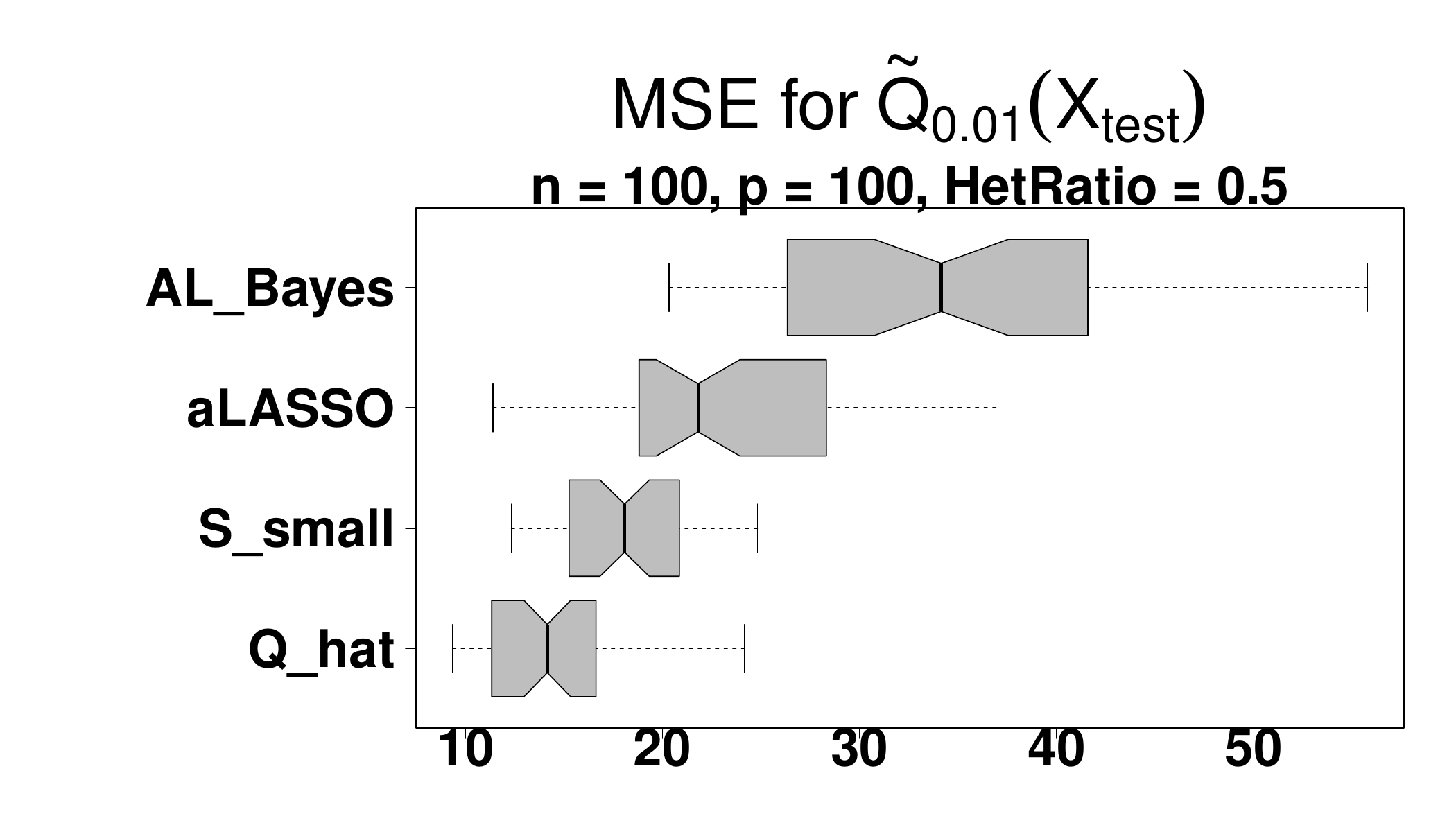}
    \includegraphics[width = .32\textwidth,keepaspectratio]{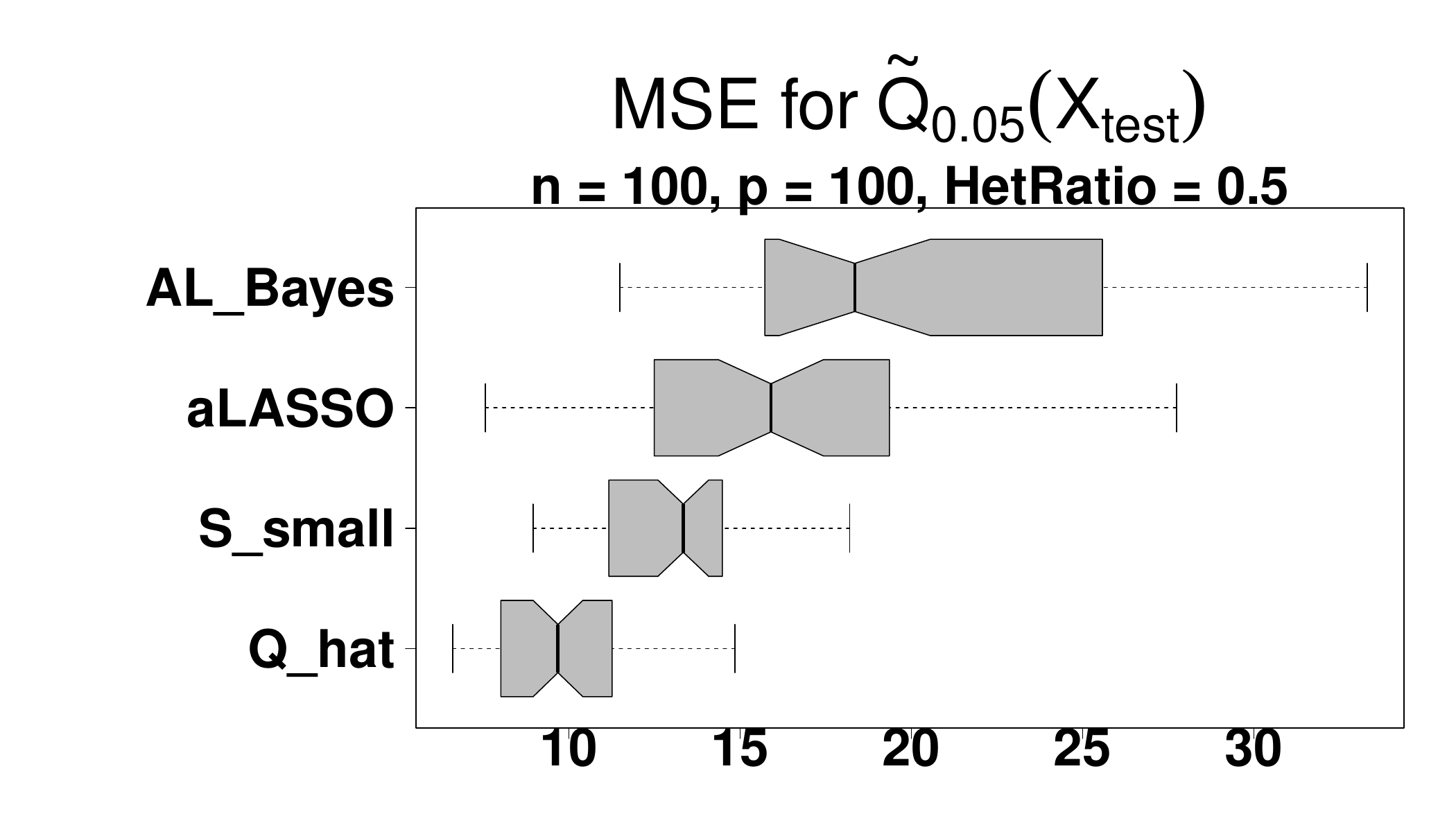}
    \includegraphics[width = .32\textwidth,keepaspectratio]{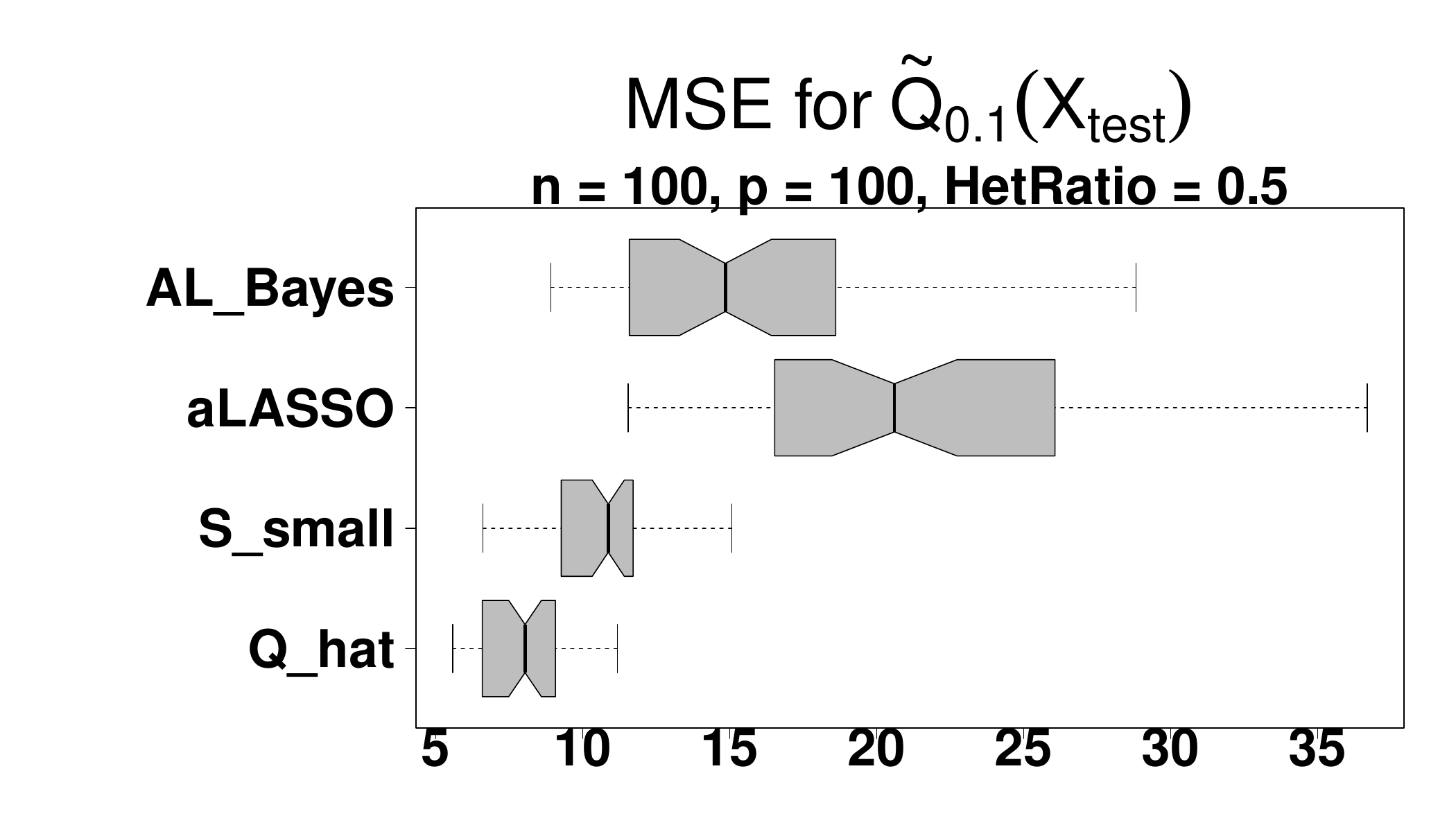}
    \includegraphics[width = .32\textwidth,keepaspectratio]{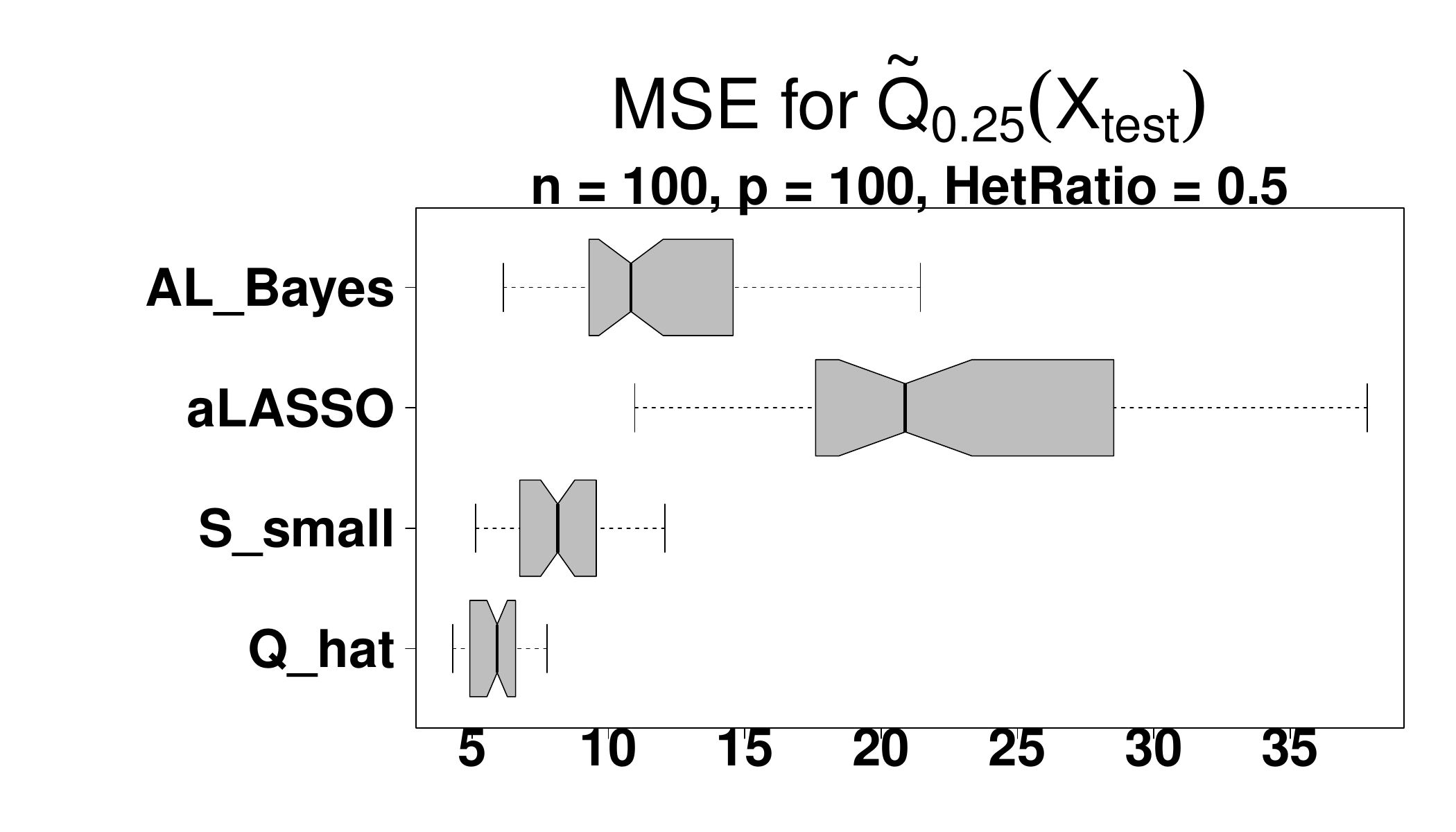}
    \includegraphics[width = .32\textwidth,keepaspectratio]{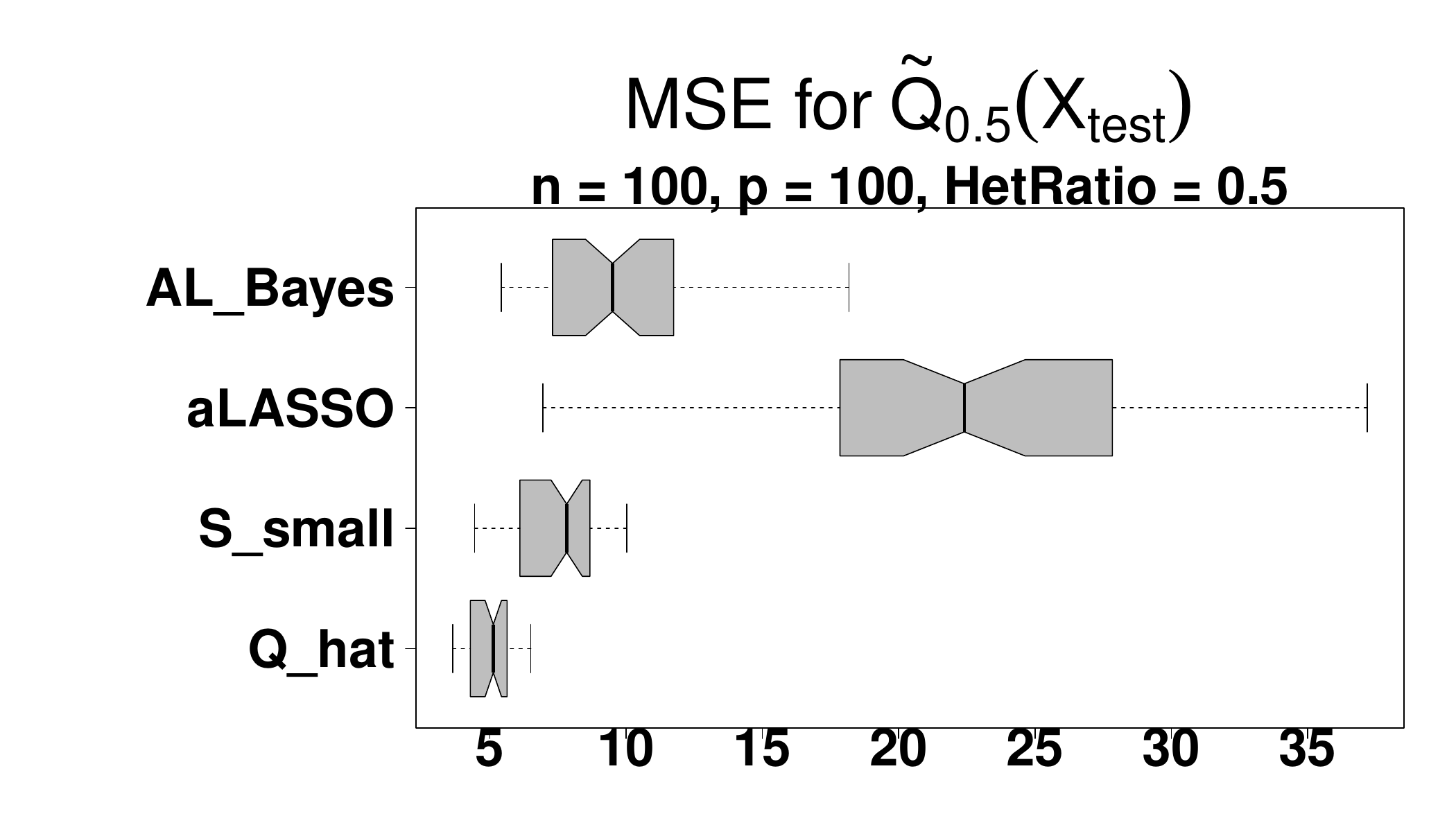}
    \includegraphics[width = .32\textwidth,keepaspectratio]{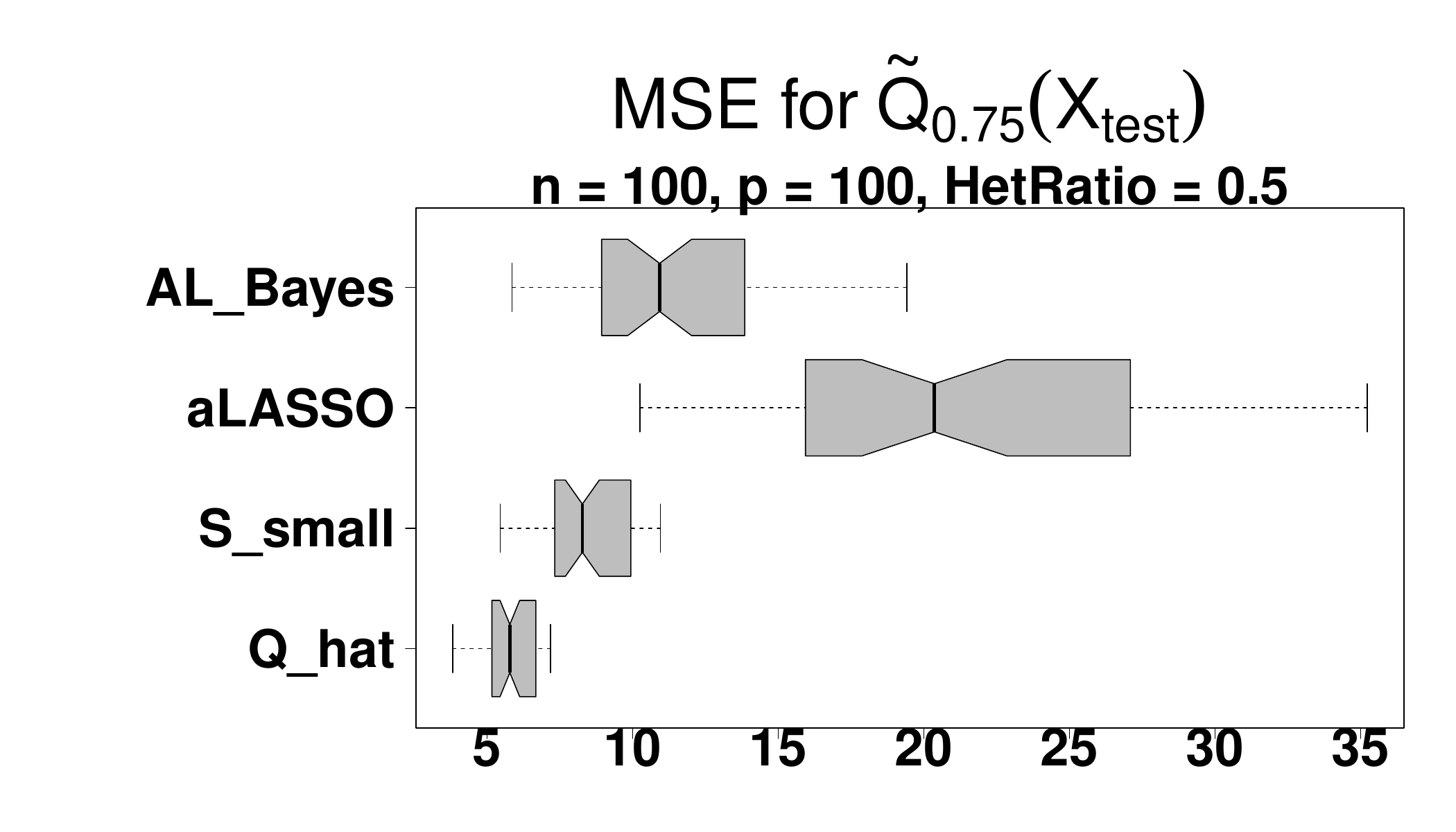}
   \includegraphics[width = .32\textwidth,keepaspectratio]{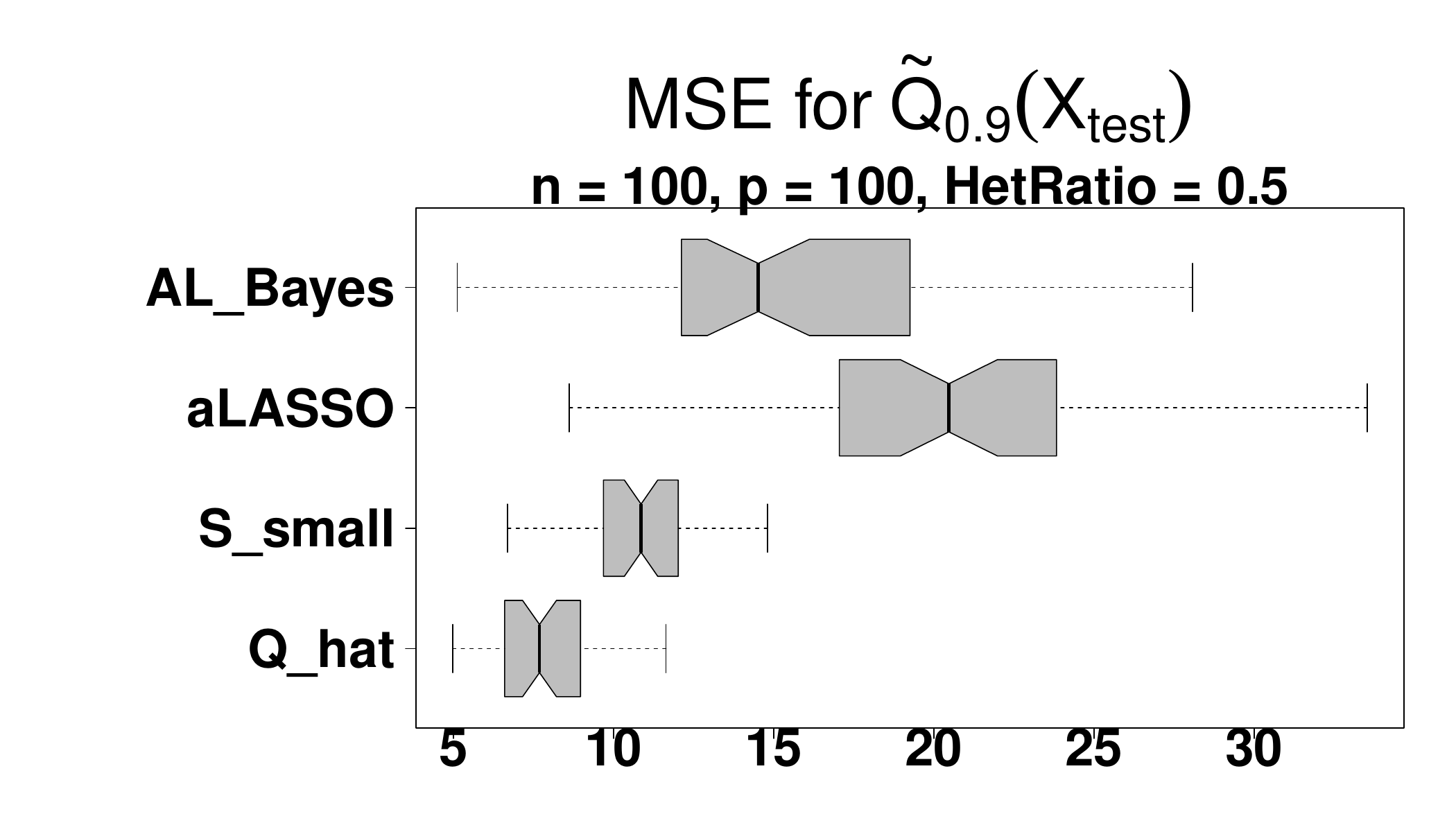}
    \includegraphics[width = .32\textwidth,keepaspectratio]{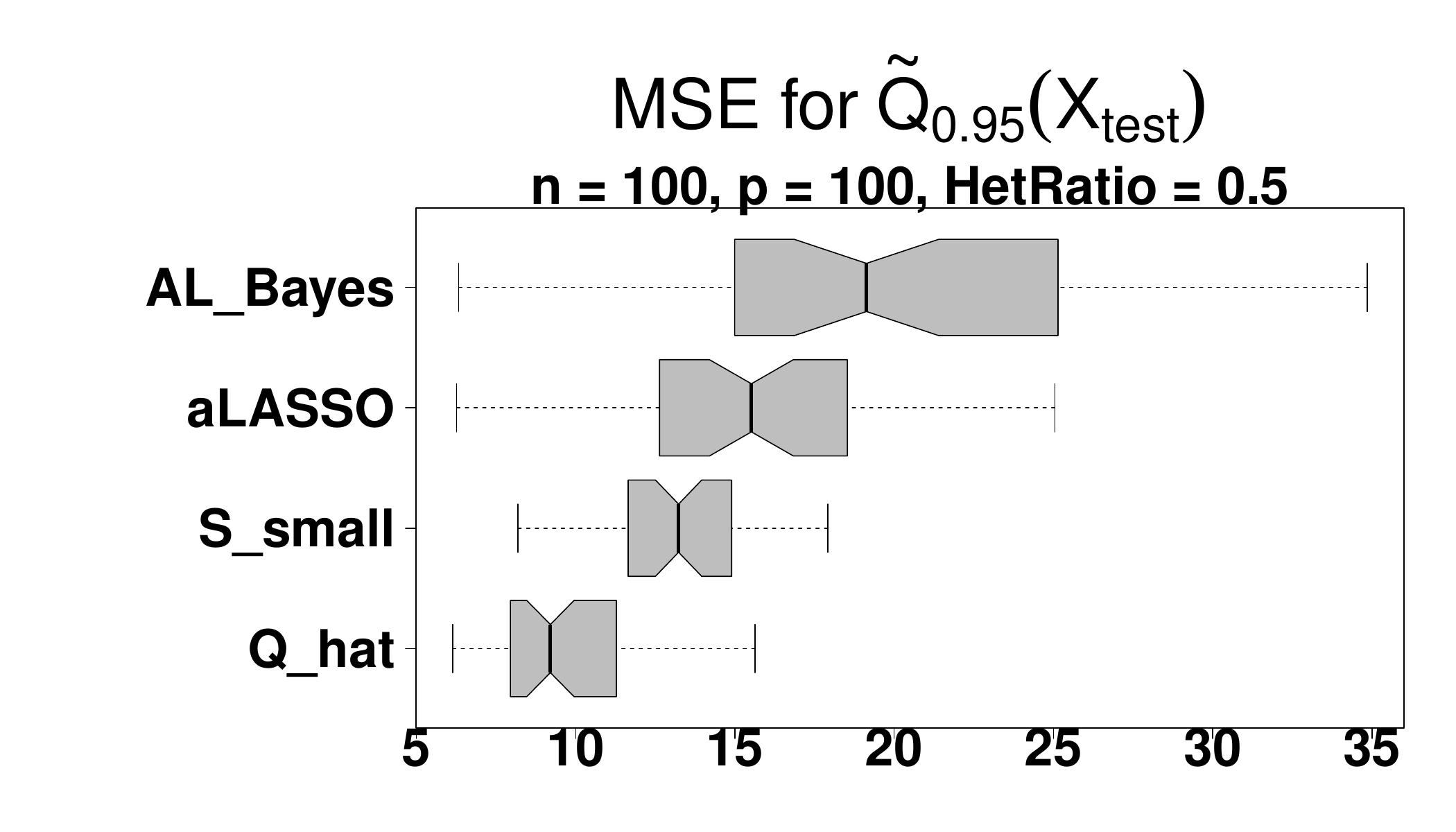}
   \includegraphics[width = .32\textwidth,keepaspectratio]{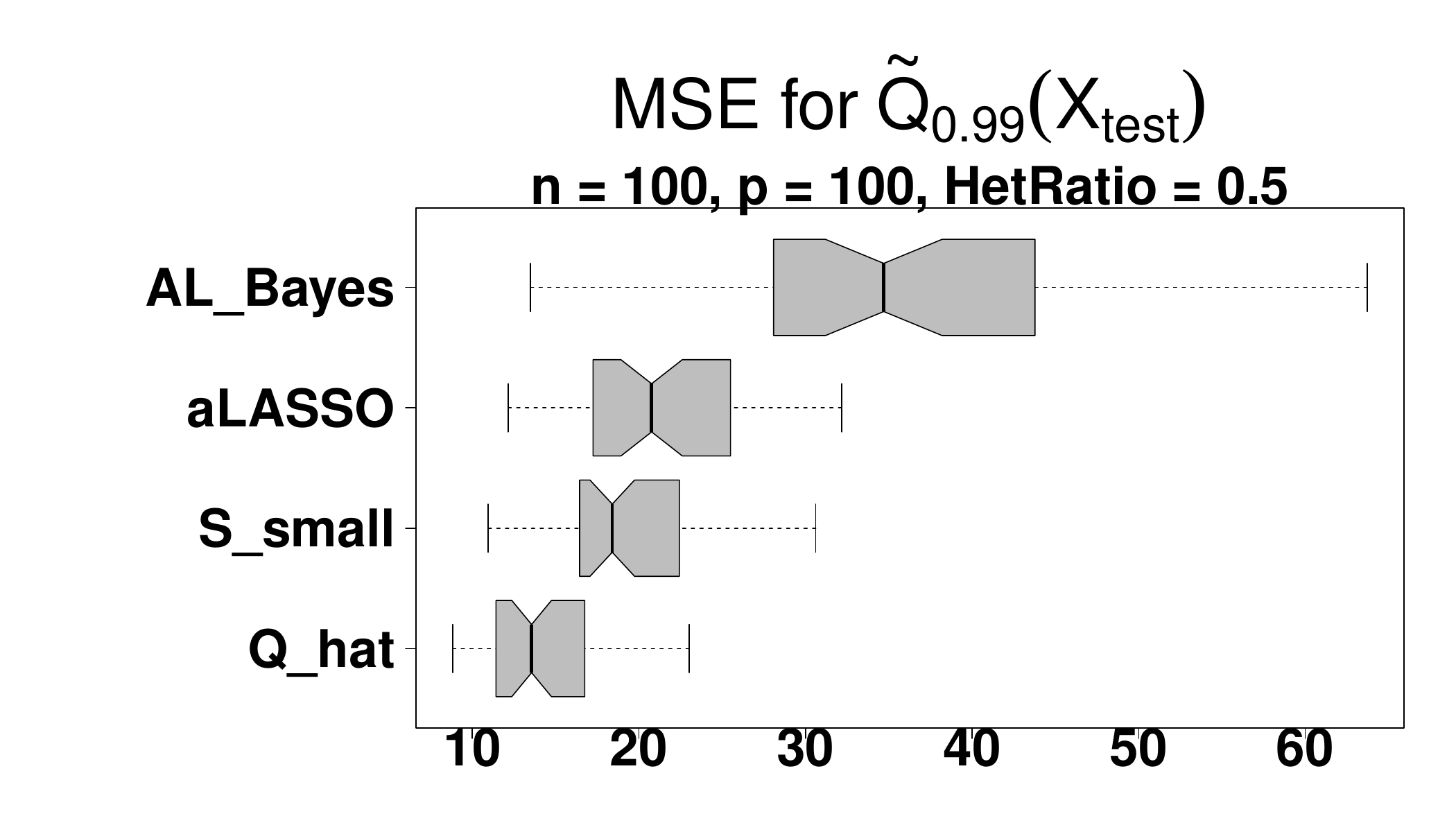}

\end{figure}

\begin{figure}[H]
    \centering
        \caption{\textbf{MSE}: $\boldsymbol{n = 100, p= 100, \mbox{\textbf{HetRatio} }= 1}$}
    \includegraphics[width = .32\textwidth,keepaspectratio]{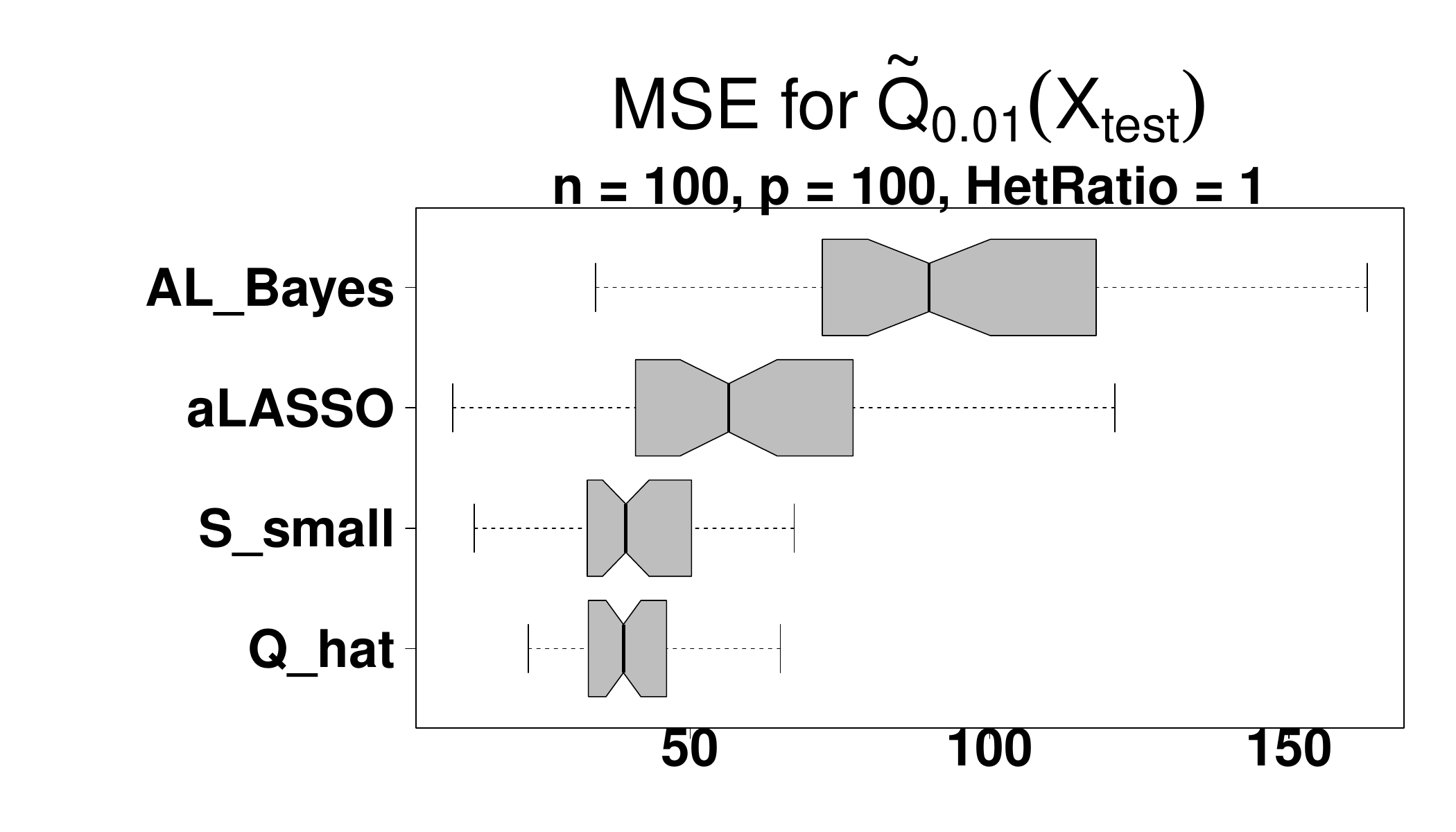}
    \includegraphics[width = .32\textwidth,keepaspectratio]{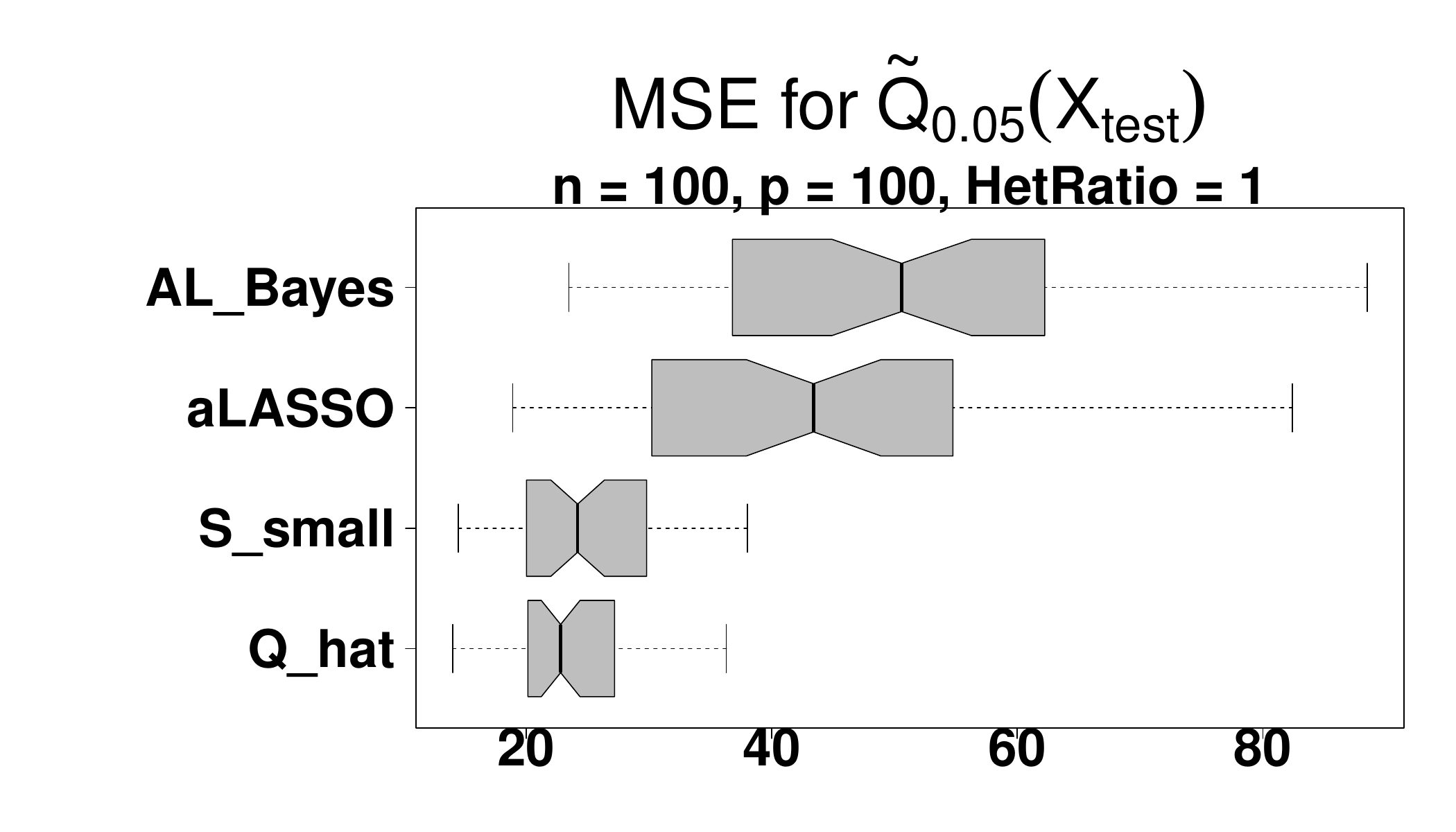}
    \includegraphics[width = .32\textwidth,keepaspectratio]{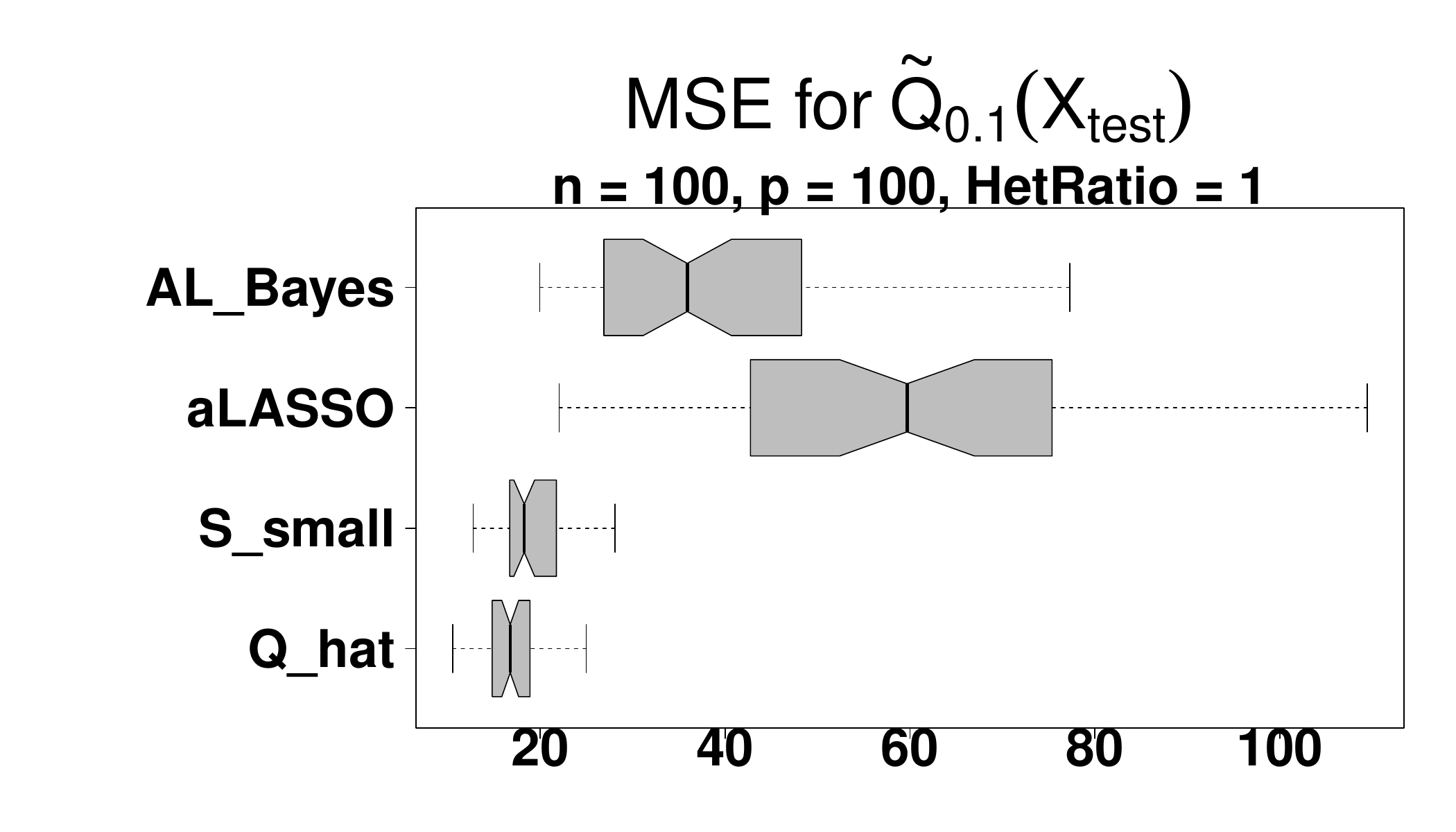}
    \includegraphics[width = .32\textwidth,keepaspectratio]{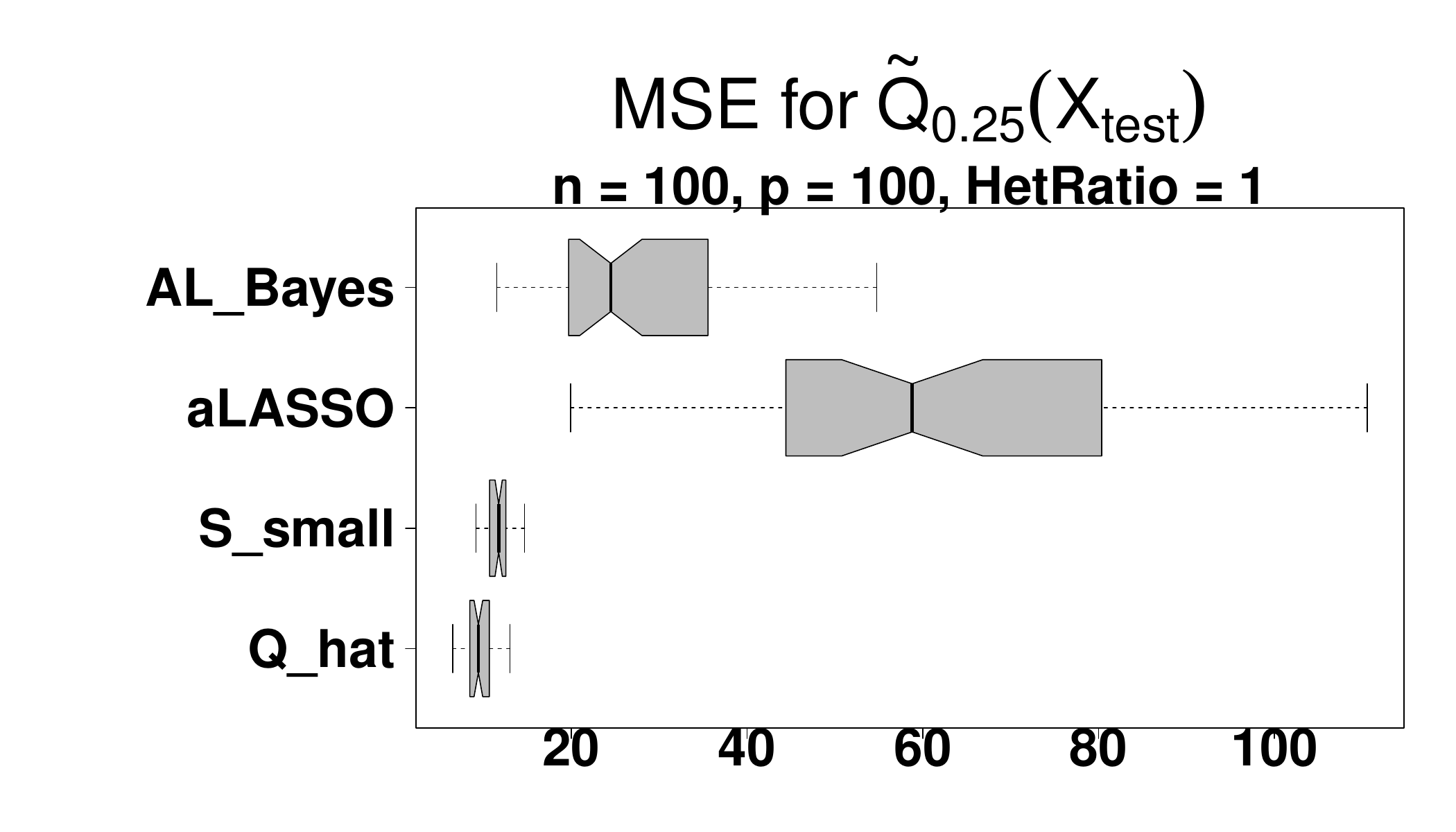}
    \includegraphics[width = .32\textwidth,keepaspectratio]{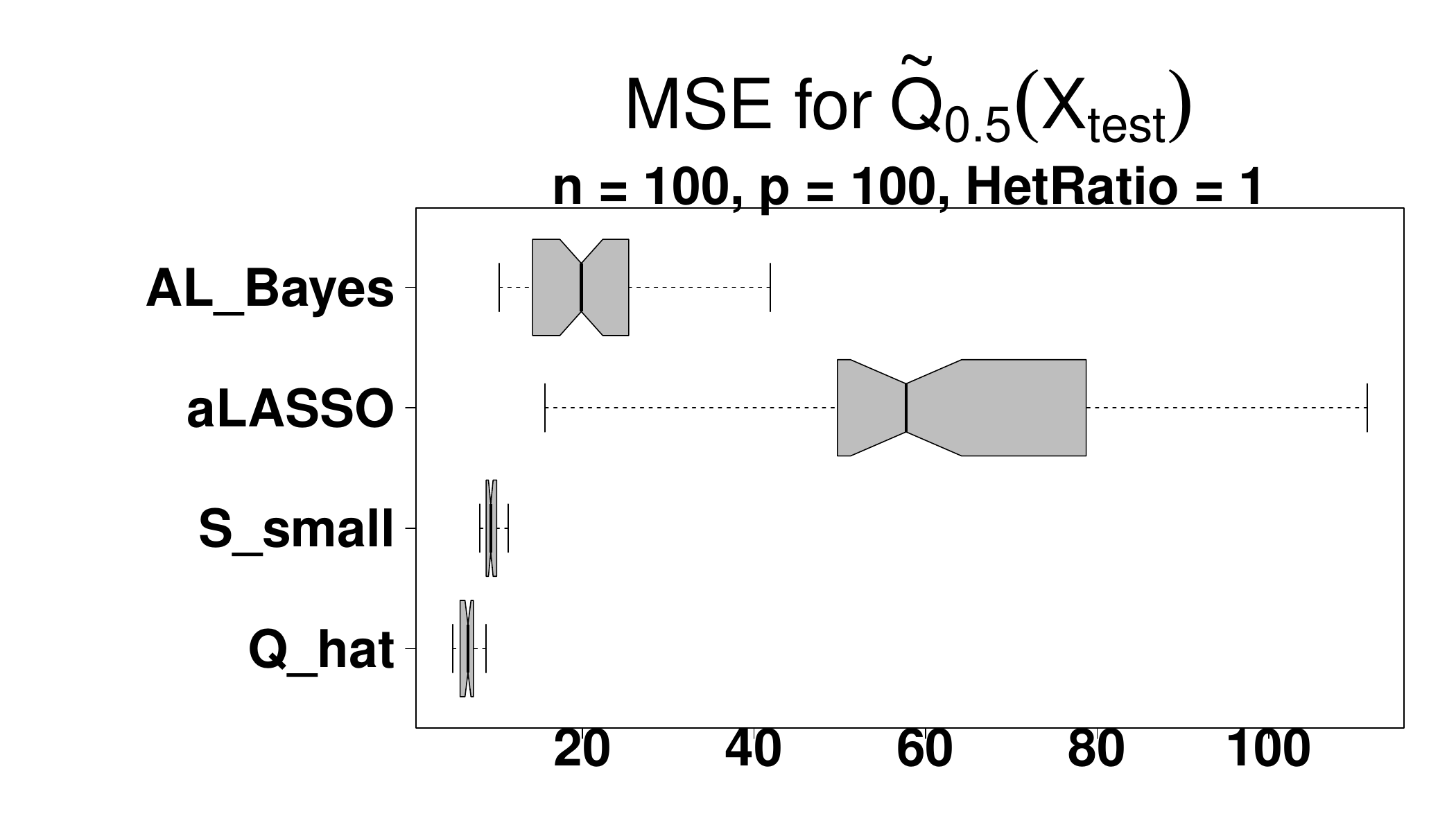}
    \includegraphics[width = .32\textwidth,keepaspectratio]{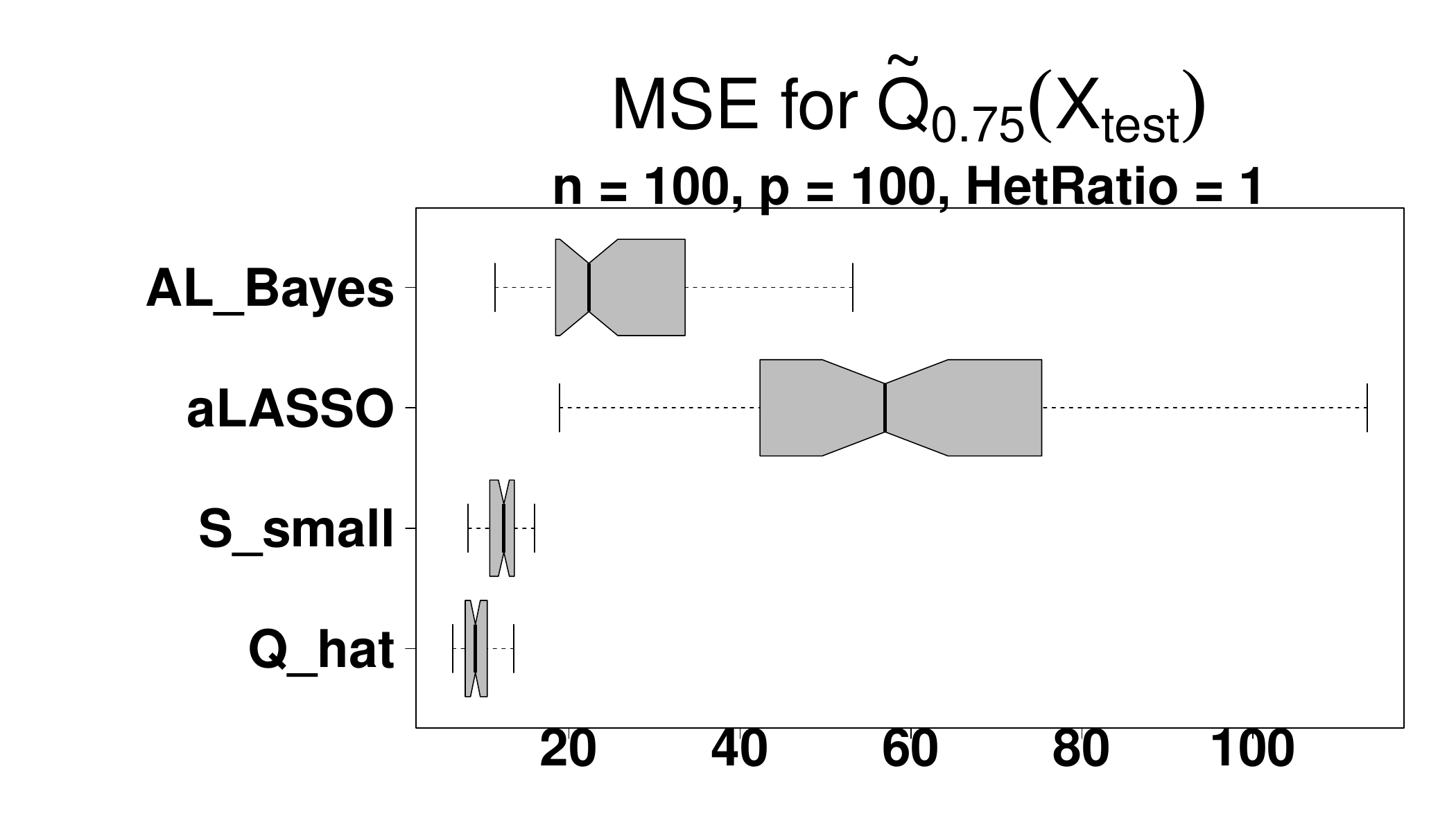}
   \includegraphics[width = .32\textwidth,keepaspectratio]{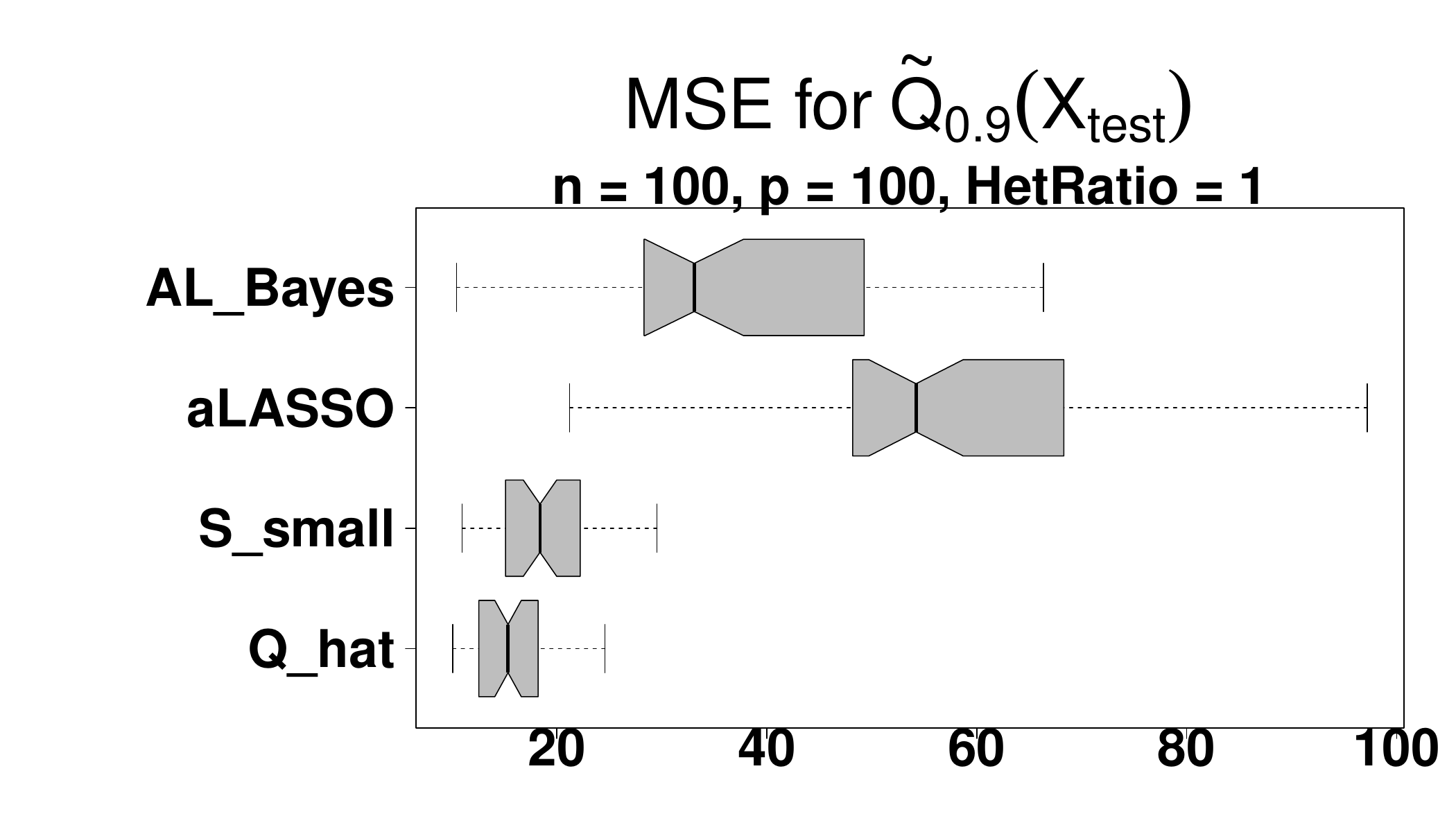}
    \includegraphics[width = .32\textwidth,keepaspectratio]{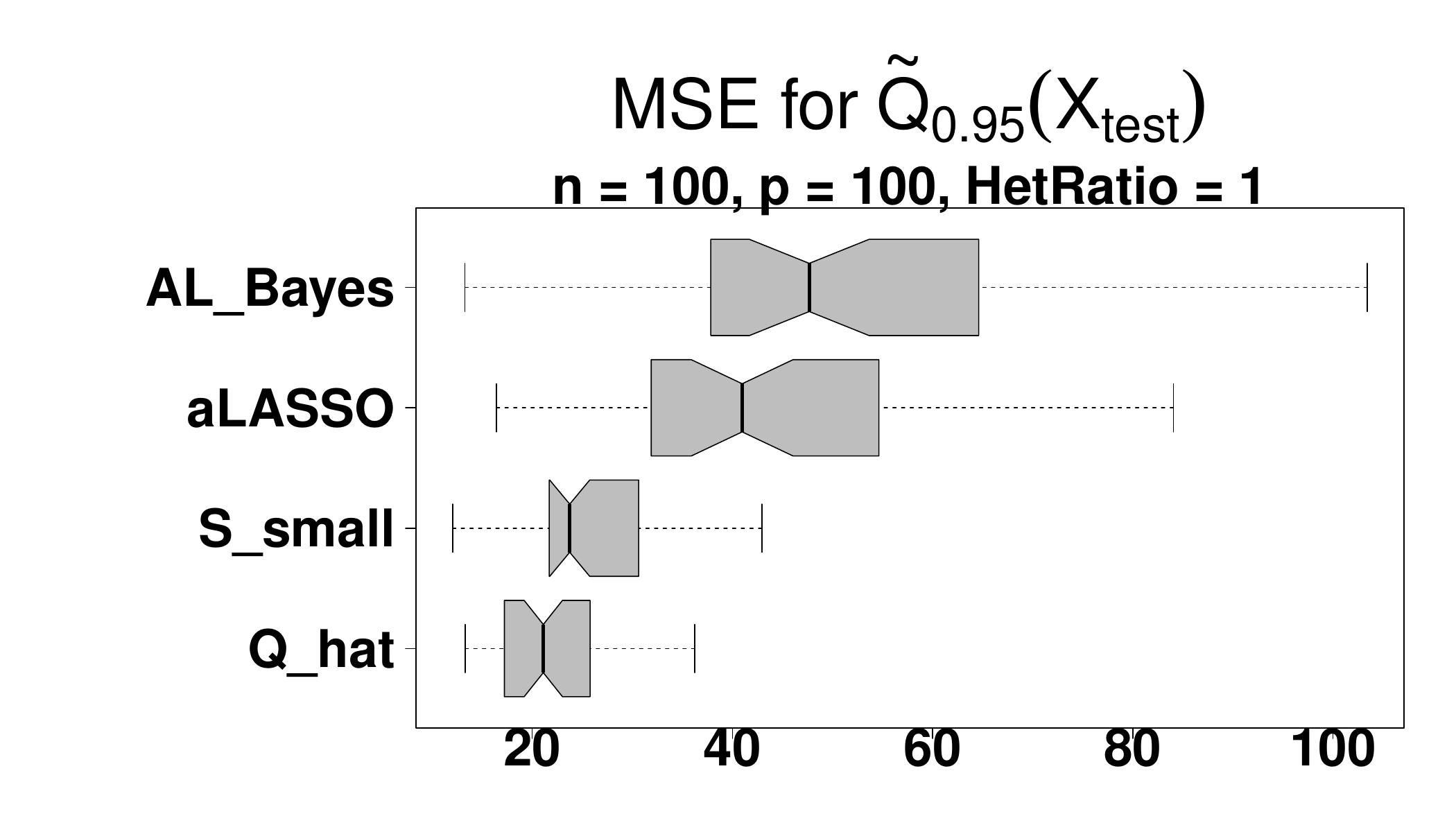}
   \includegraphics[width = .32\textwidth,keepaspectratio]{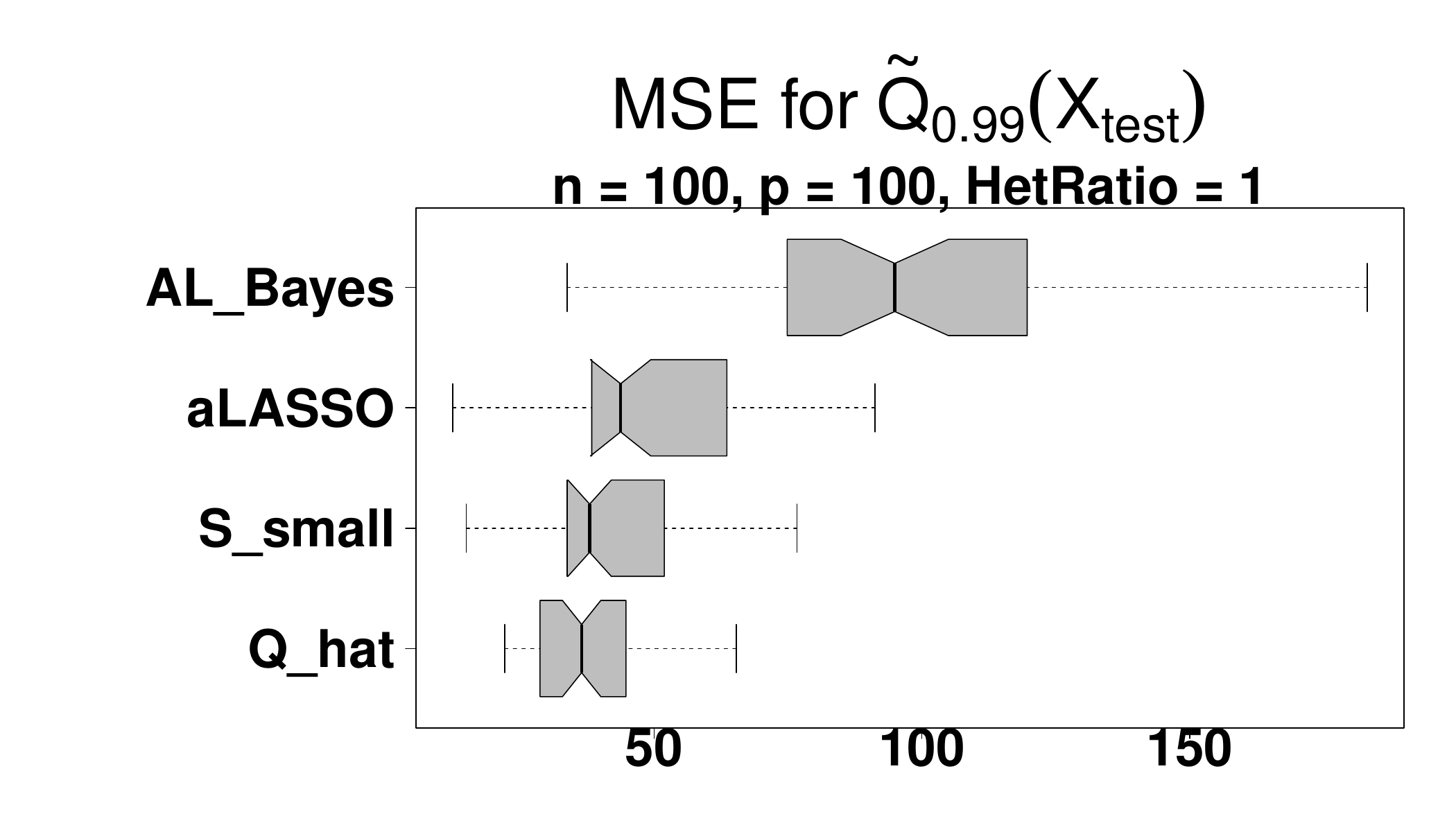}

    \label{fig:enter-label}
\end{figure}

\begin{figure}[H]
    \centering
        \caption{\textbf{MSE}: $\boldsymbol{n = 200, p= 50, \mbox{\textbf{HetRatio} }= 0.5}$}
  
    \includegraphics[width = .32\textwidth,keepaspectratio]{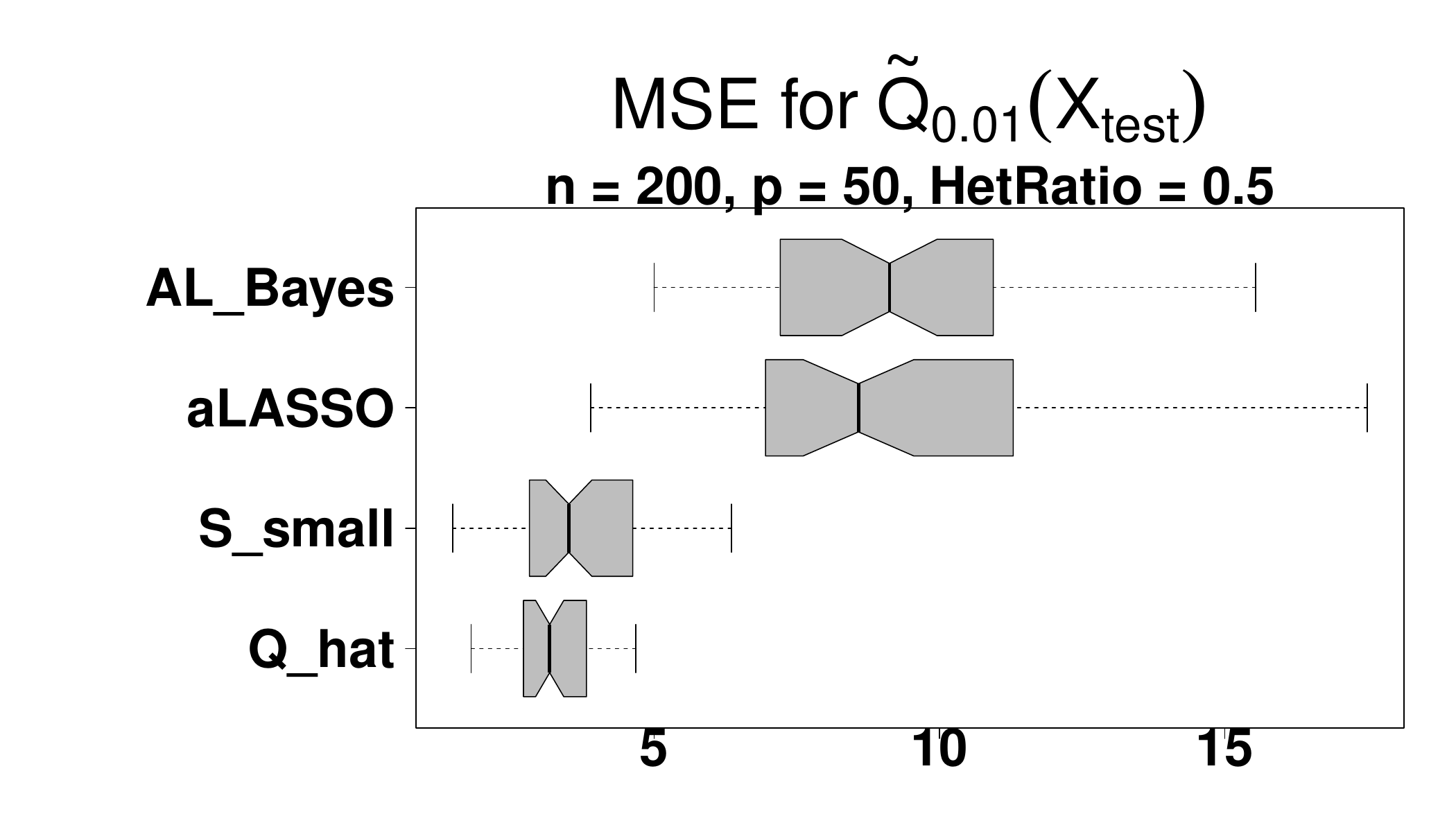}
    \includegraphics[width = .32\textwidth,keepaspectratio]{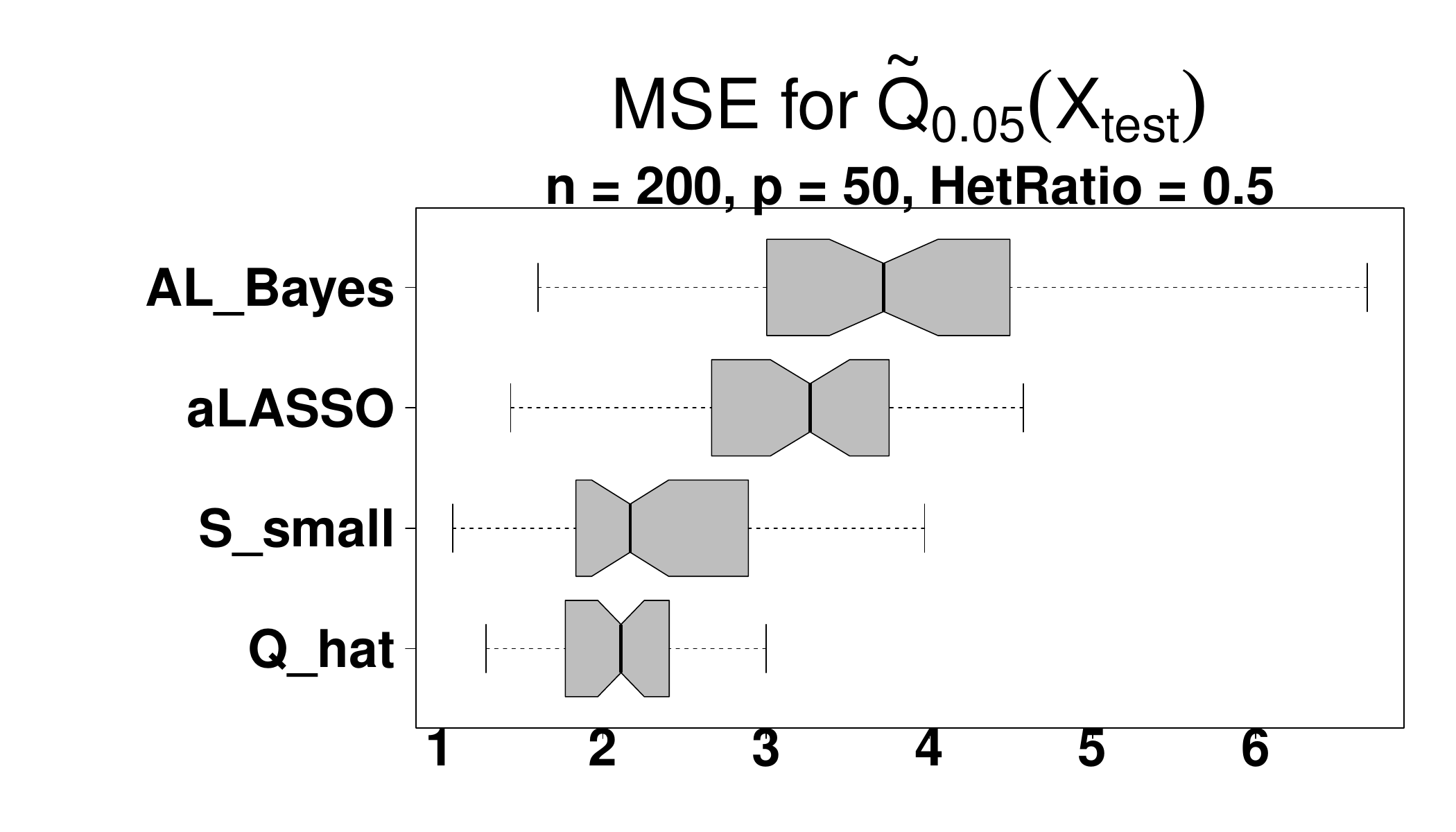}
    \includegraphics[width = .32\textwidth,keepaspectratio]{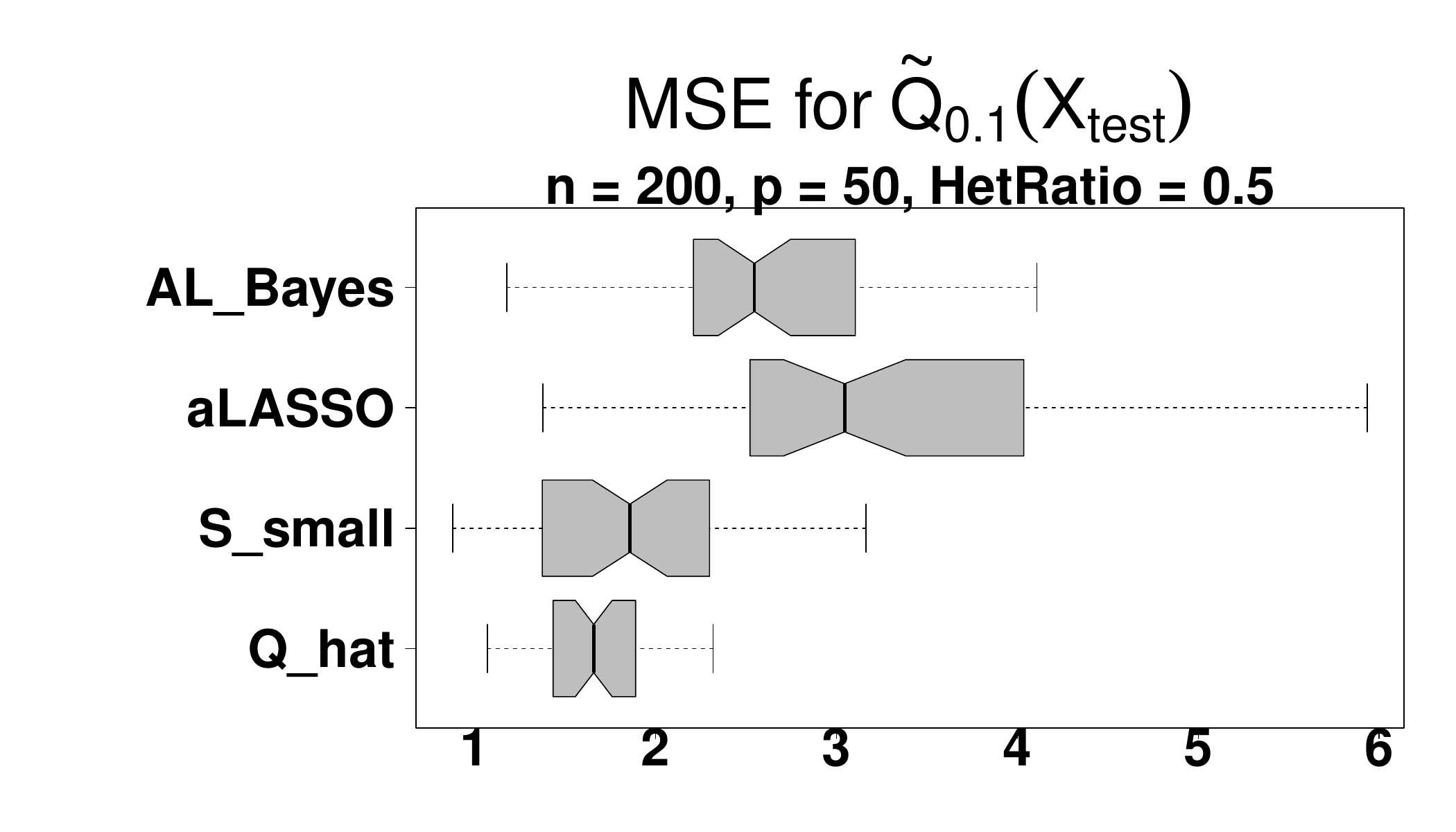}
    \includegraphics[width = .32\textwidth,keepaspectratio]{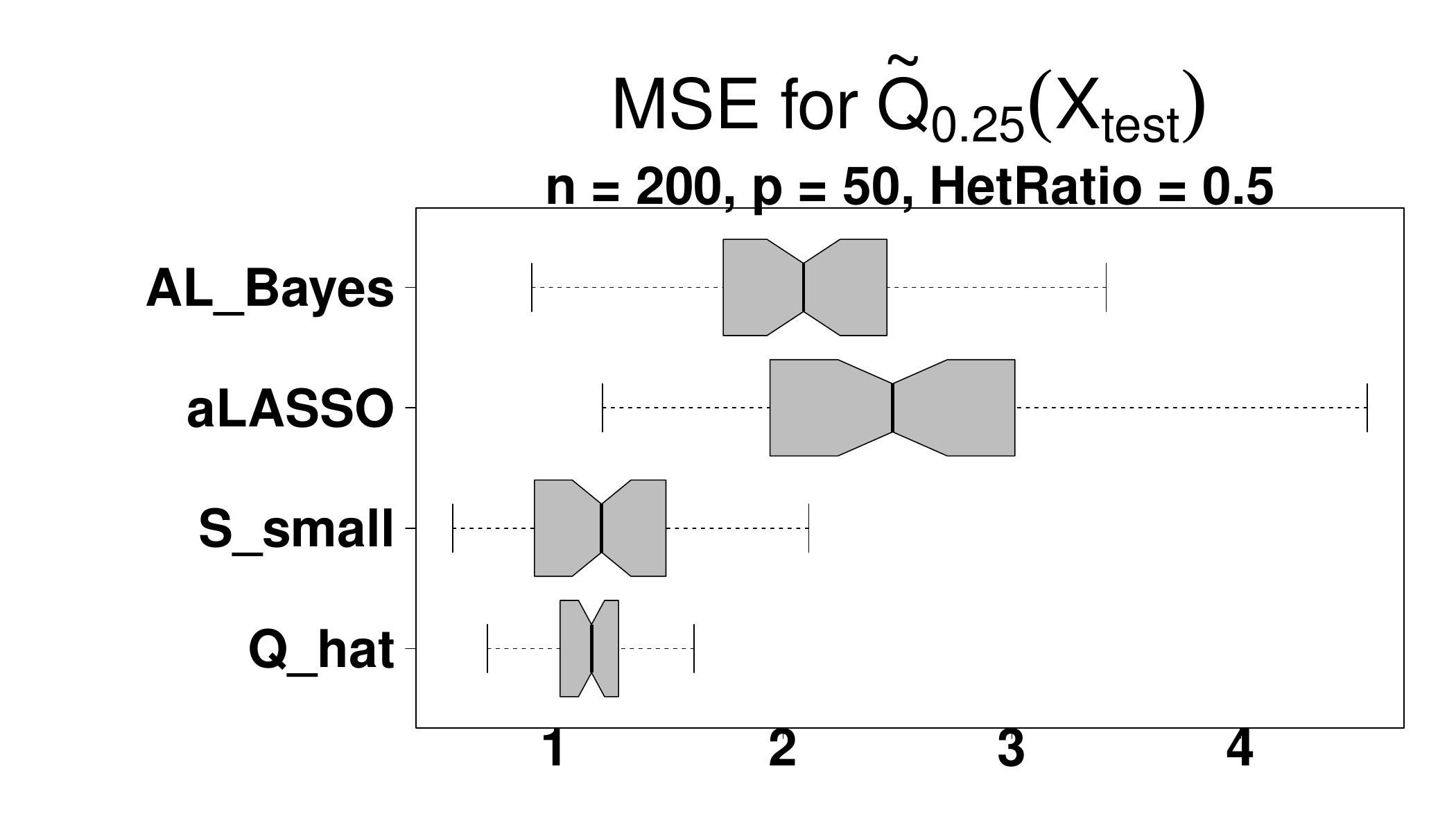}
    \includegraphics[width = .32\textwidth,keepaspectratio]{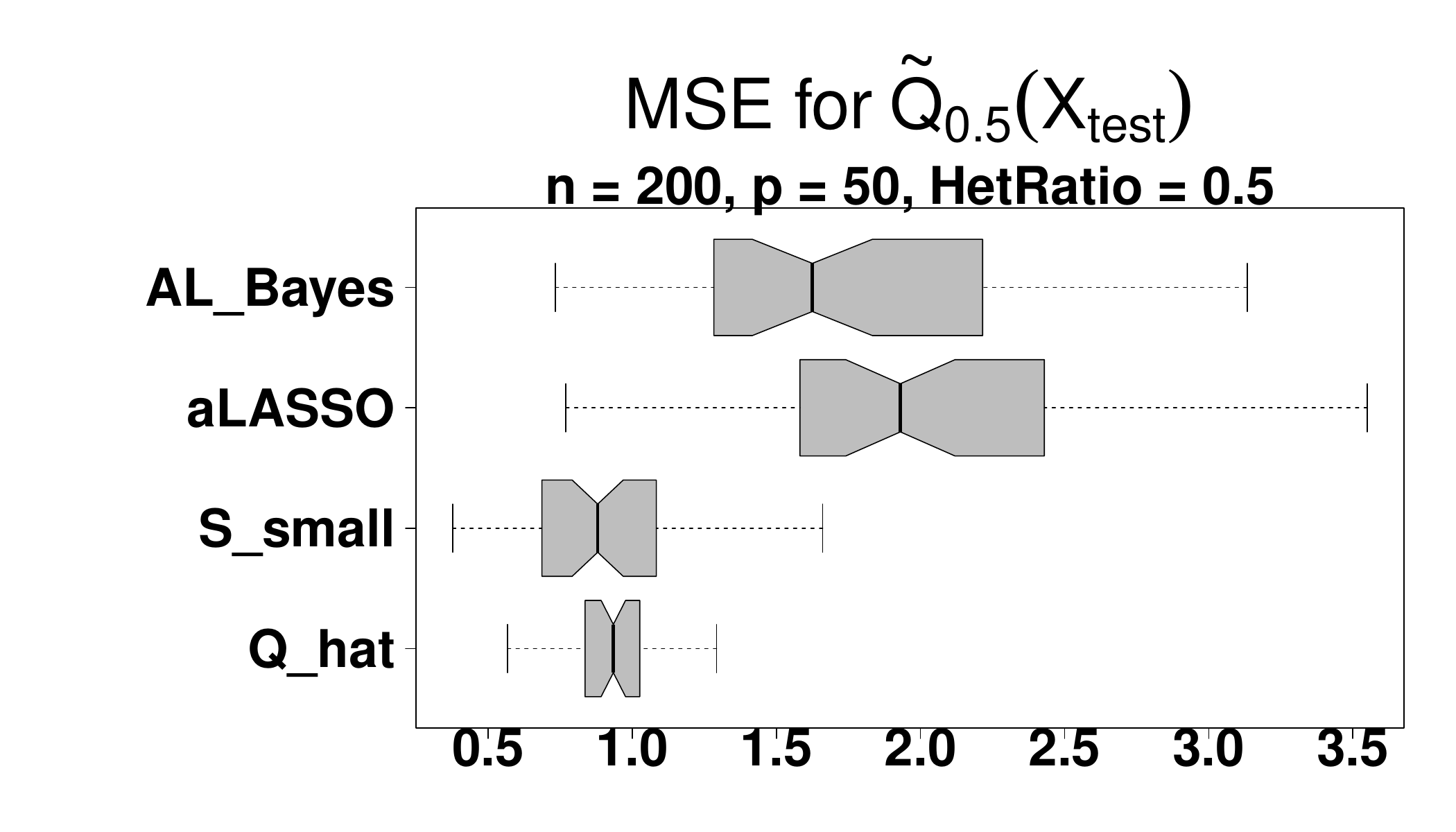}
    \includegraphics[width = .32\textwidth,keepaspectratio]{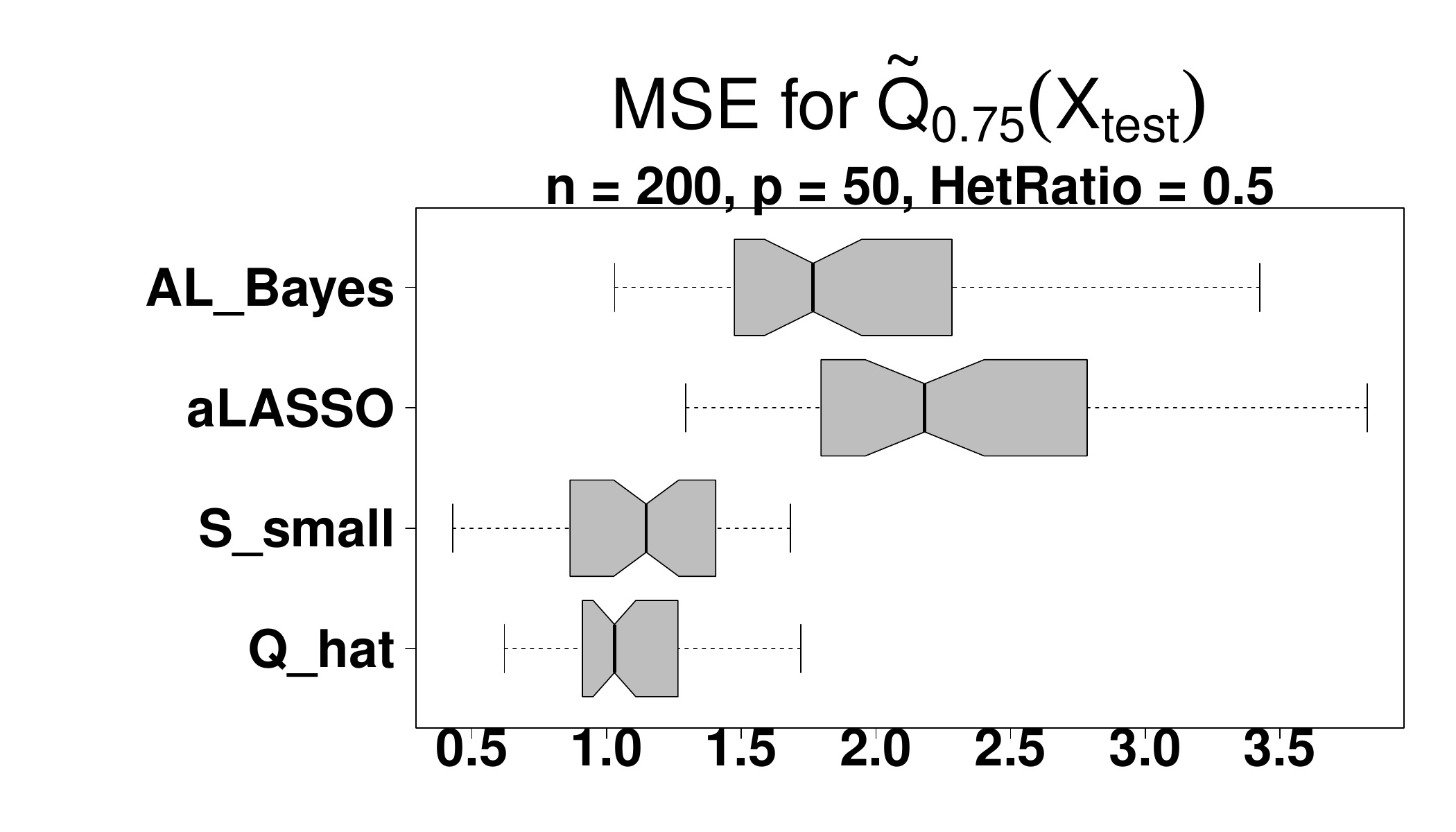}
   \includegraphics[width = .32\textwidth,keepaspectratio]{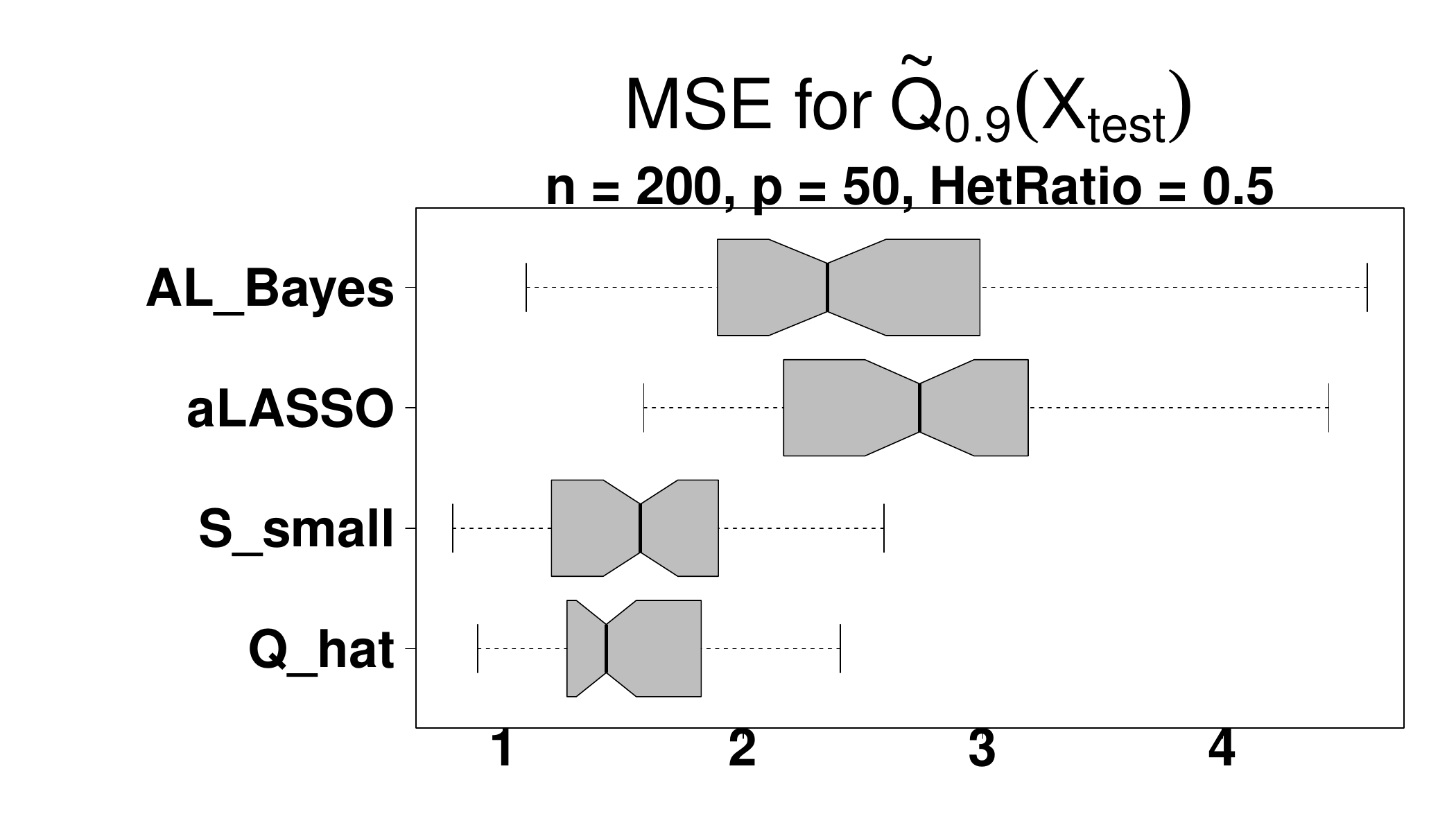}
    \includegraphics[width = .32\textwidth,keepaspectratio]{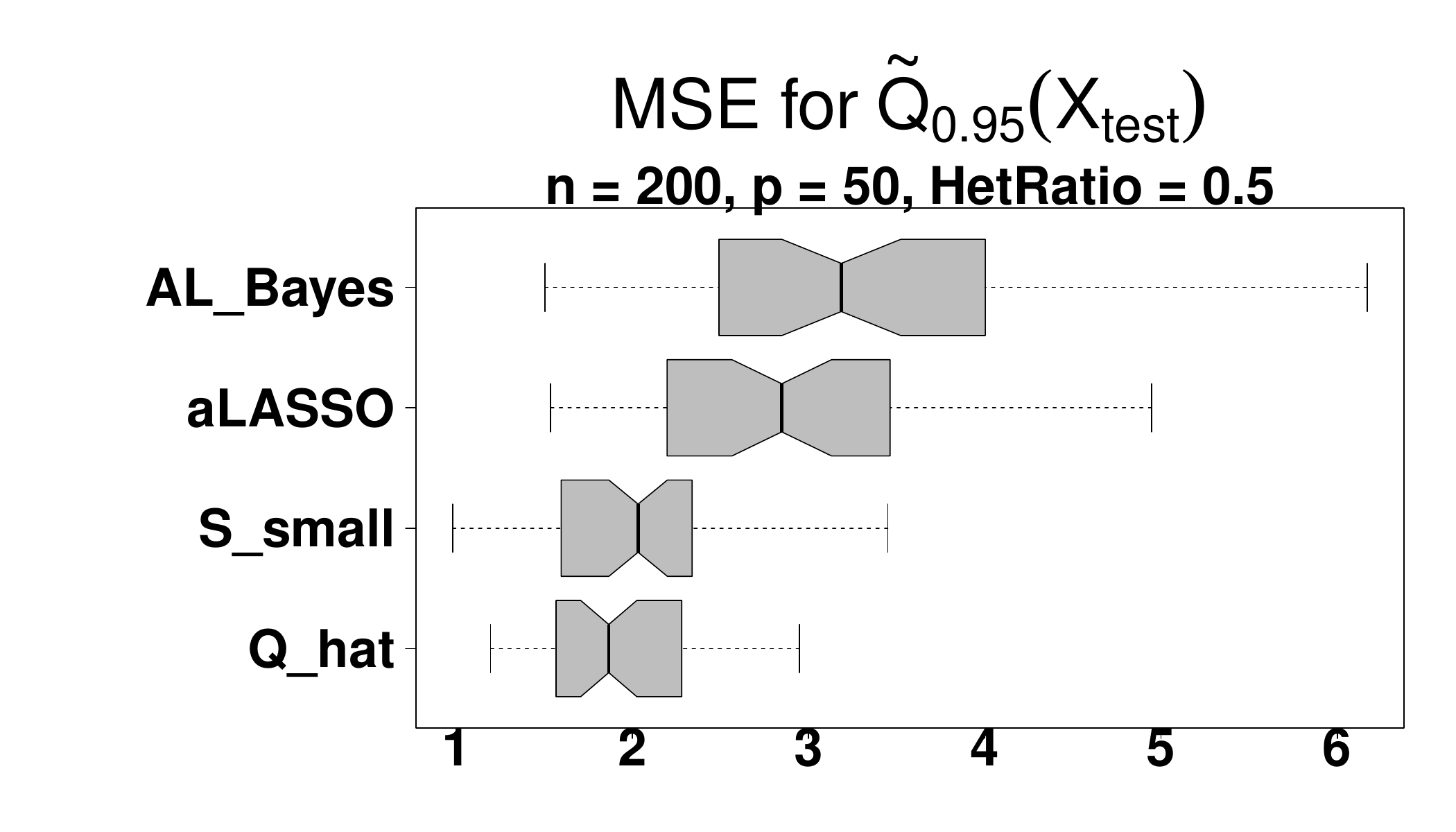}
   \includegraphics[width = .32\textwidth,keepaspectratio]{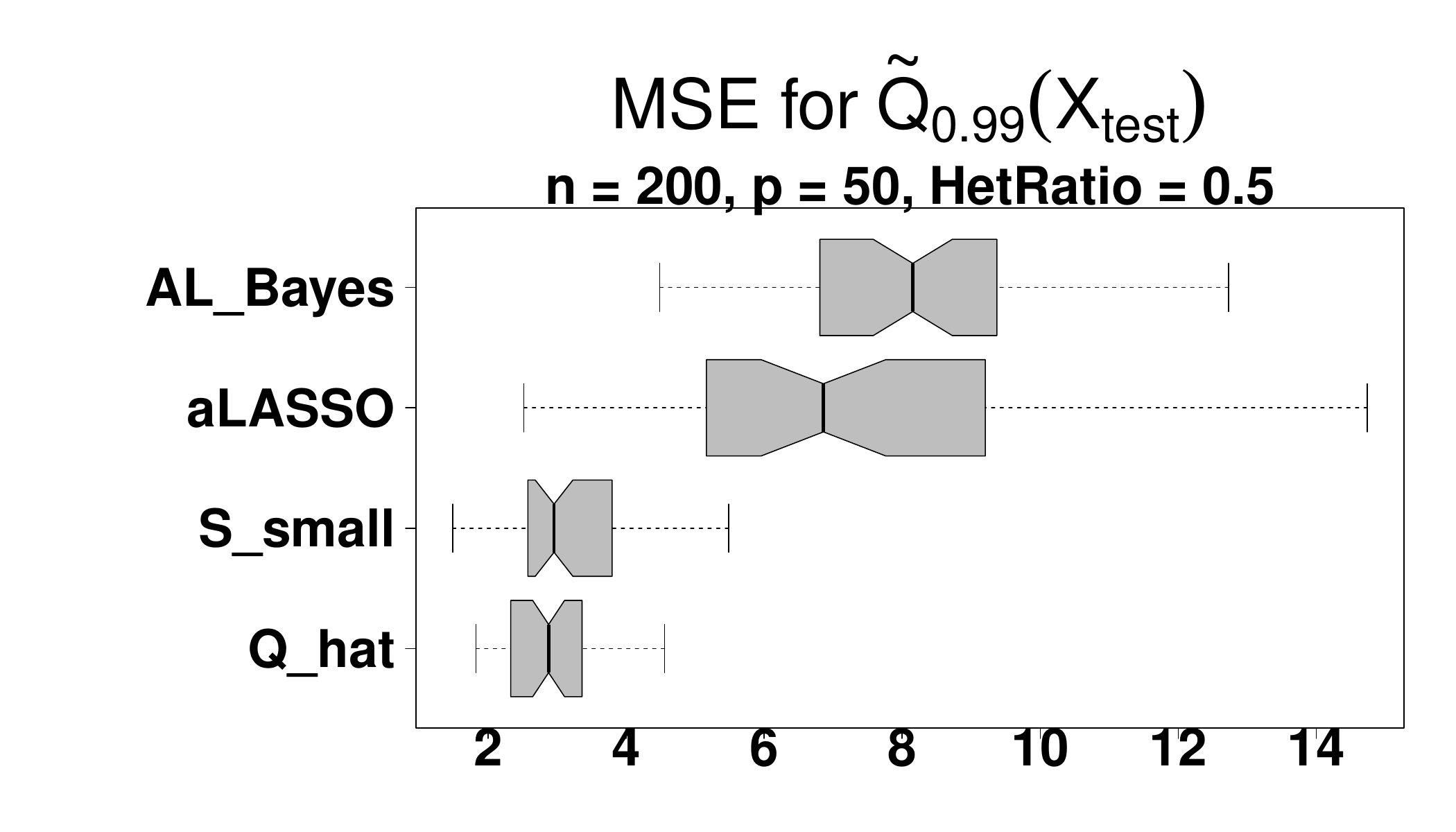}

    \label{fig:enter-label}
\end{figure}

\begin{figure}[H]
    \centering
        \caption{\textbf{MSE}: $\boldsymbol{n = 200, p= 50, \mbox{\textbf{HetRatio} }= 1}$}
  
    \includegraphics[width = .32\textwidth,keepaspectratio]{images/n200_p50_MSE_SNR1_1.pdf}
    \includegraphics[width = .32\textwidth,keepaspectratio]{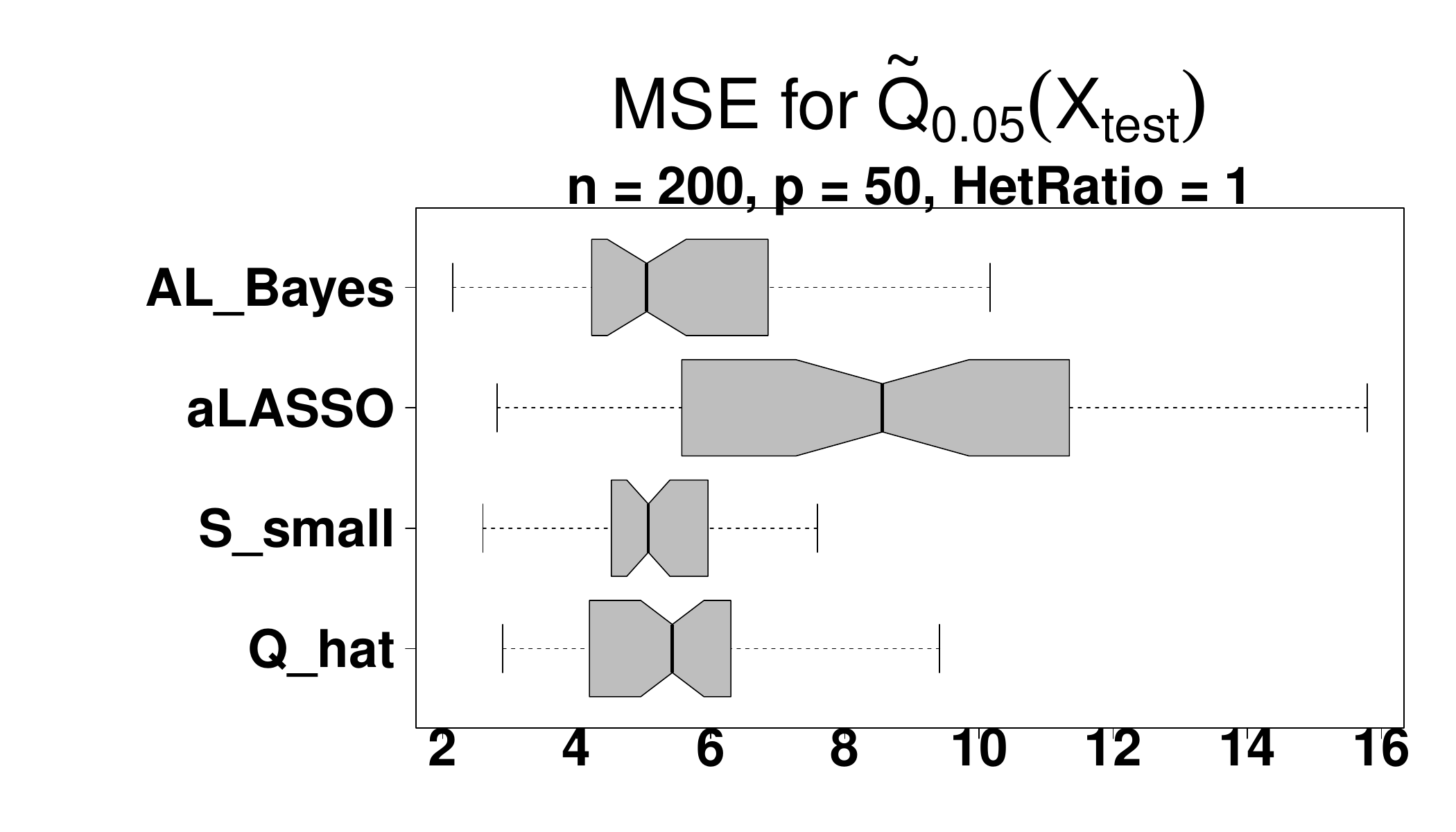}
    \includegraphics[width = .32\textwidth,keepaspectratio]{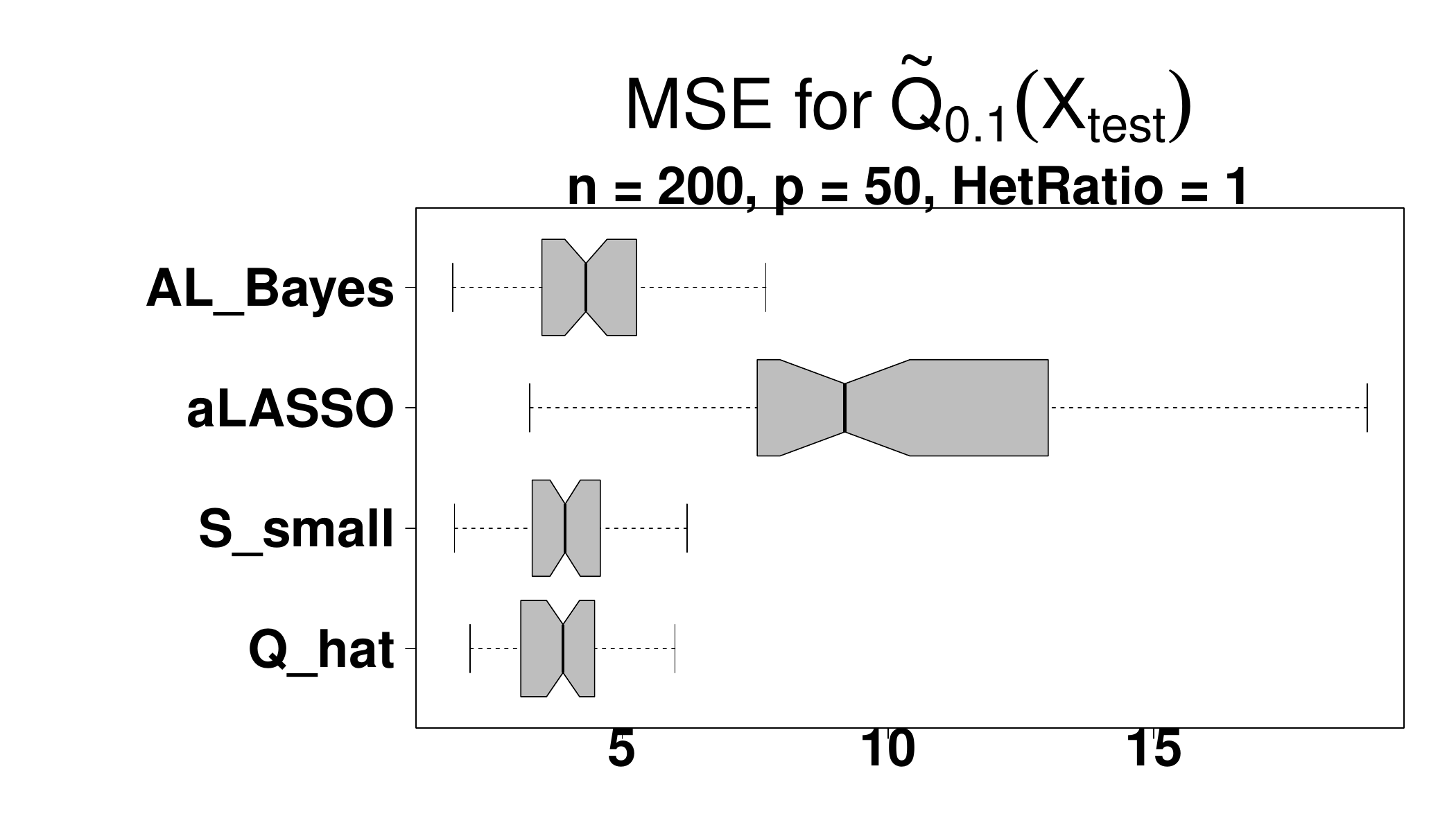}
    \includegraphics[width = .32\textwidth,keepaspectratio]{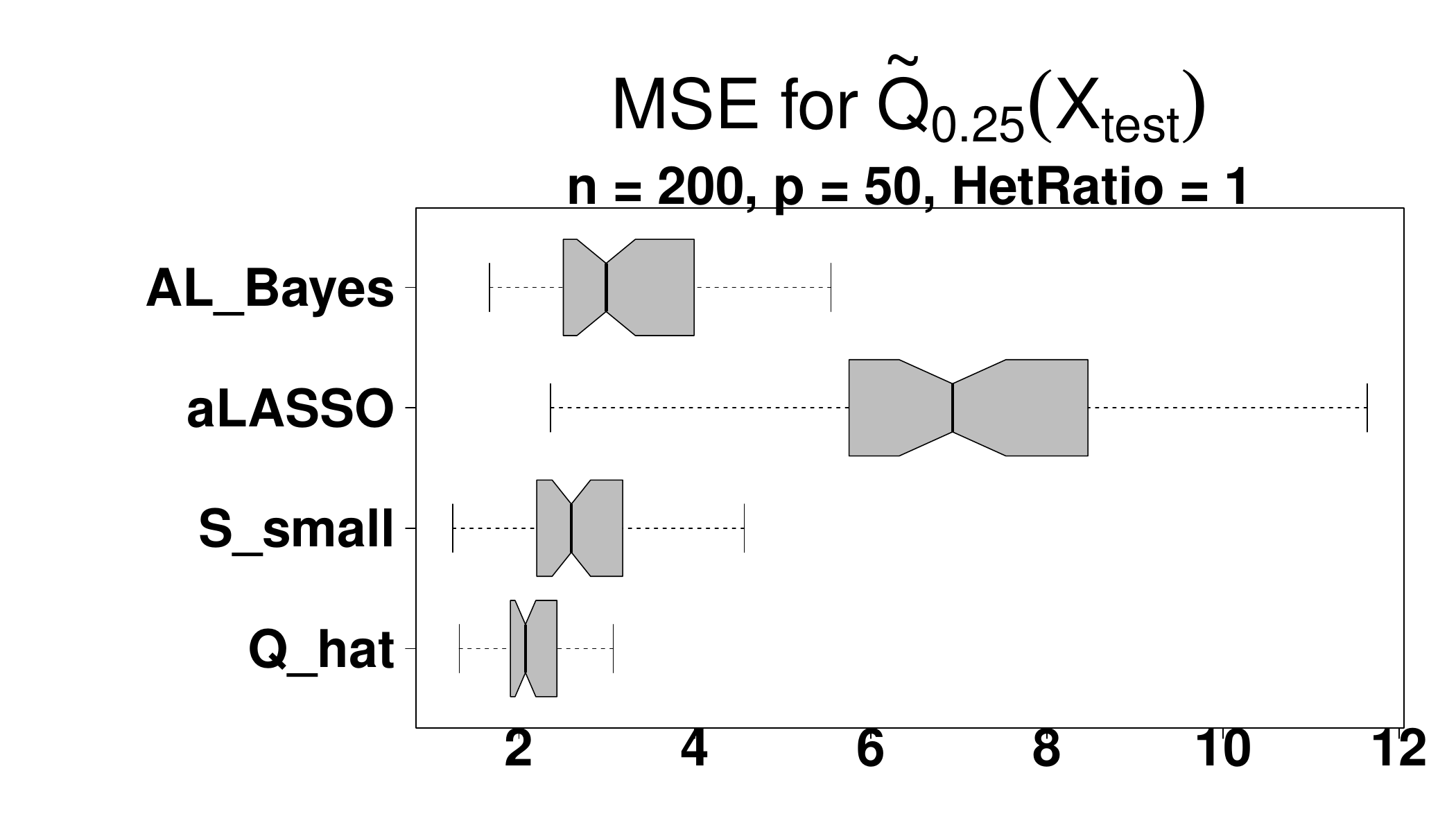}
    \includegraphics[width = .32\textwidth,keepaspectratio]{images/n200_p50_MSE_SNR1_50.pdf}
    \includegraphics[width = .32\textwidth,keepaspectratio]{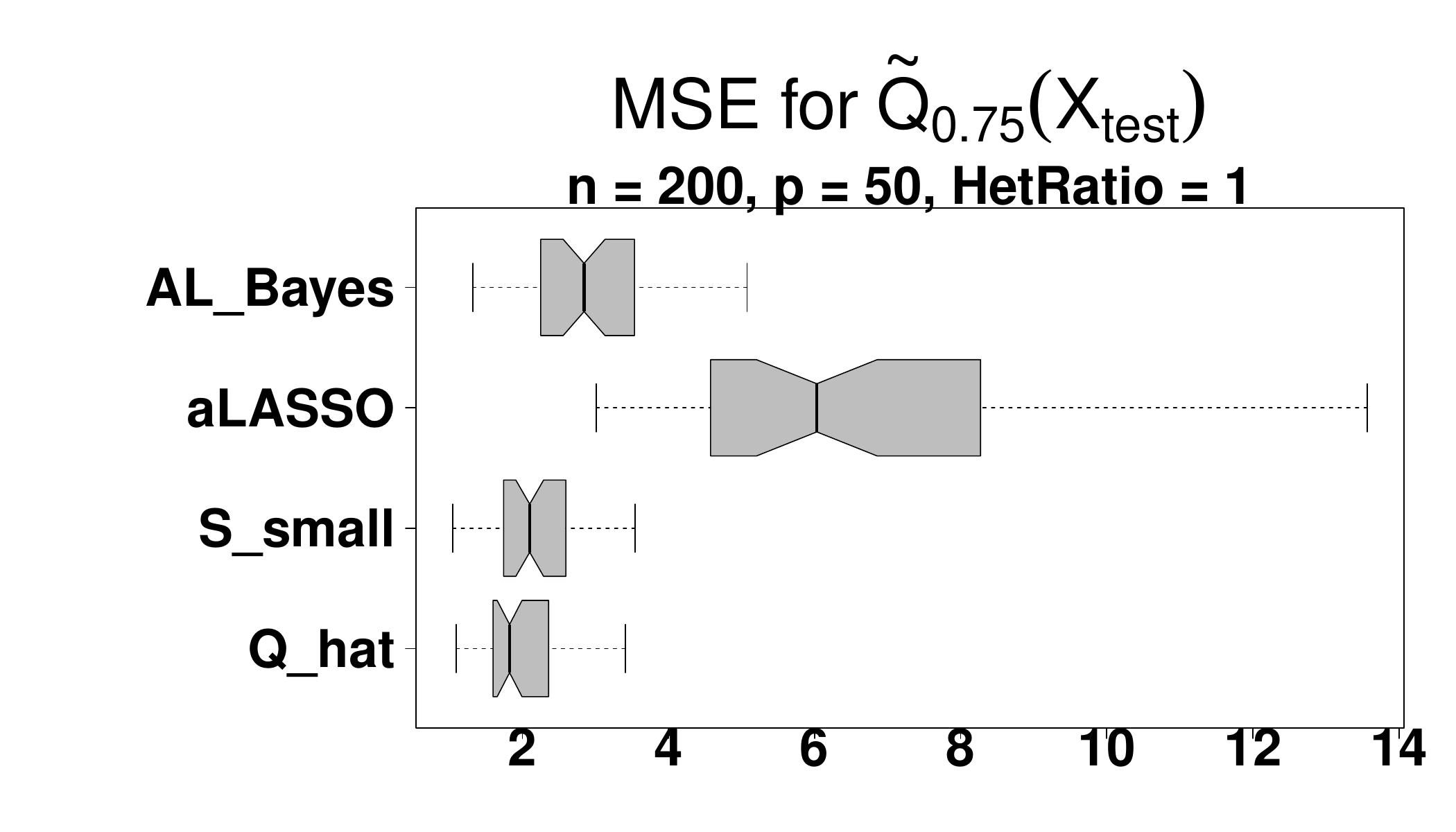}
   \includegraphics[width = .32\textwidth,keepaspectratio]{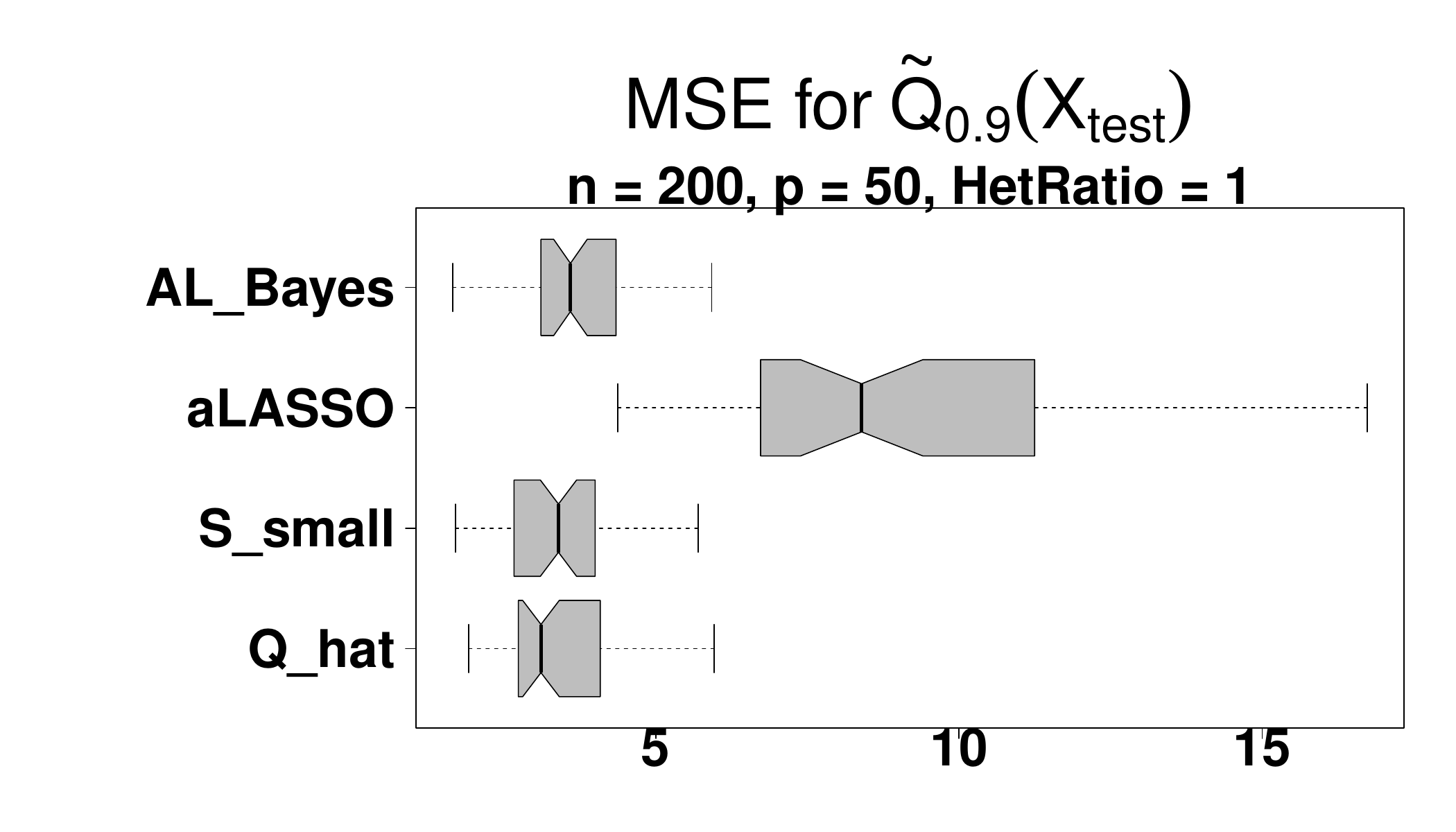}
    \includegraphics[width = .32\textwidth,keepaspectratio]{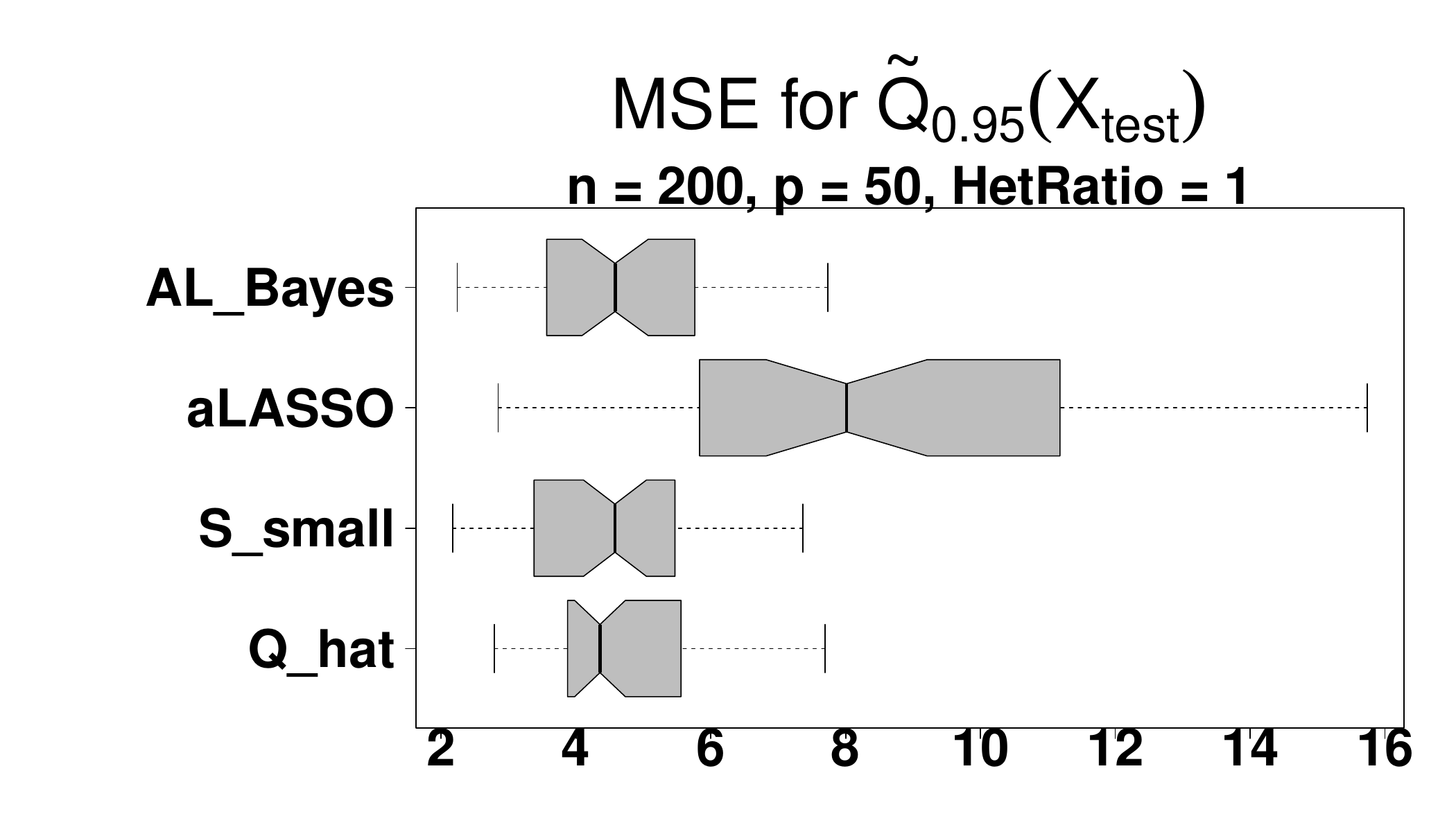}
   \includegraphics[width = .32\textwidth,keepaspectratio]{images/n200_p50_MSE_SNR1_99.pdf}
\end{figure}

\begin{figure}[H]
    \centering
        \caption{\textbf{Check Loss}: $\boldsymbol{n = 500, p= 20, \mbox{\textbf{HetRatio} }= 0.5}$}
  \includegraphics[width = .32\textwidth,keepaspectratio]{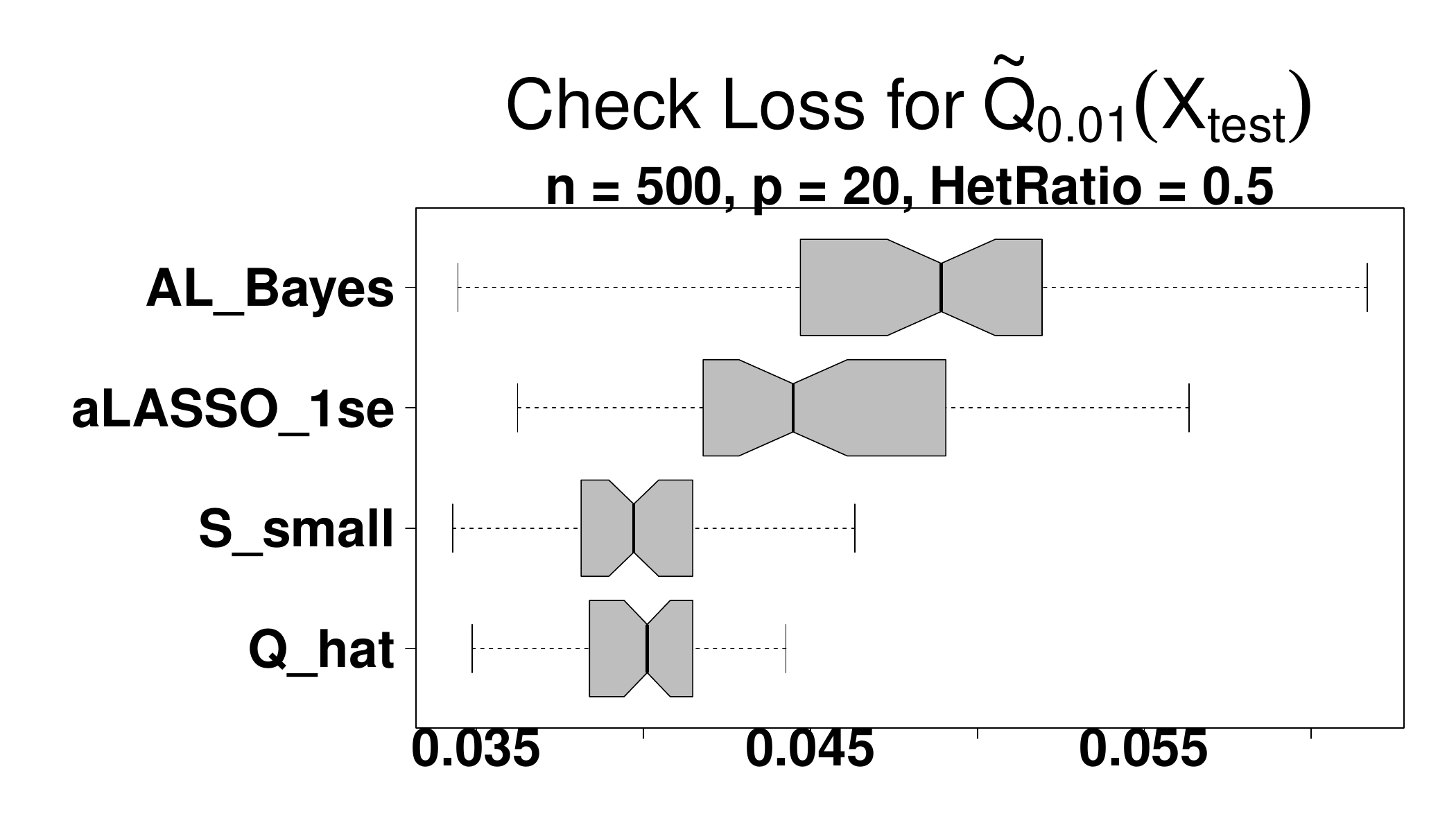}
    \includegraphics[width = .32\textwidth,keepaspectratio]{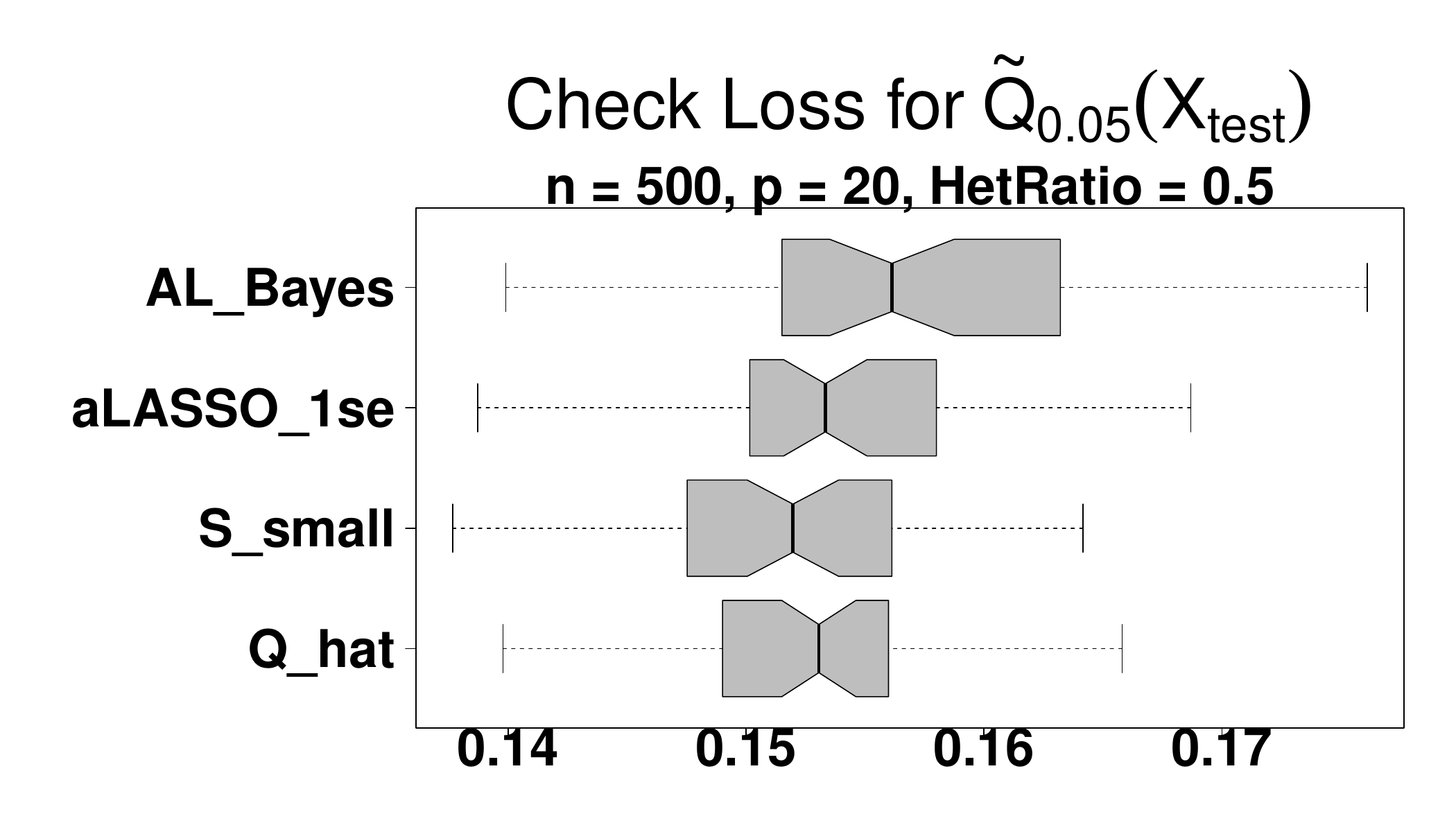}
    \includegraphics[width = .32\textwidth,keepaspectratio]{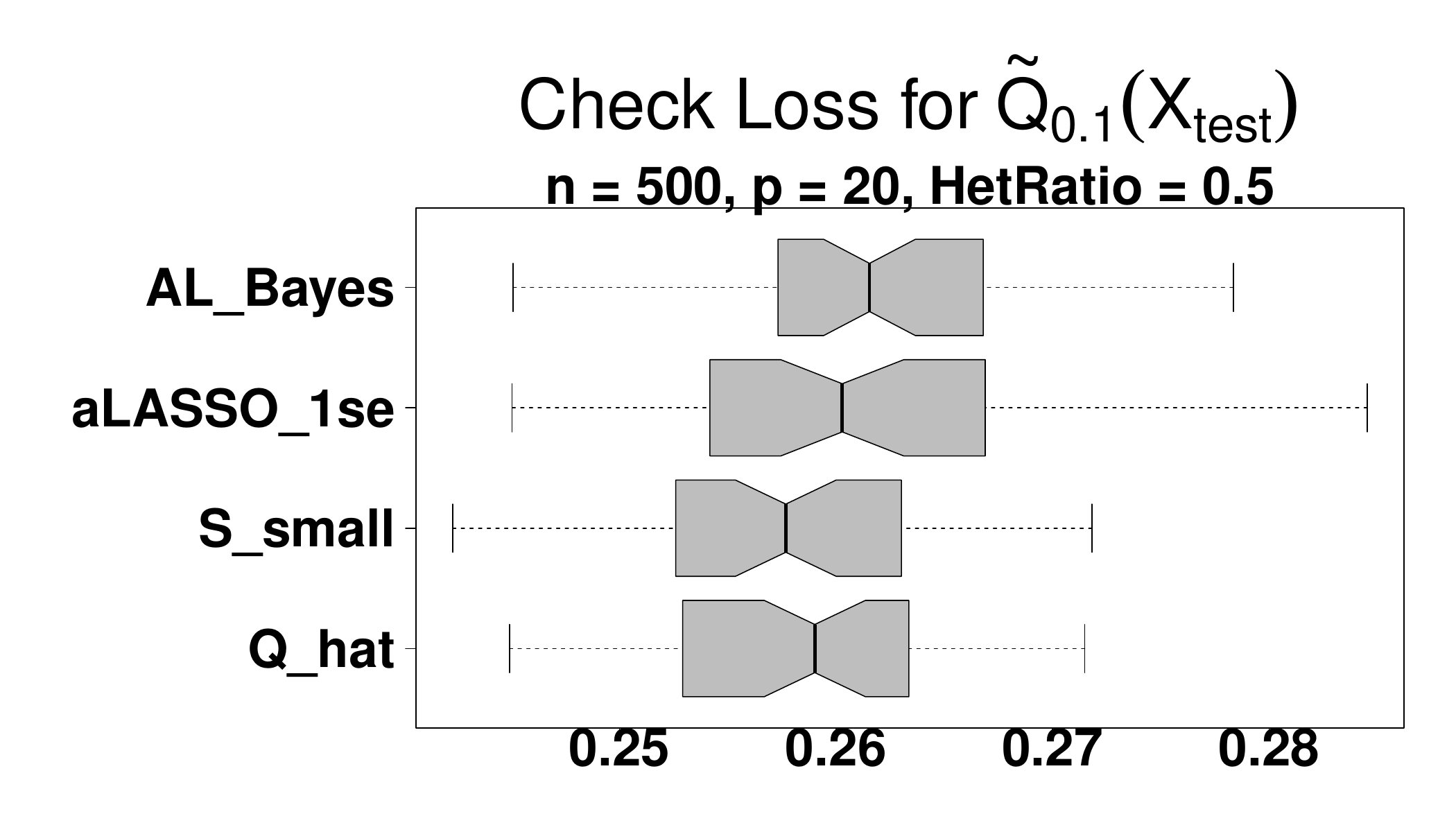}
    \includegraphics[width = .32\textwidth,keepaspectratio]{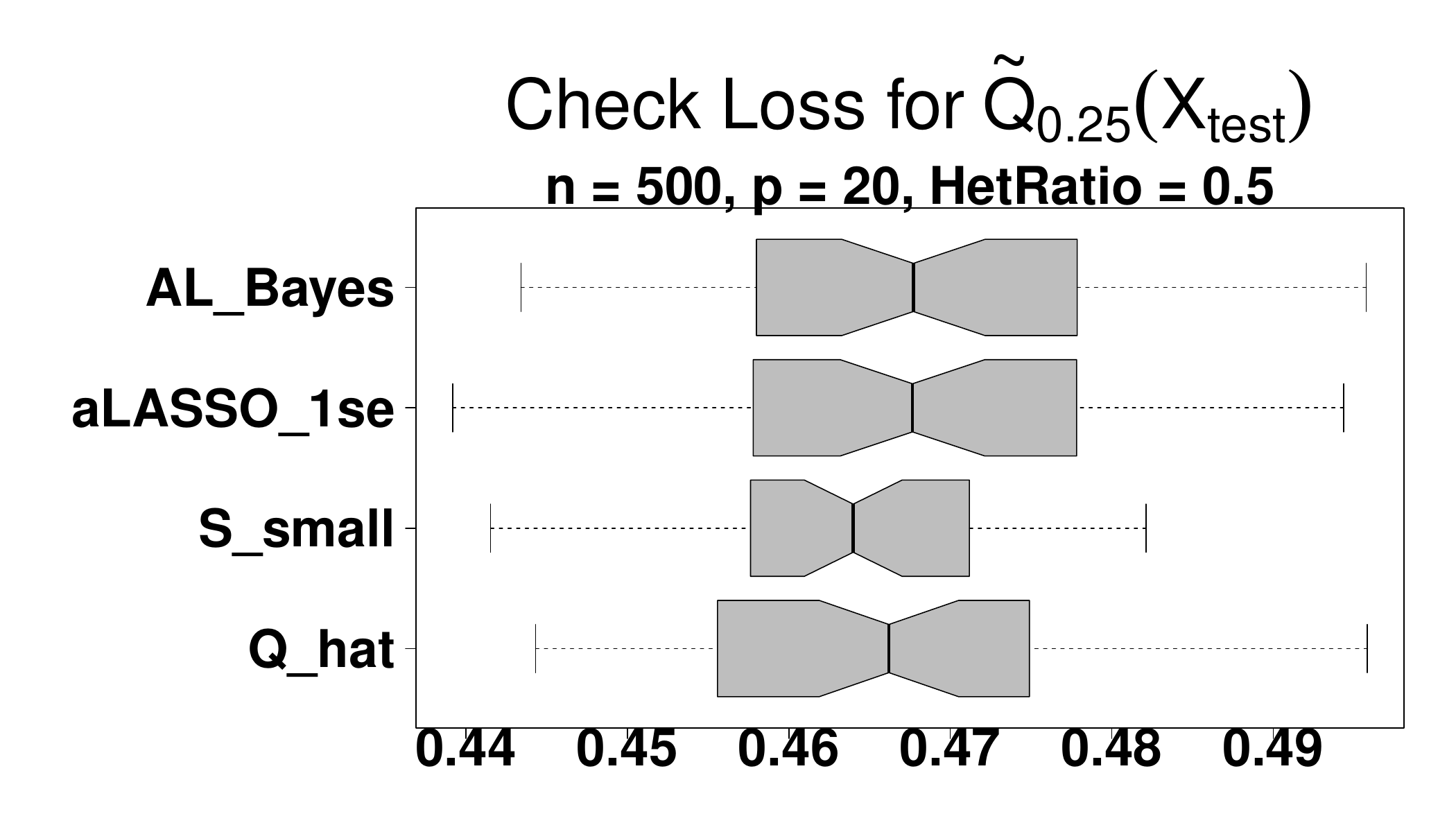}
    \includegraphics[width = .32\textwidth,keepaspectratio]{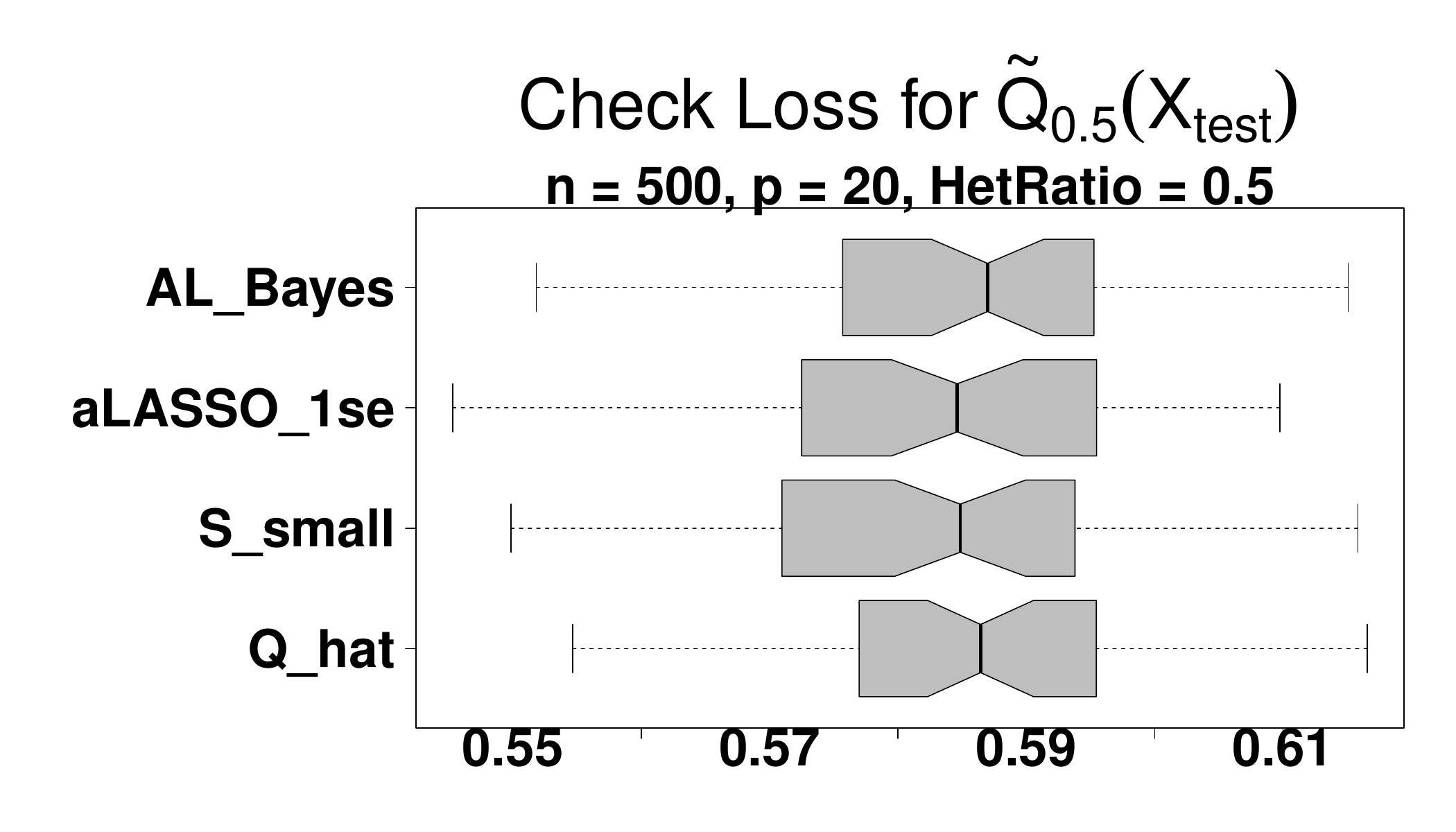}
    \includegraphics[width = .32\textwidth,keepaspectratio]{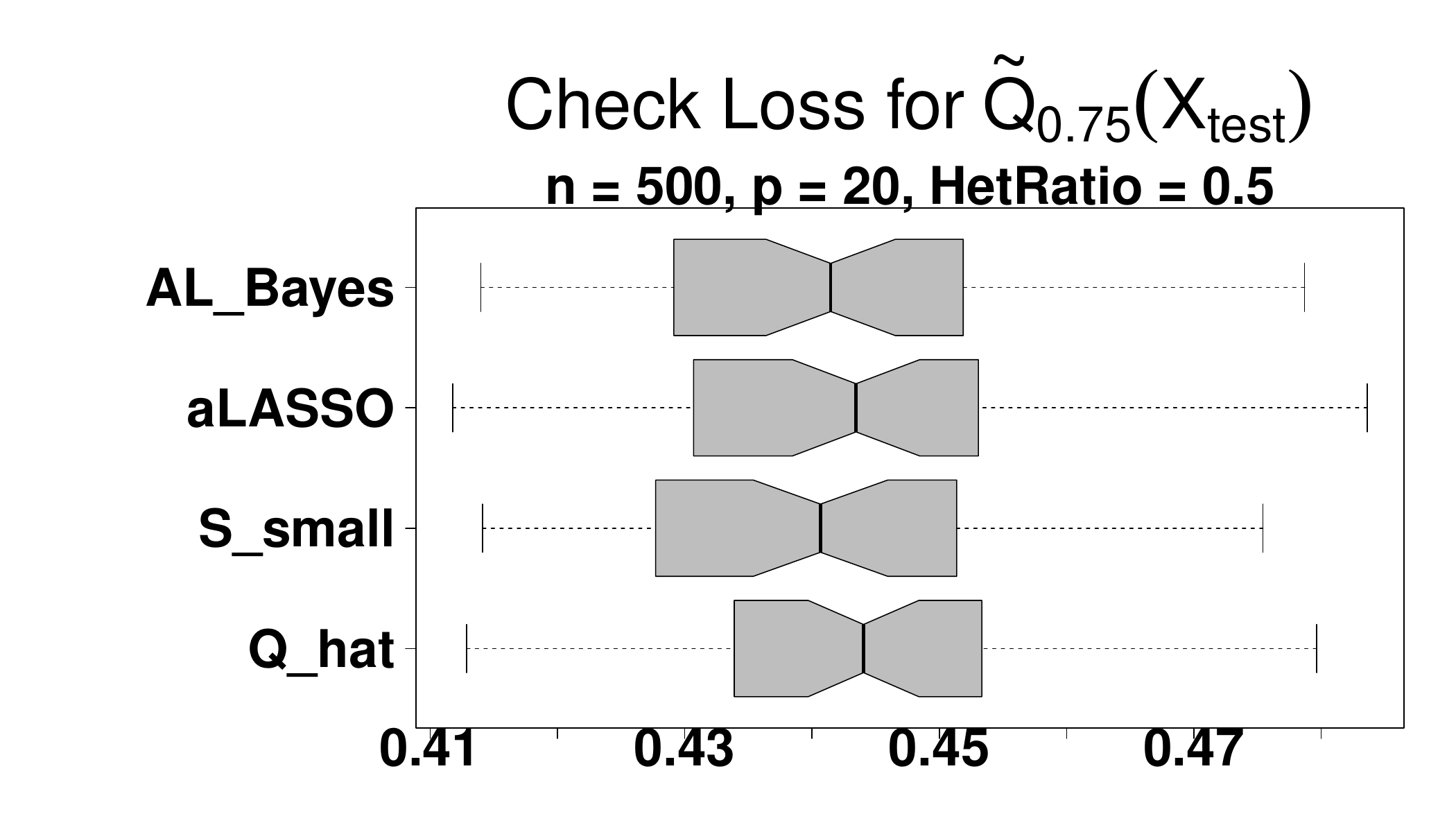}
   \includegraphics[width = .32\textwidth,keepaspectratio]{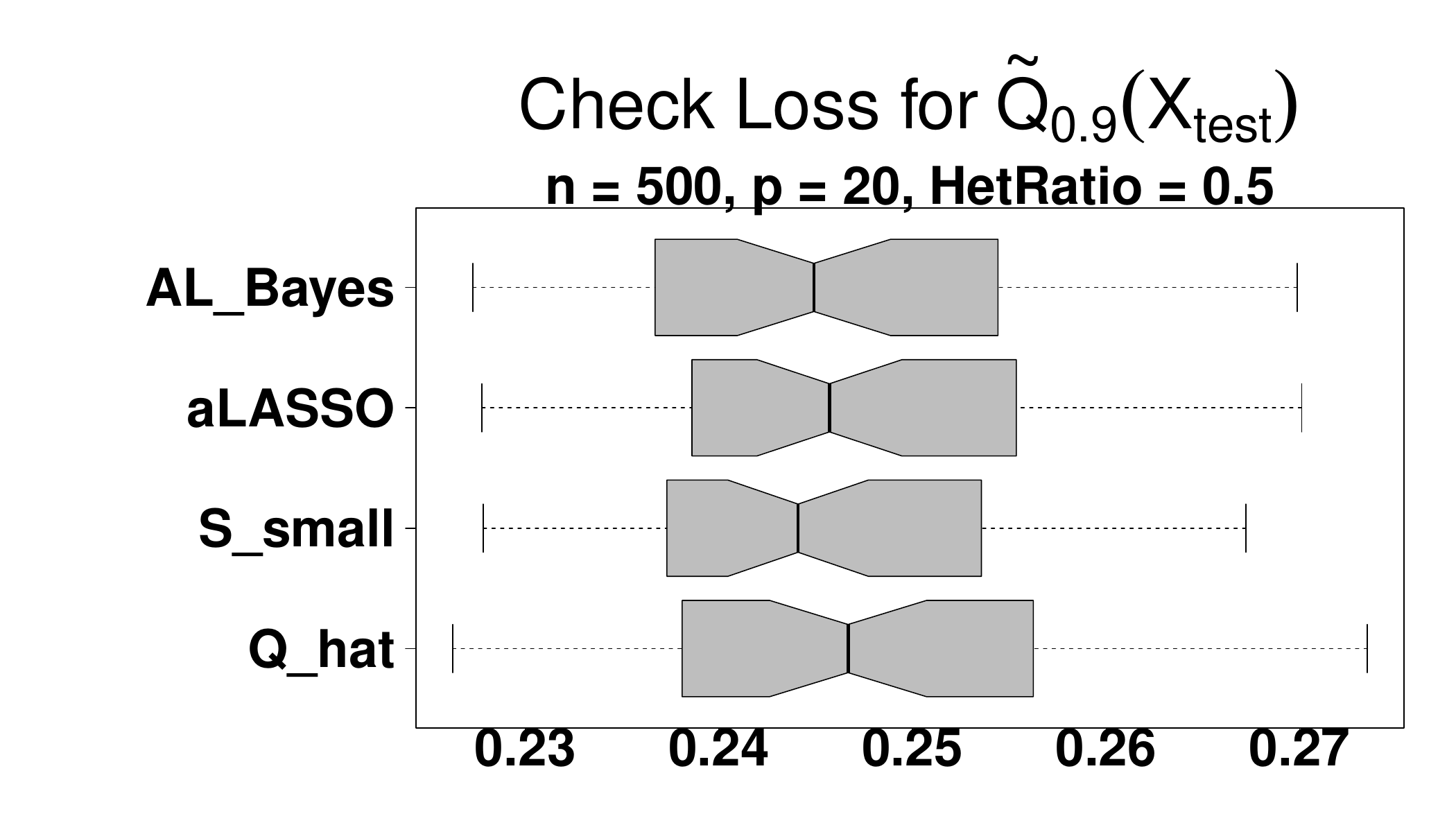}
    \includegraphics[width = .32\textwidth,keepaspectratio]{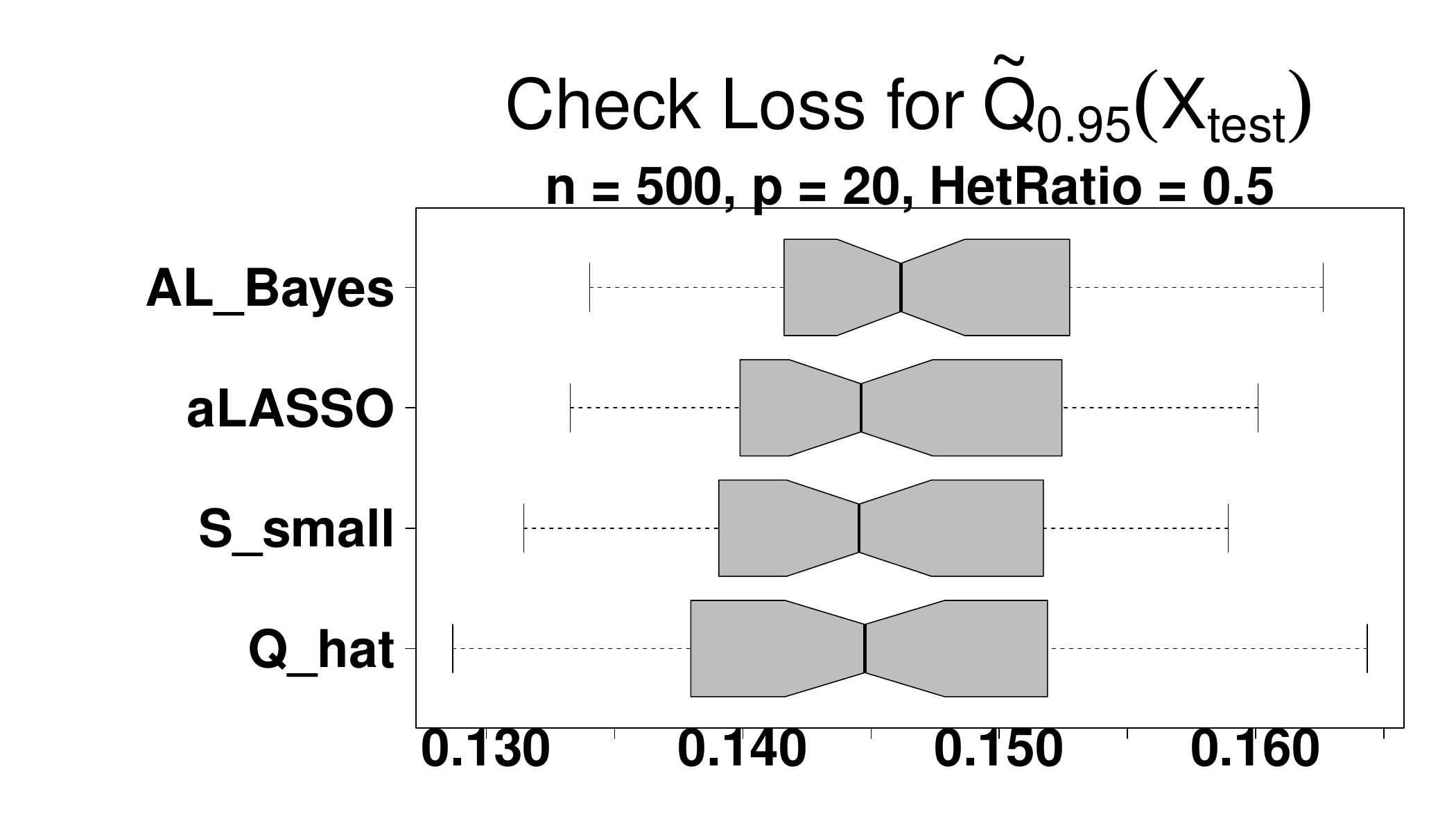}
   \includegraphics[width = .32\textwidth,keepaspectratio]{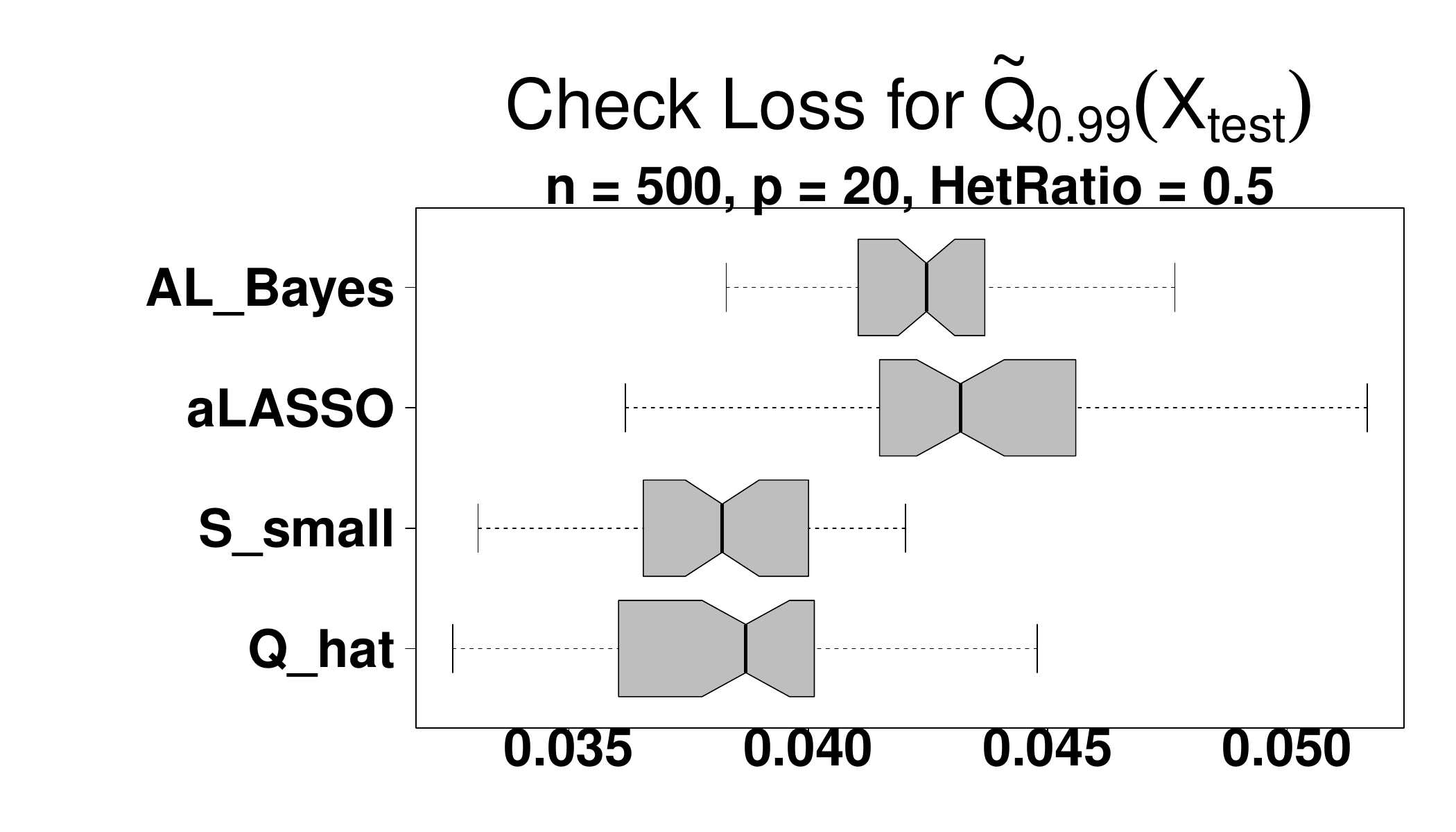}

\end{figure}

\begin{figure}[H]
    \centering
        \caption{\textbf{Check Loss}: $\boldsymbol{n = 500, p= 20, \mbox{\textbf{HetRatio} }= 1}$}
    \includegraphics[width = .32\textwidth,keepaspectratio]{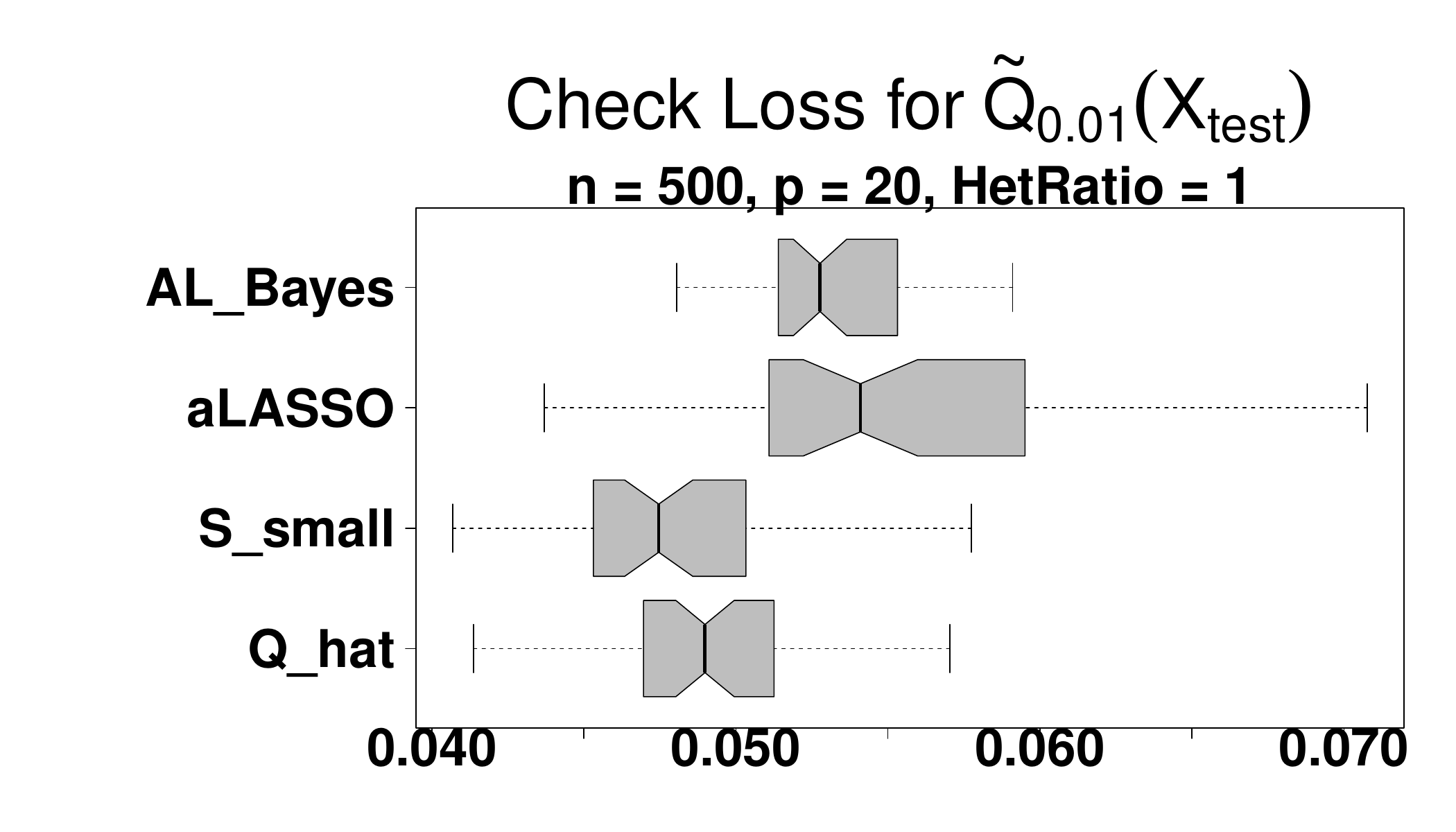}
    \includegraphics[width = .32\textwidth,keepaspectratio]{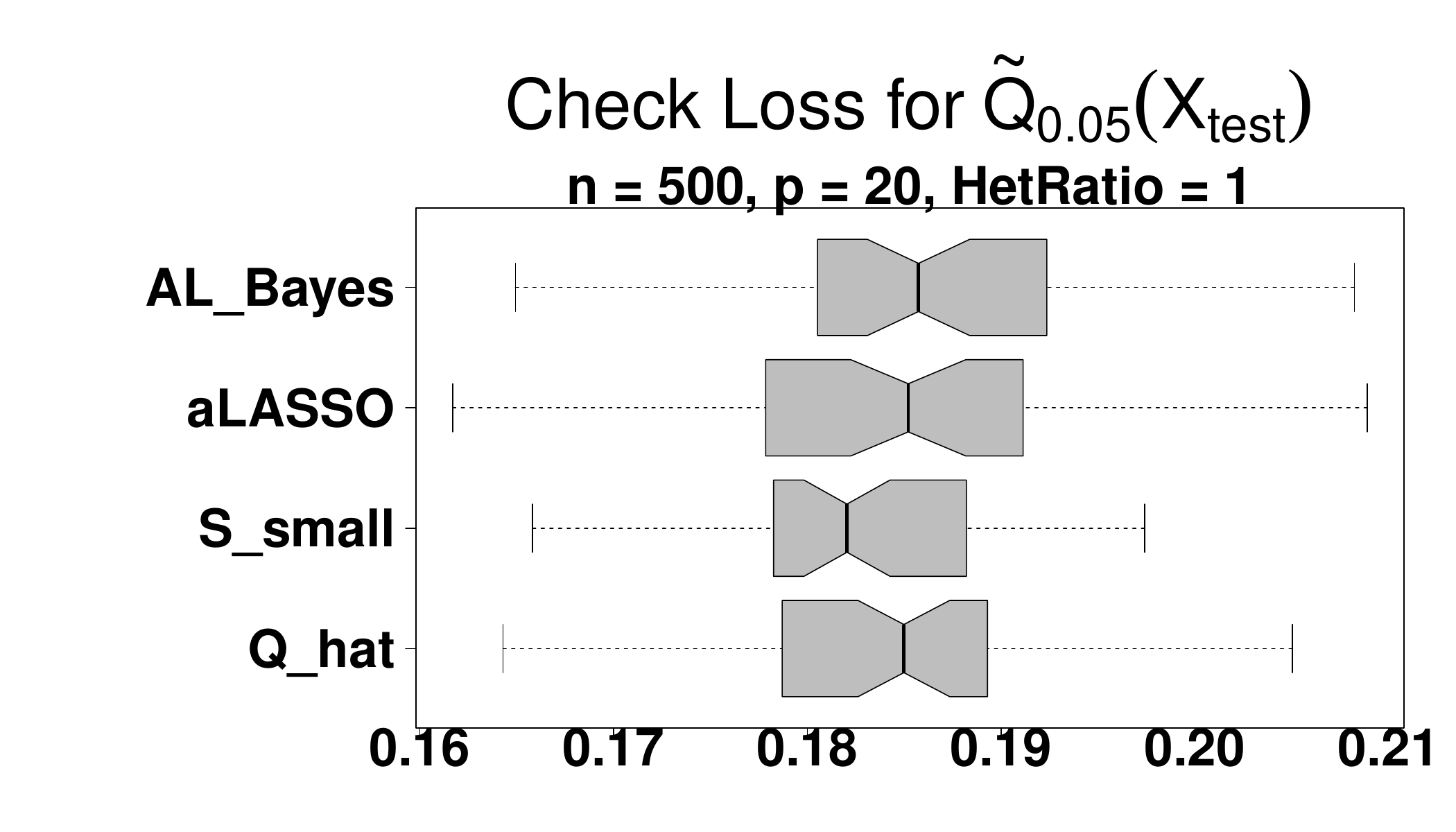}
    \includegraphics[width = .32\textwidth,keepaspectratio]{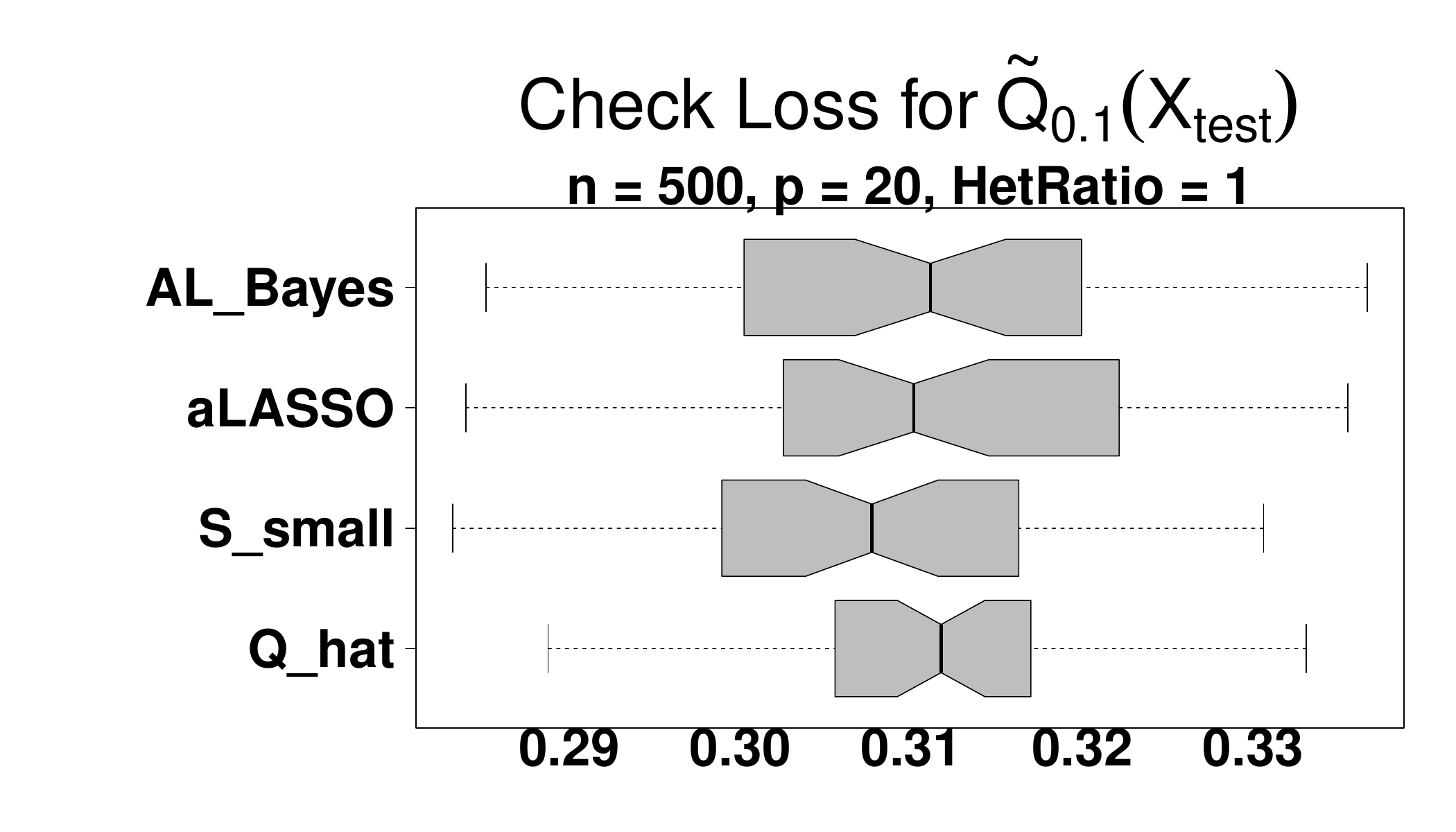}
    \includegraphics[width = .32\textwidth,keepaspectratio]{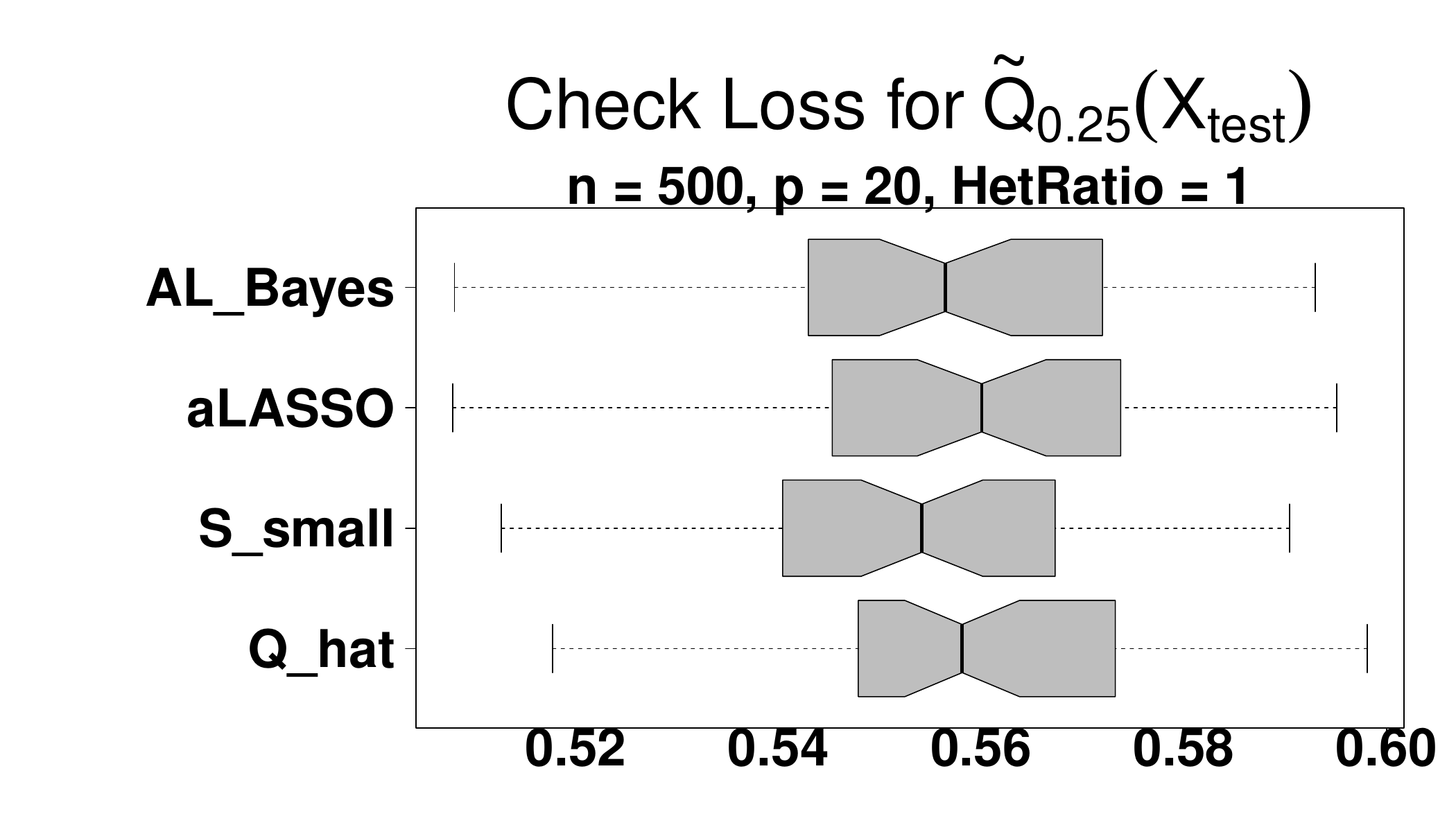}
    \includegraphics[width = .32\textwidth,keepaspectratio]{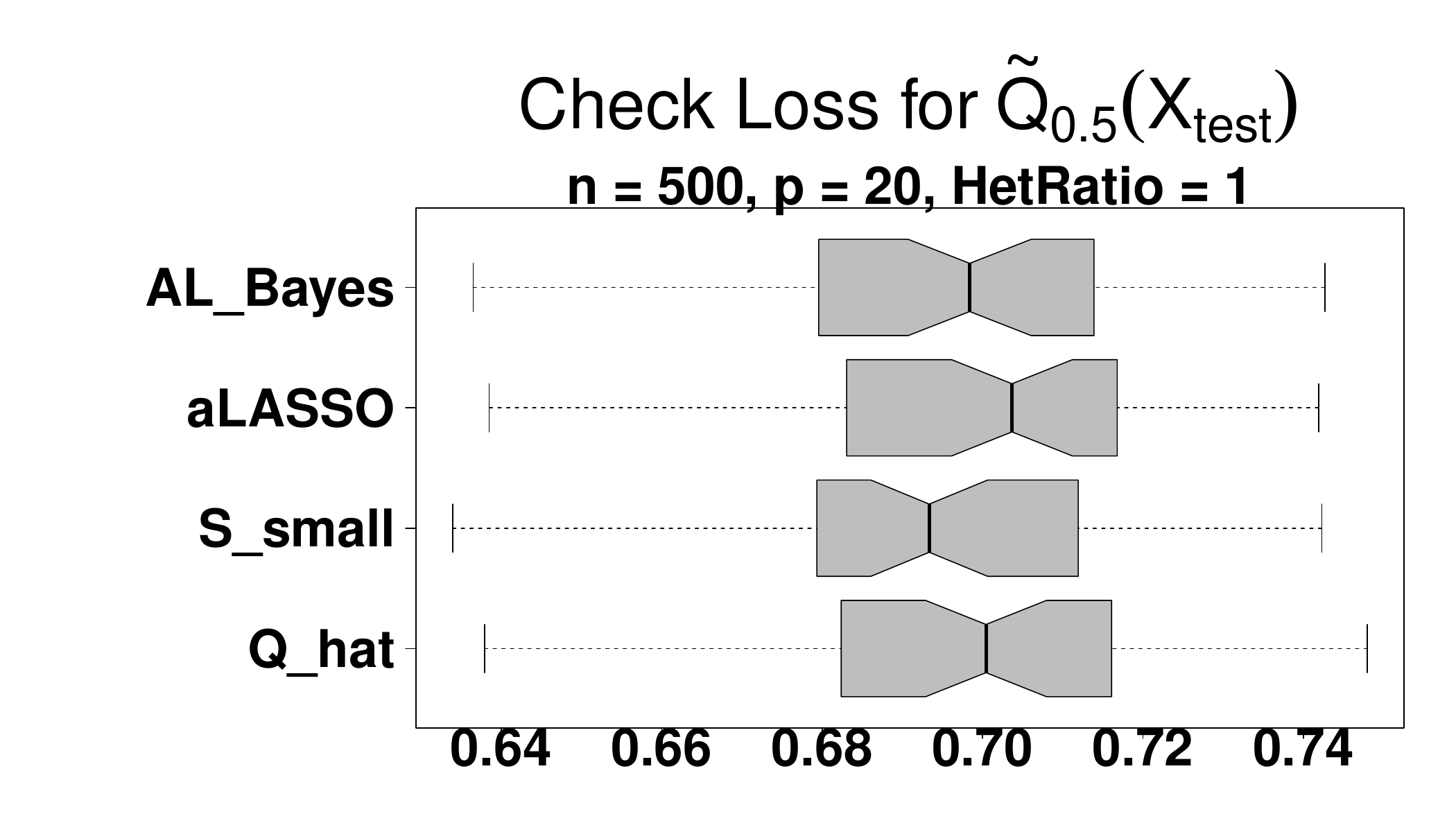}
    \includegraphics[width = .32\textwidth,keepaspectratio]{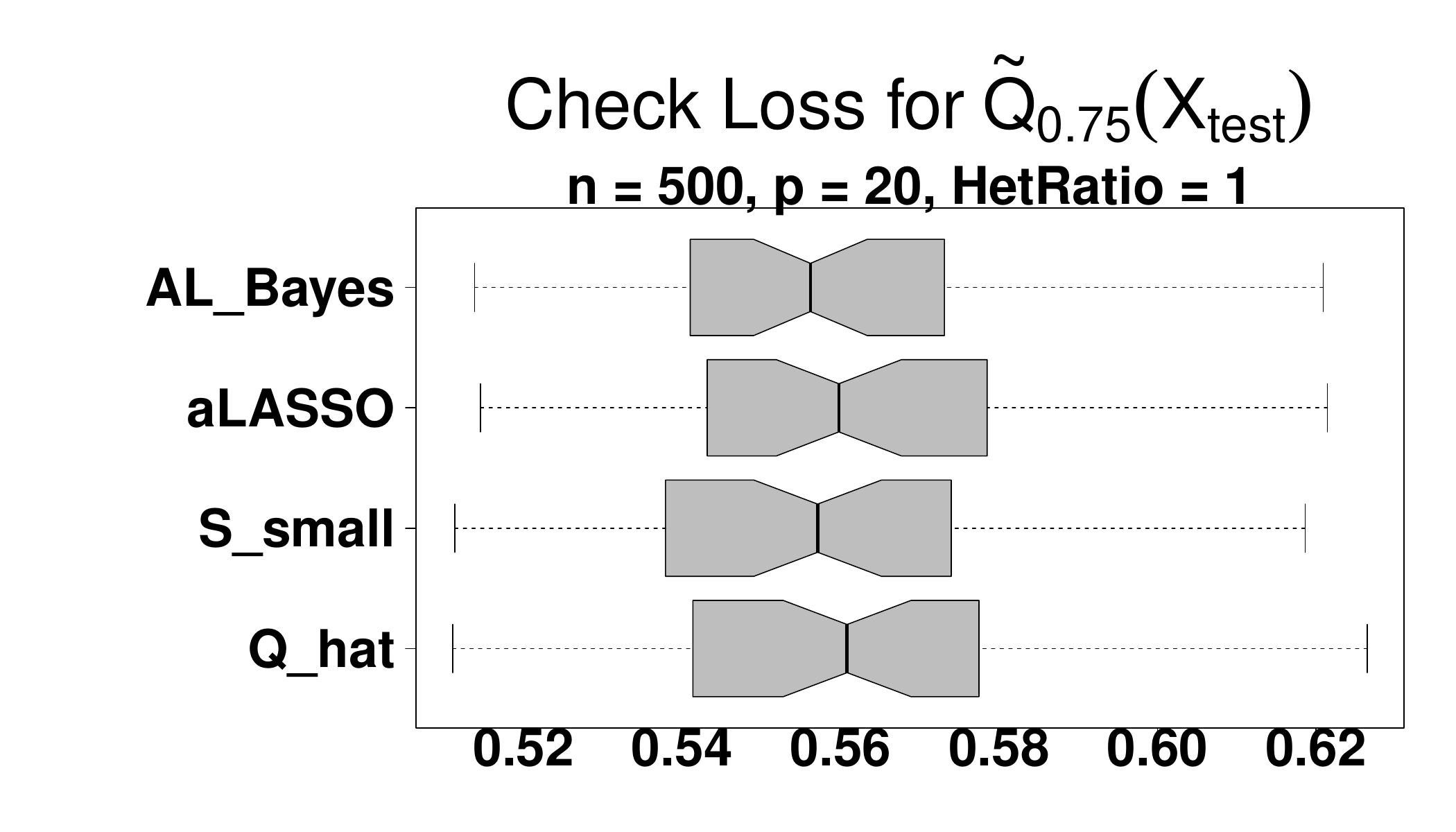}
   \includegraphics[width = .32\textwidth,keepaspectratio]{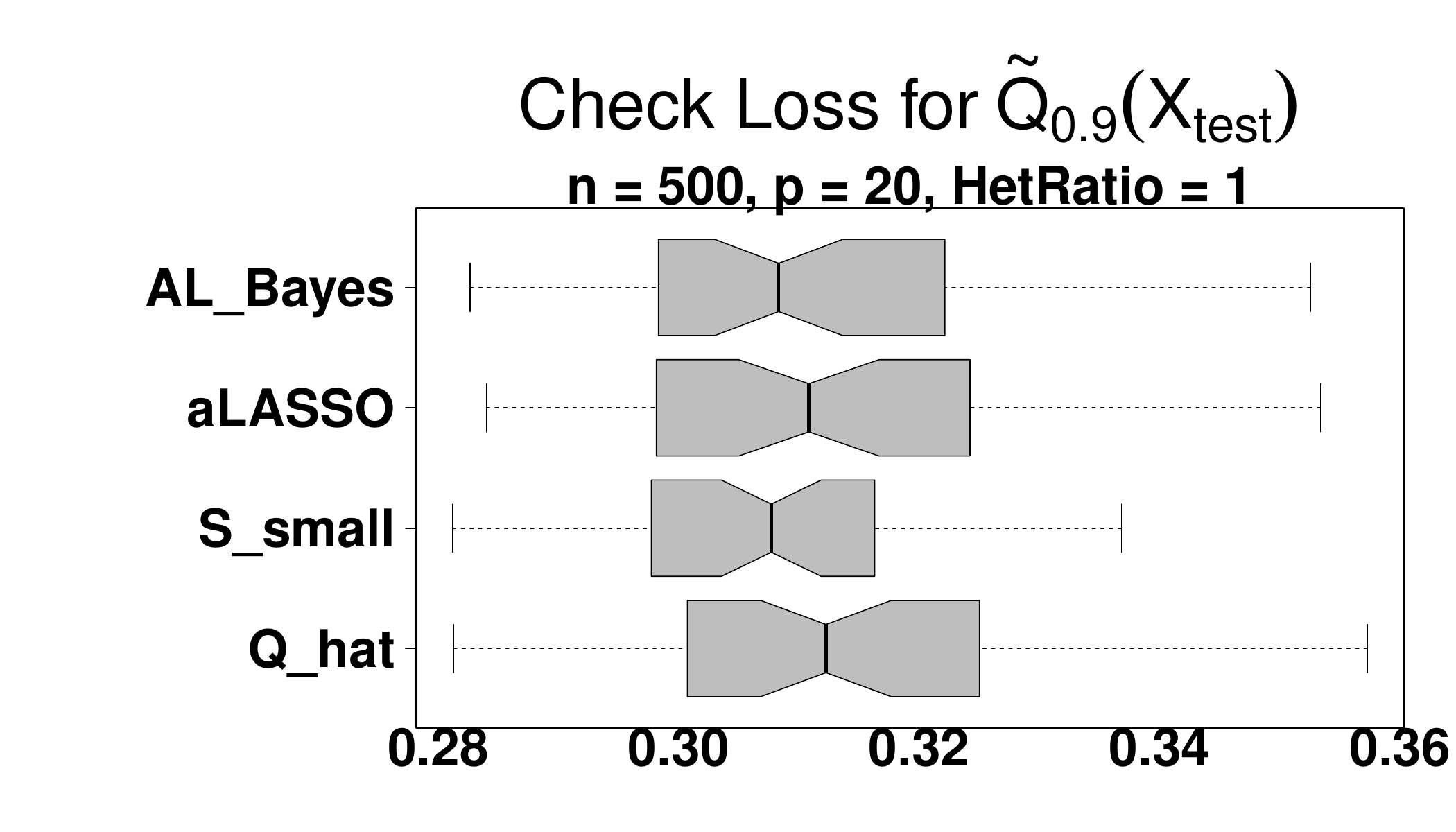}
    \includegraphics[width = .32\textwidth,keepaspectratio]{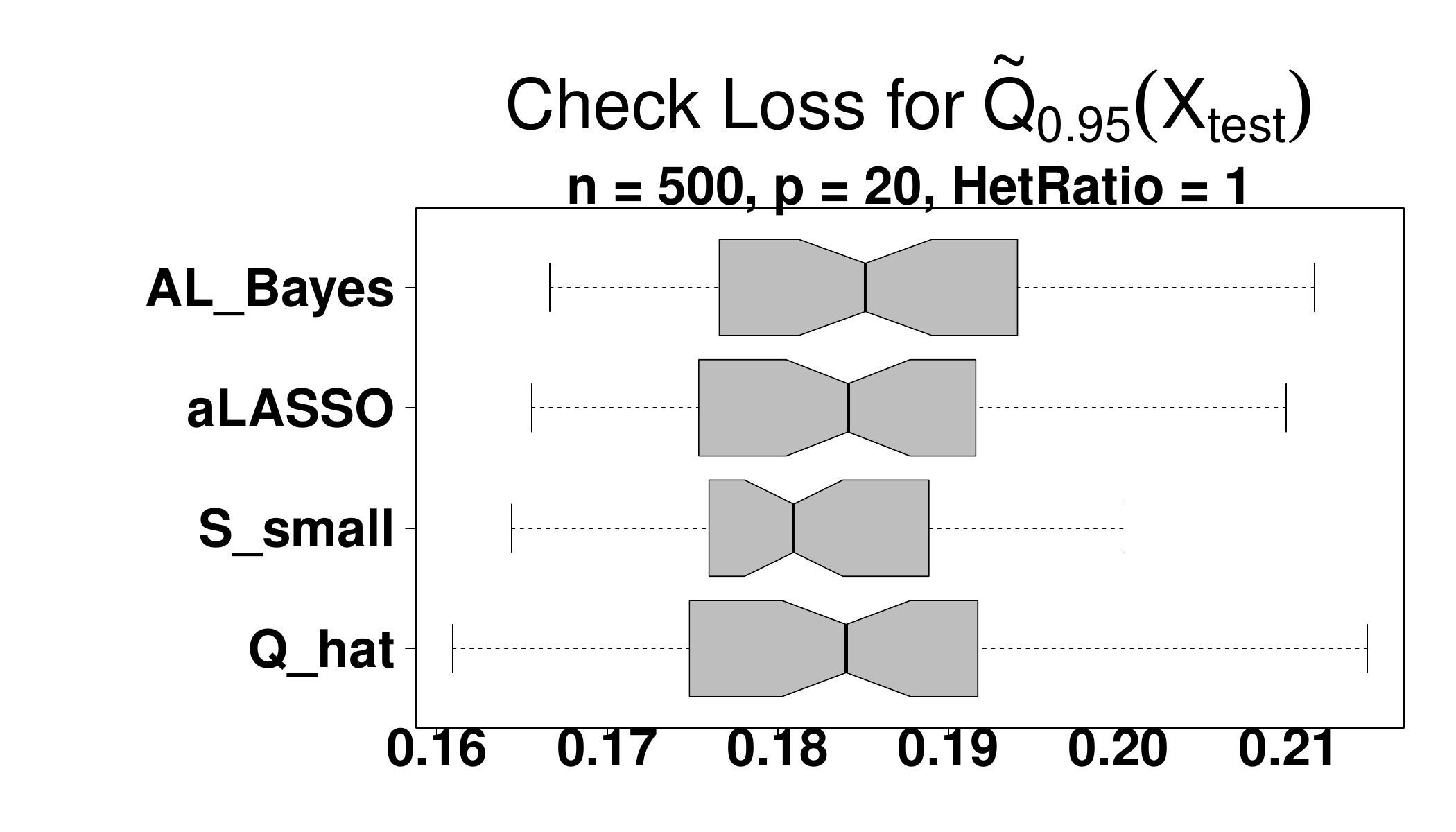}
   \includegraphics[width = .32\textwidth,keepaspectratio]{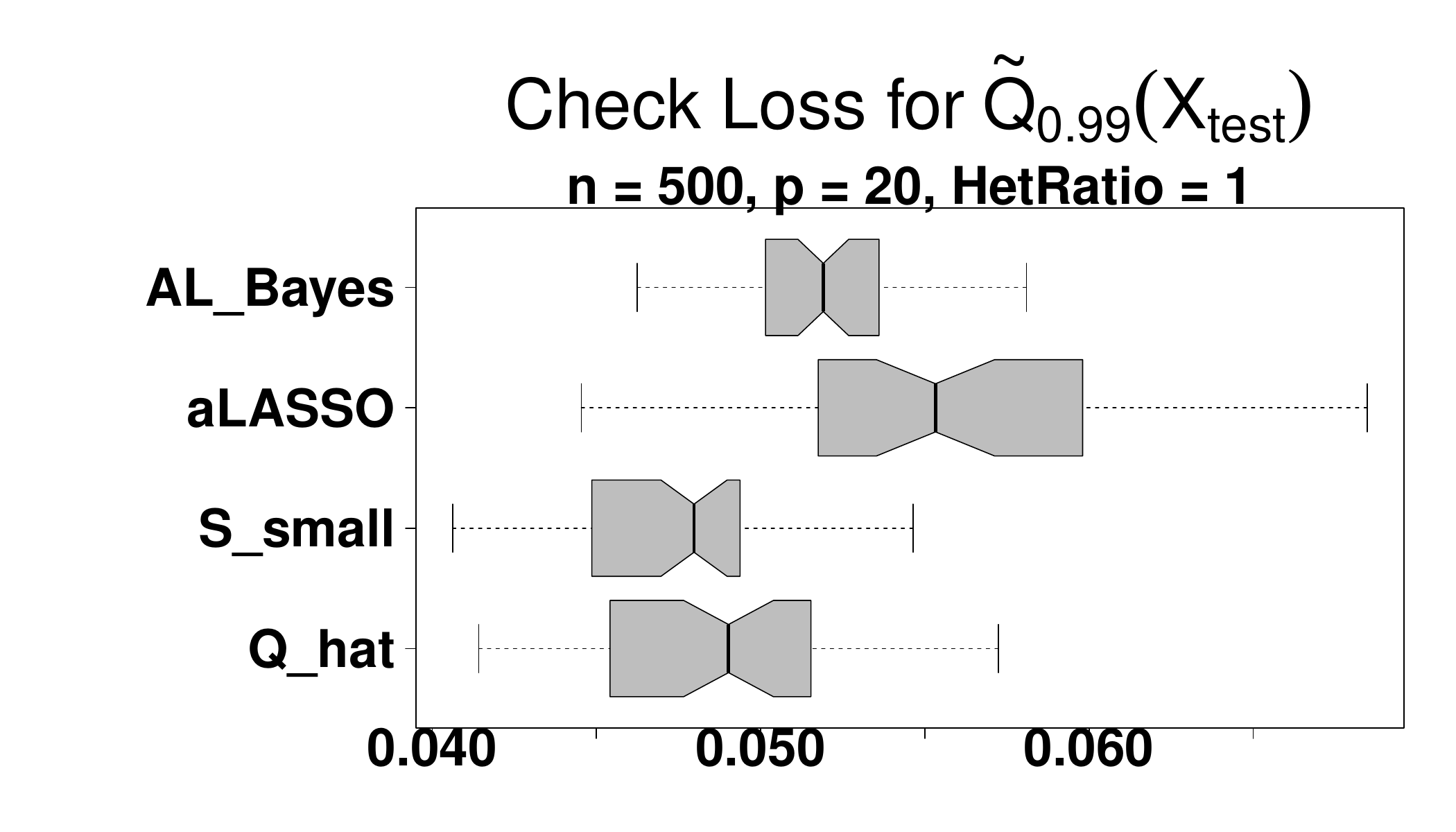}

\end{figure}
\begin{figure}[H]
    \centering
        \caption{\textbf{Check Loss}: $\boldsymbol{n = 100, p= 100, \mbox{\textbf{HetRatio} }= 0.5}$}
  \includegraphics[width = .32\textwidth,keepaspectratio]{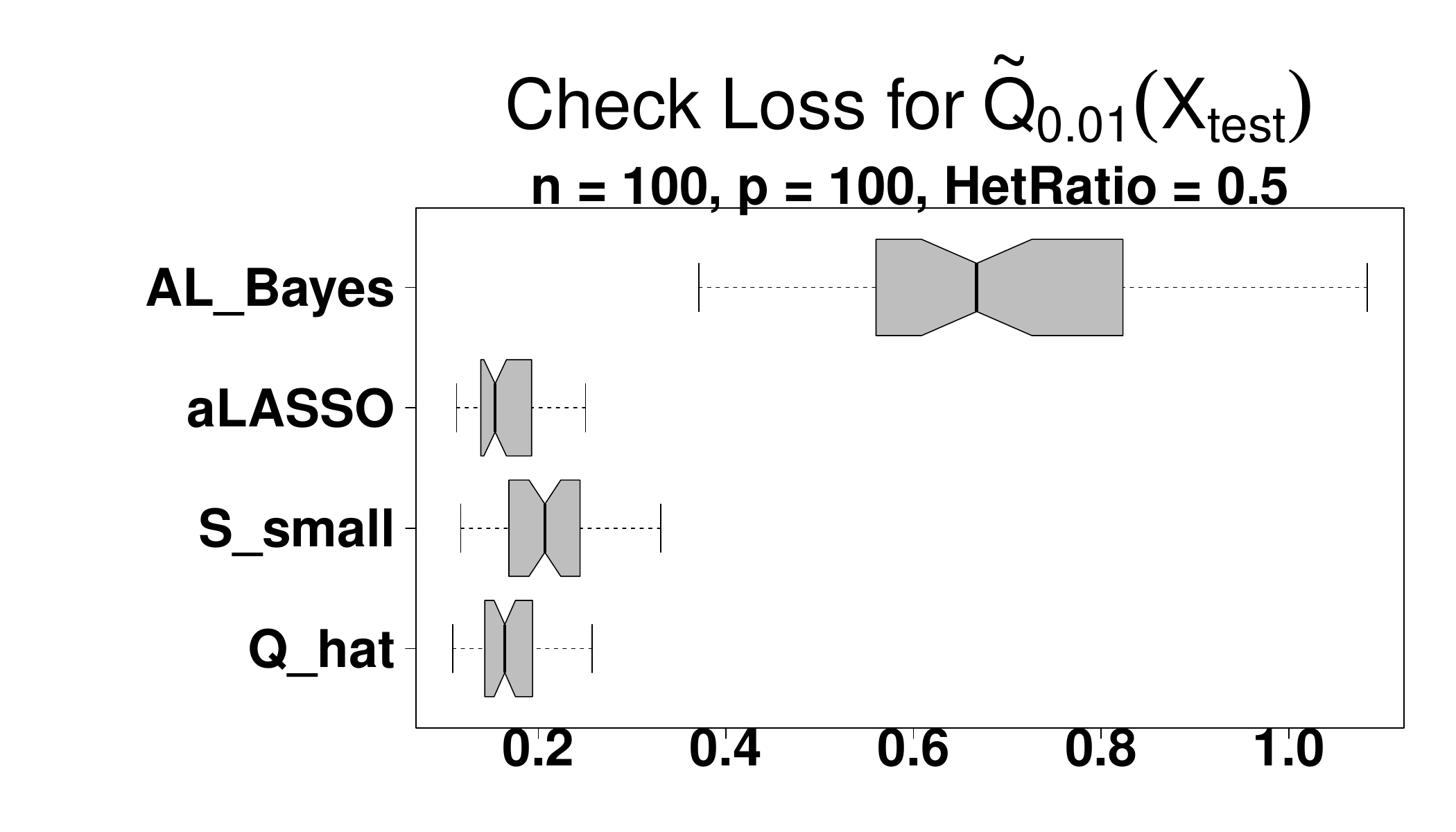}
    \includegraphics[width = .32\textwidth,keepaspectratio]{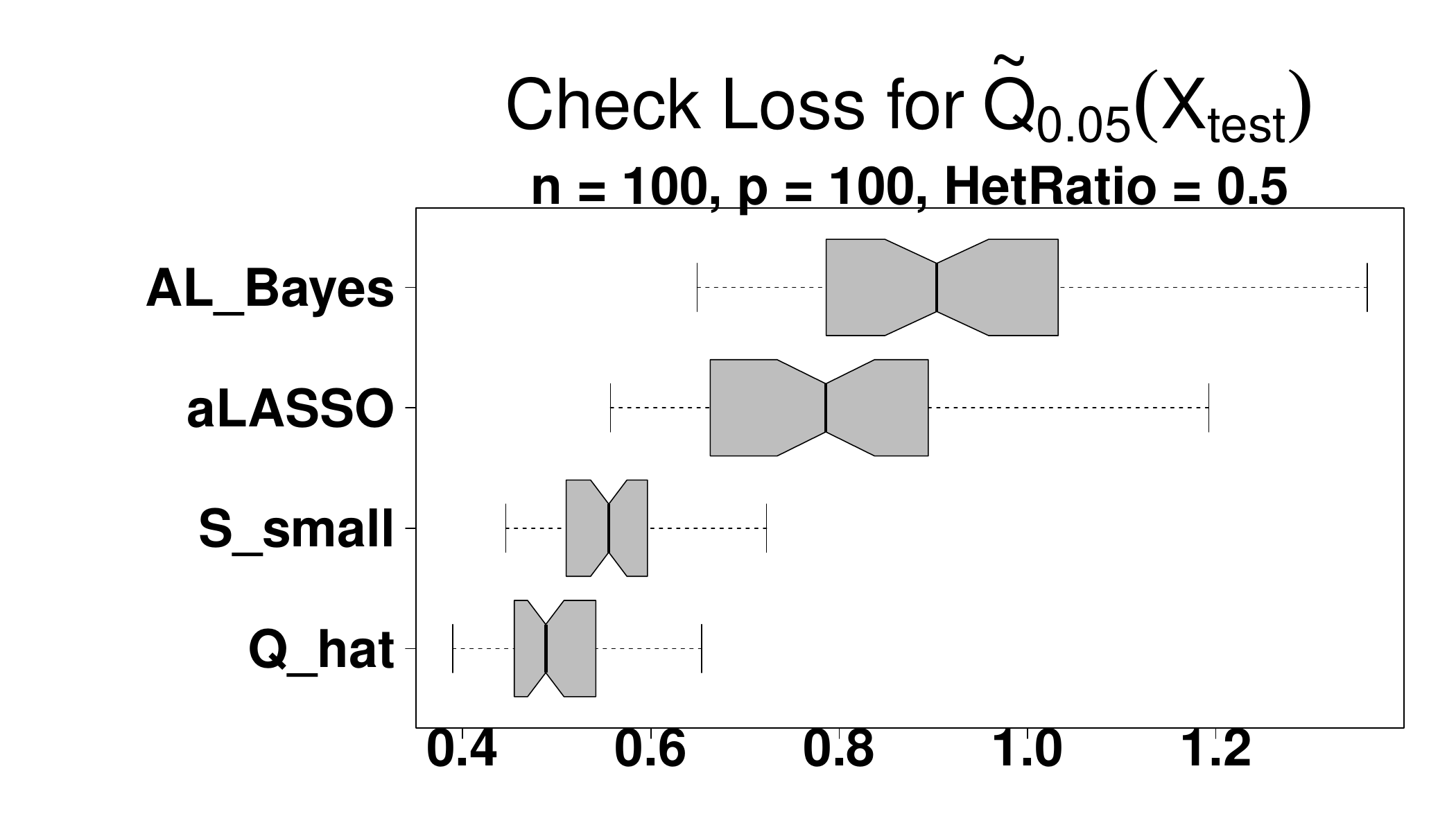}
    \includegraphics[width = .32\textwidth,keepaspectratio]{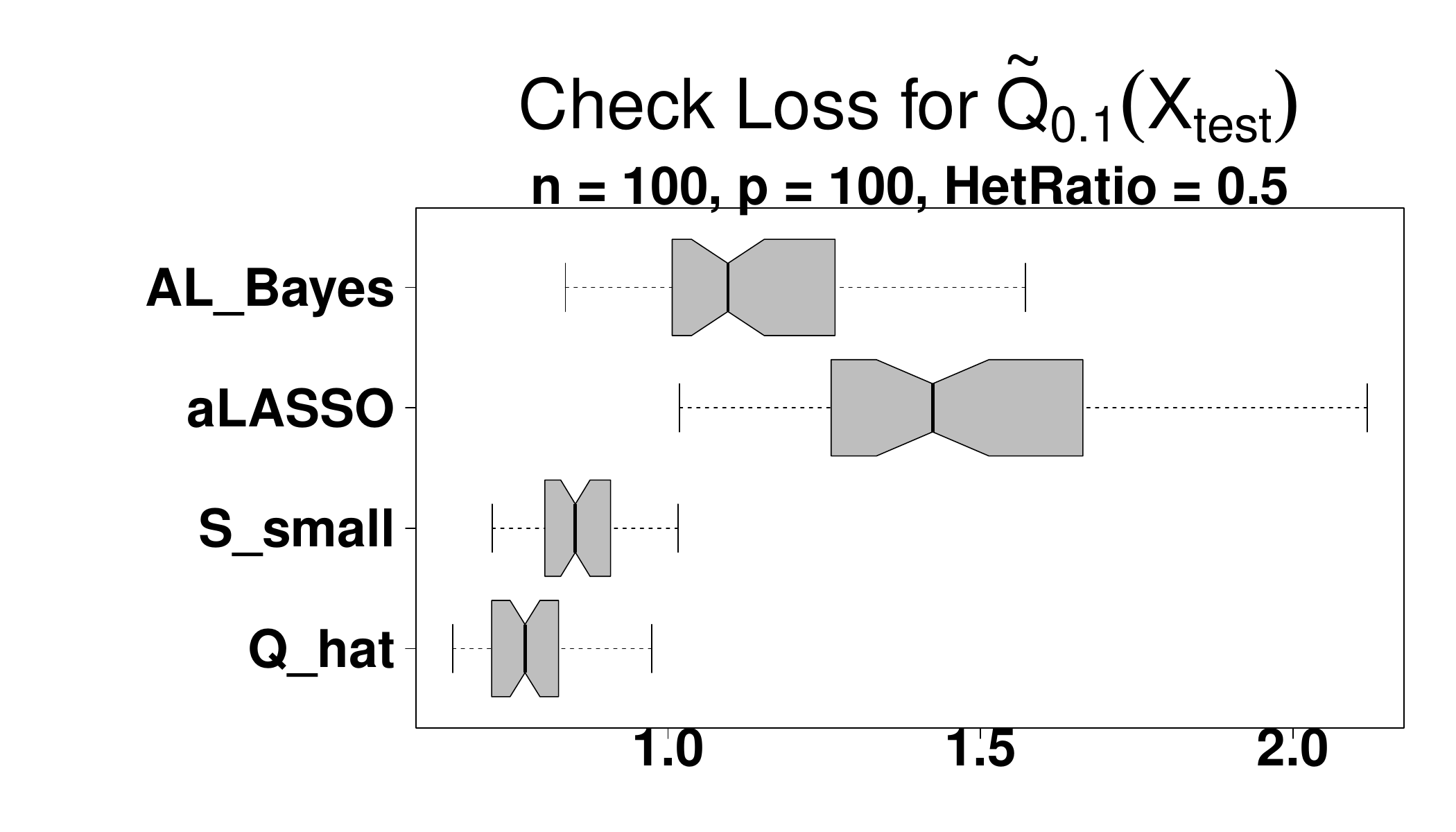}
    \includegraphics[width = .32\textwidth,keepaspectratio]{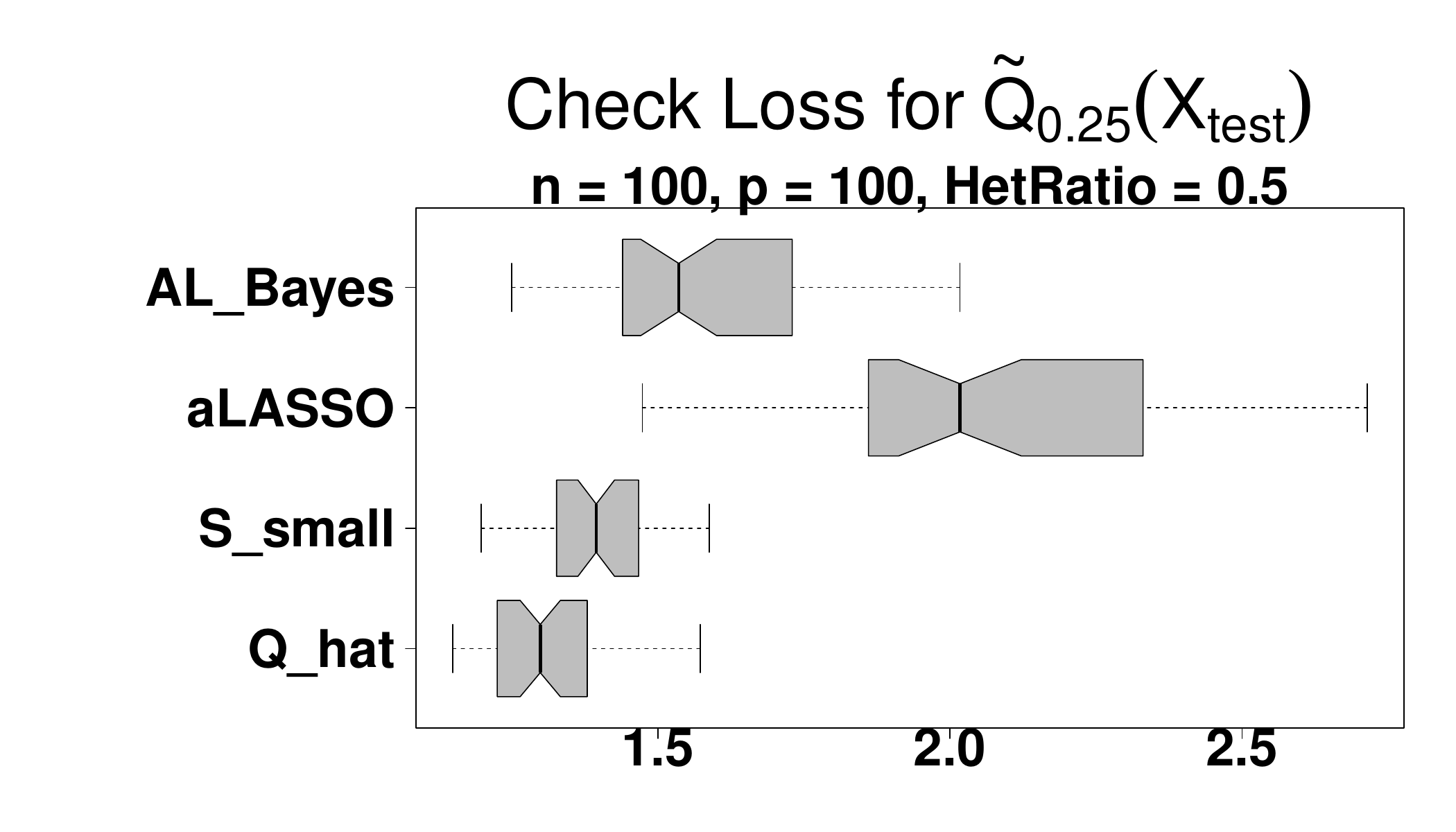}
    \includegraphics[width = .32\textwidth,keepaspectratio]{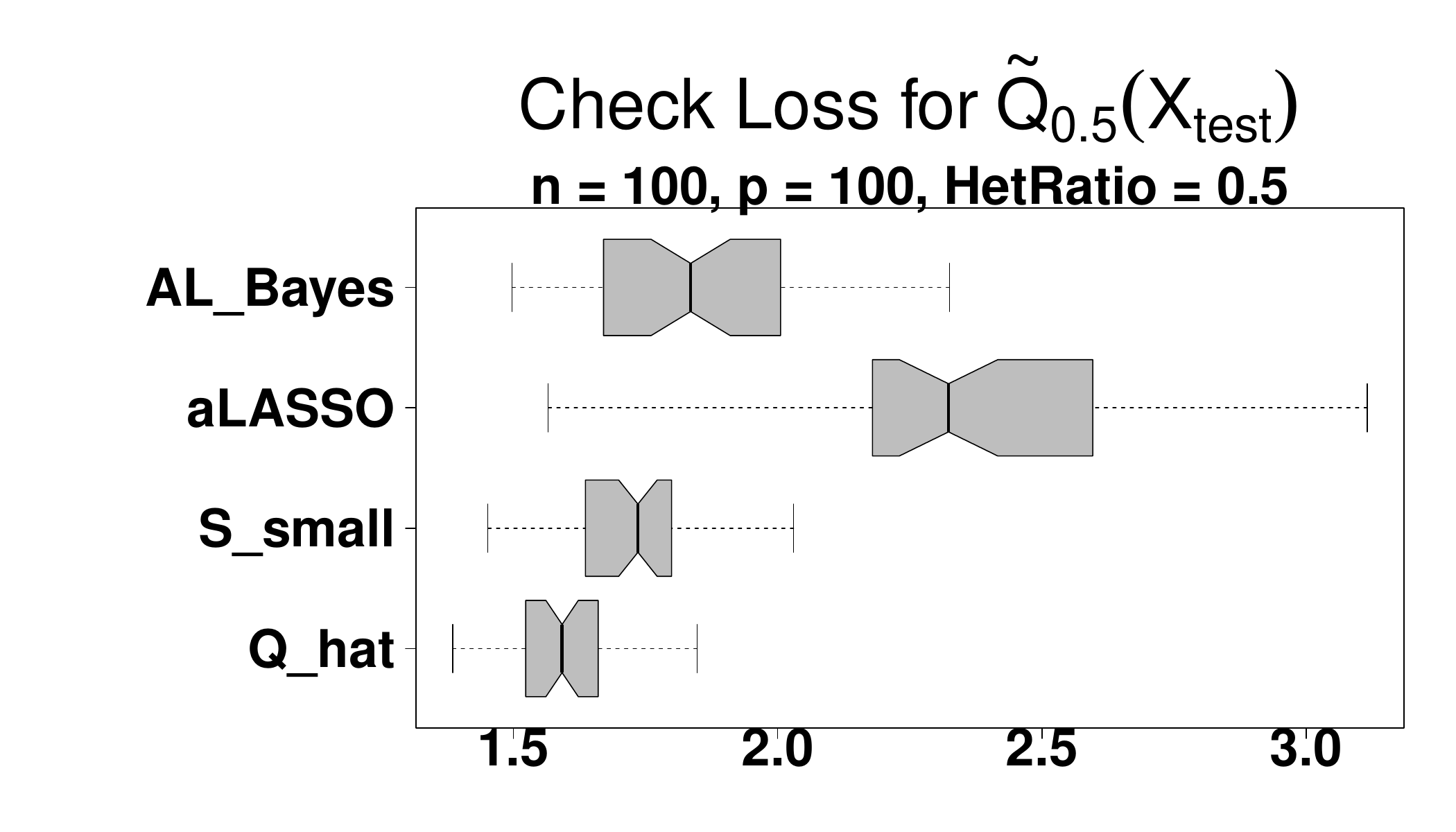}
    \includegraphics[width = .32\textwidth,keepaspectratio]{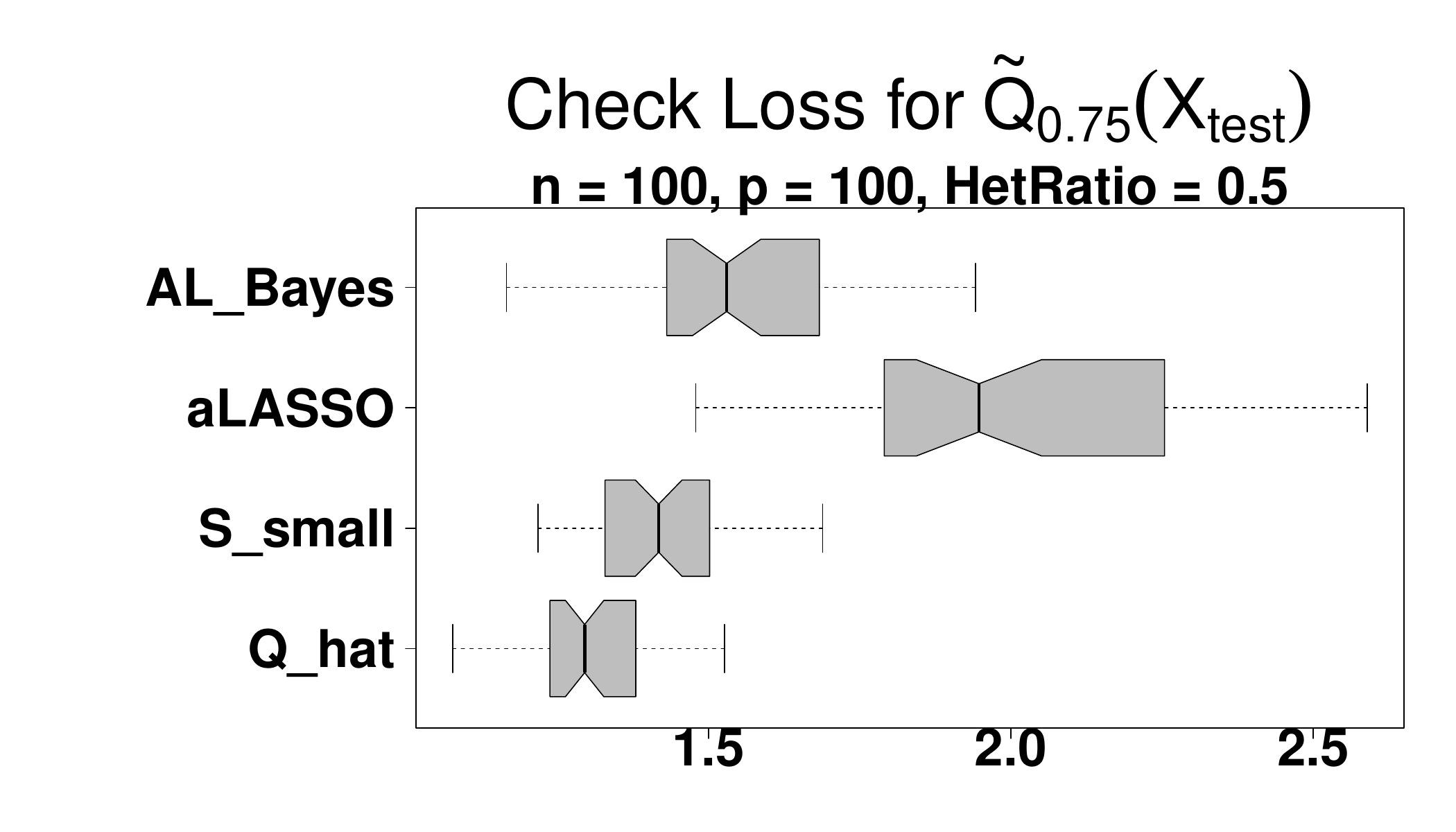}
   \includegraphics[width = .32\textwidth,keepaspectratio]{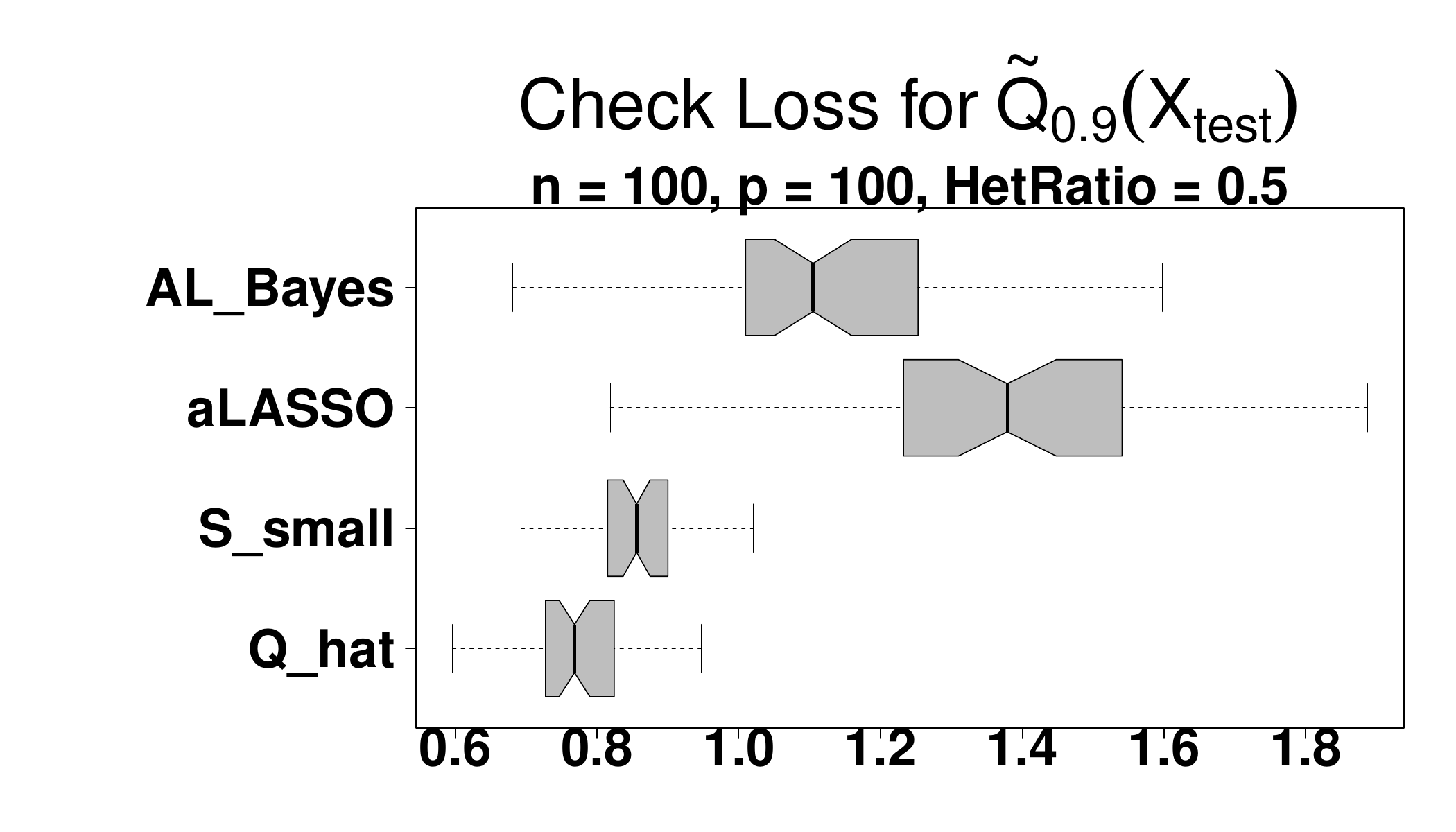}
    \includegraphics[width = .32\textwidth,keepaspectratio]{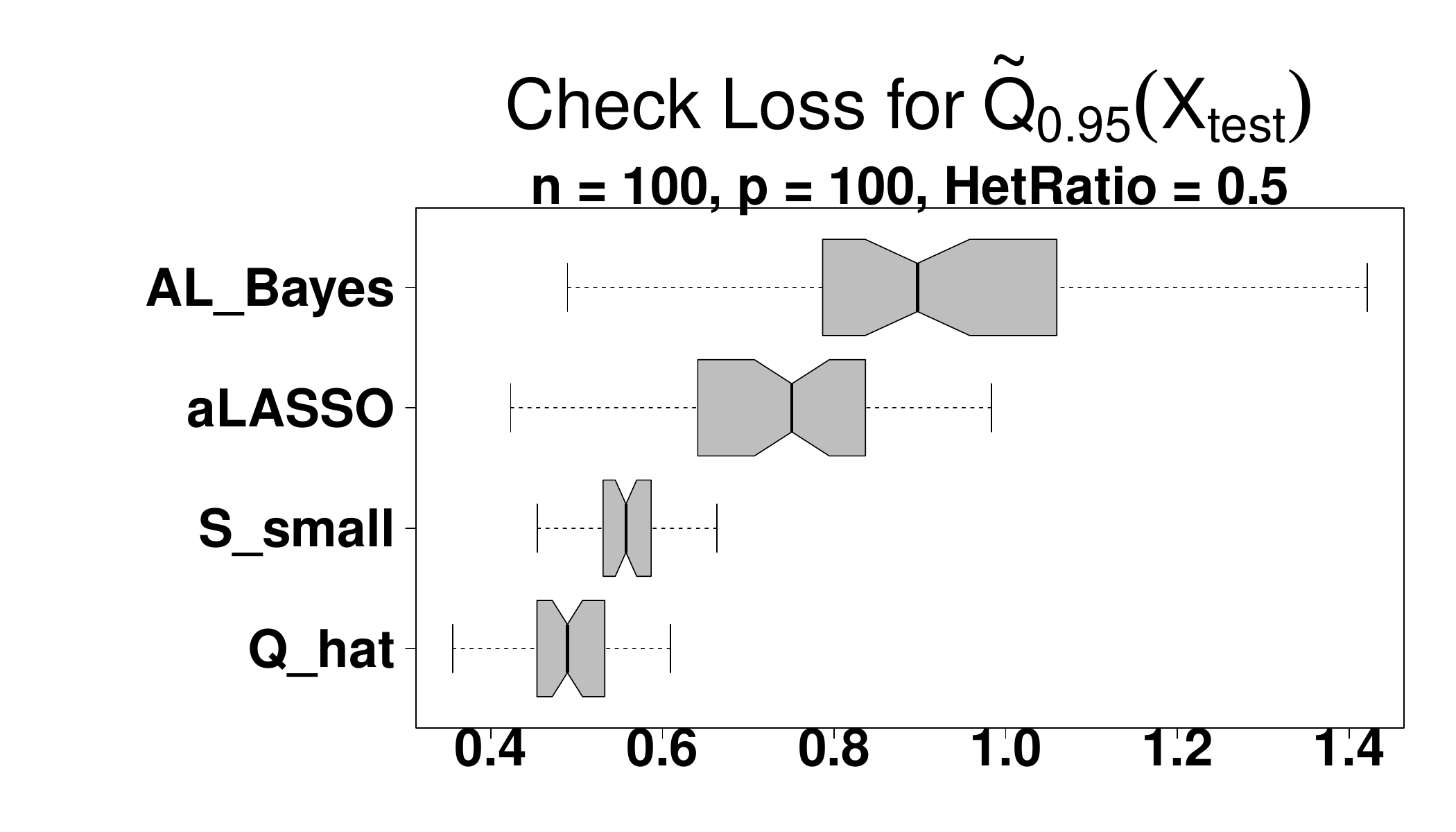}
   \includegraphics[width = .32\textwidth,keepaspectratio]{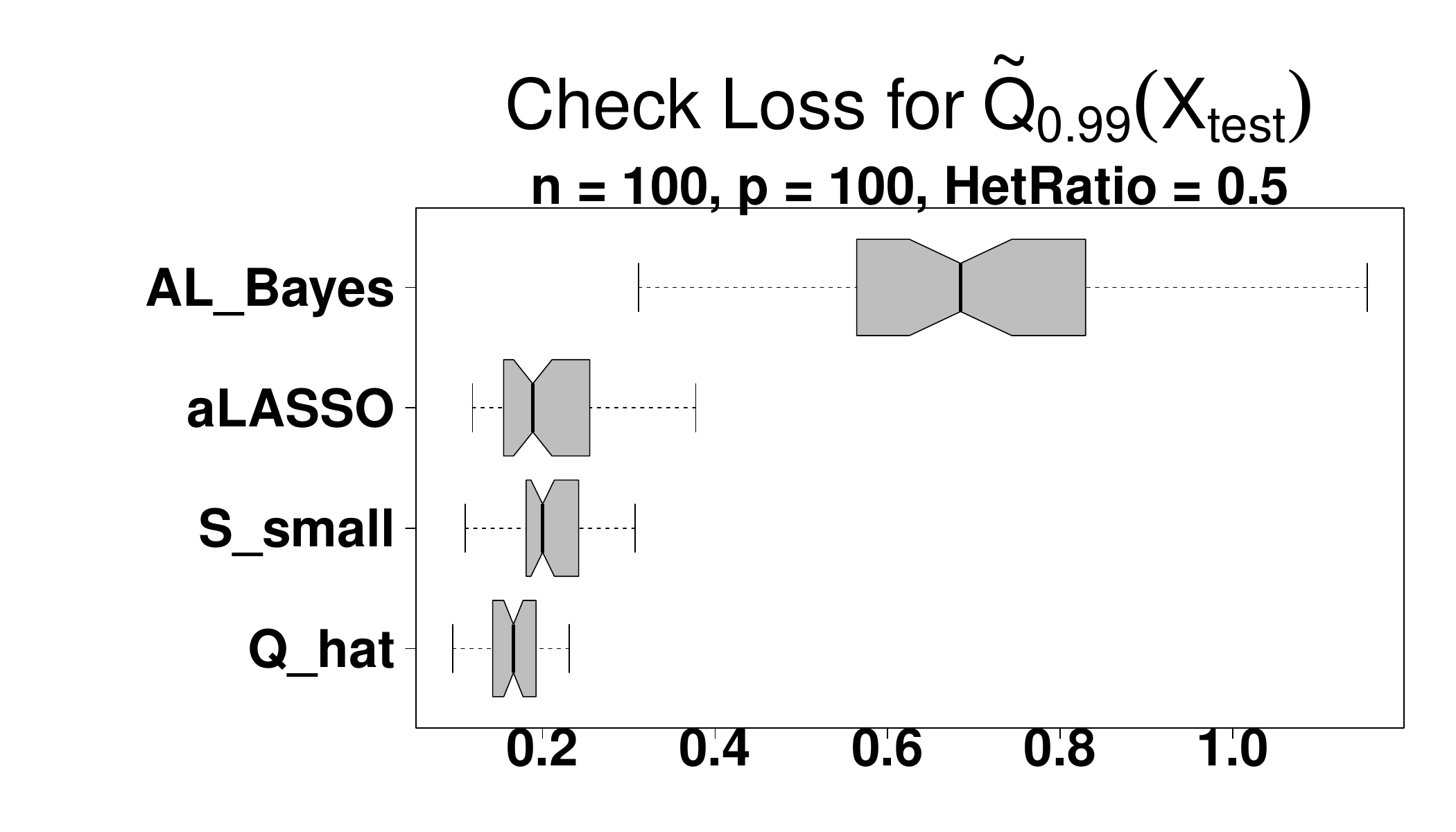}

\end{figure}

\begin{figure}[H]
    \centering
        \caption{\textbf{Check Loss}: $\boldsymbol{n = 100, p= 100, \mbox{\textbf{HetRatio} }= 1}$}
    \includegraphics[width = .32\textwidth,keepaspectratio]{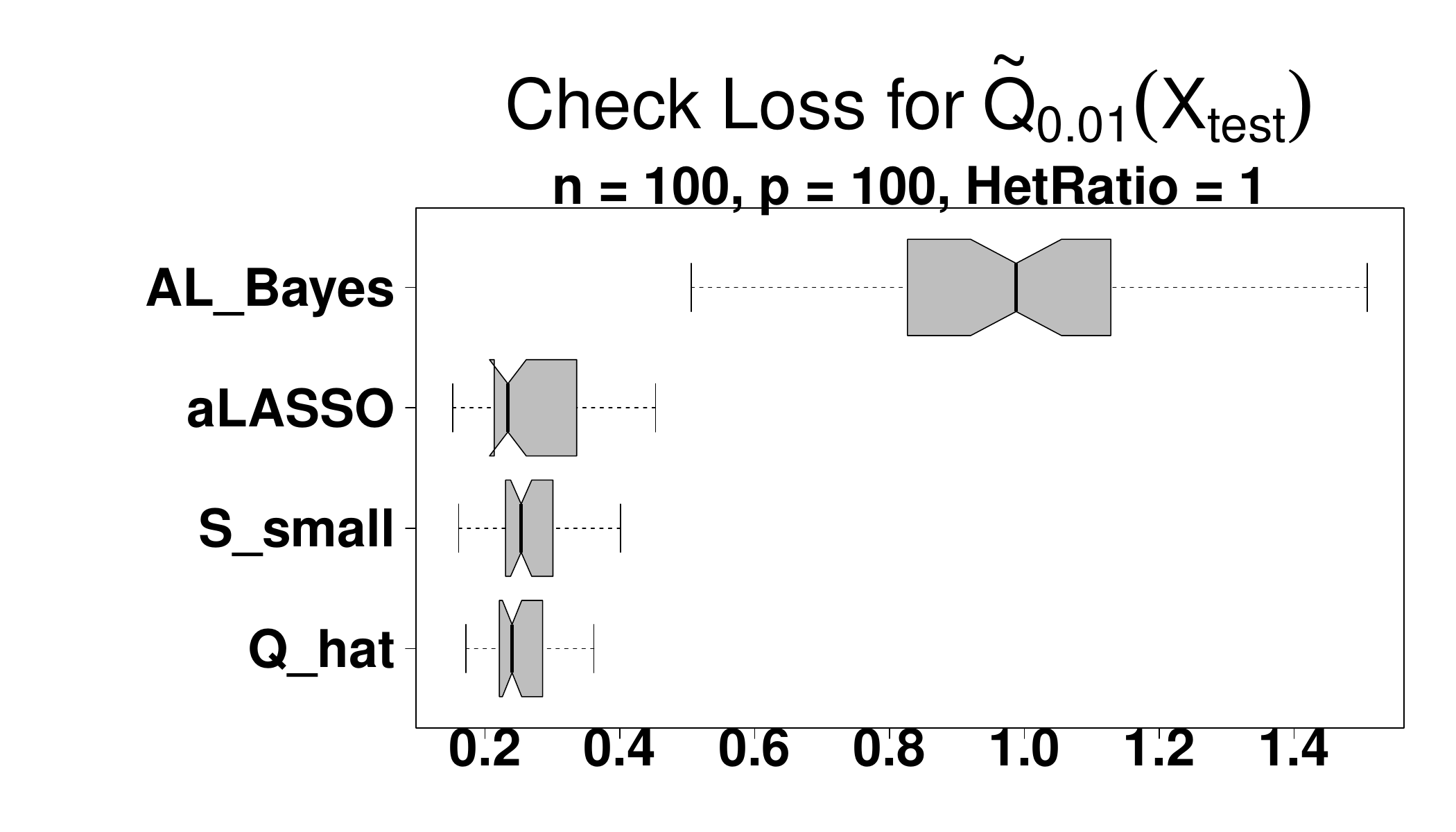}
    \includegraphics[width = .32\textwidth,keepaspectratio]{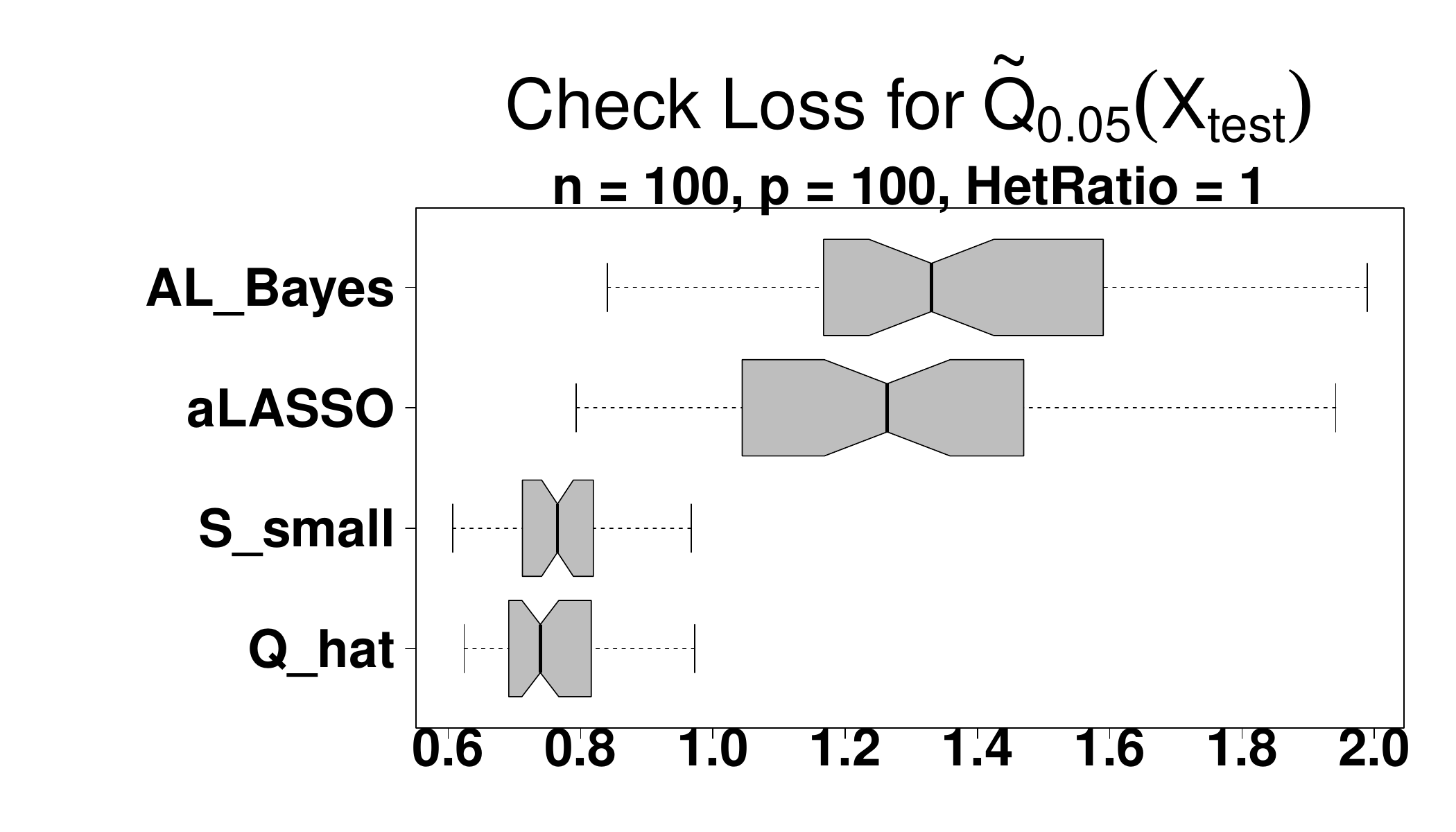}
    \includegraphics[width = .32\textwidth,keepaspectratio]{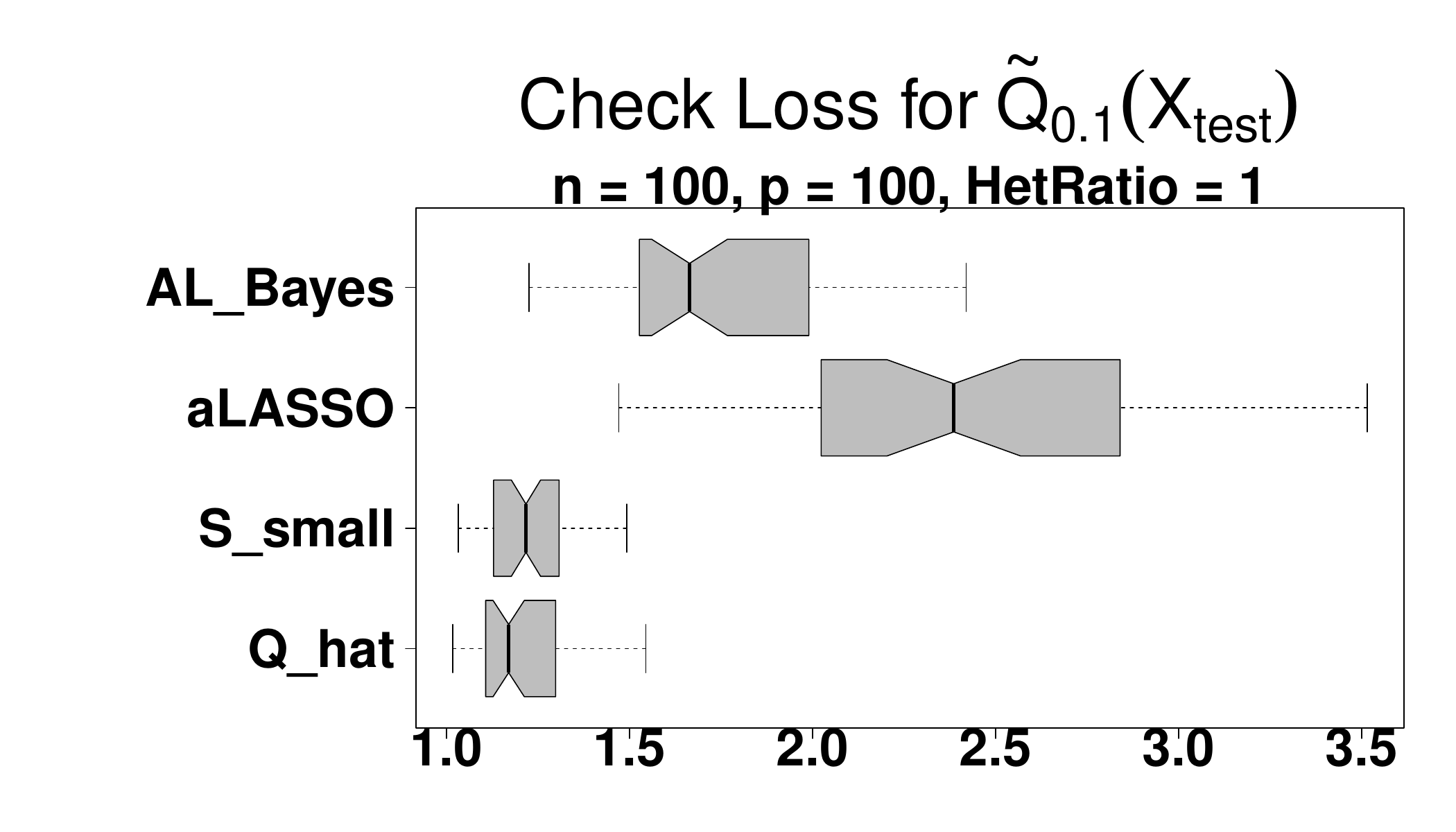}
    \includegraphics[width = .32\textwidth,keepaspectratio]{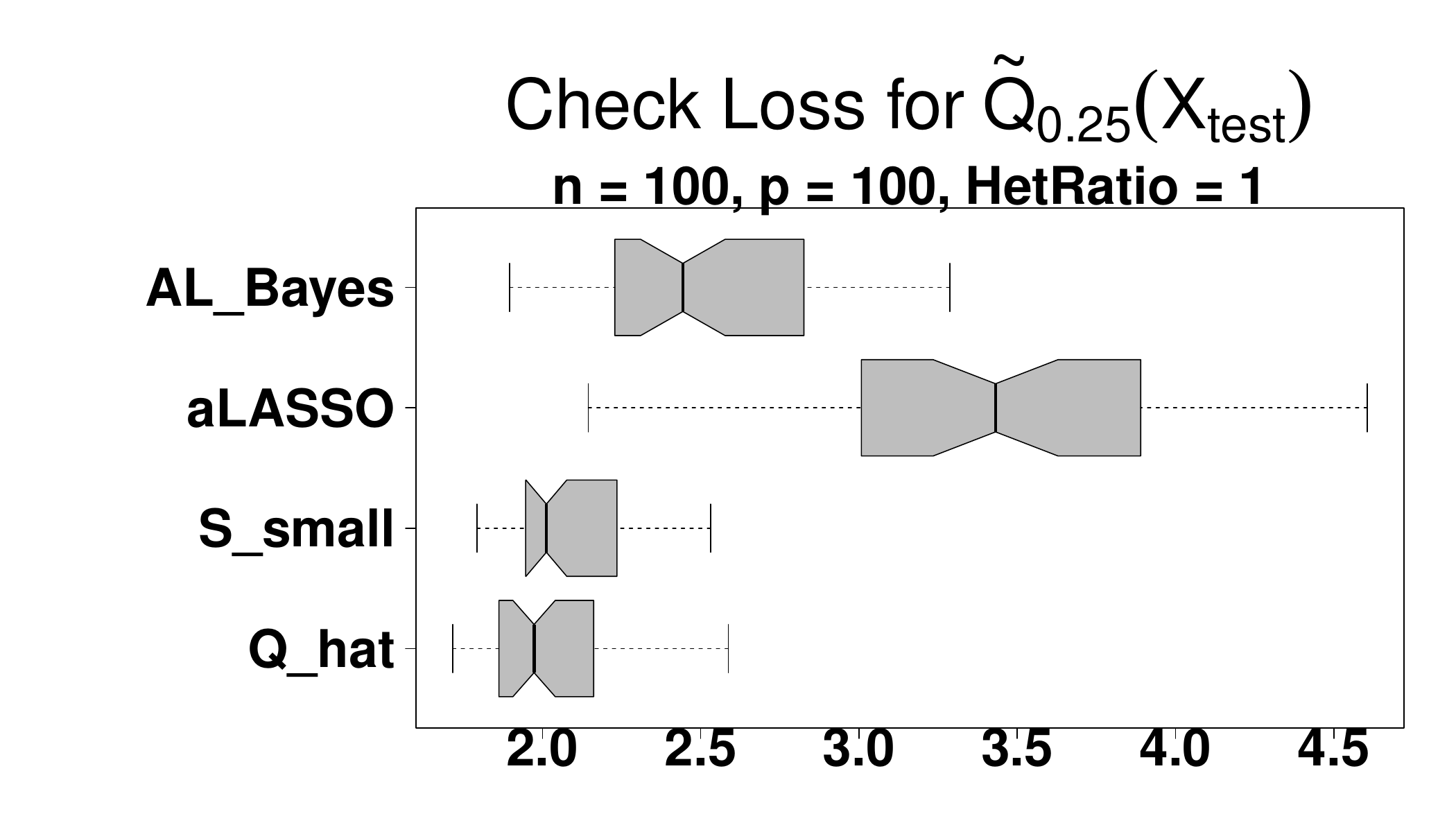}
    \includegraphics[width = .32\textwidth,keepaspectratio]{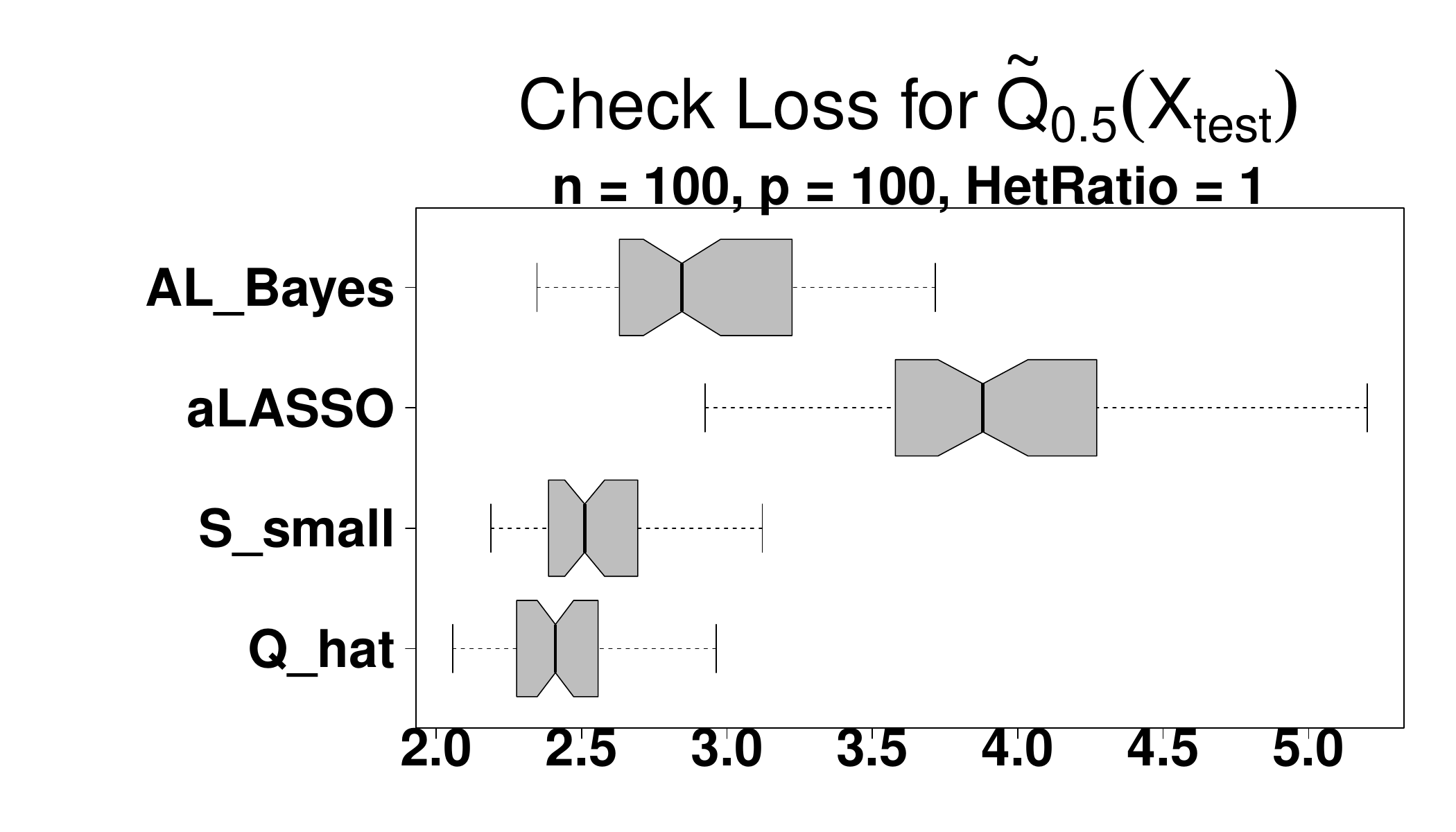}
    \includegraphics[width = .32\textwidth,keepaspectratio]{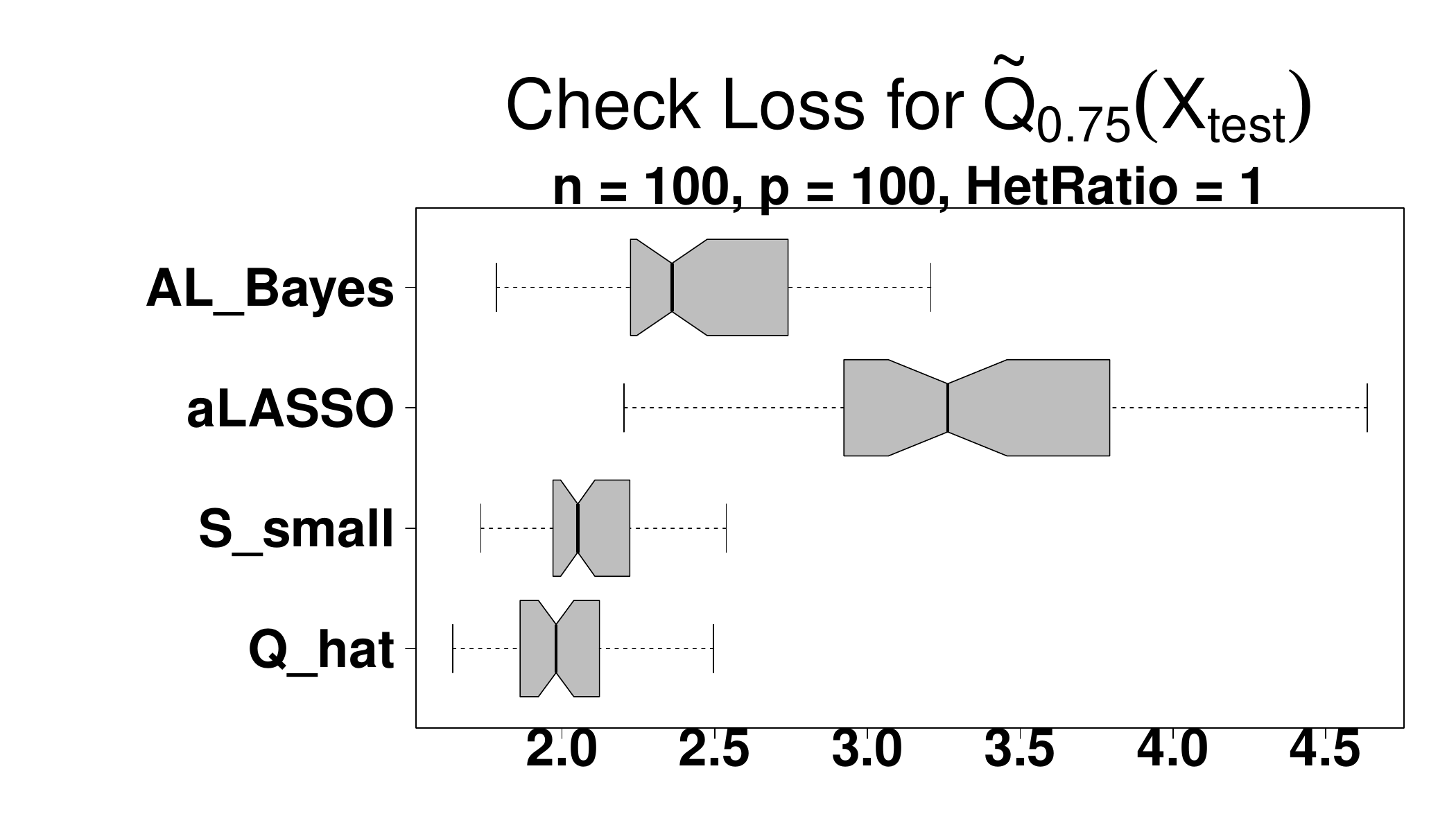}
   \includegraphics[width = .32\textwidth,keepaspectratio]{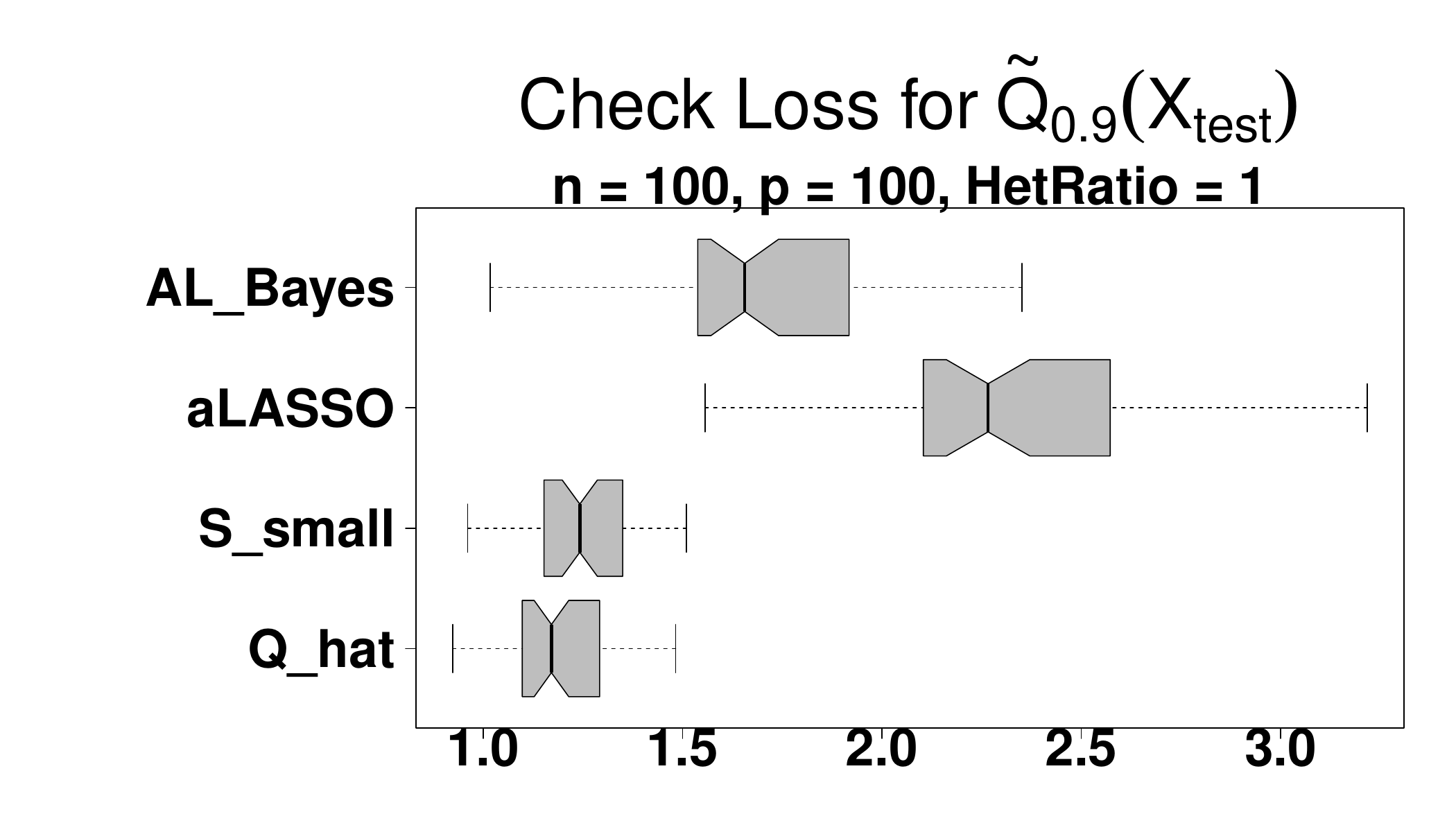}
    \includegraphics[width = .32\textwidth,keepaspectratio]{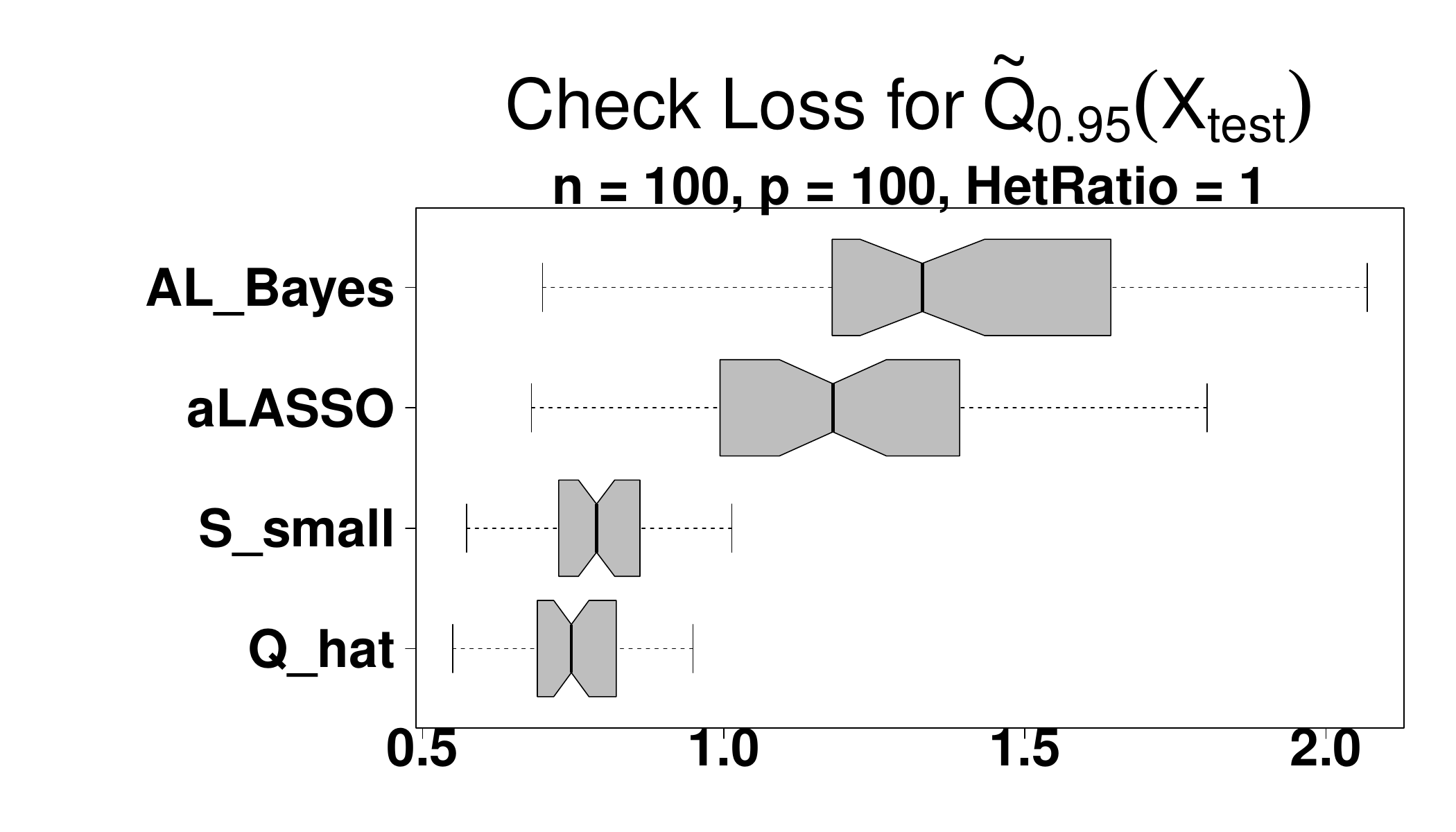}
   \includegraphics[width = .32\textwidth,keepaspectratio]{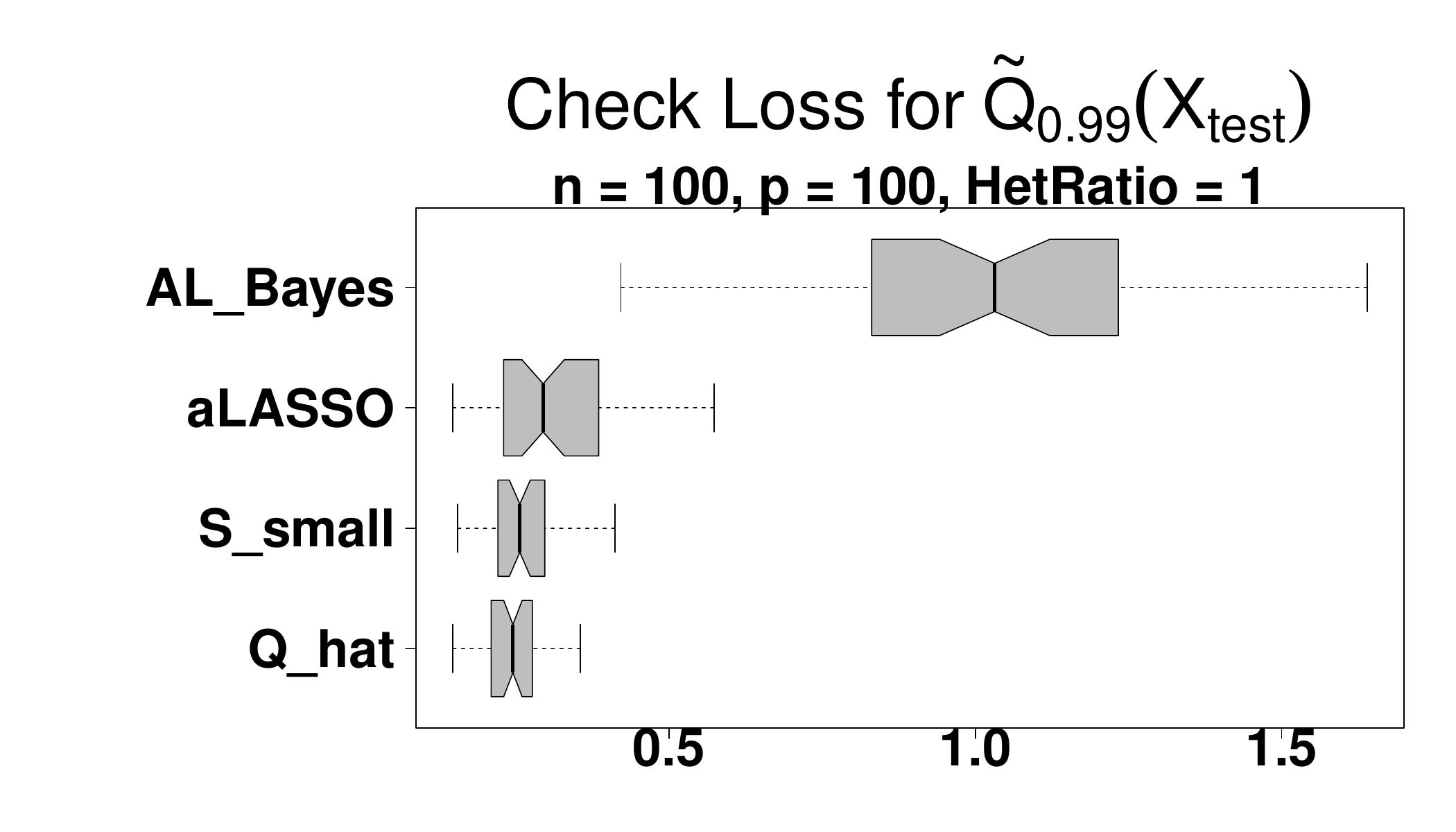}

\end{figure}

\begin{figure}[H]
    \centering
        \caption{\textbf{Check Loss}: $\boldsymbol{n = 200, p= 50, \mbox{\textbf{HetRatio} }= 0.5}$}
  
    \includegraphics[width = .32\textwidth,keepaspectratio]{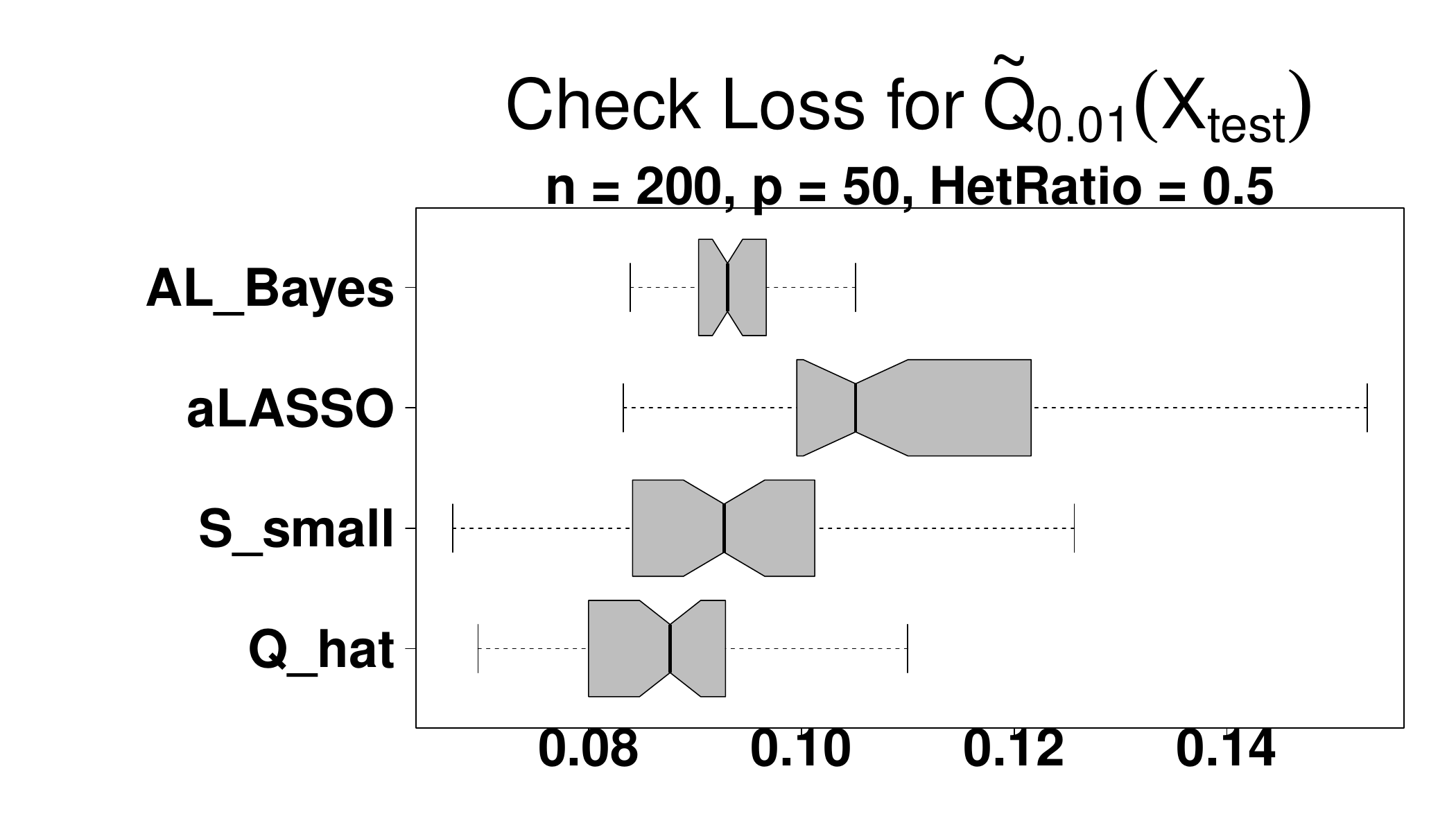}
    \includegraphics[width = .32\textwidth,keepaspectratio]{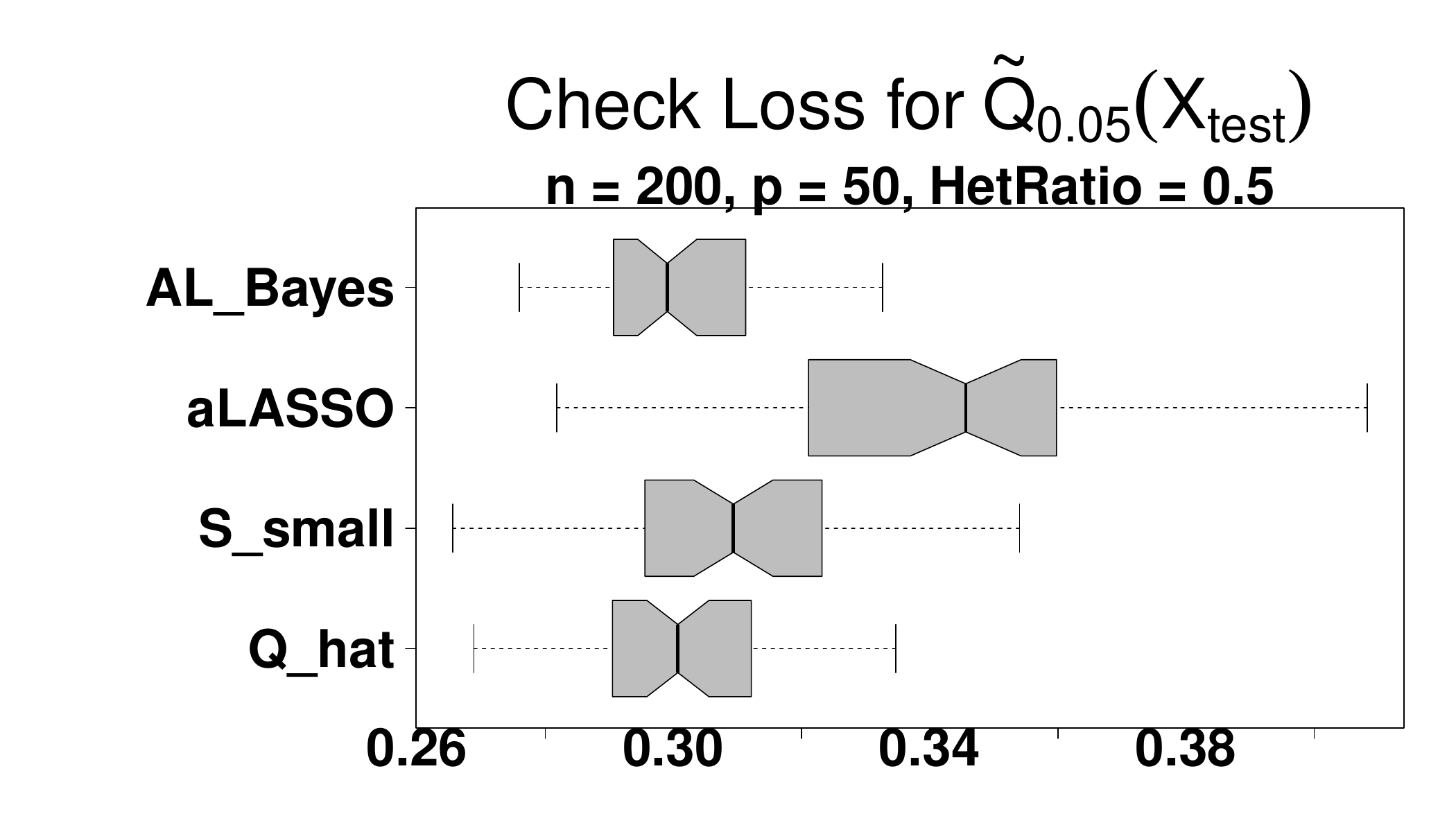}
    \includegraphics[width = .32\textwidth,keepaspectratio]{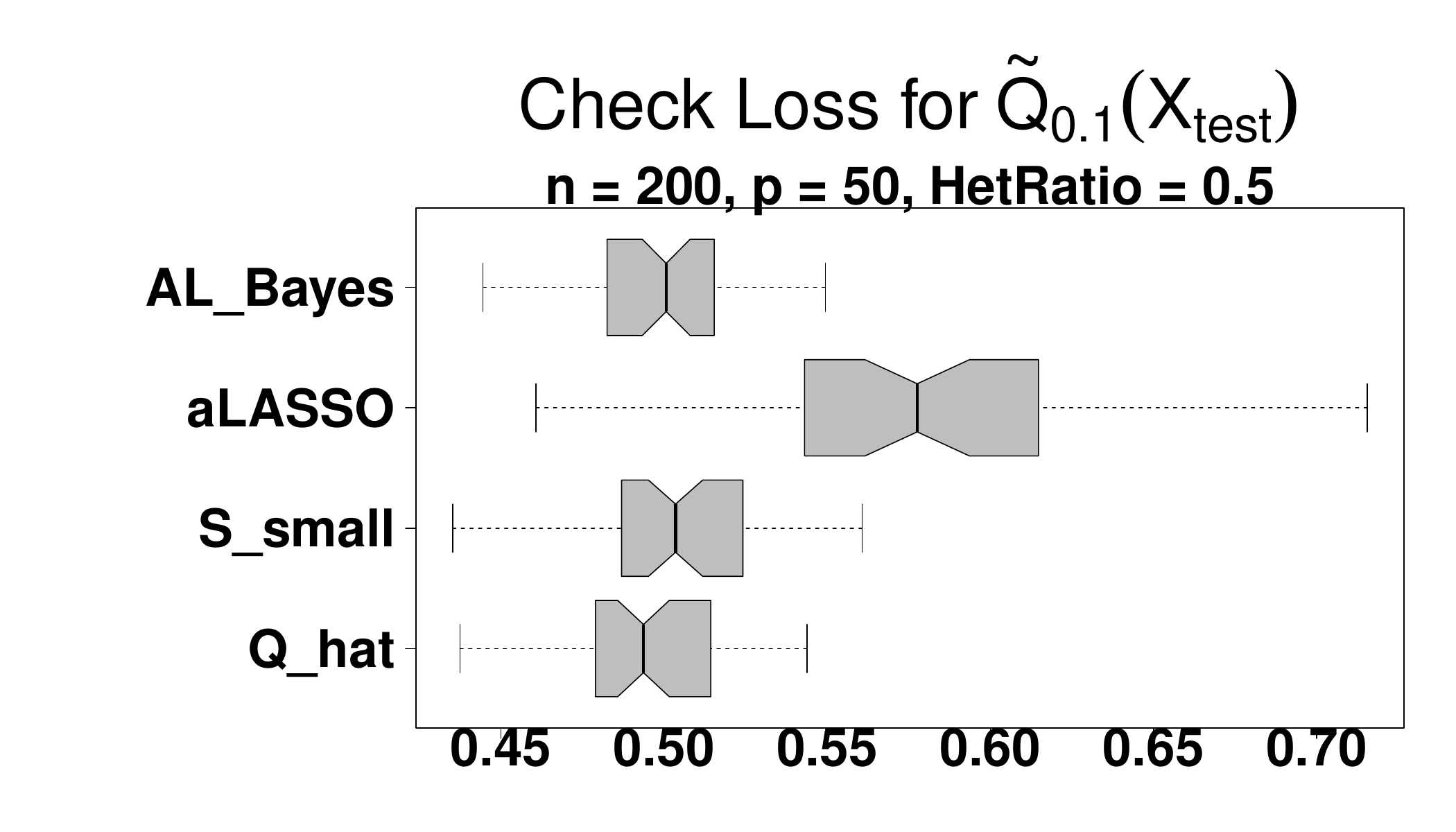}
    \includegraphics[width = .32\textwidth,keepaspectratio]{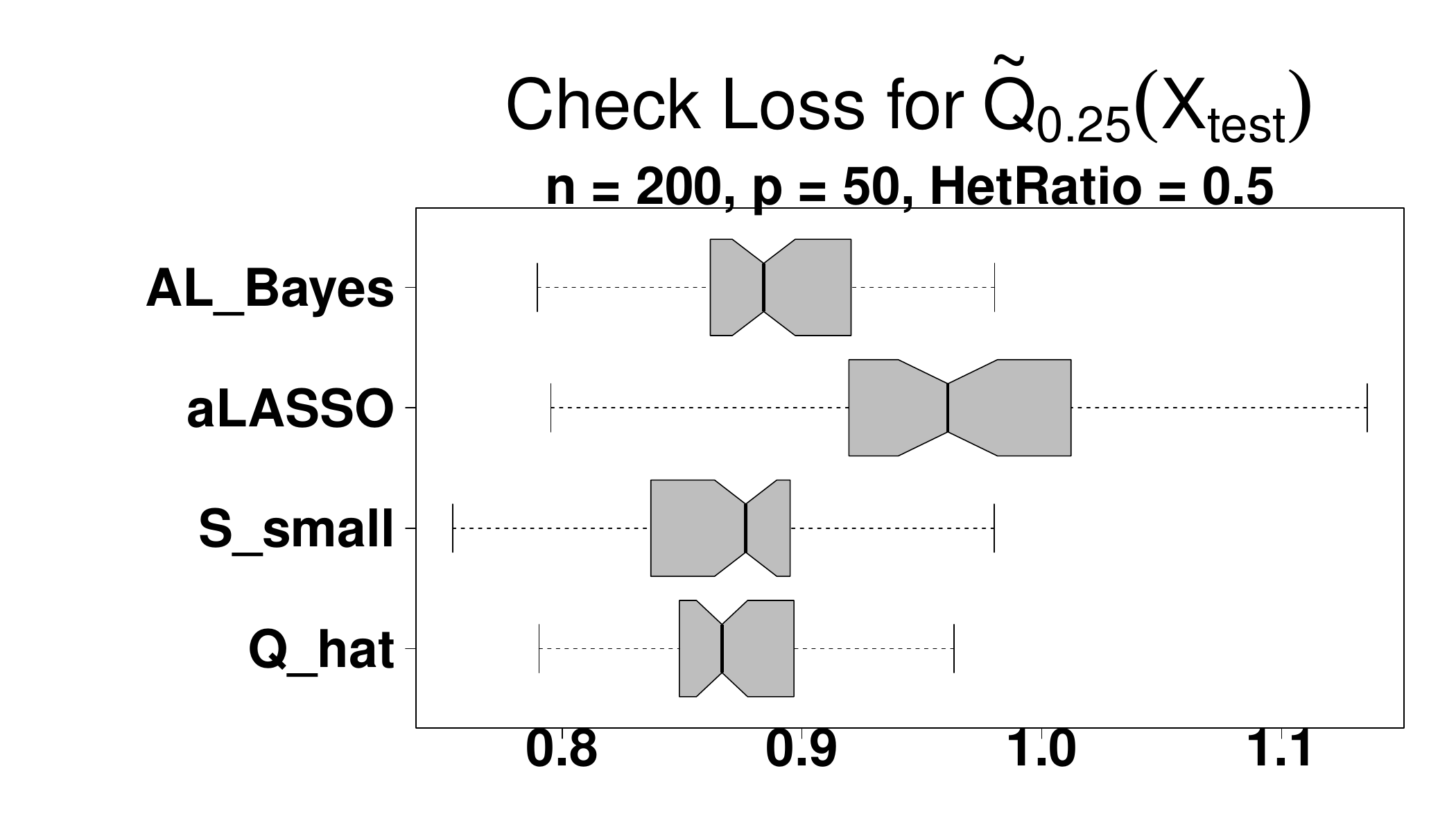}
    \includegraphics[width = .32\textwidth,keepaspectratio]{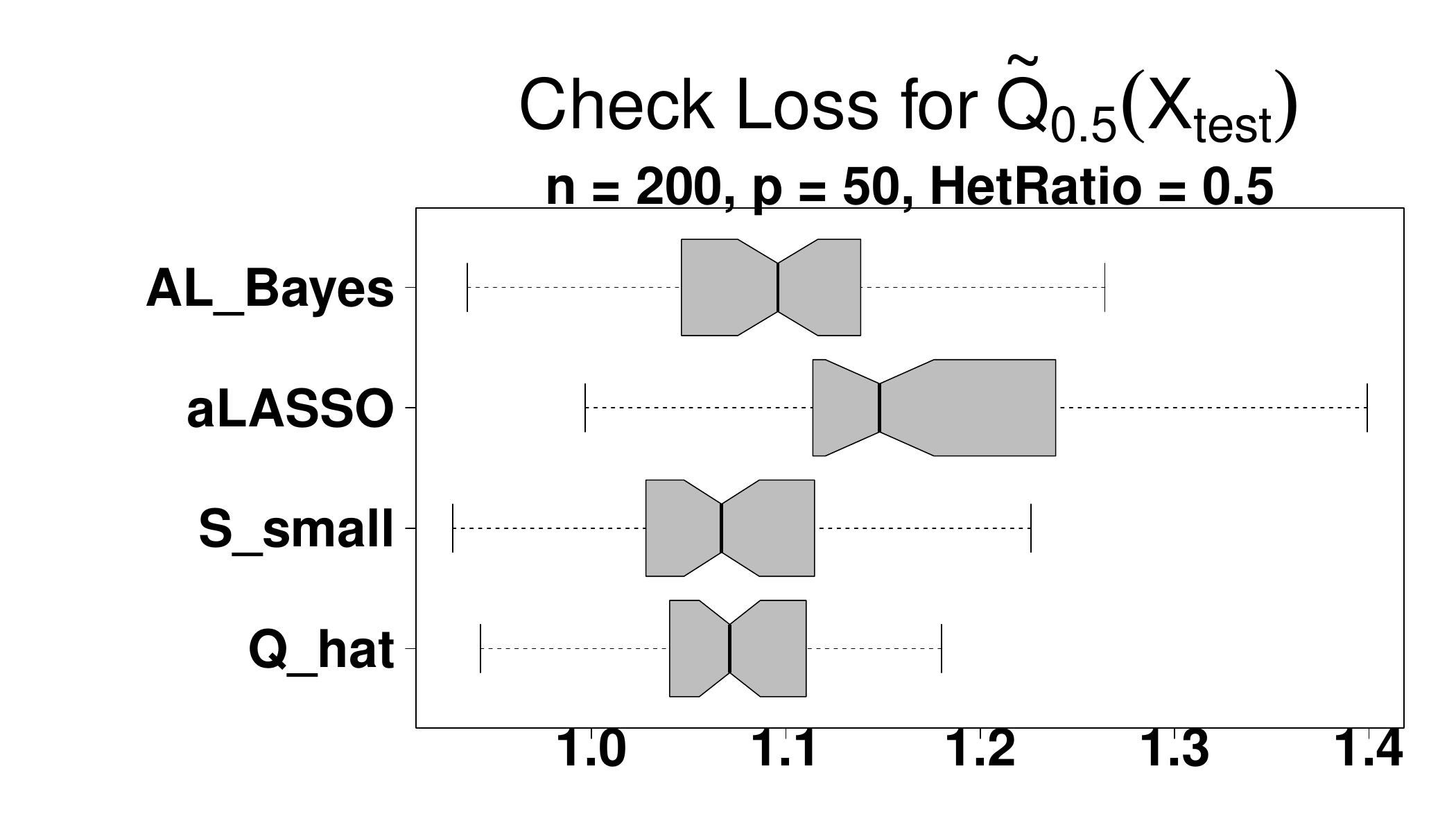}
    \includegraphics[width = .32\textwidth,keepaspectratio]{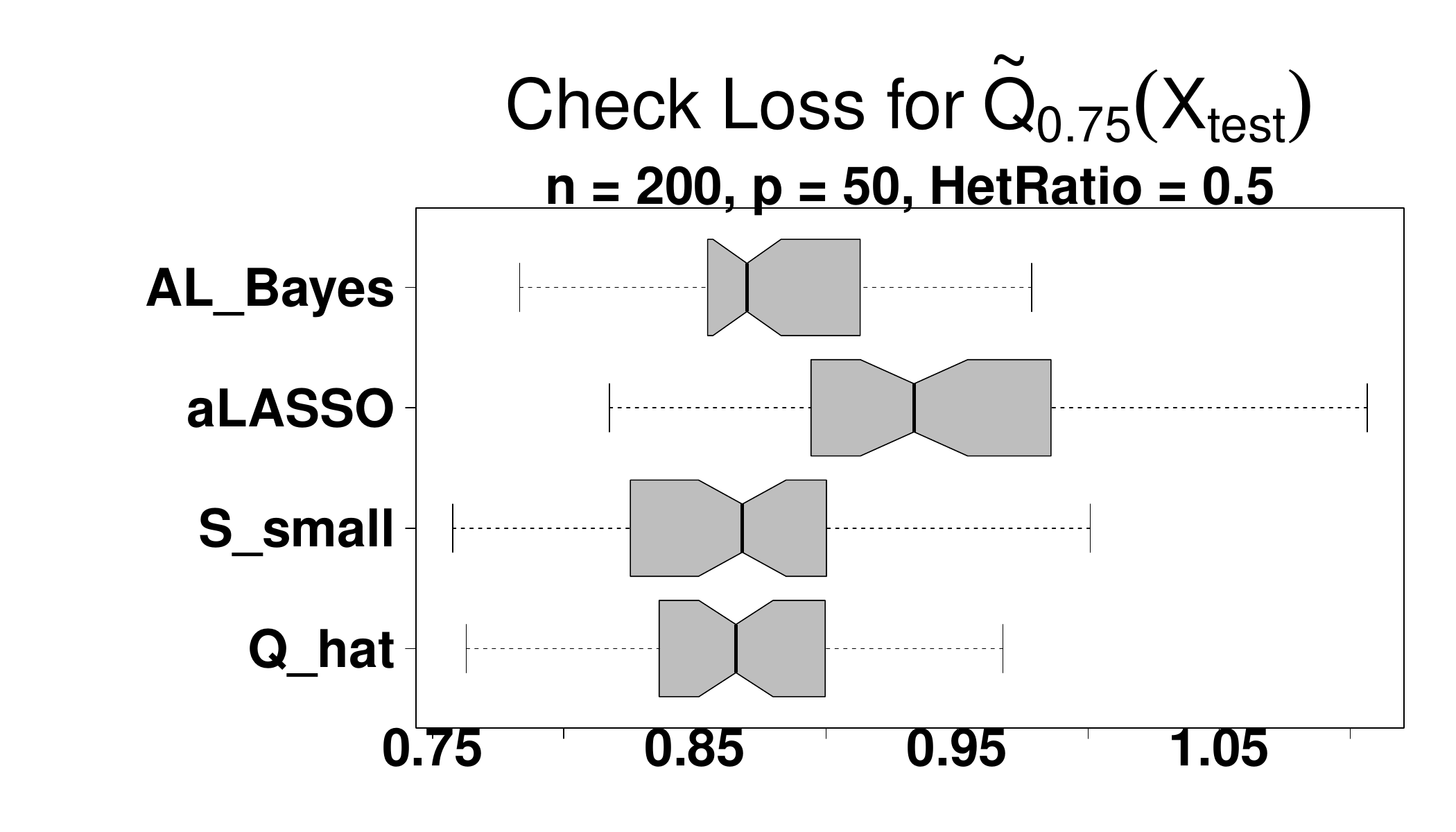}
   \includegraphics[width = .32\textwidth,keepaspectratio]{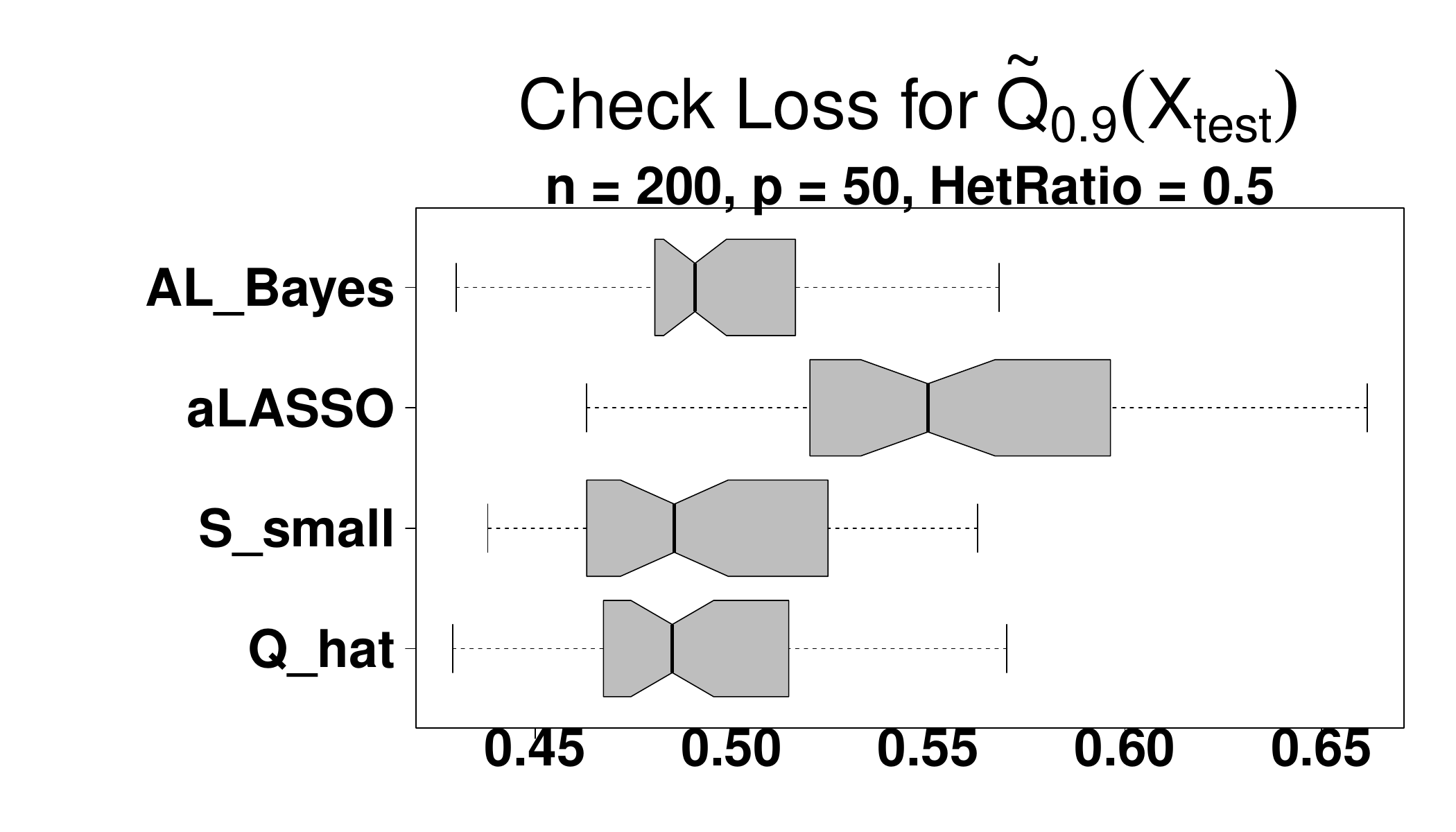}
    \includegraphics[width = .32\textwidth,keepaspectratio]{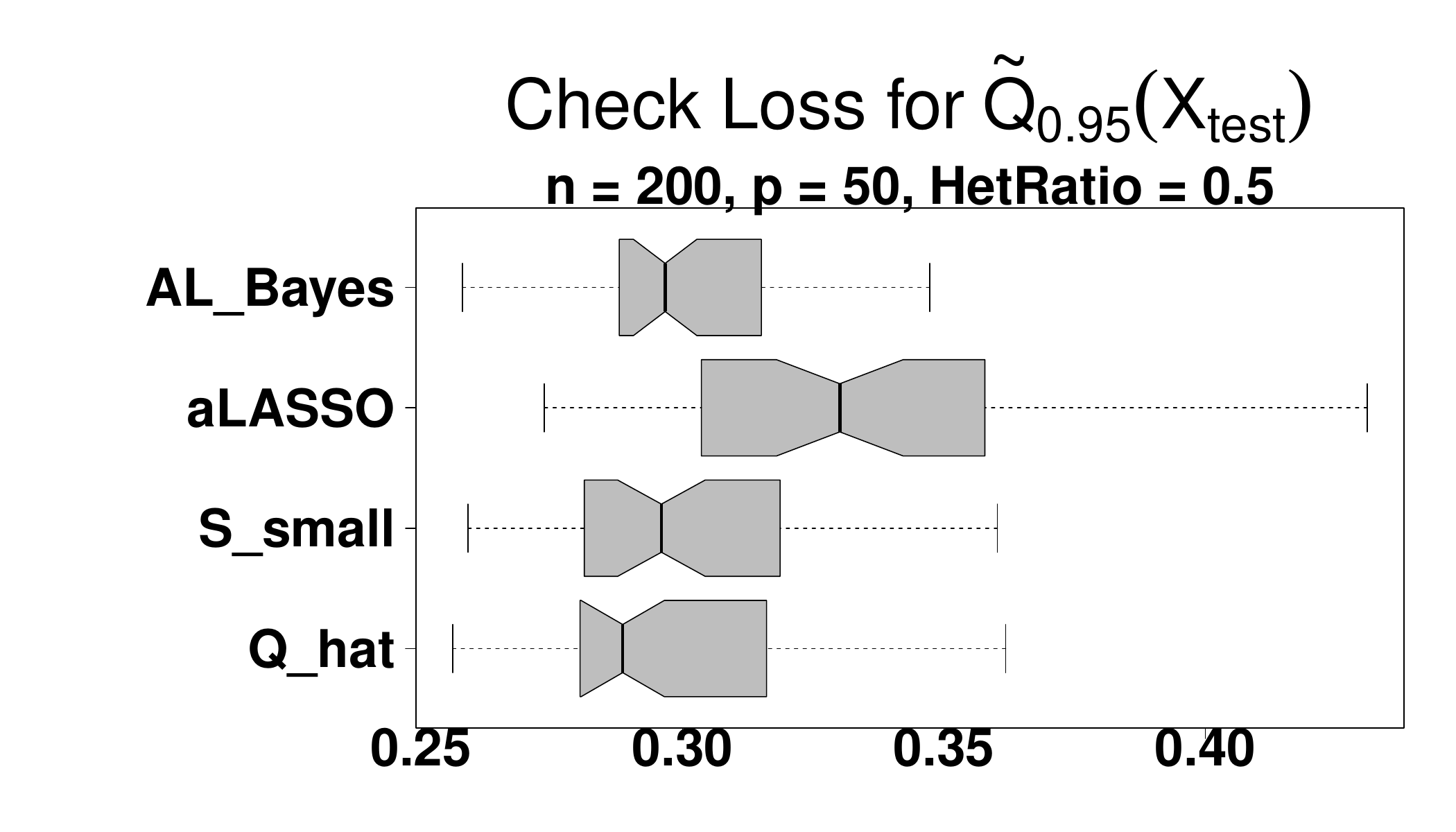}
   \includegraphics[width = .32\textwidth,keepaspectratio]{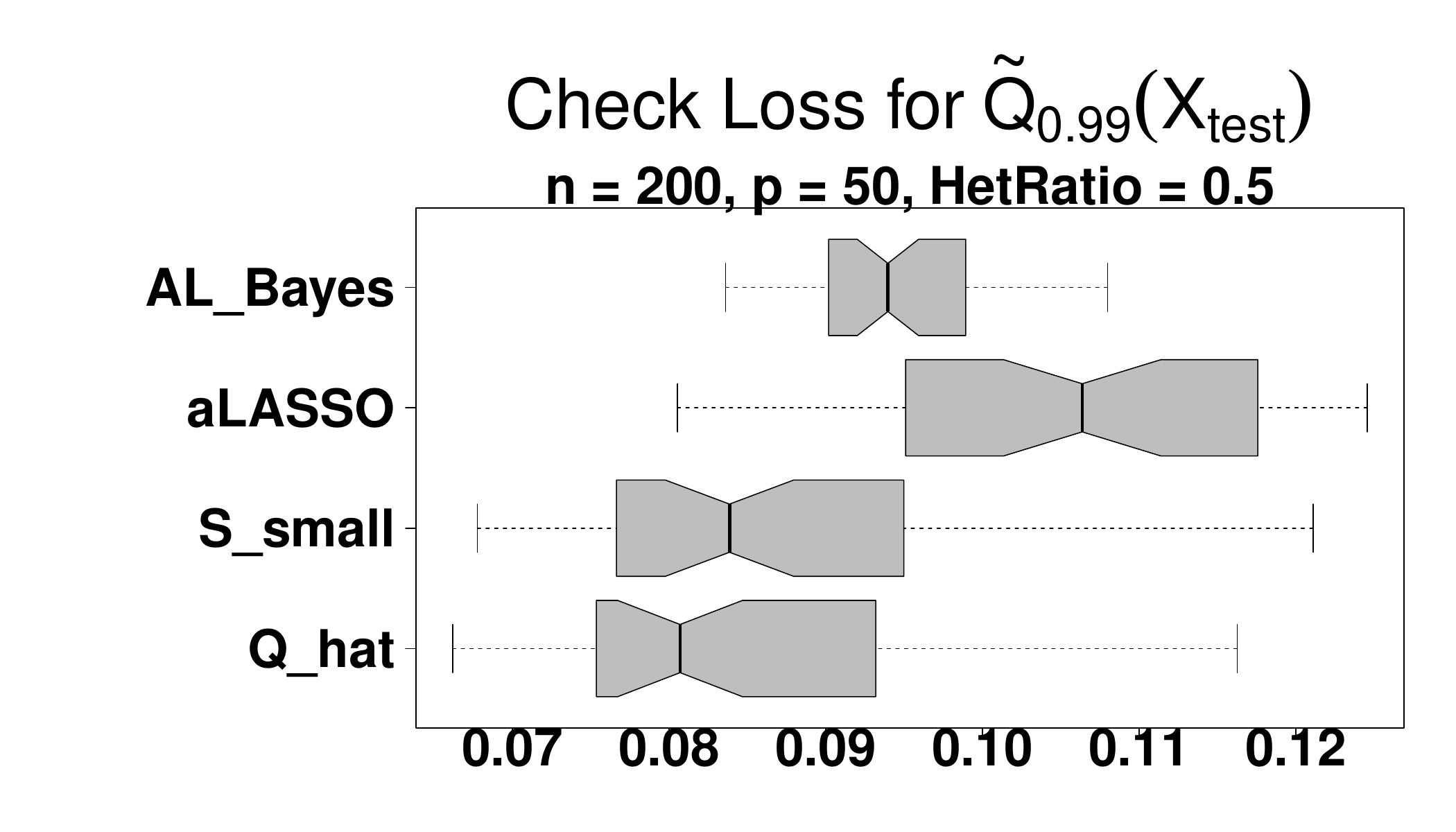}
\end{figure}

\begin{figure}[H]
    \centering
        \caption{\textbf{Check Loss}: $\boldsymbol{n = 200, p= 50, \mbox{\textbf{HetRatio} }= 1}$}
  
    \includegraphics[width = .32\textwidth,keepaspectratio]{images/n200_p50_cl_SNR1_1.pdf}
    \includegraphics[width = .32\textwidth,keepaspectratio]{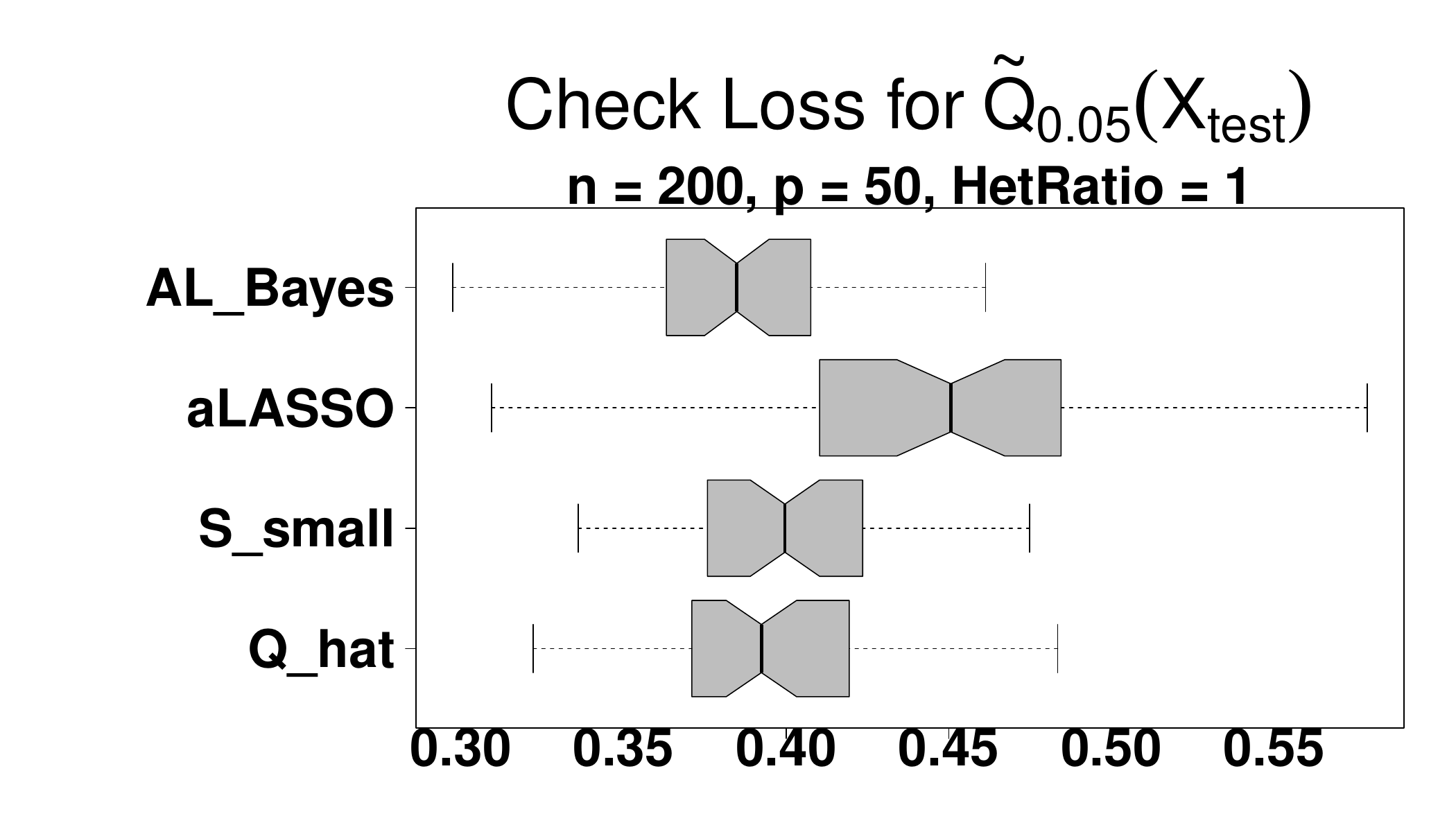}
    \includegraphics[width = .32\textwidth,keepaspectratio]{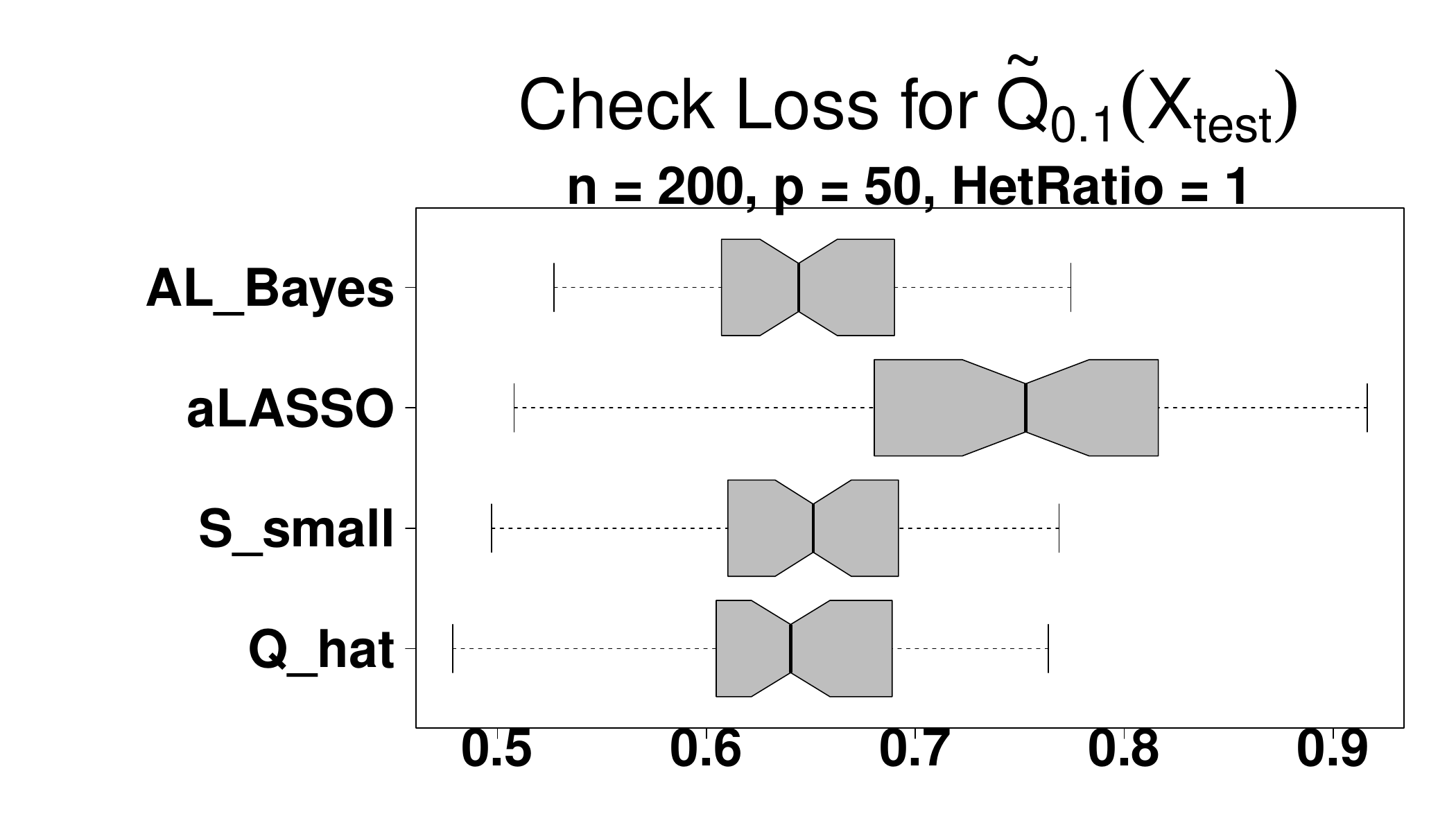}
    \includegraphics[width = .32\textwidth,keepaspectratio]{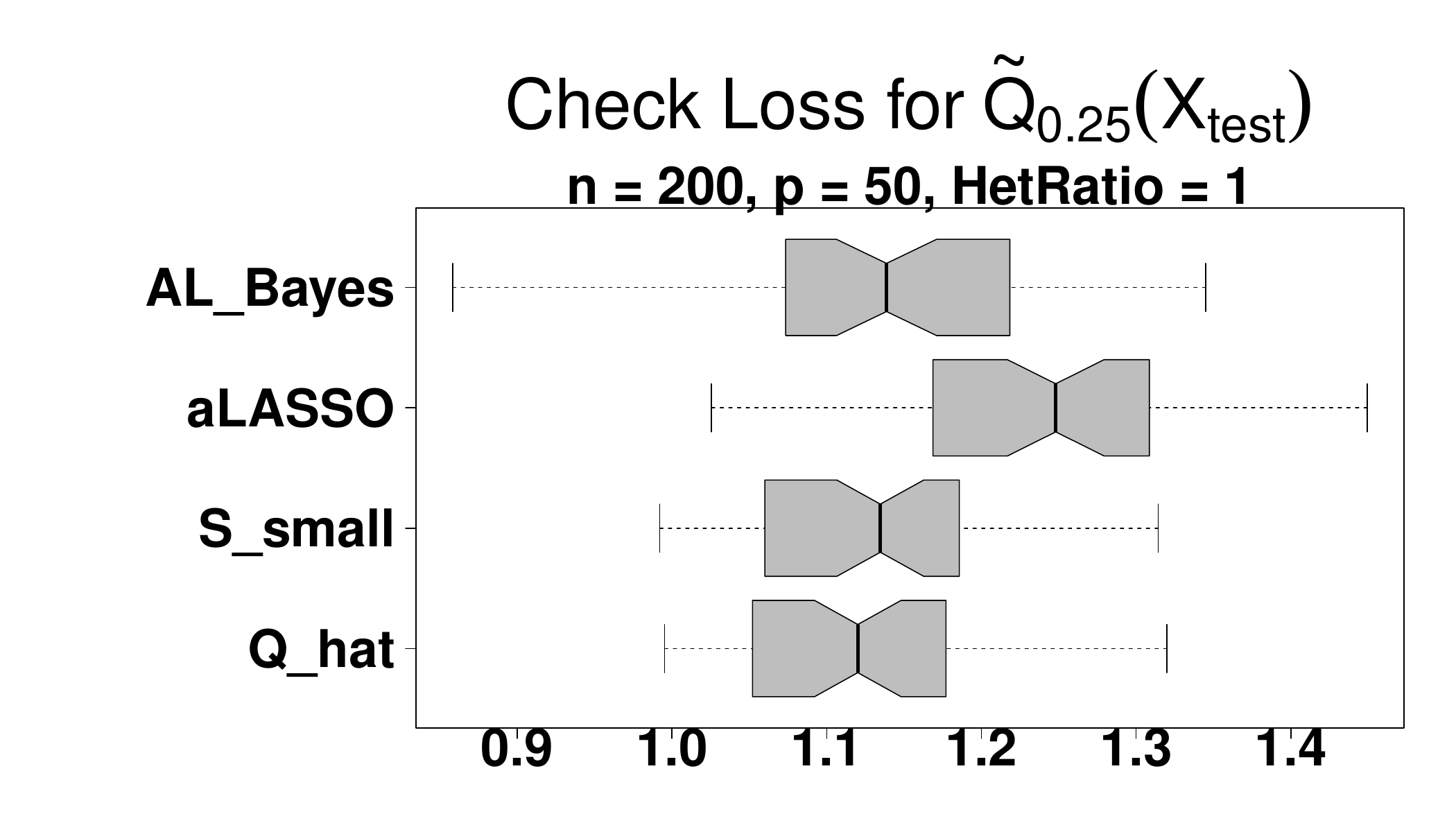}
    \includegraphics[width = .32\textwidth,keepaspectratio]{images/n200_p50_cl_SNR1_50.pdf}
    \includegraphics[width = .32\textwidth,keepaspectratio]{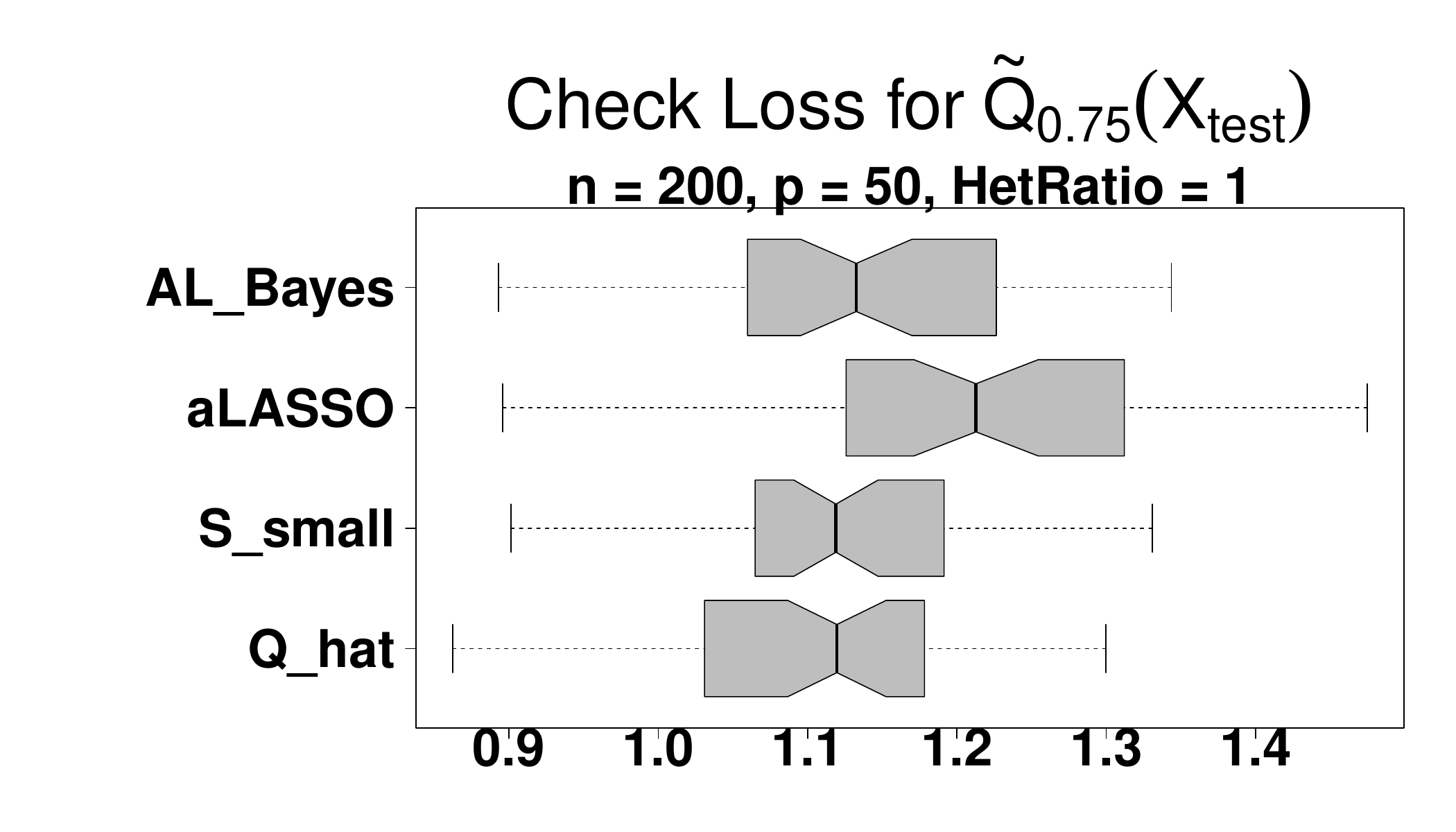}
   \includegraphics[width = .32\textwidth,keepaspectratio]{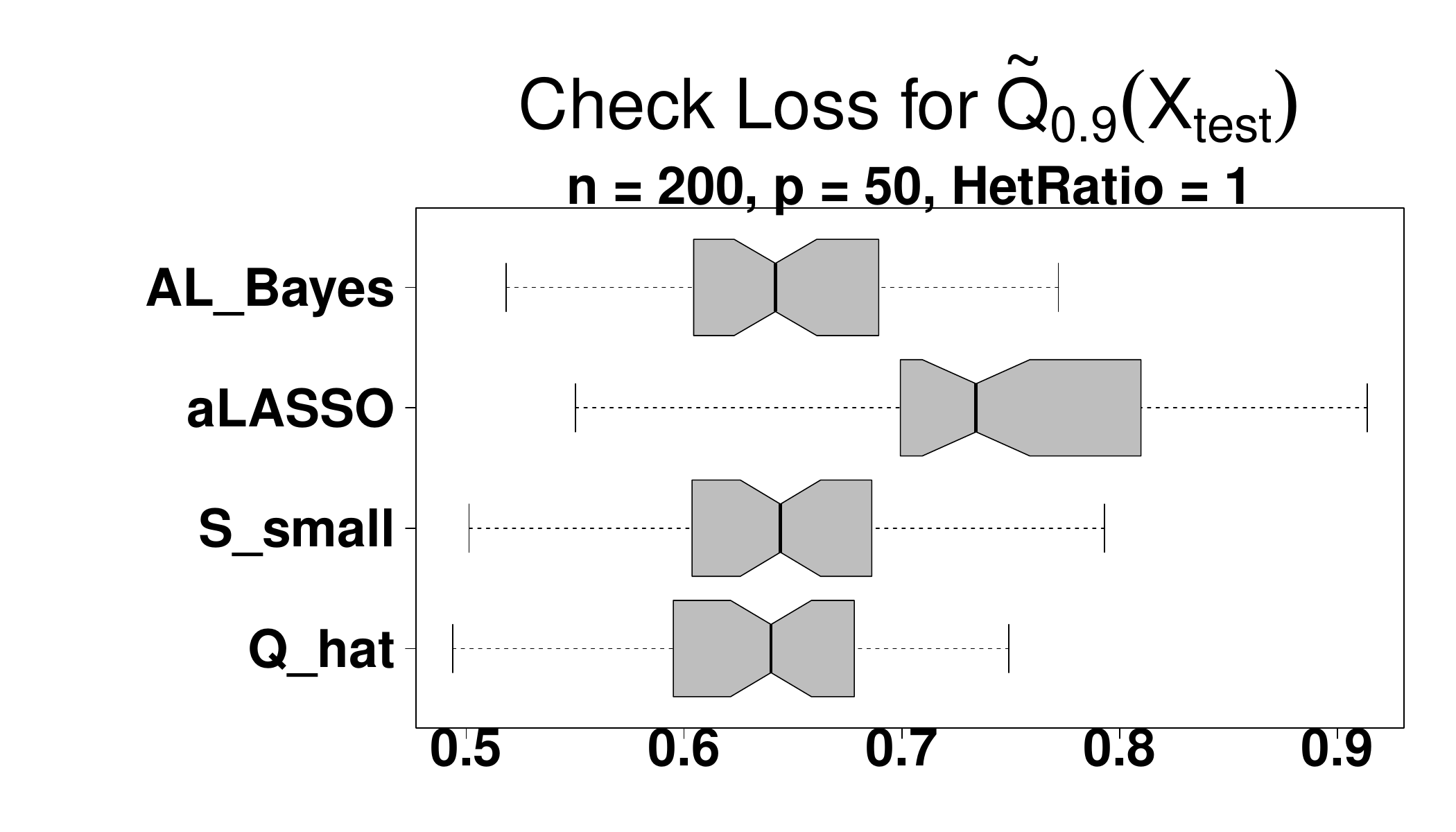}
    \includegraphics[width = .32\textwidth,keepaspectratio]{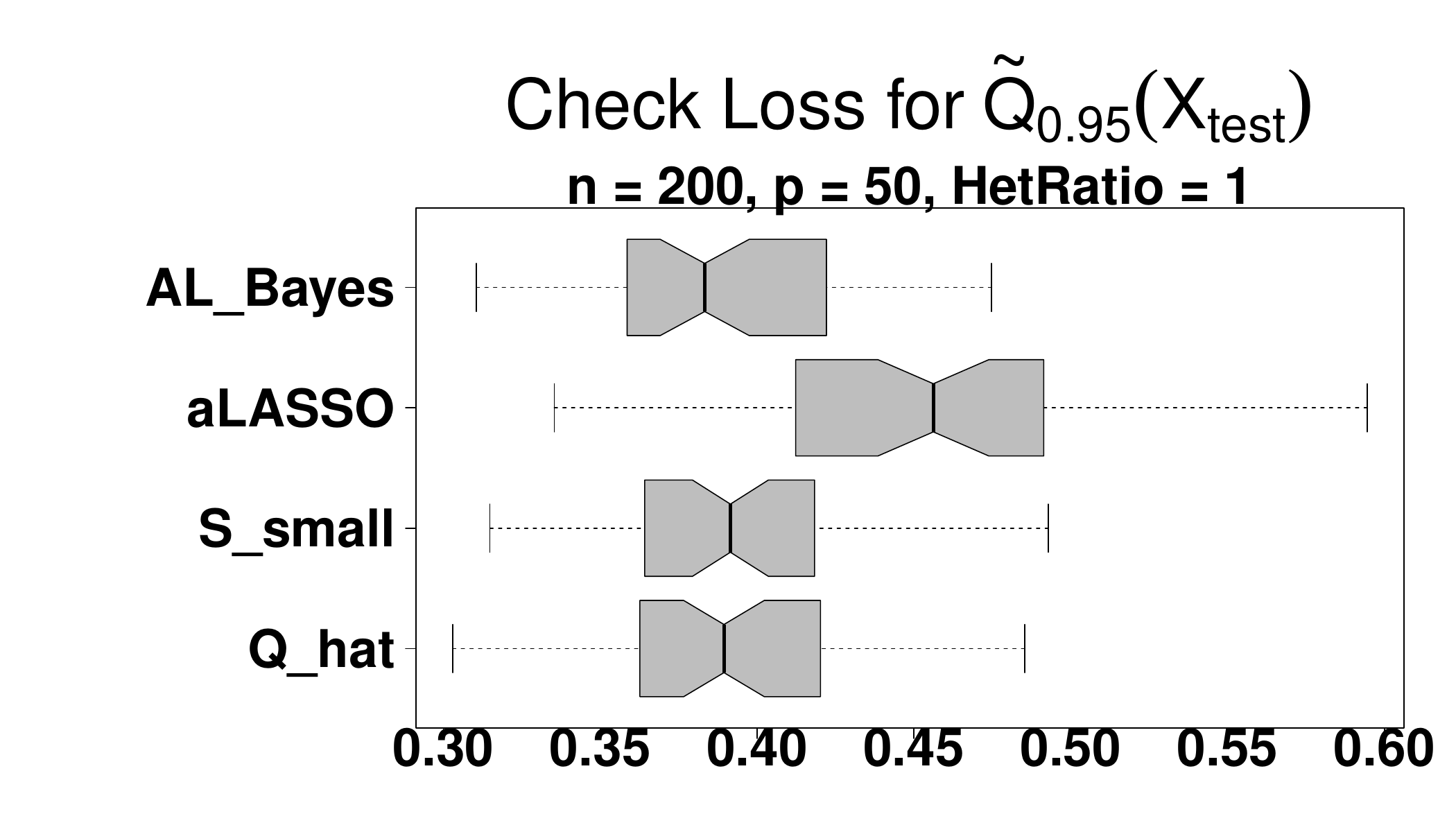}
   \includegraphics[width = .32\textwidth,keepaspectratio]{images/n200_p50_cl_SNR1_99.pdf}

    \label{fig:enter-label}
\end{figure}
\begin{figure}[H]
    \centering
        \caption*{\textbf{Calibration}: $\boldsymbol{n = 500, p= 20, \mbox{\textbf{HetRatio} }= 0.5}$}
  
    \includegraphics[width = .32\textwidth,keepaspectratio]{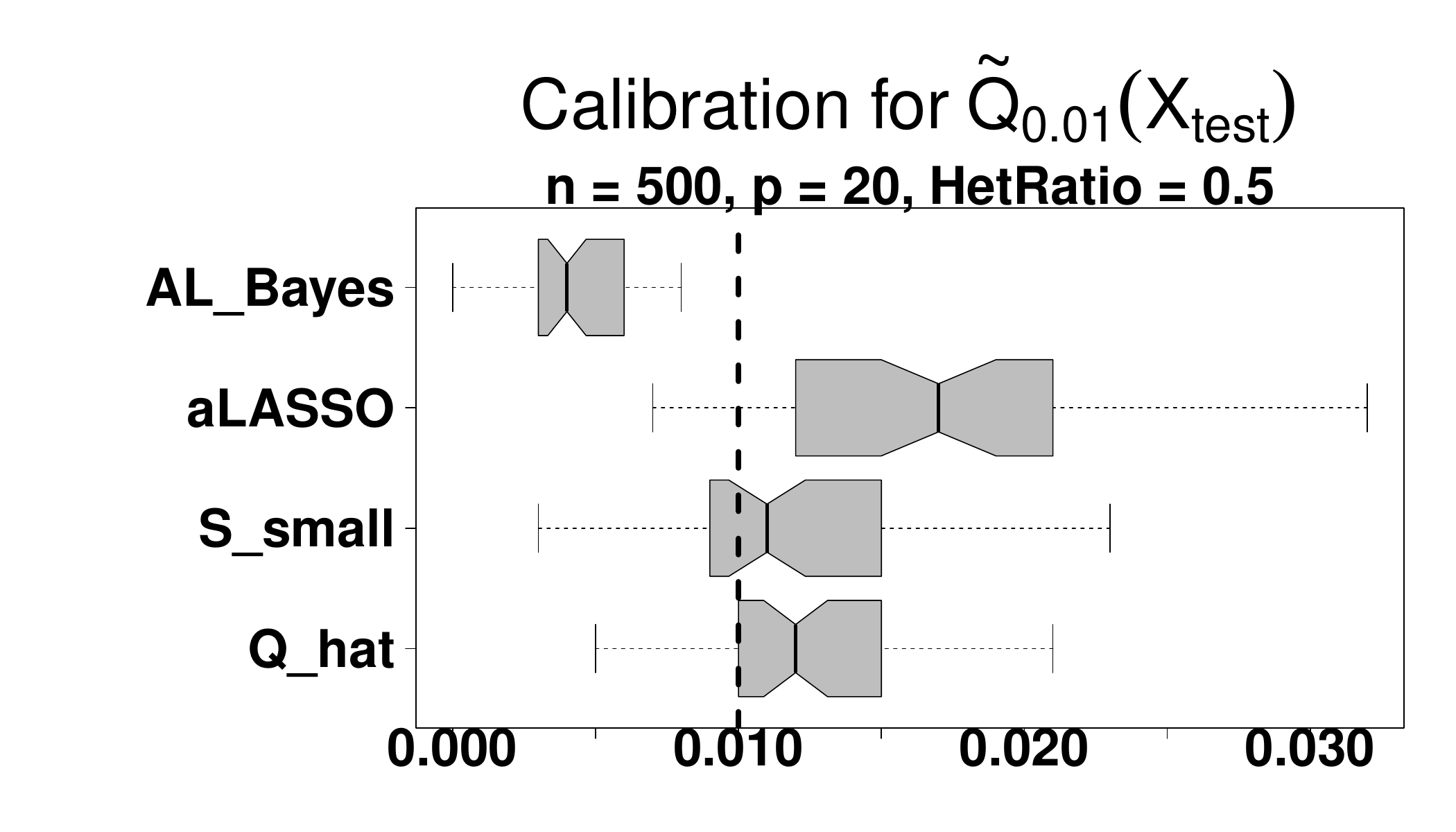}
    \includegraphics[width = .32\textwidth,keepaspectratio]{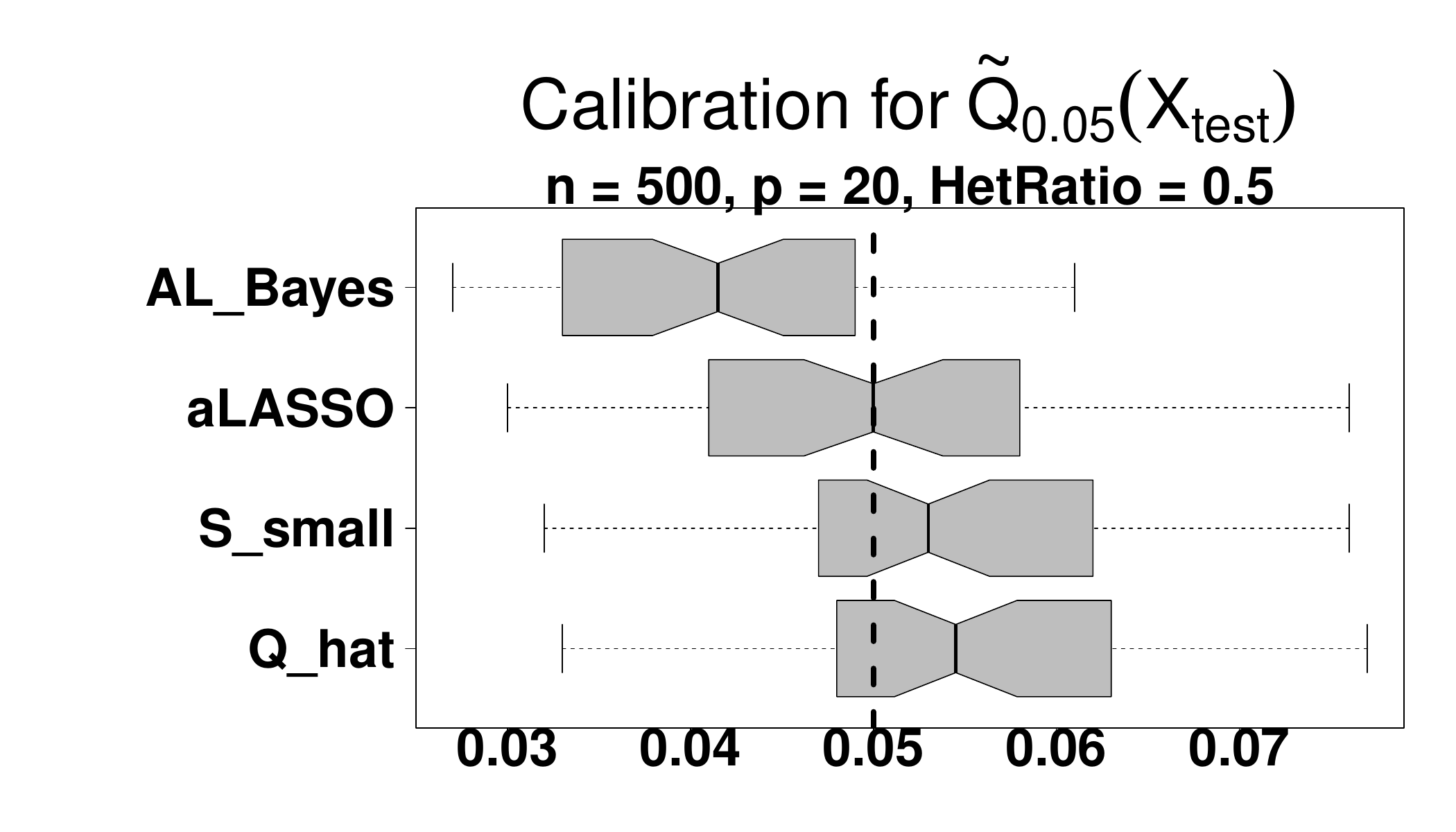}
    \includegraphics[width = .32\textwidth,keepaspectratio]{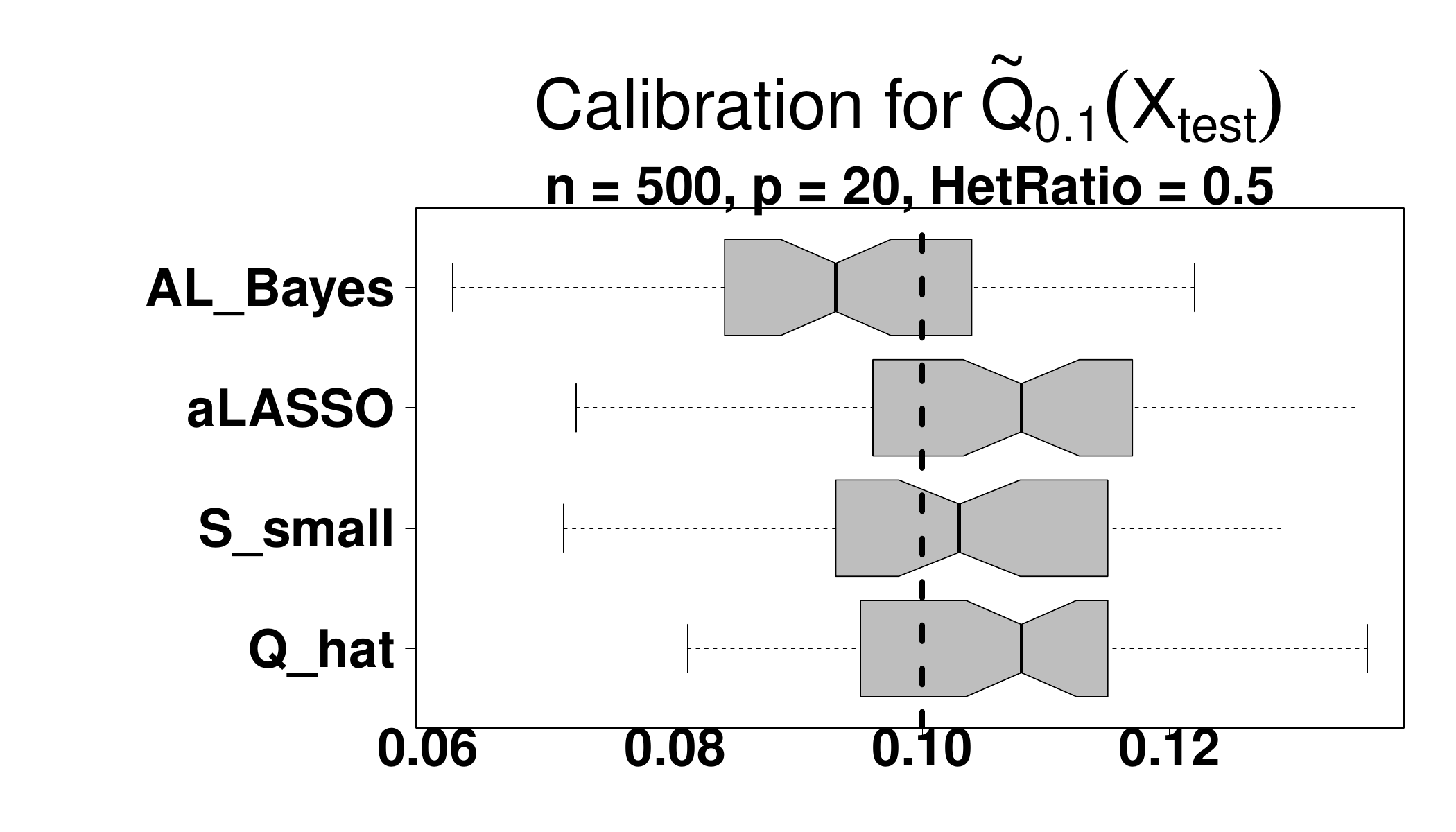}
    \includegraphics[width = .32\textwidth,keepaspectratio]{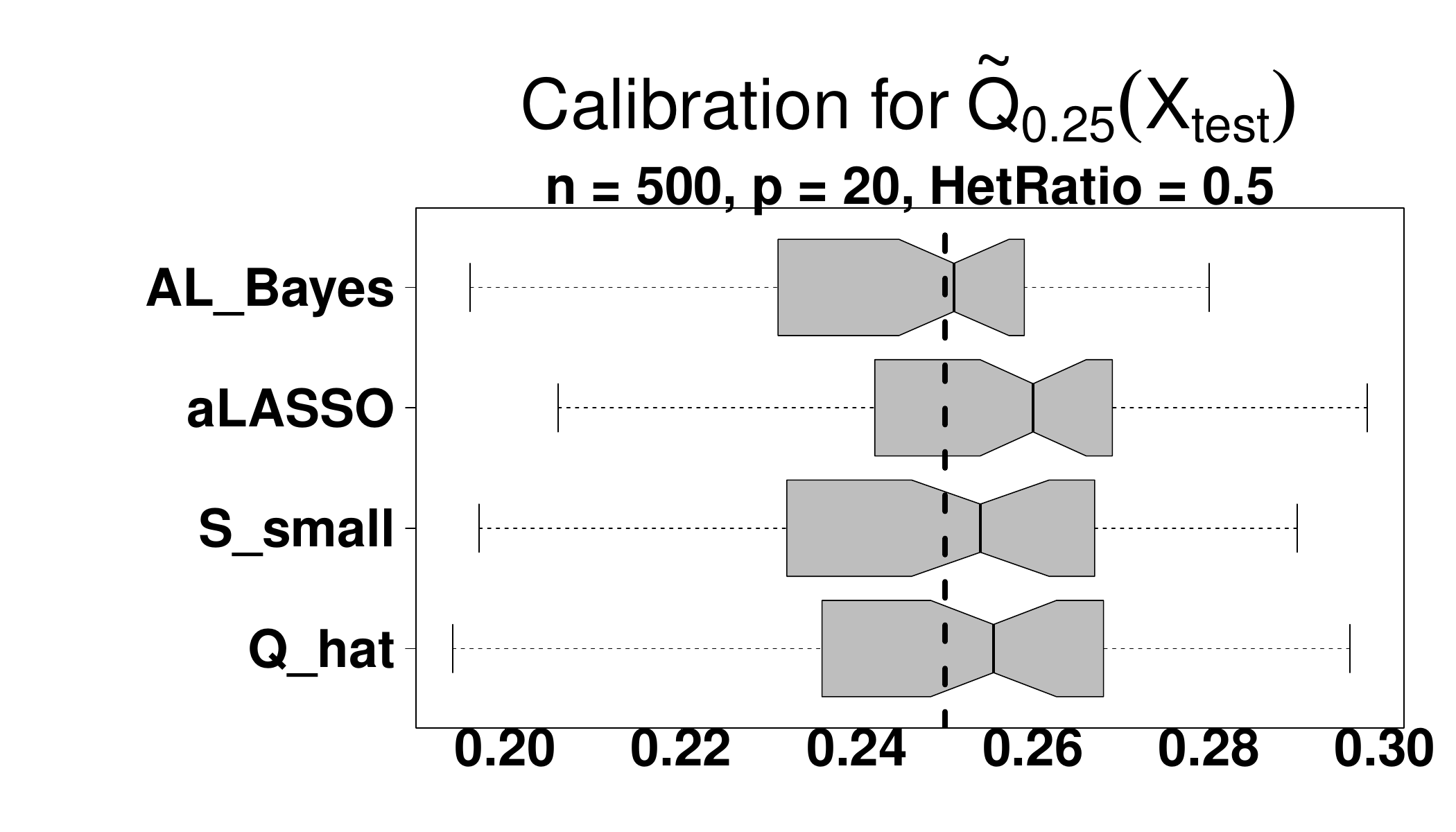}
    \includegraphics[width = .32\textwidth,keepaspectratio]{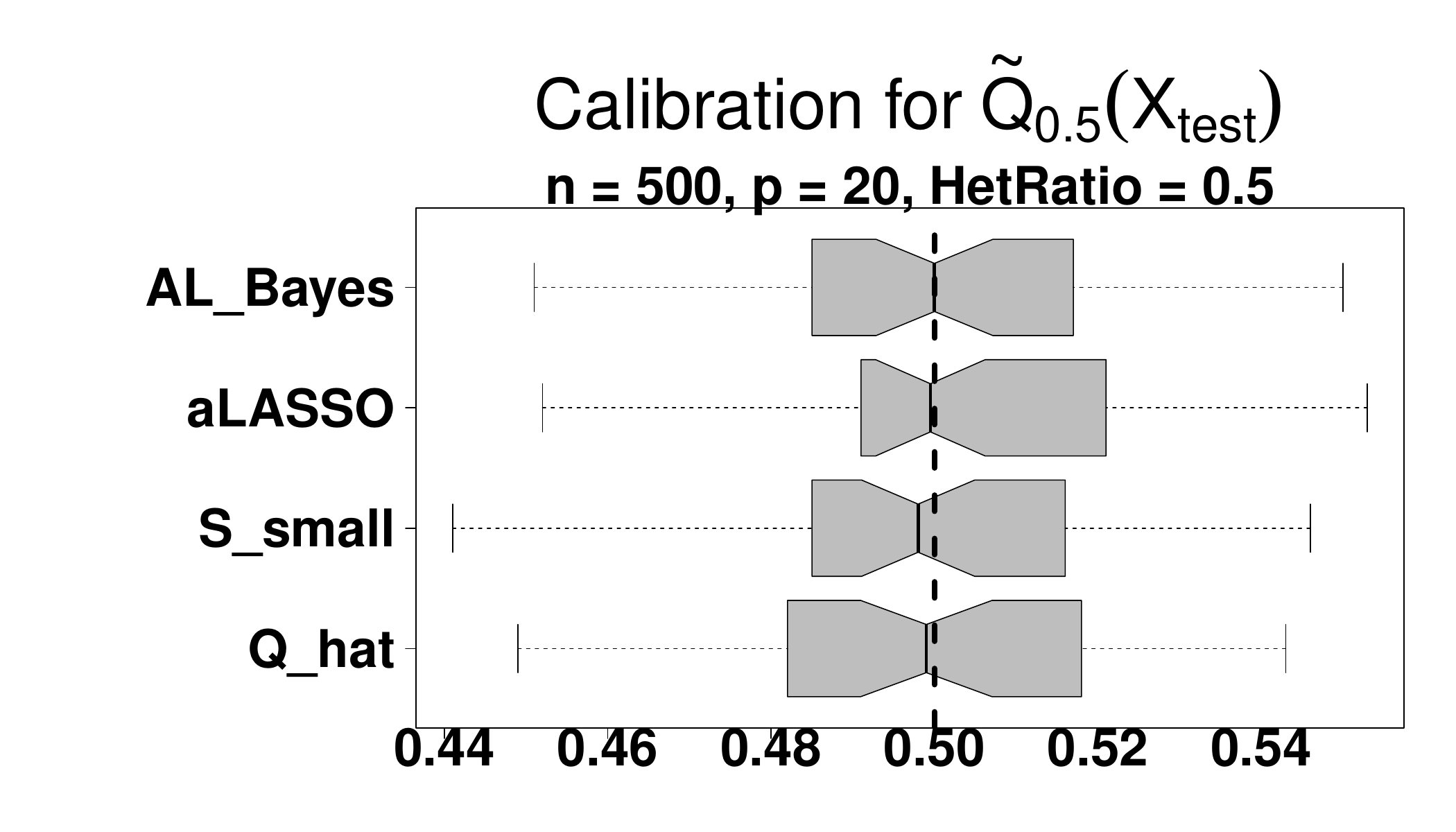}
    \includegraphics[width = .32\textwidth,keepaspectratio]{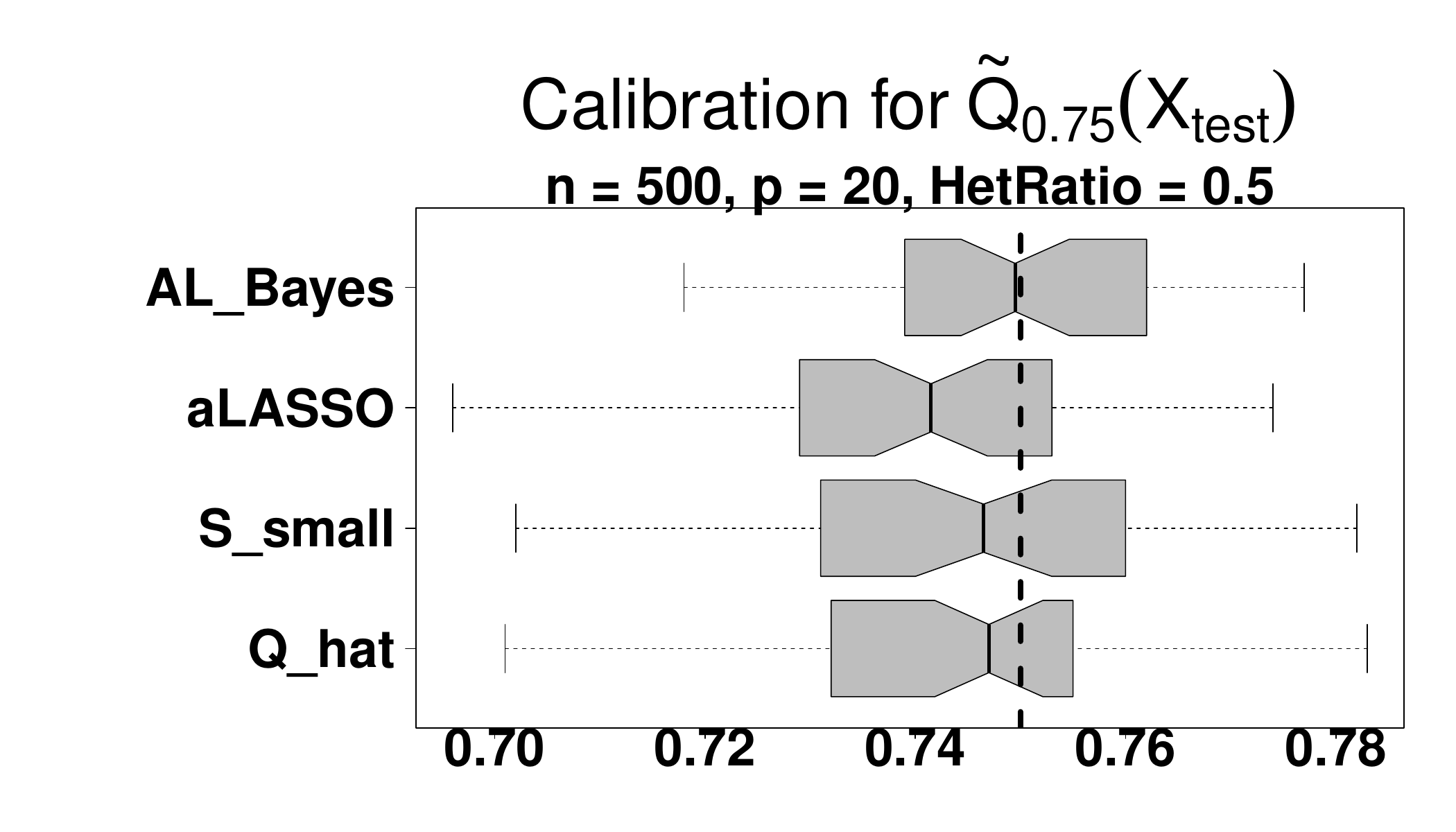}
   \includegraphics[width = .32\textwidth,keepaspectratio]{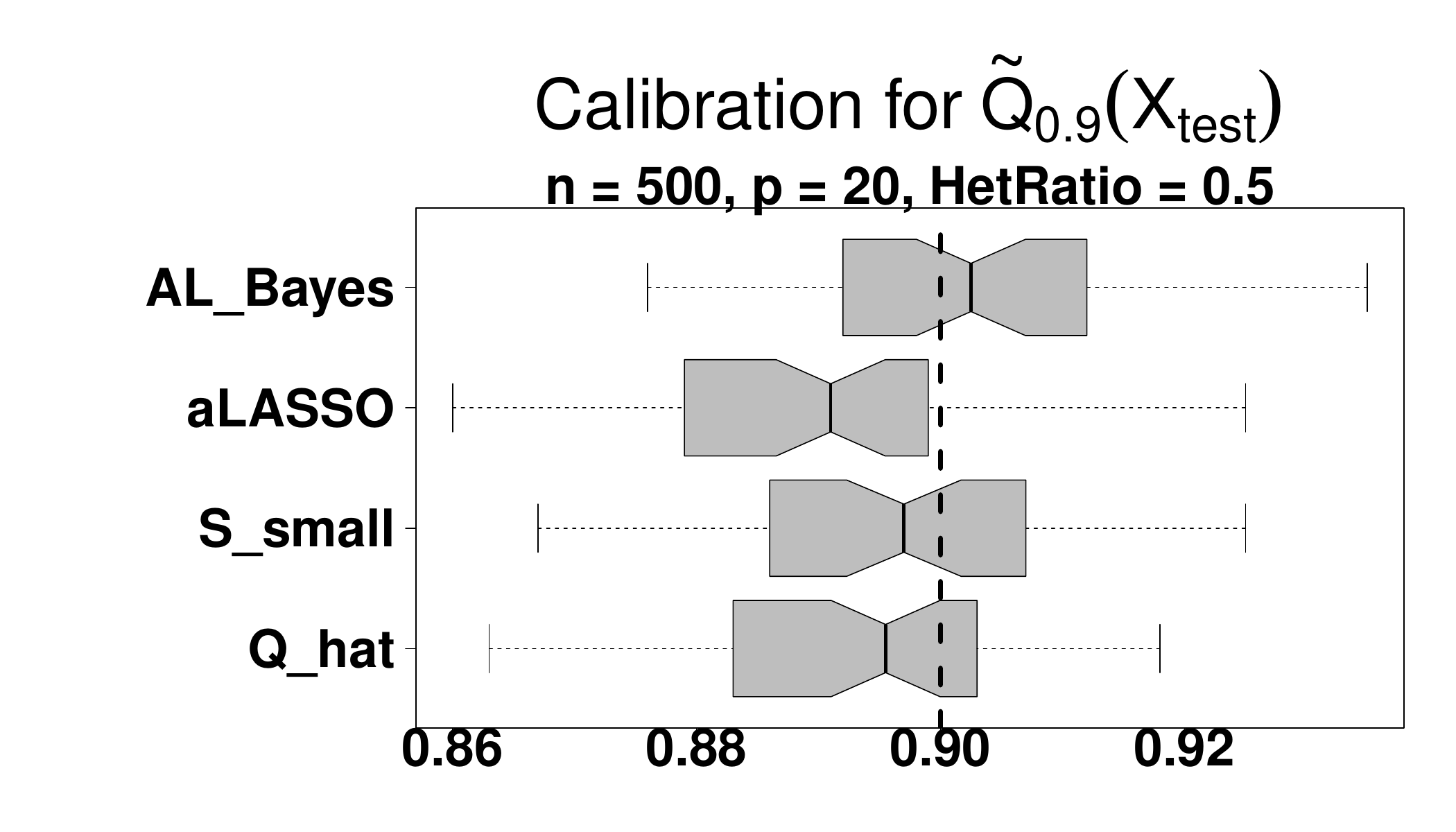}
    \includegraphics[width = .32\textwidth,keepaspectratio]{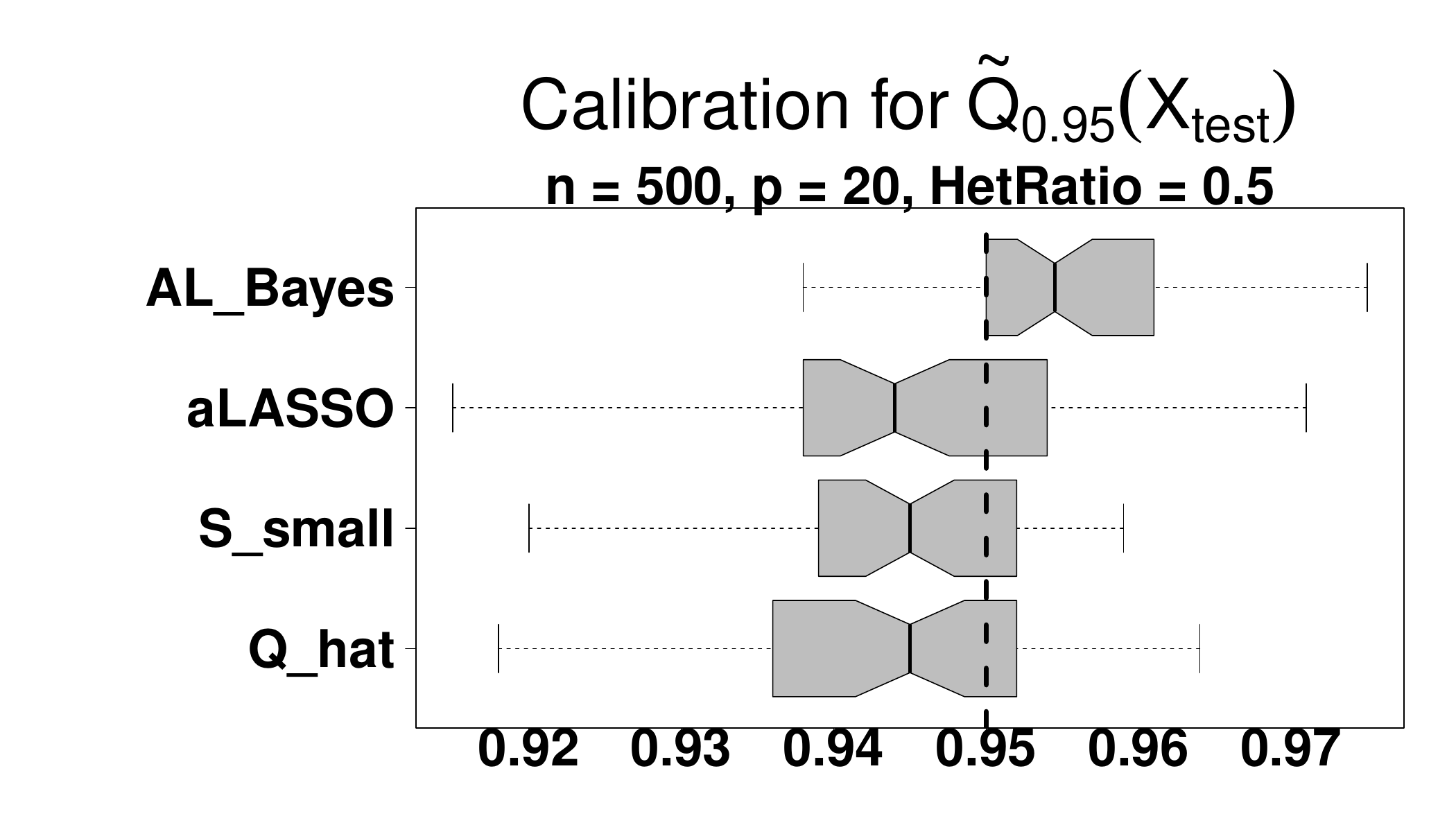}
   \includegraphics[width = .32\textwidth,keepaspectratio]{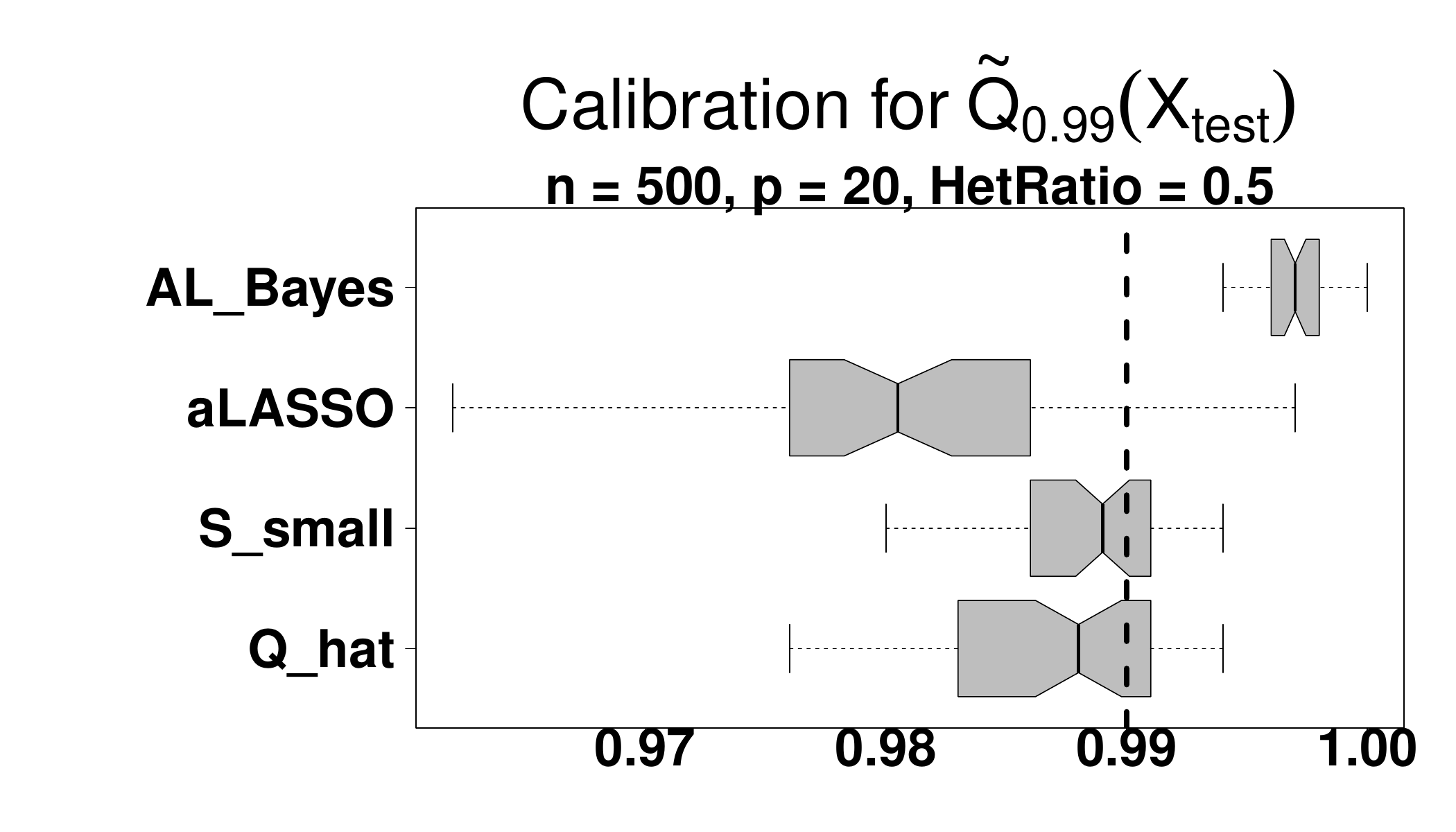}

\end{figure}

\begin{figure}[H]
    \centering
        \caption{\textbf{Calibration}: $\boldsymbol{n = 500, p= 20, \mbox{\textbf{HetRatio} }= 1}$}
  
    \includegraphics[width = .32\textwidth,keepaspectratio]{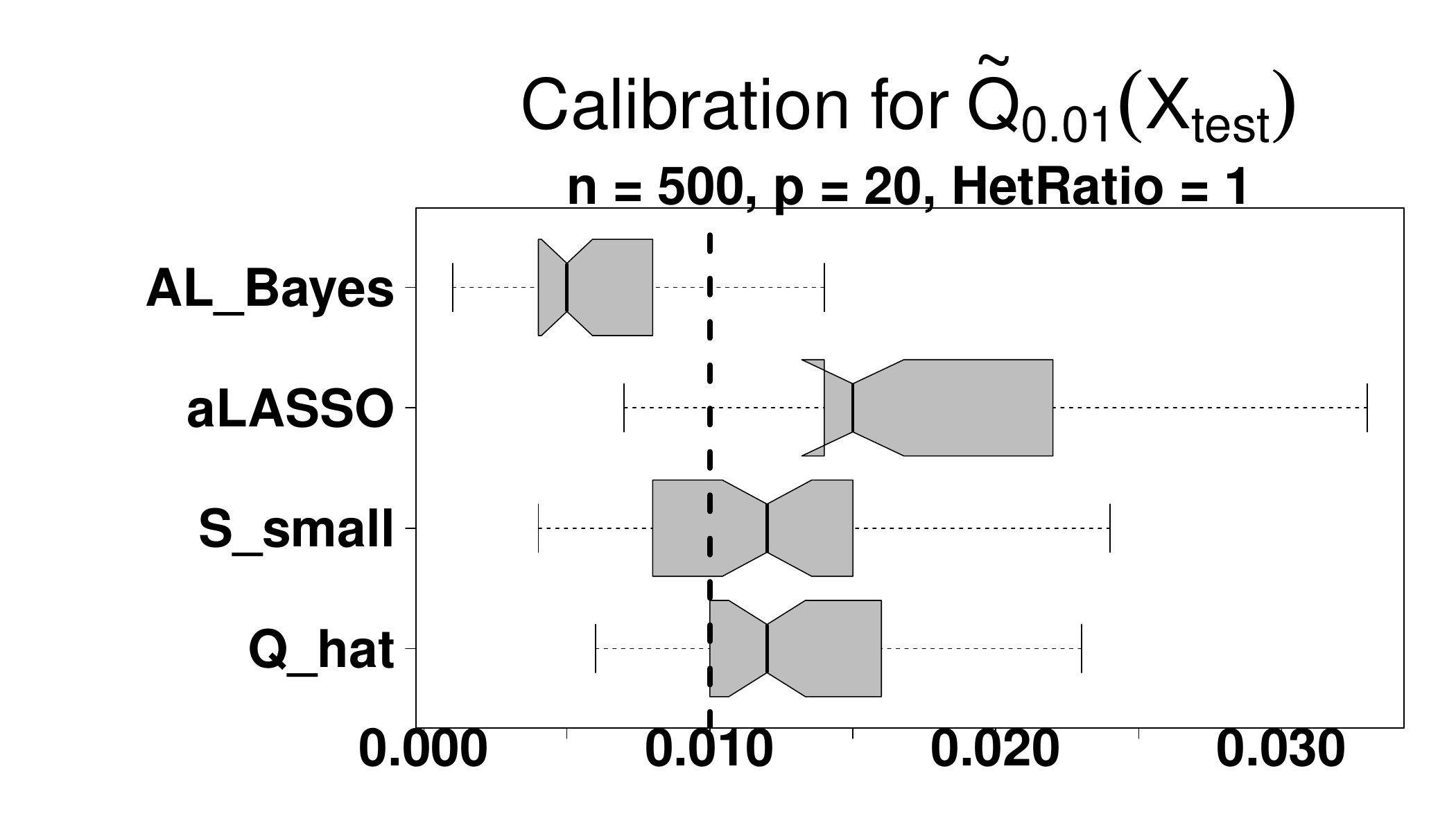}
    \includegraphics[width = .32\textwidth,keepaspectratio]{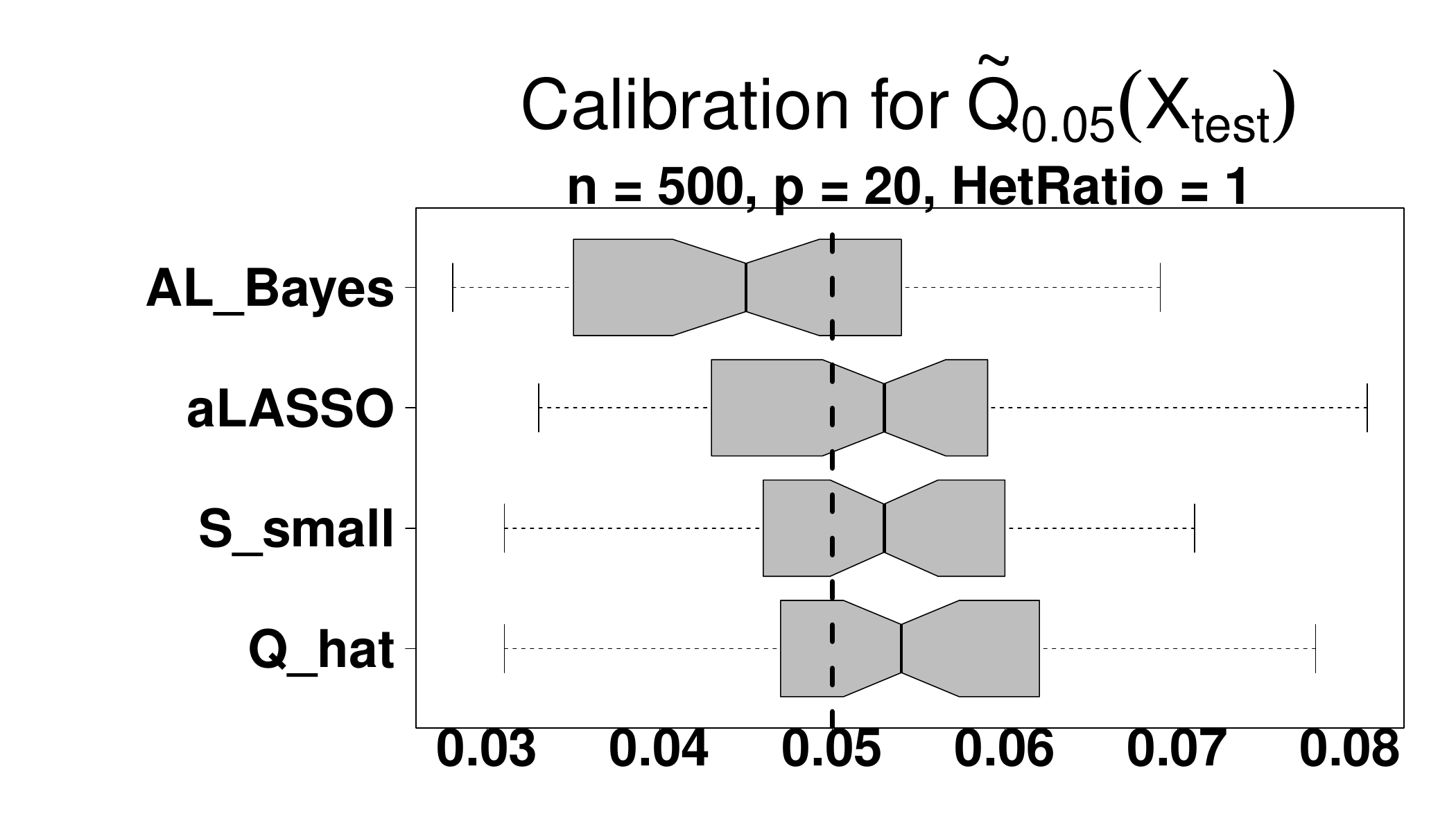}
    \includegraphics[width = .32\textwidth,keepaspectratio]{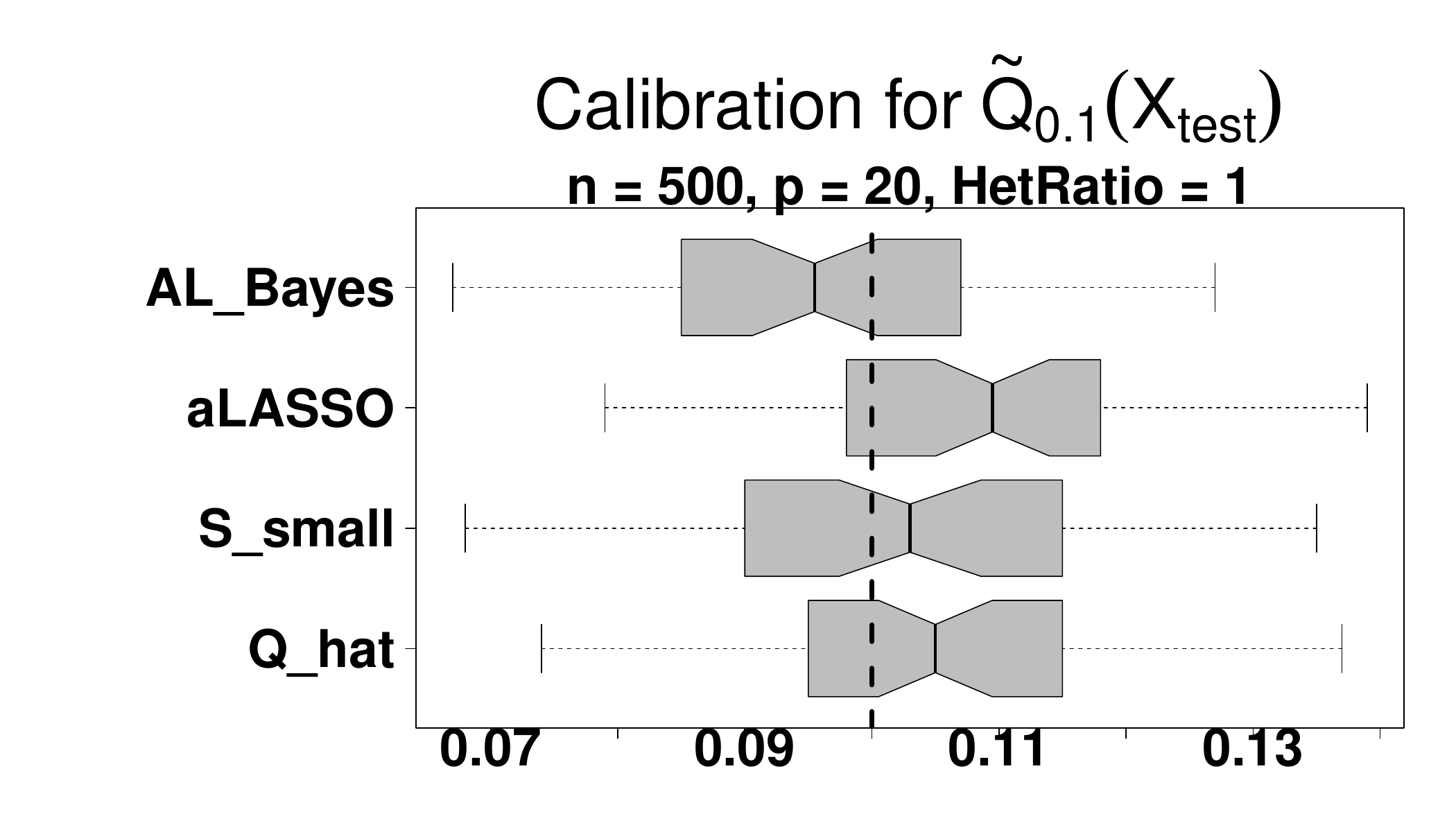}
    \includegraphics[width = .32\textwidth,keepaspectratio]{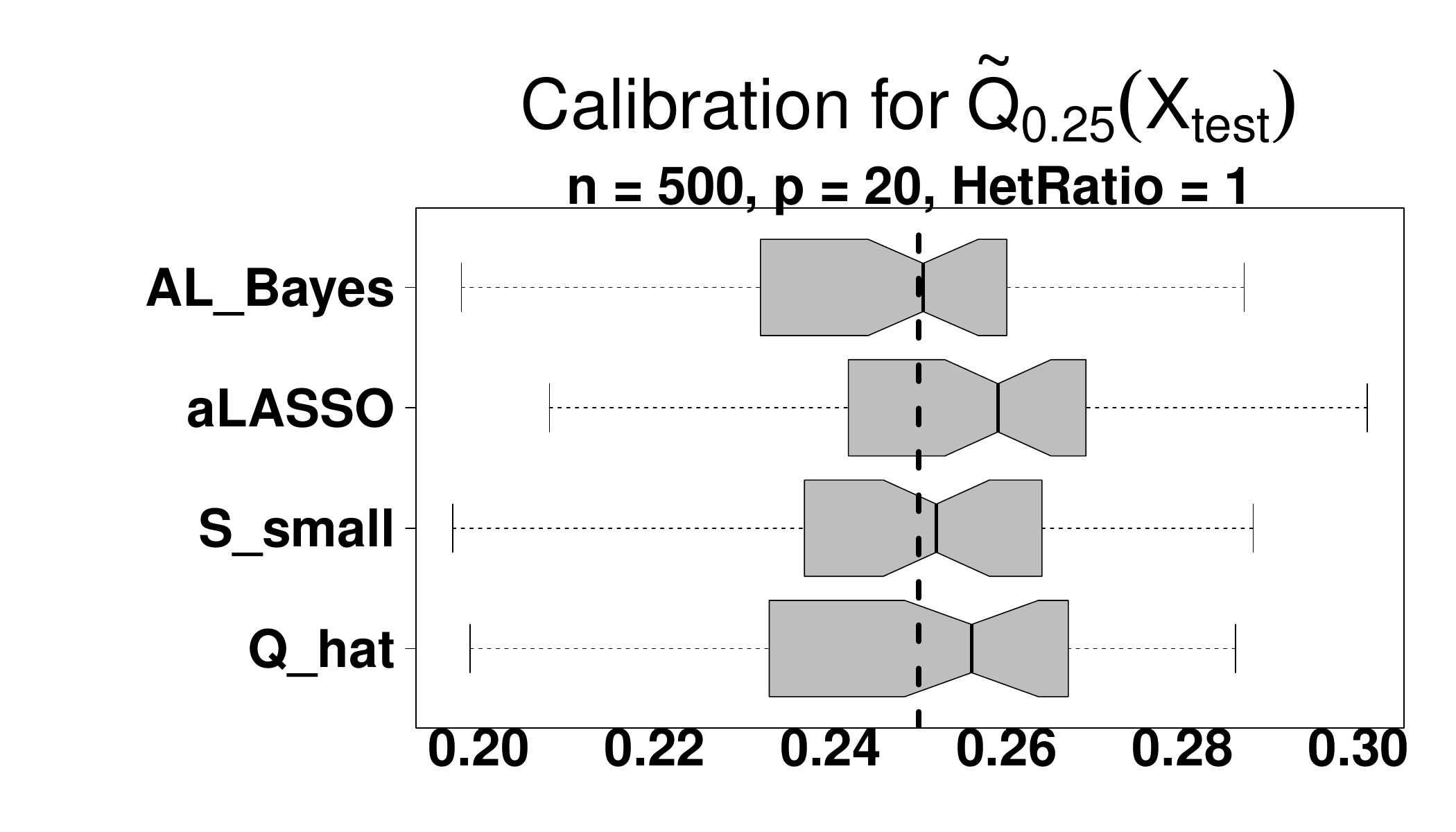}
    \includegraphics[width = .32\textwidth,keepaspectratio]{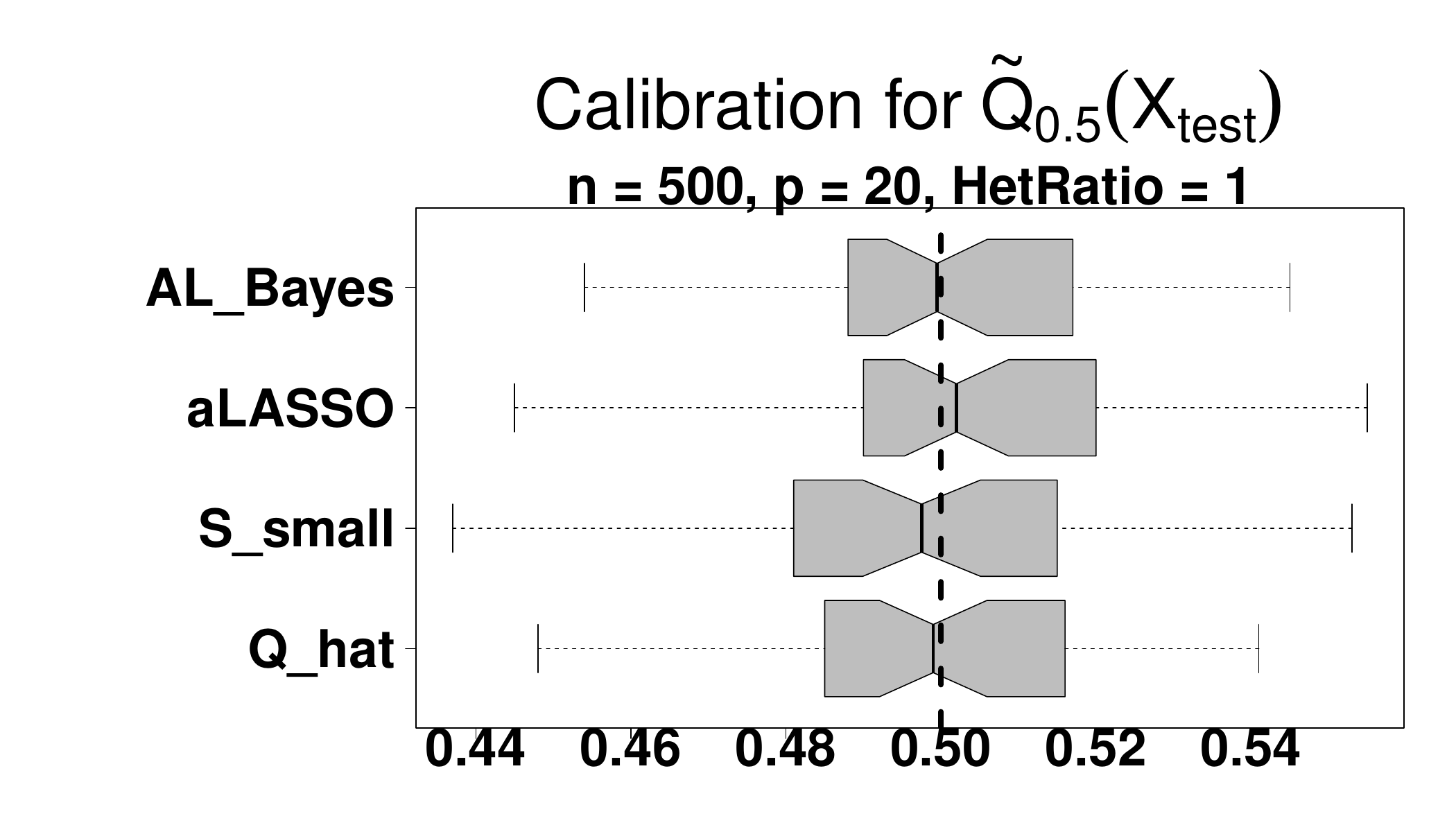}
    \includegraphics[width = .32\textwidth,keepaspectratio]{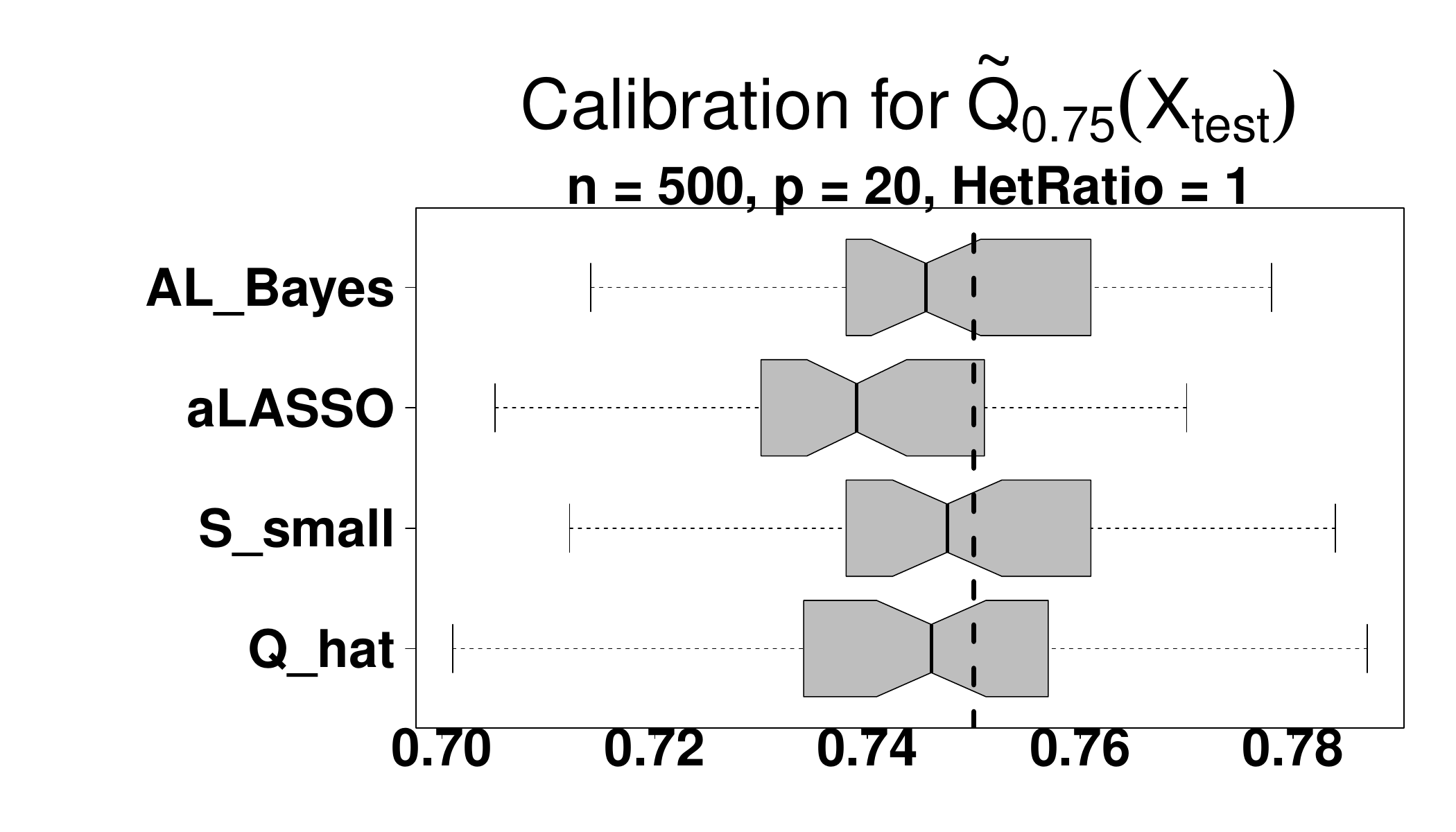}
   \includegraphics[width = .32\textwidth,keepaspectratio]{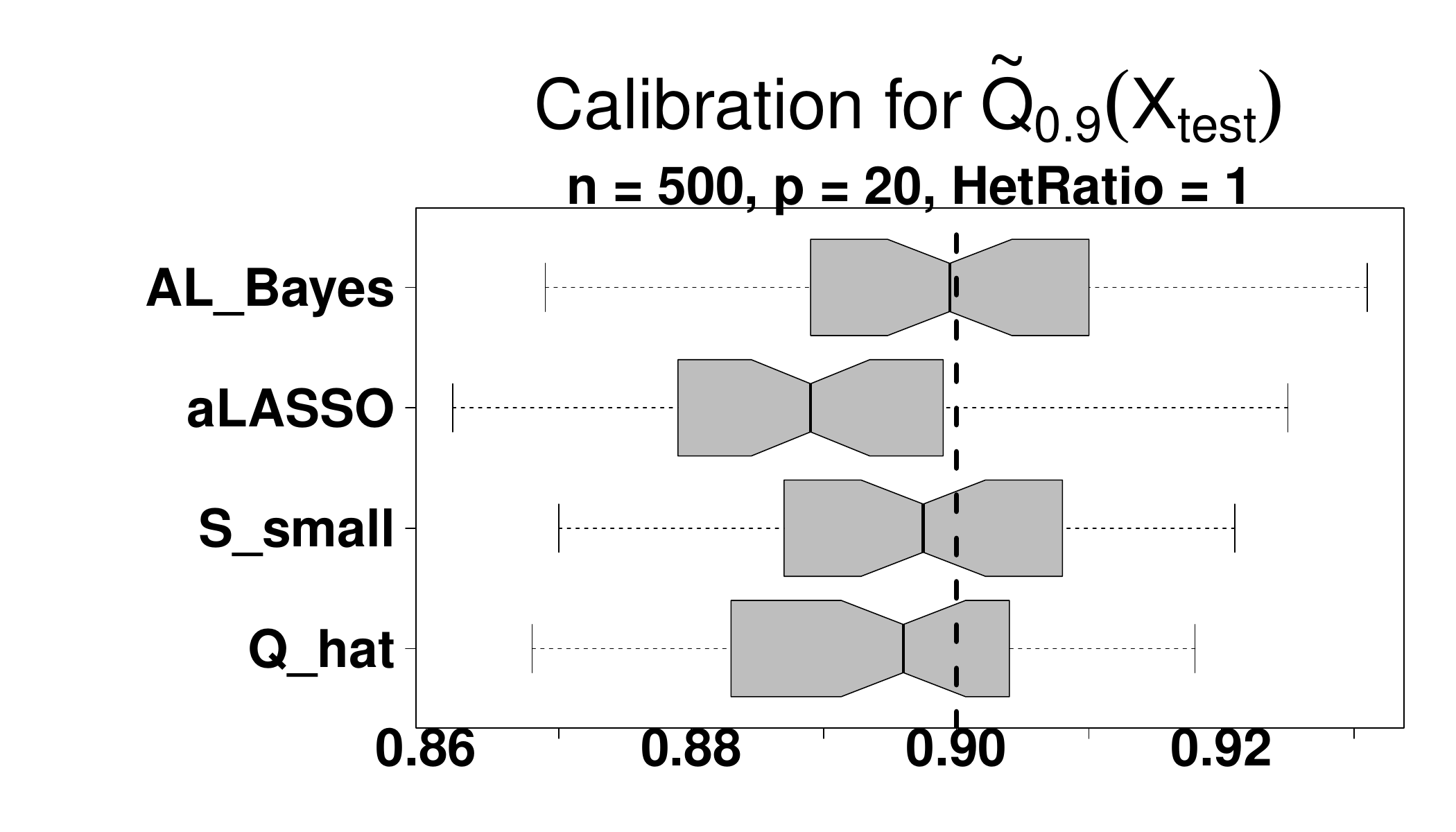}
    \includegraphics[width = .32\textwidth,keepaspectratio]{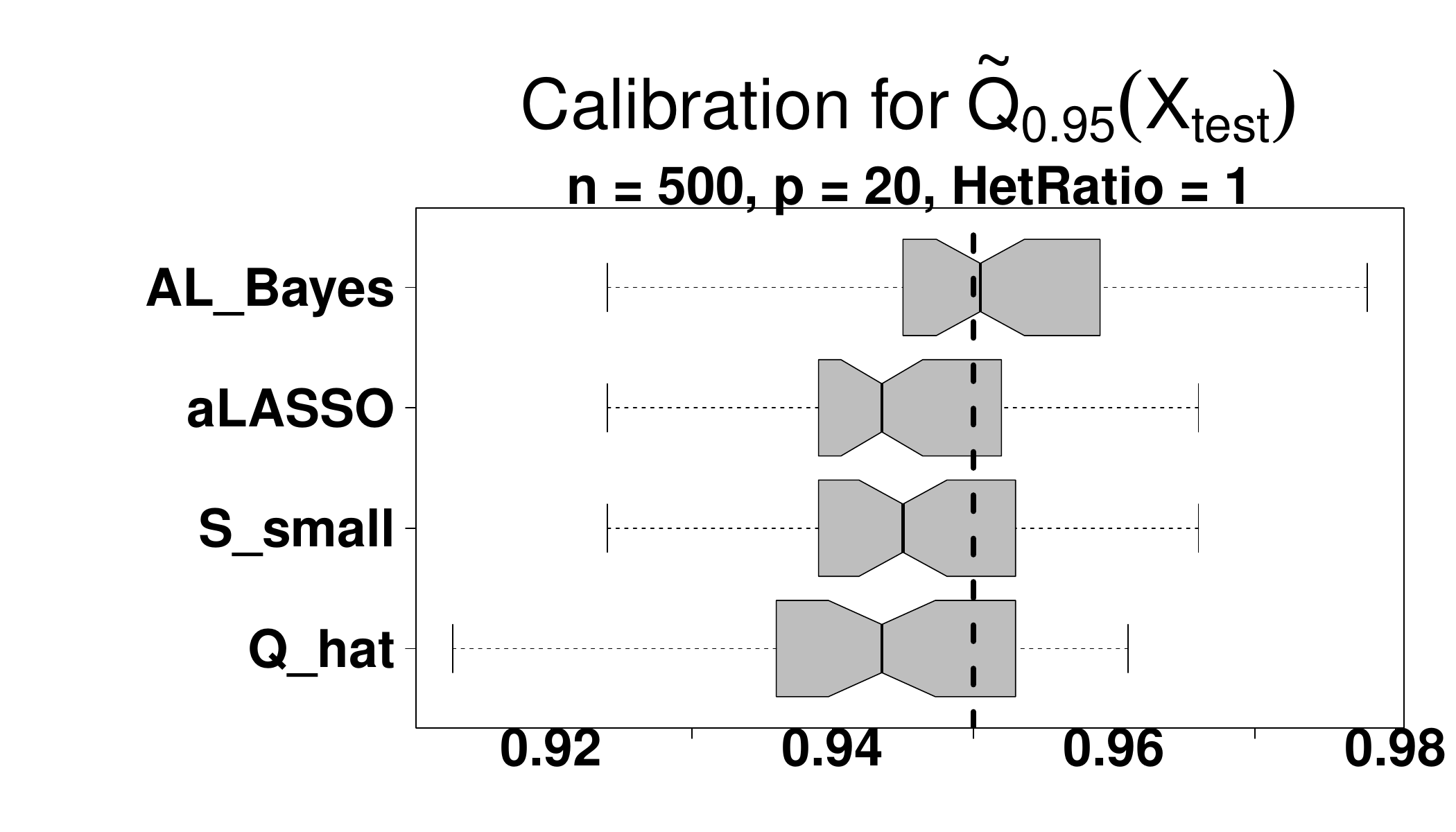}
   \includegraphics[width = .32\textwidth,keepaspectratio]{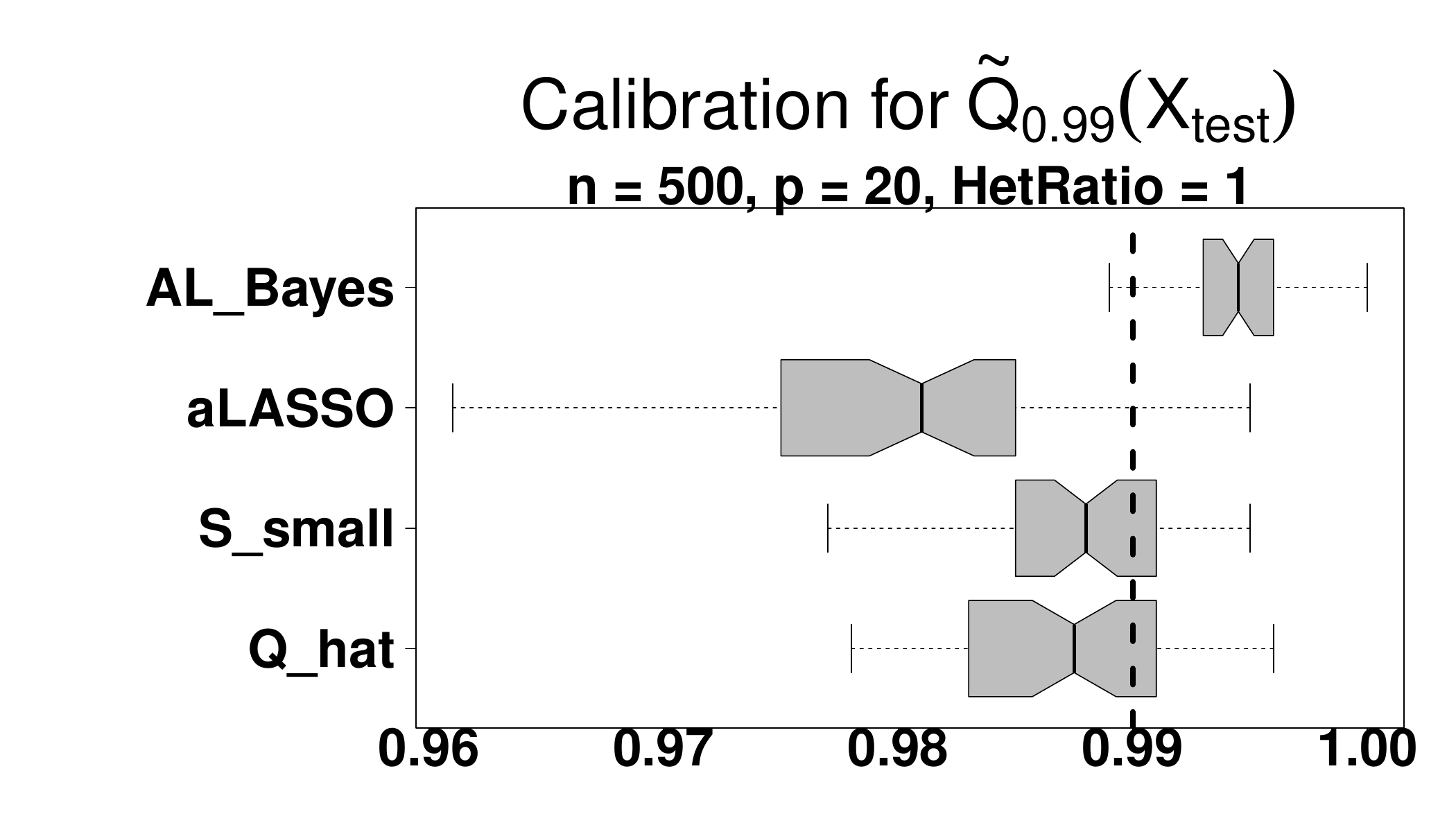}

\end{figure}

\begin{figure}[H]
    \centering
        \caption*{\textbf{Calibration}: $\boldsymbol{n = 100, p= 100, \mbox{\textbf{HetRatio} }= 0.5}$}
  
    \includegraphics[width = .32\textwidth,keepaspectratio]{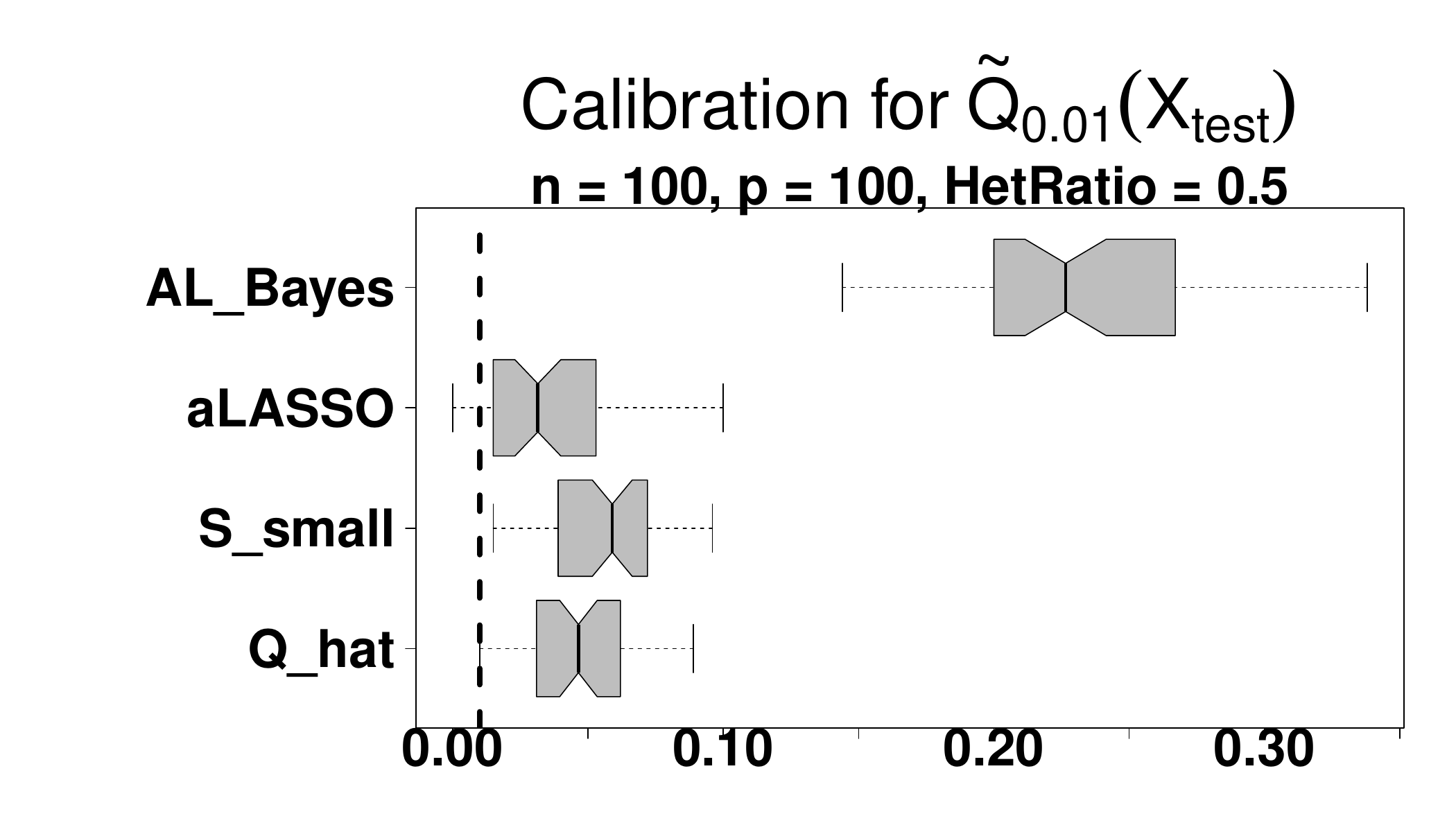}
    \includegraphics[width = .32\textwidth,keepaspectratio]{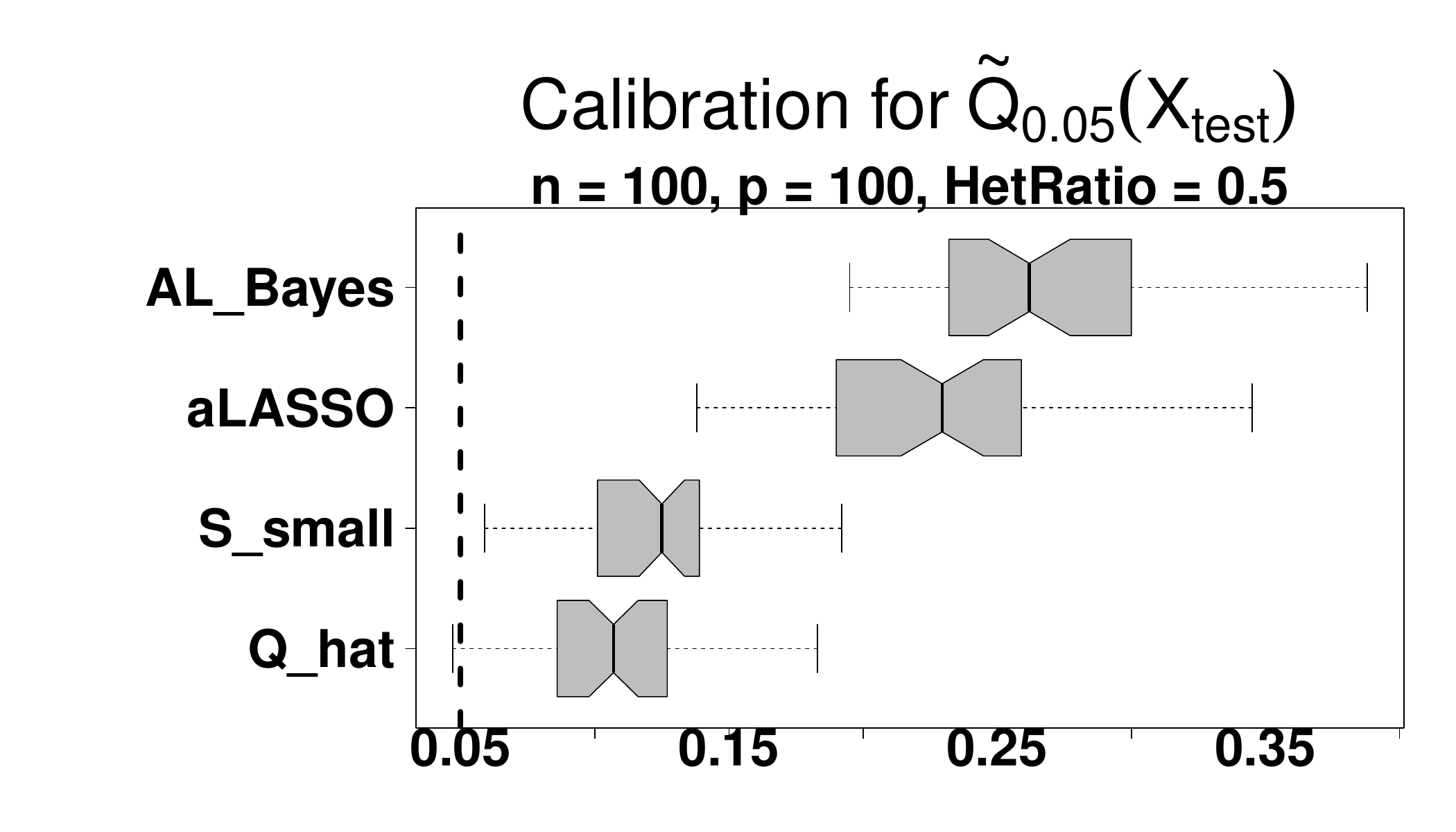}
    \includegraphics[width = .32\textwidth,keepaspectratio]{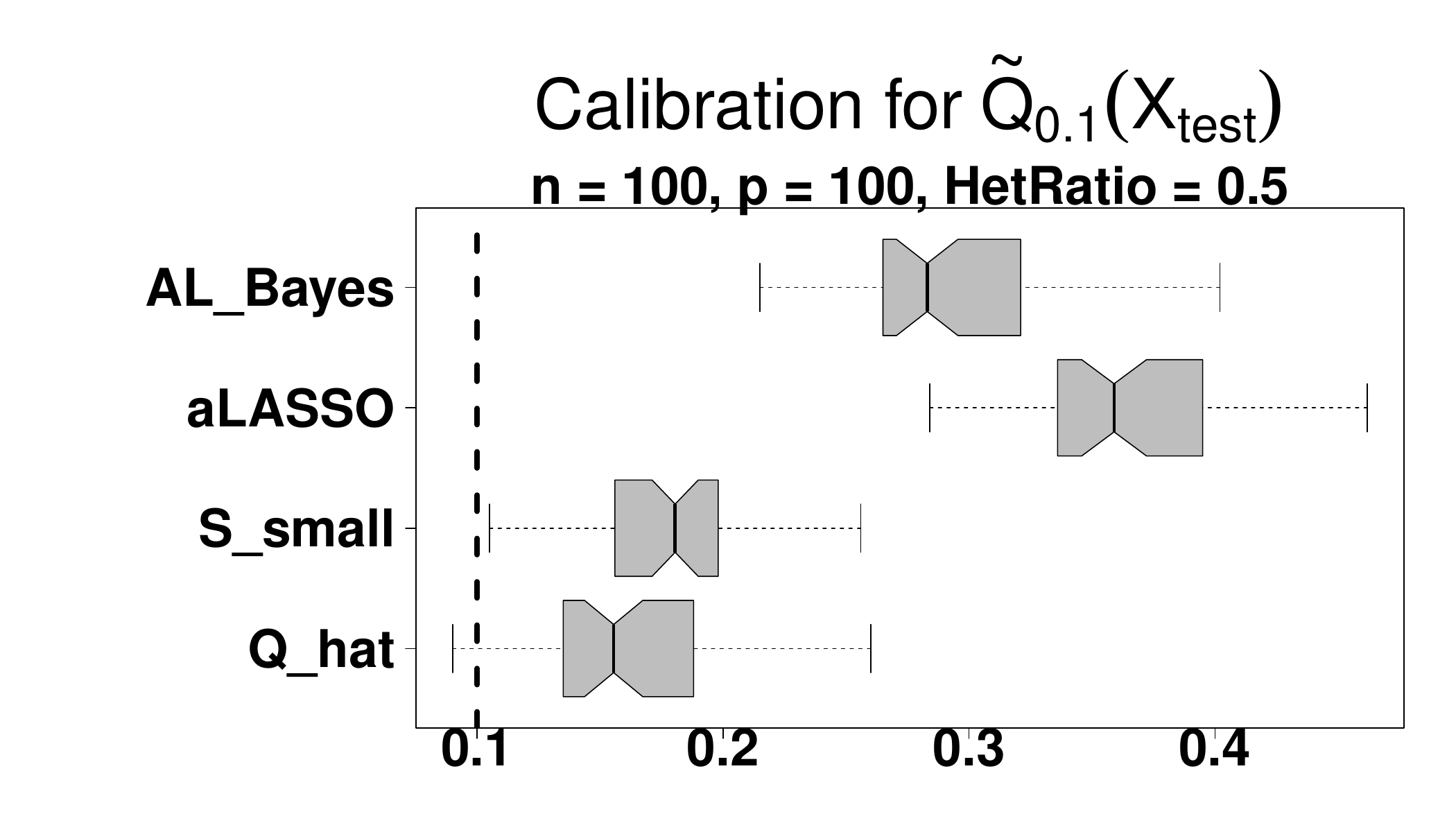}
    \includegraphics[width = .32\textwidth,keepaspectratio]{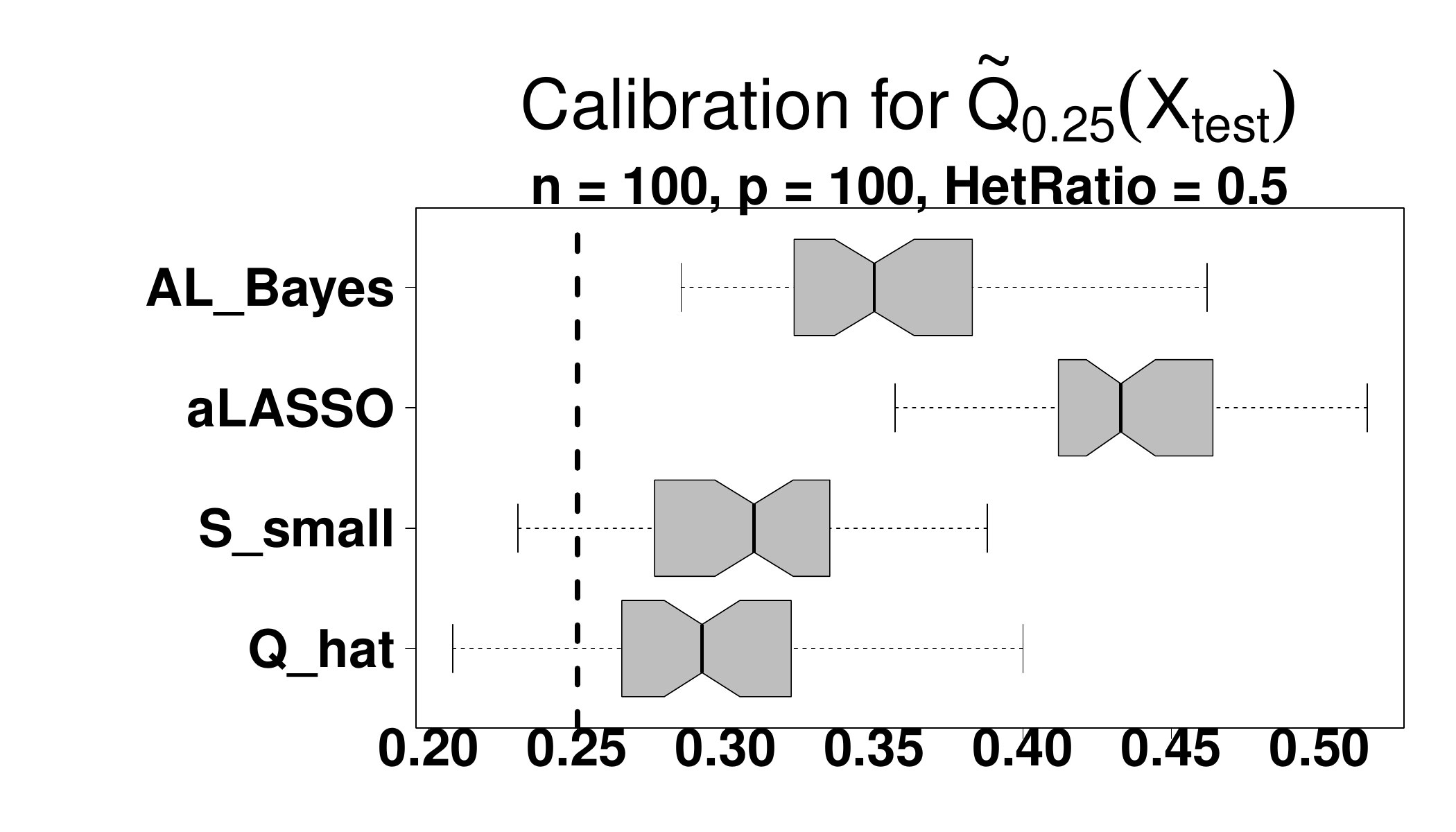}
    \includegraphics[width = .32\textwidth,keepaspectratio]{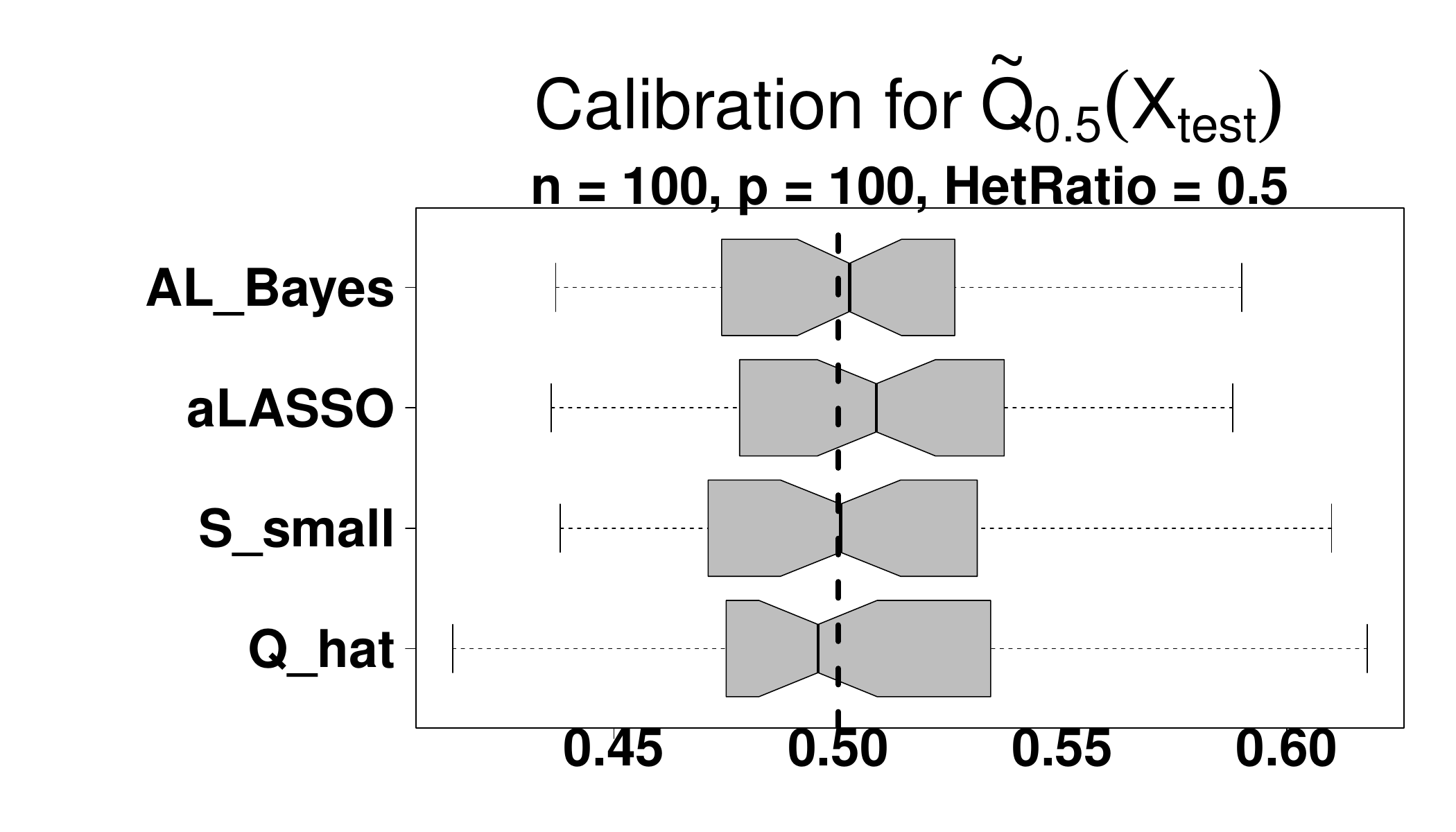}
    \includegraphics[width = .32\textwidth,keepaspectratio]{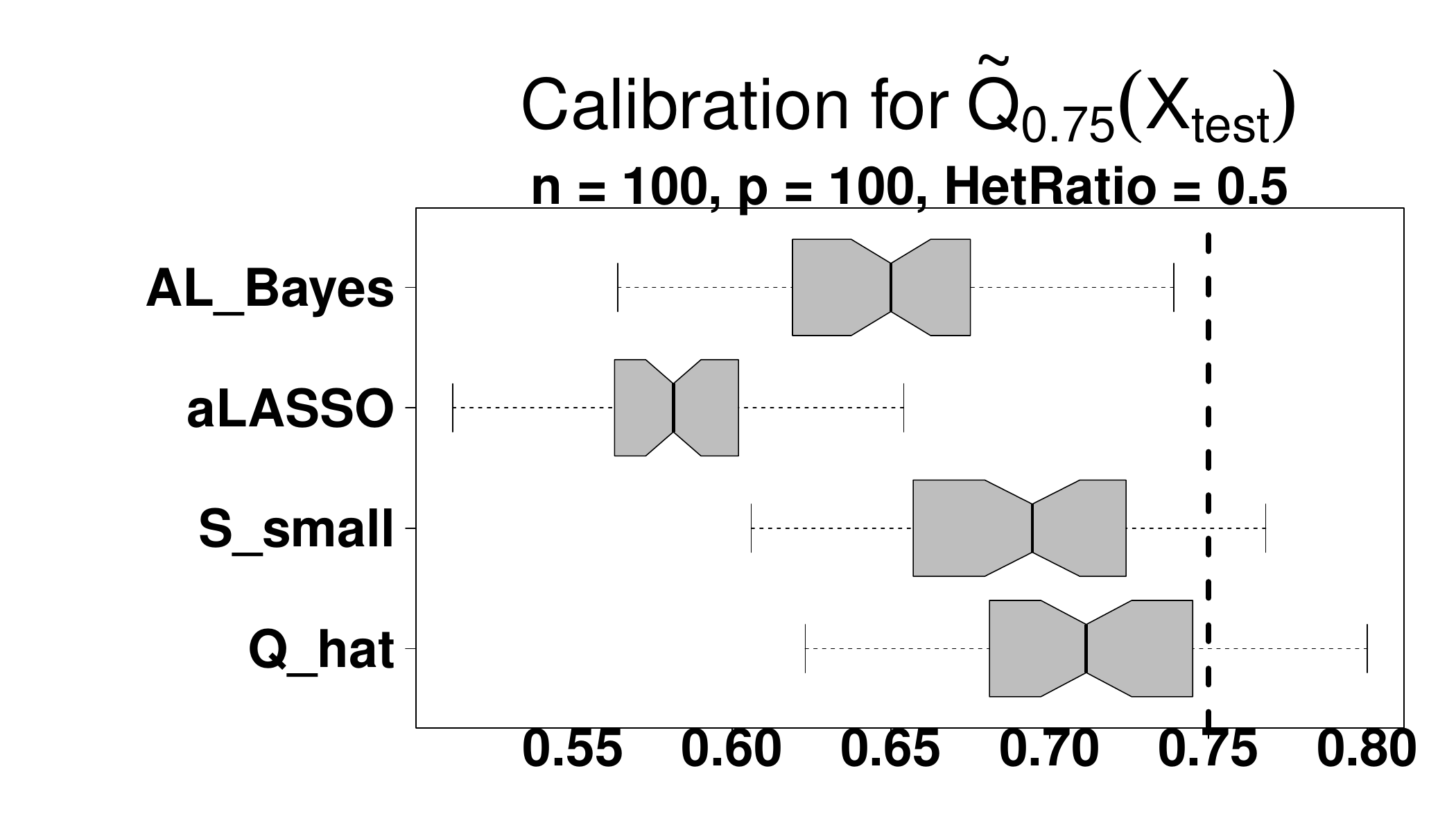}
   \includegraphics[width = .32\textwidth,keepaspectratio]{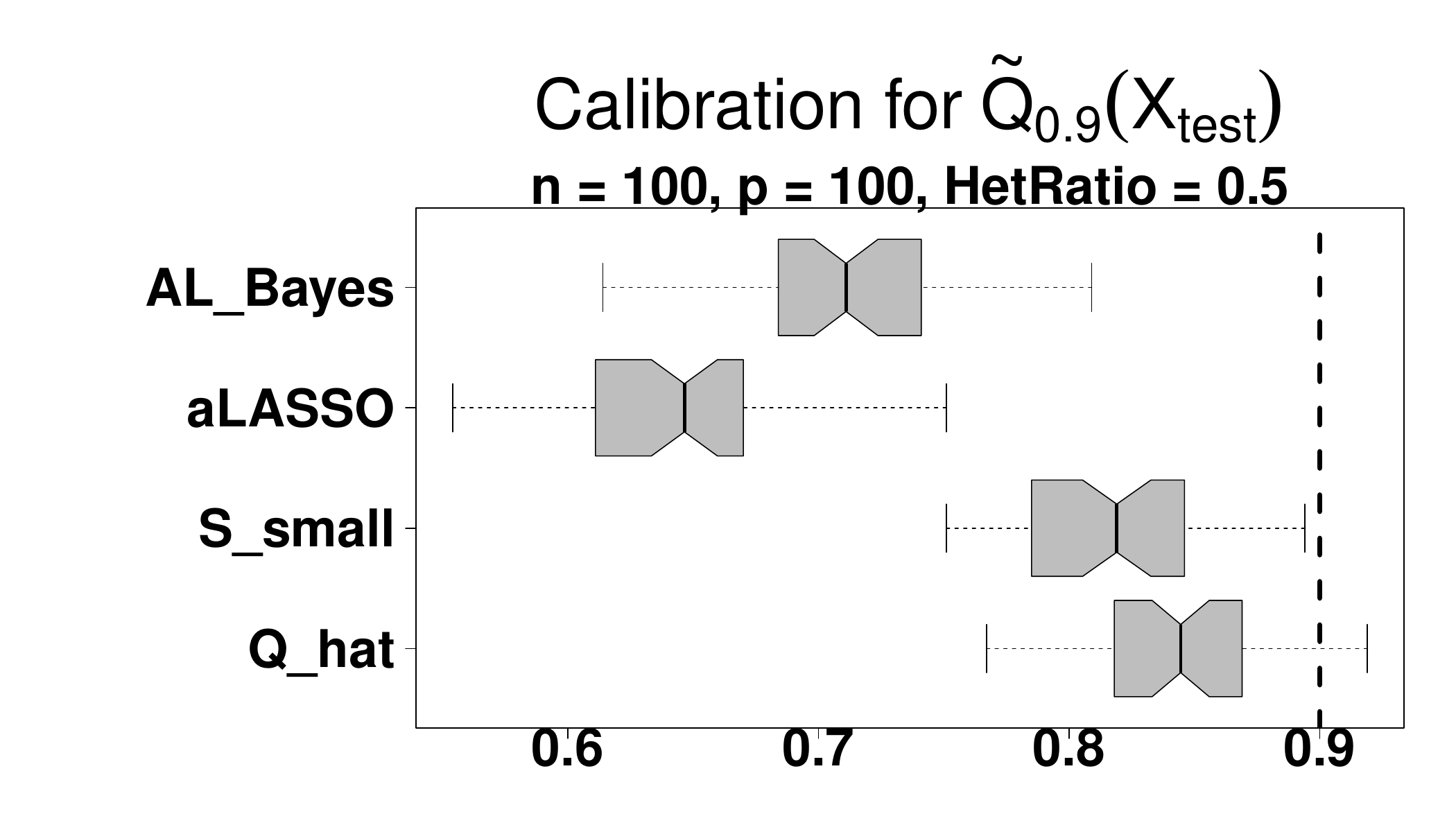}
    \includegraphics[width = .32\textwidth,keepaspectratio]{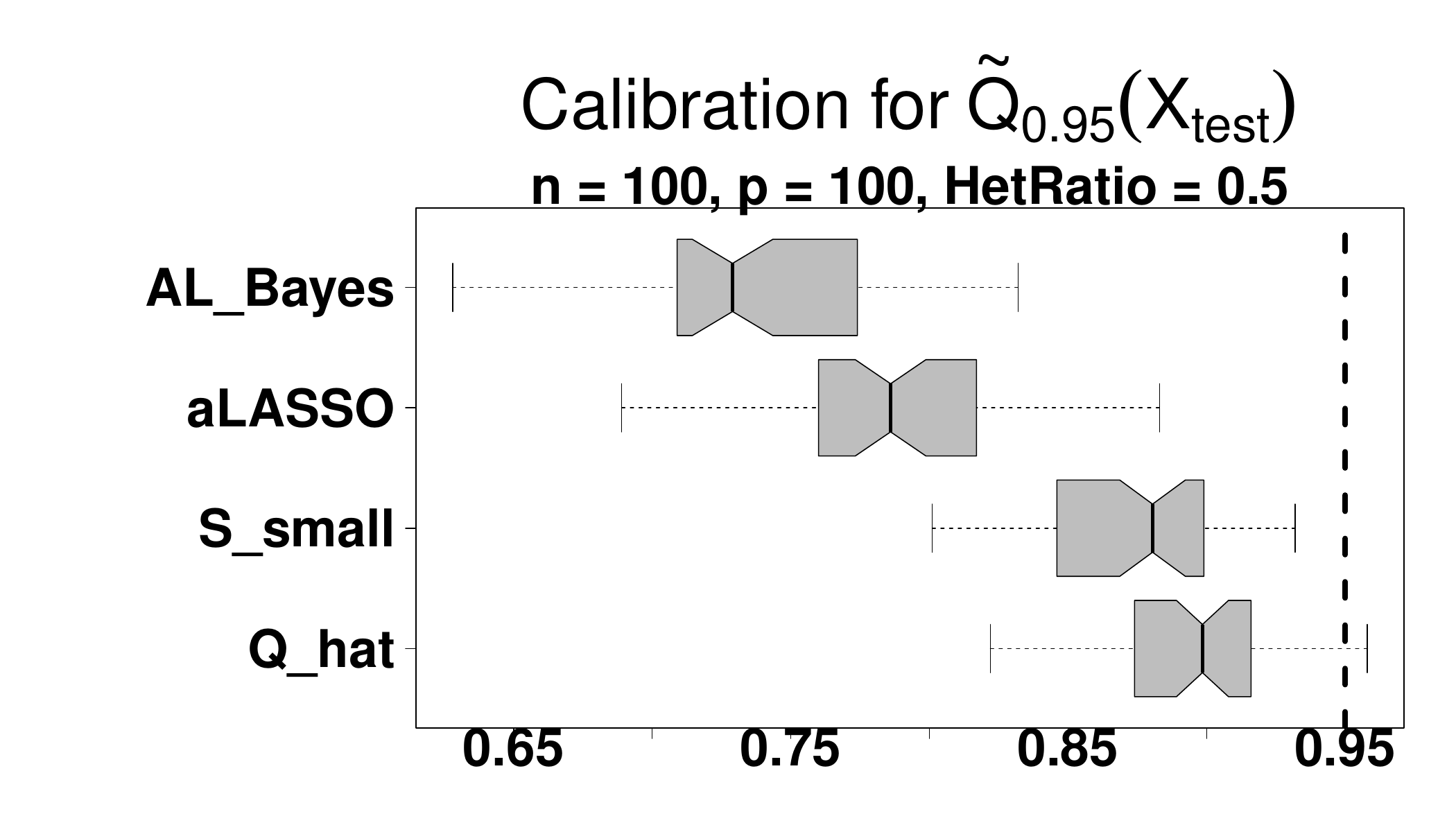}
   \includegraphics[width = .32\textwidth,keepaspectratio]{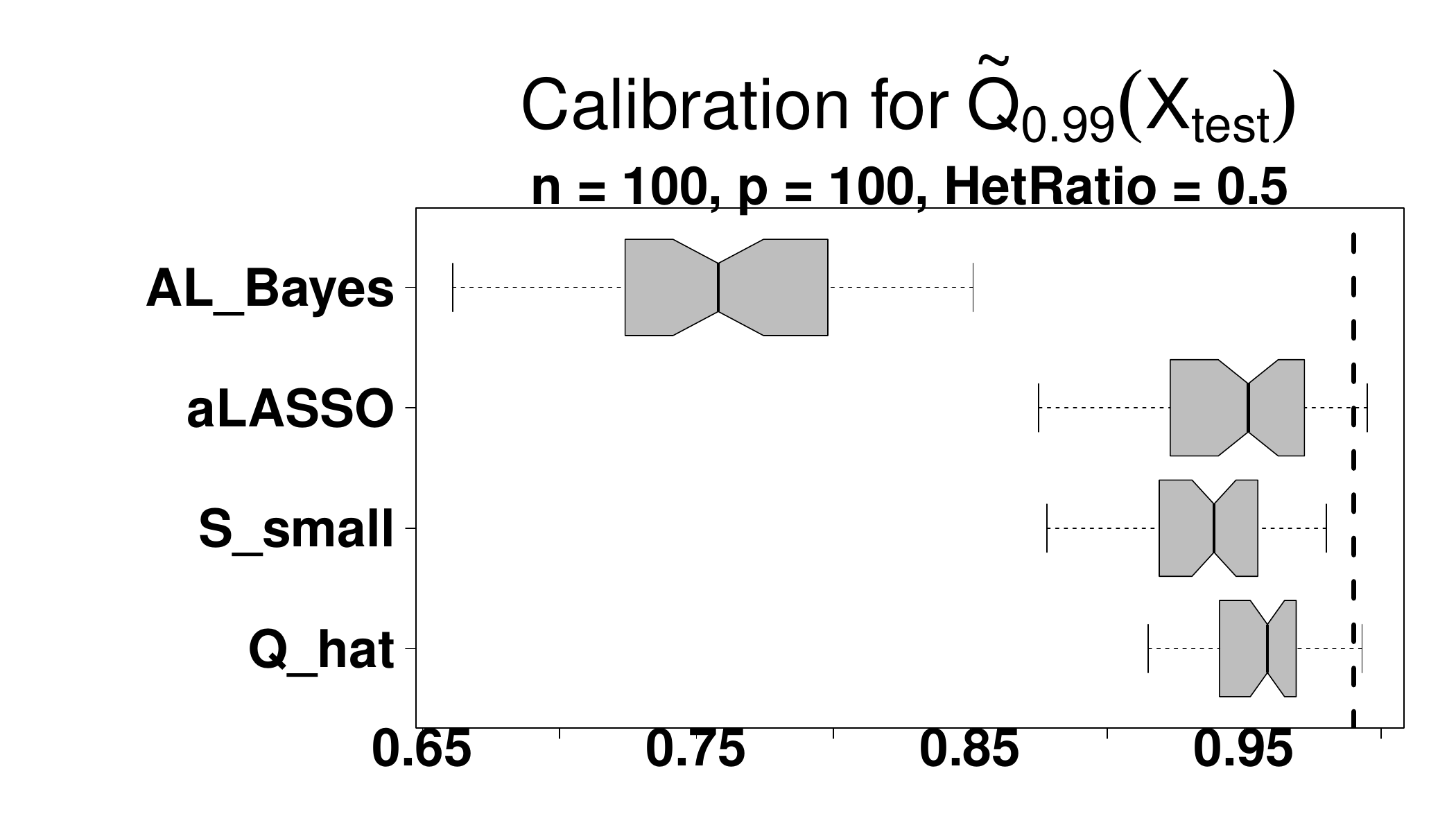}

\end{figure}

\begin{figure}[H]
    \centering
        \caption{\textbf{Calibration}: $\boldsymbol{n = 100, p= 100, \mbox{\textbf{HetRatio} }= 1}$}
  
    \includegraphics[width = .32\textwidth,keepaspectratio]{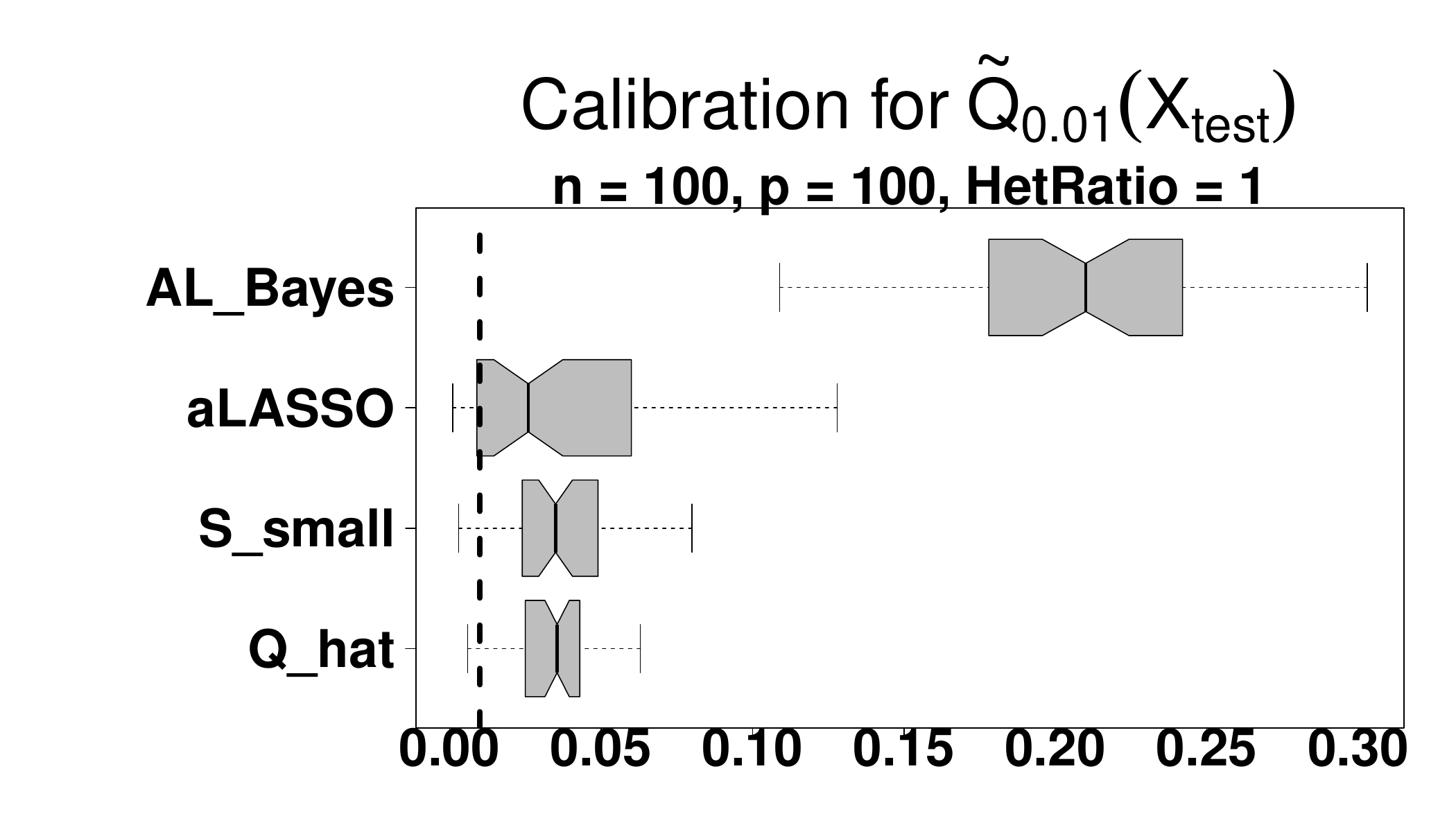}
    \includegraphics[width = .32\textwidth,keepaspectratio]{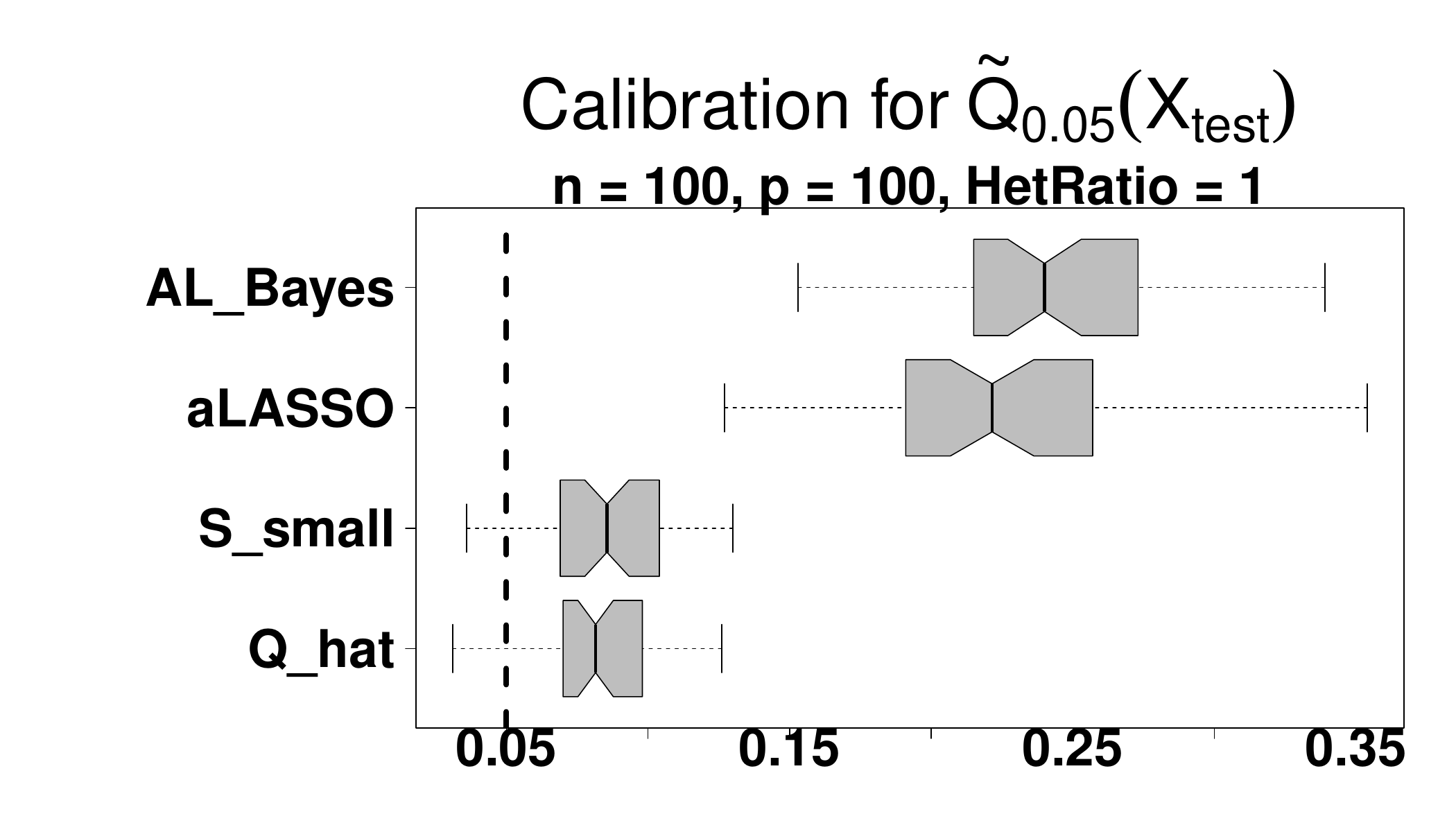}
    \includegraphics[width = .32\textwidth,keepaspectratio]{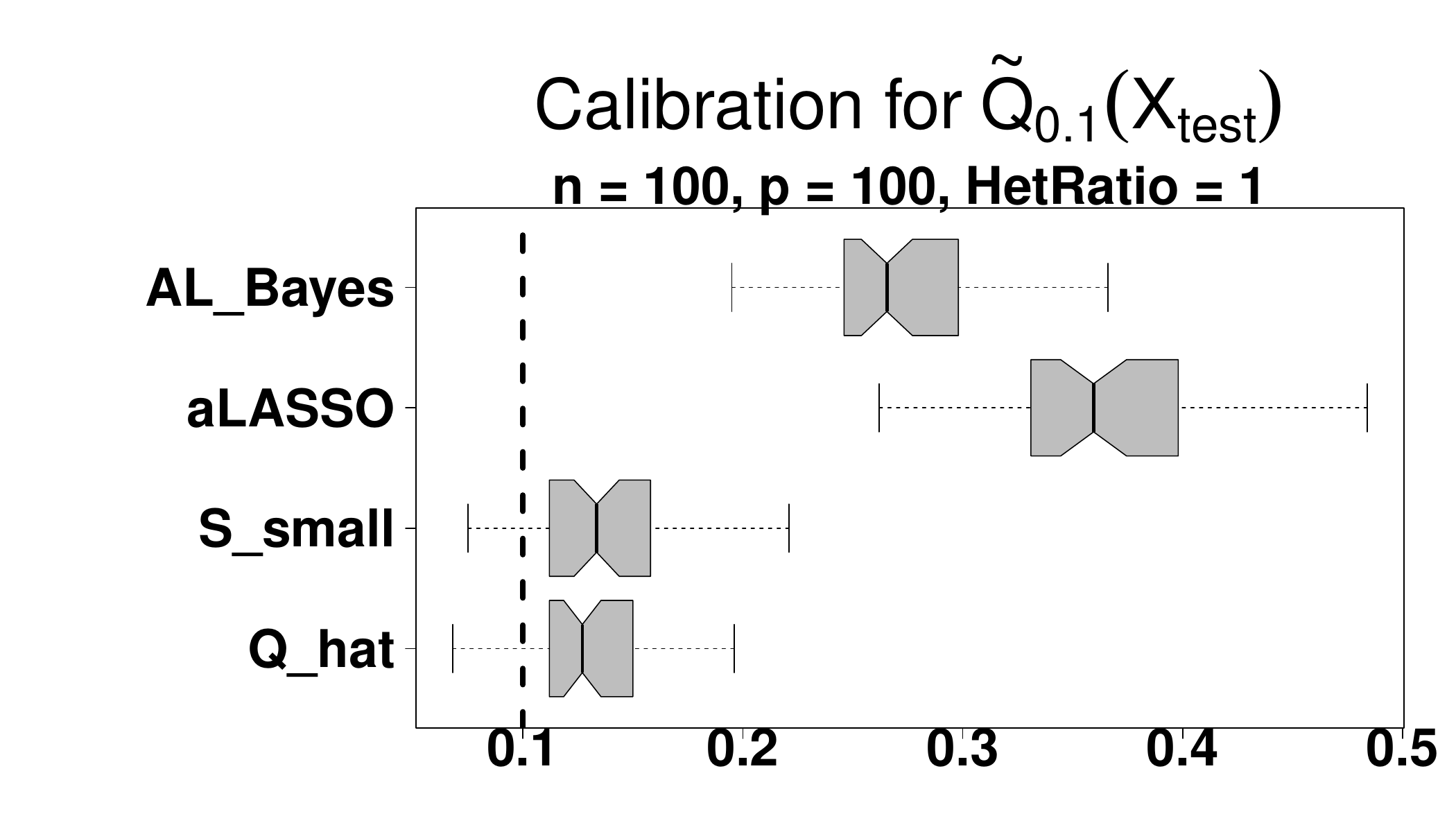}
    \includegraphics[width = .32\textwidth,keepaspectratio]{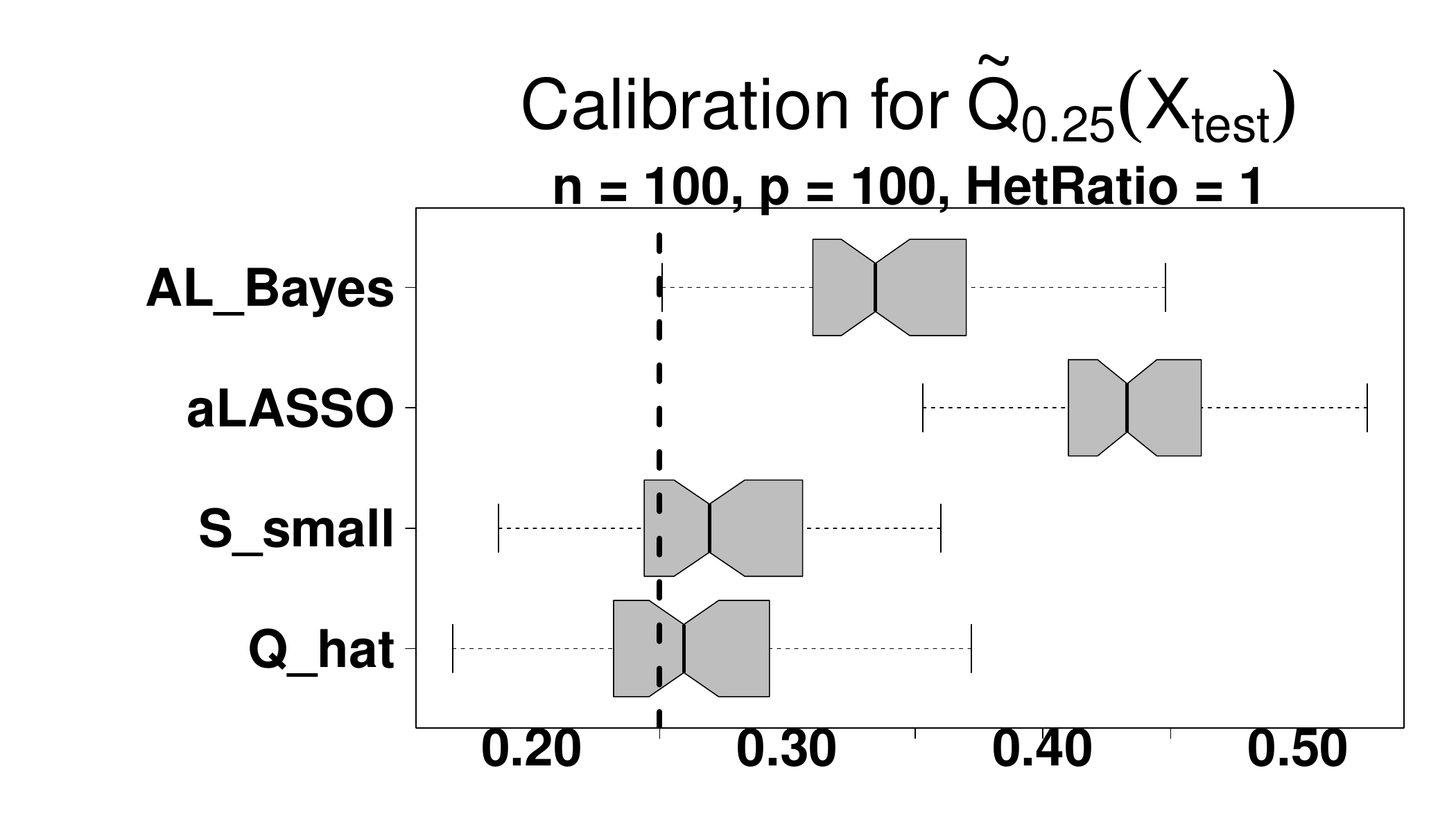}
    \includegraphics[width = .32\textwidth,keepaspectratio]{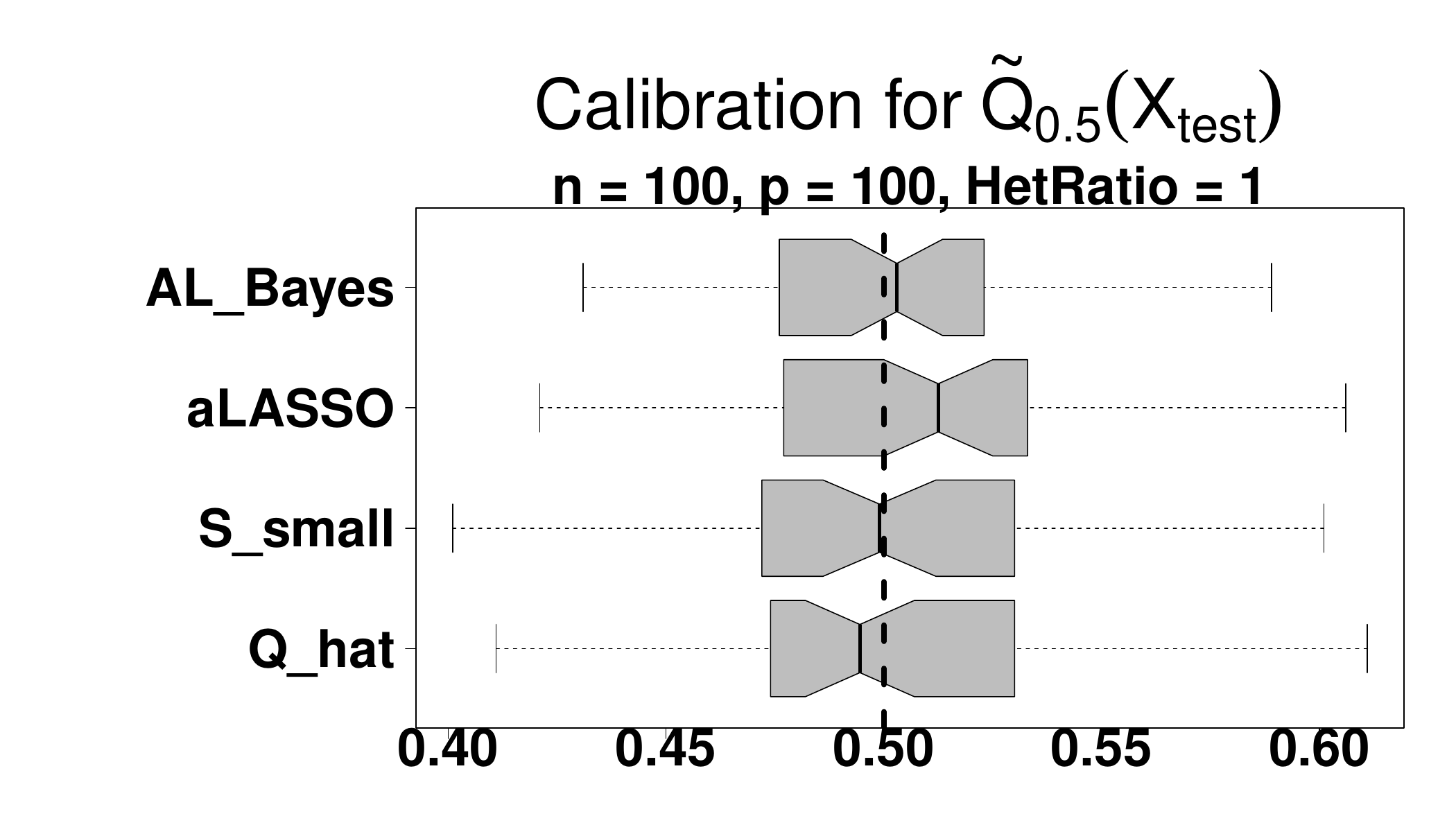}
    \includegraphics[width = .32\textwidth,keepaspectratio]{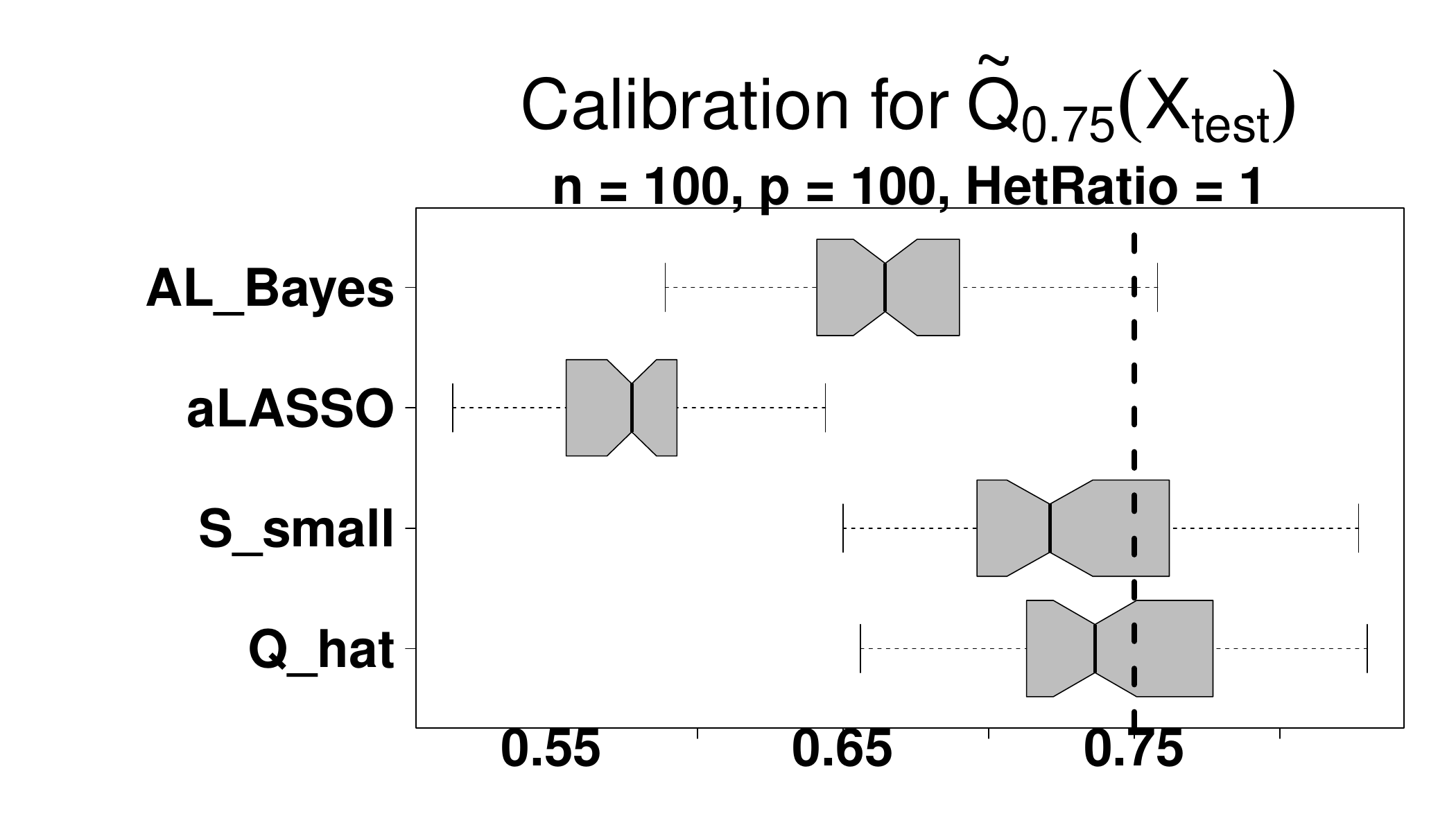}
   \includegraphics[width = .32\textwidth,keepaspectratio]{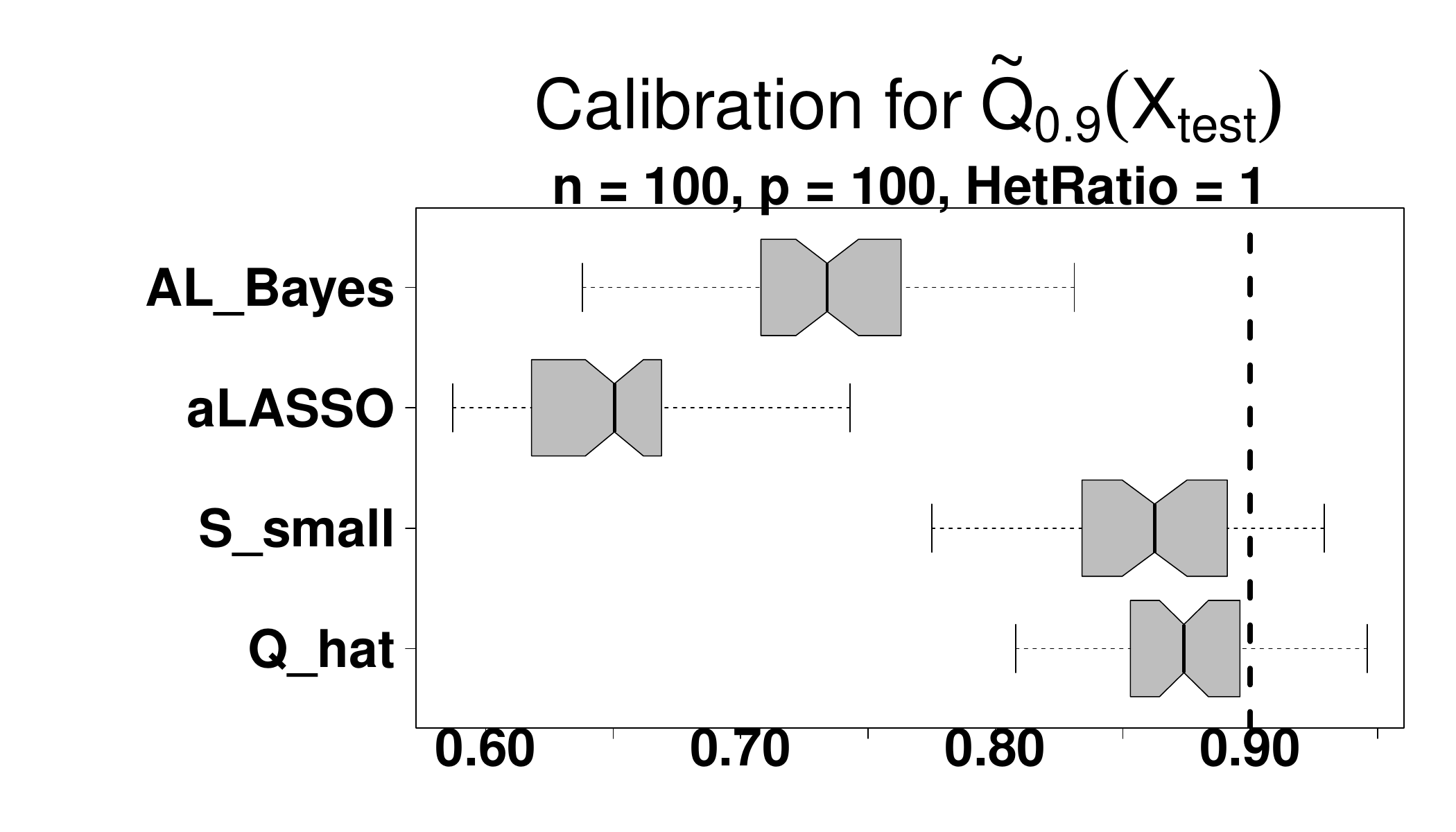}
    \includegraphics[width = .32\textwidth,keepaspectratio]{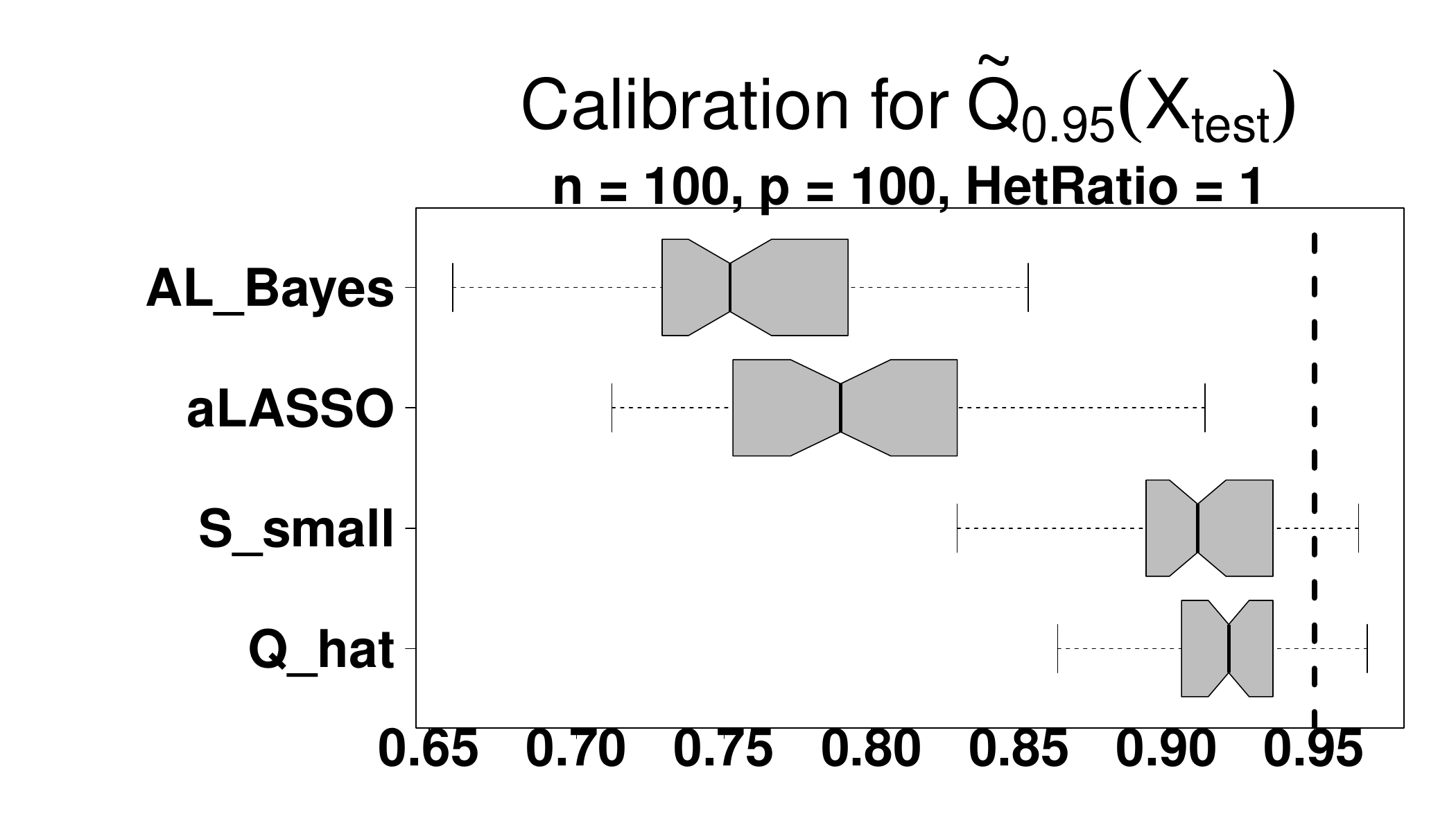}
   \includegraphics[width = .32\textwidth,keepaspectratio]{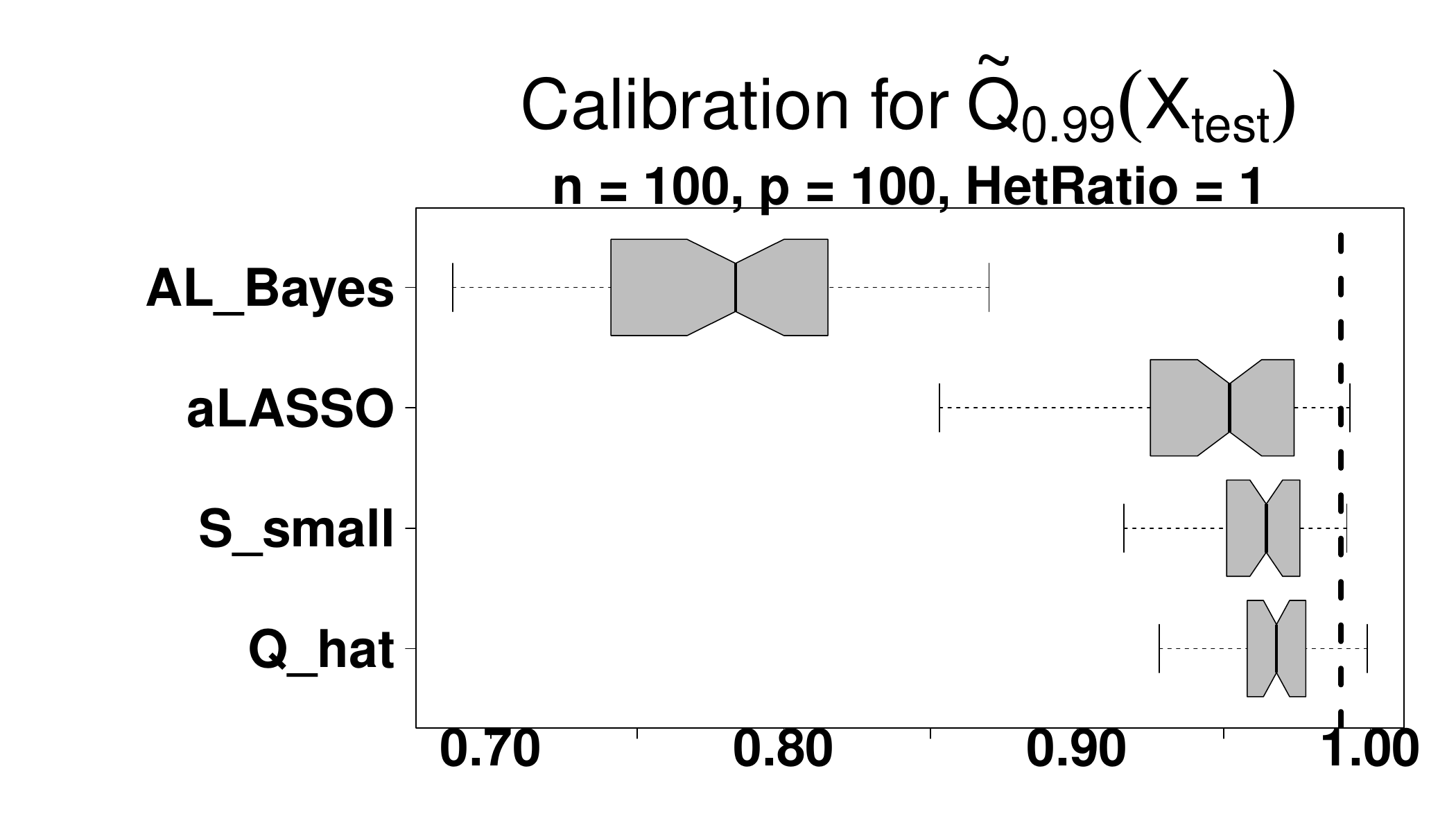}

\end{figure}
\begin{figure}[H]
    \centering
        \caption{\textbf{Calibration}: $\boldsymbol{n = 200, p= 50, \mbox{\textbf{HetRatio} }= 0.5}$}
  
    \includegraphics[width = .32\textwidth,keepaspectratio]{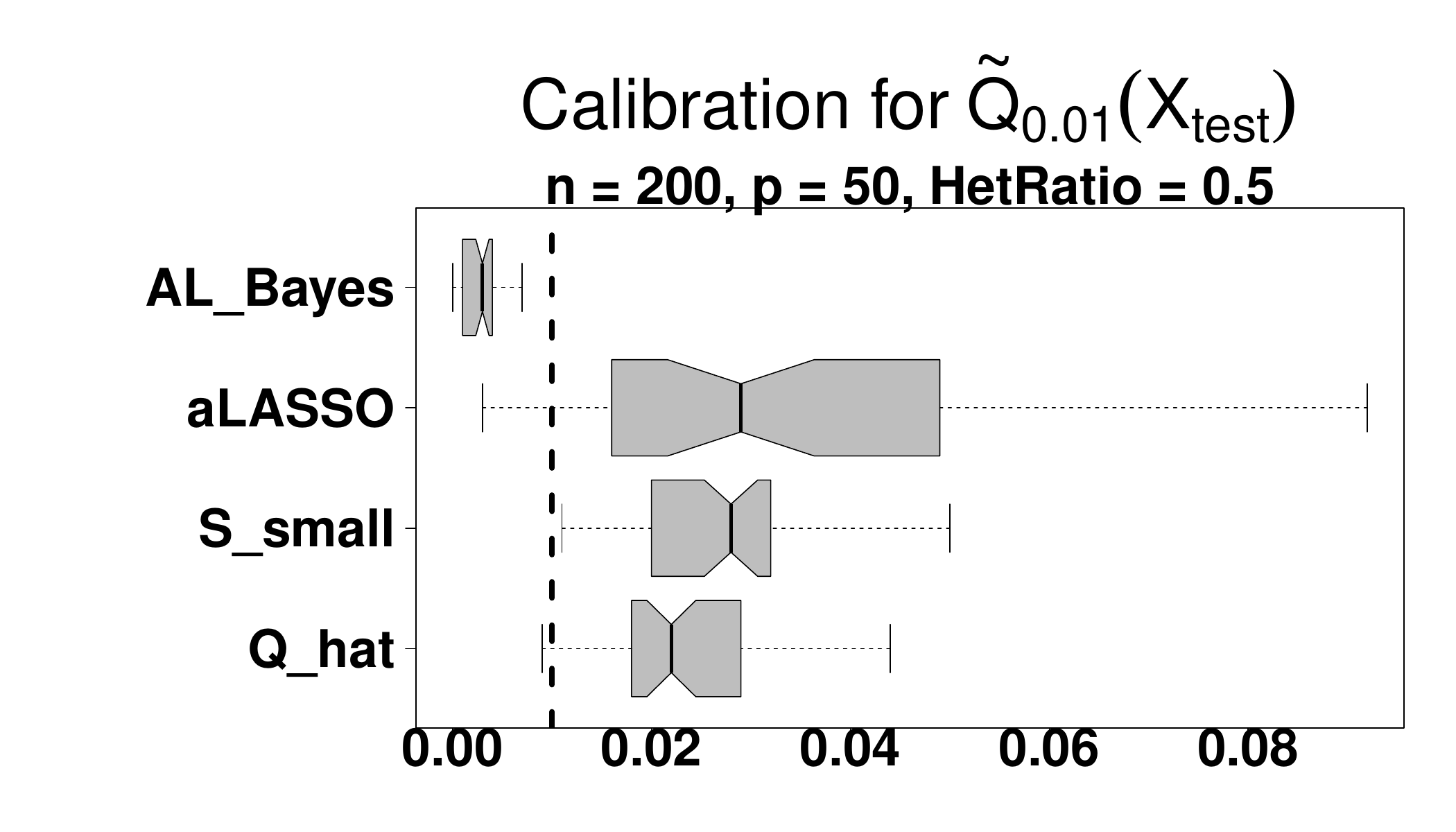}
    \includegraphics[width = .32\textwidth,keepaspectratio]{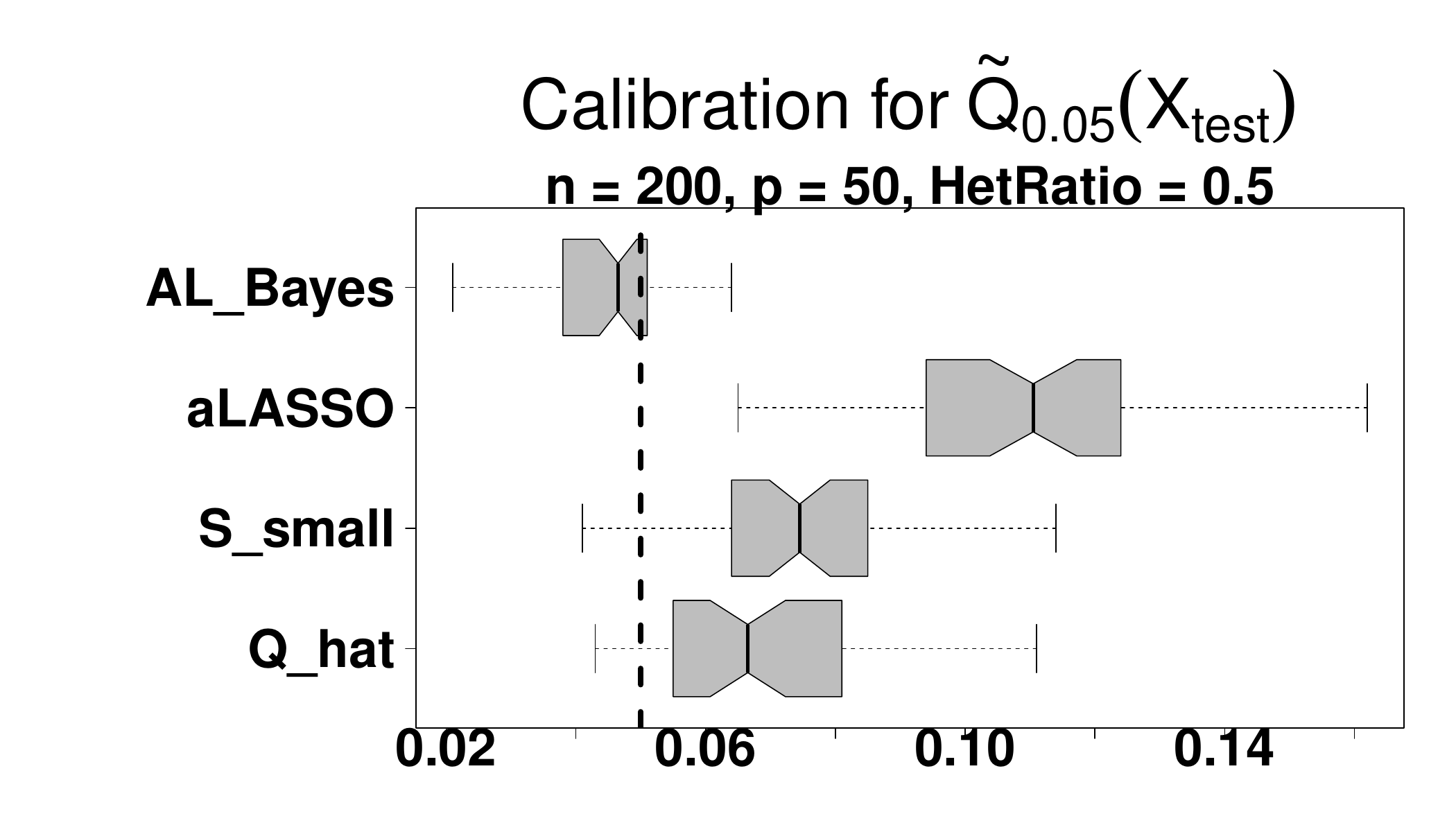}
    \includegraphics[width = .32\textwidth,keepaspectratio]{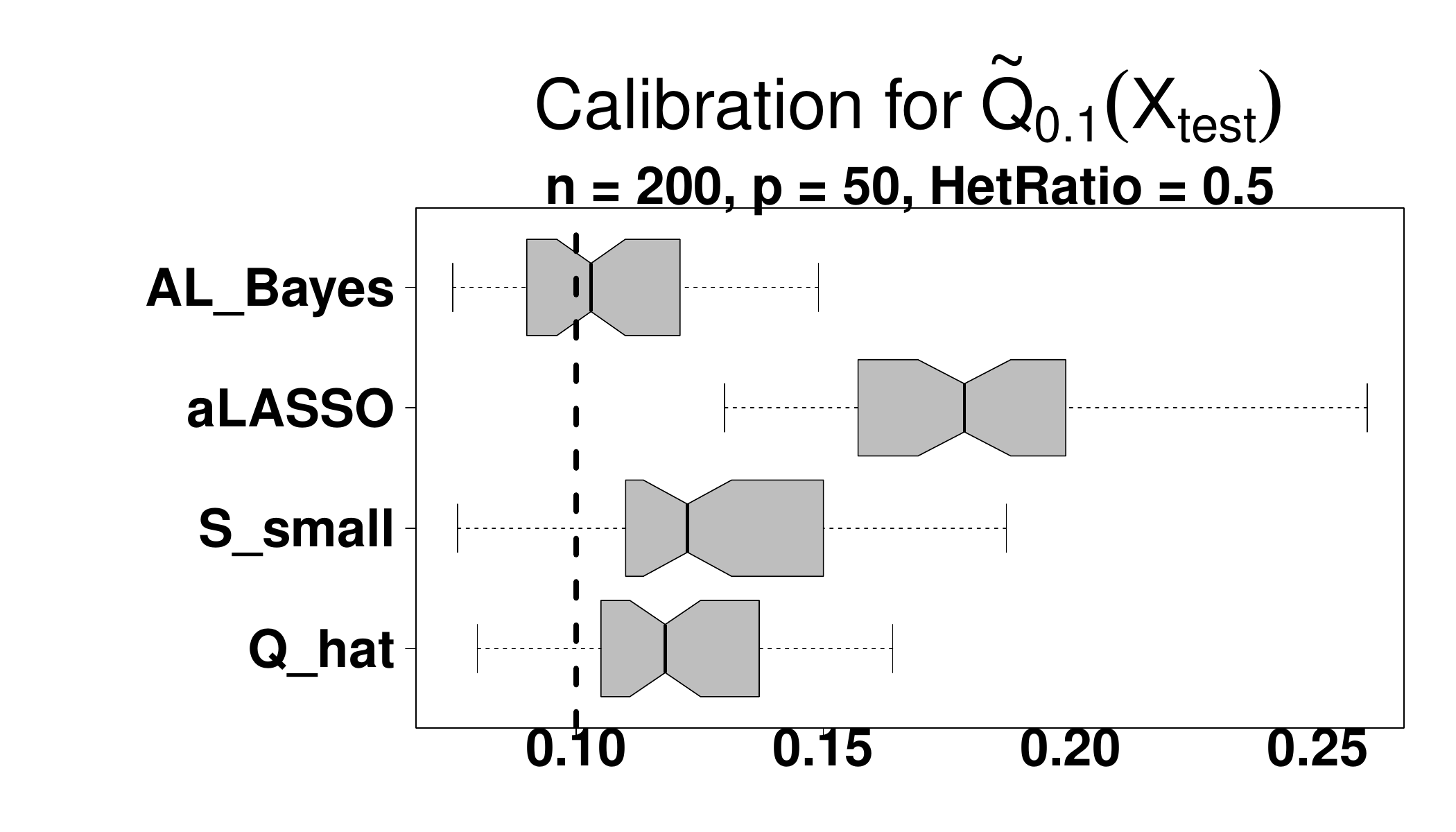}
    \includegraphics[width = .32\textwidth,keepaspectratio]{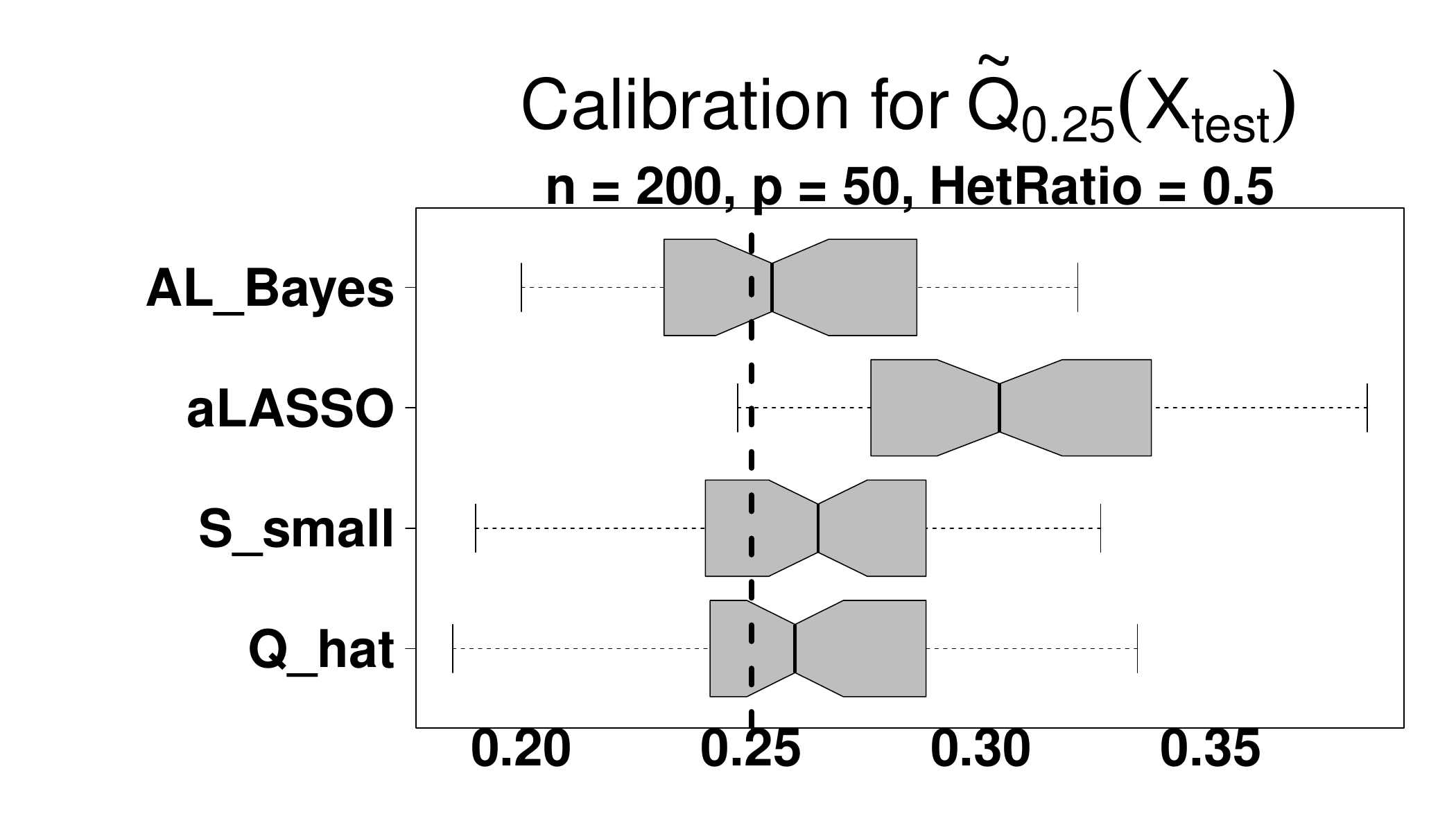}
    \includegraphics[width = .32\textwidth,keepaspectratio]{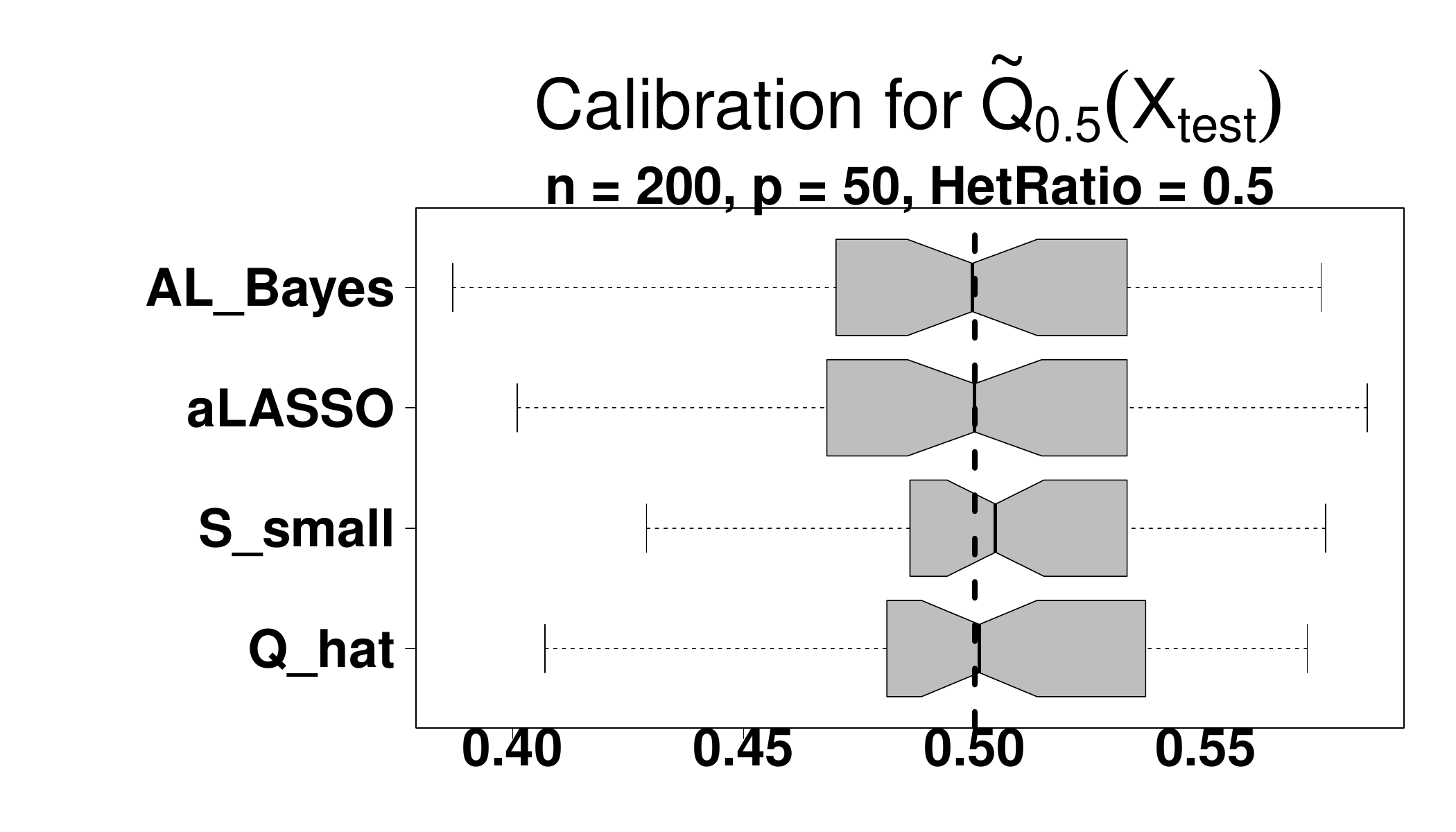}
    \includegraphics[width = .32\textwidth,keepaspectratio]{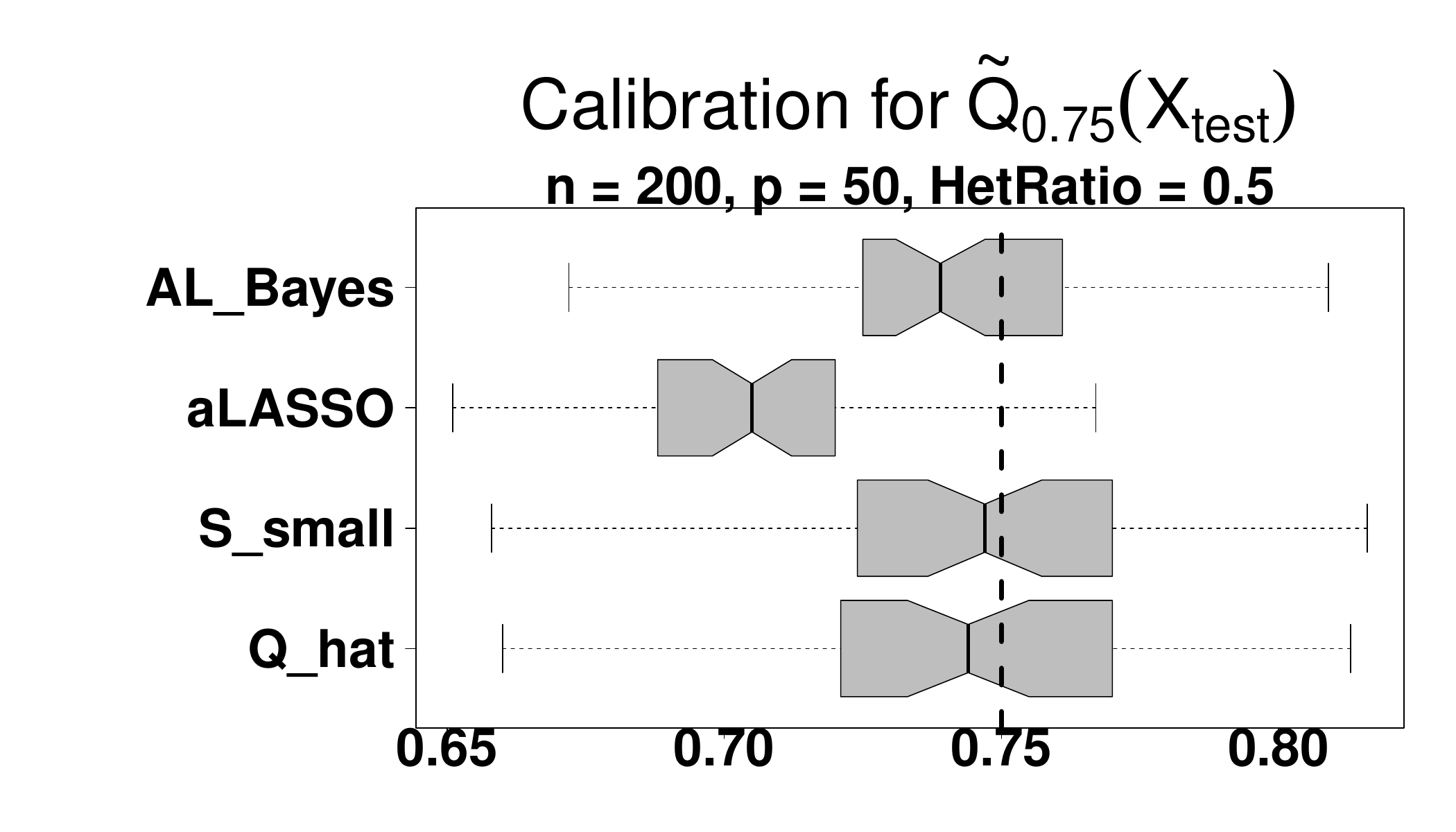}
   \includegraphics[width = .32\textwidth,keepaspectratio]{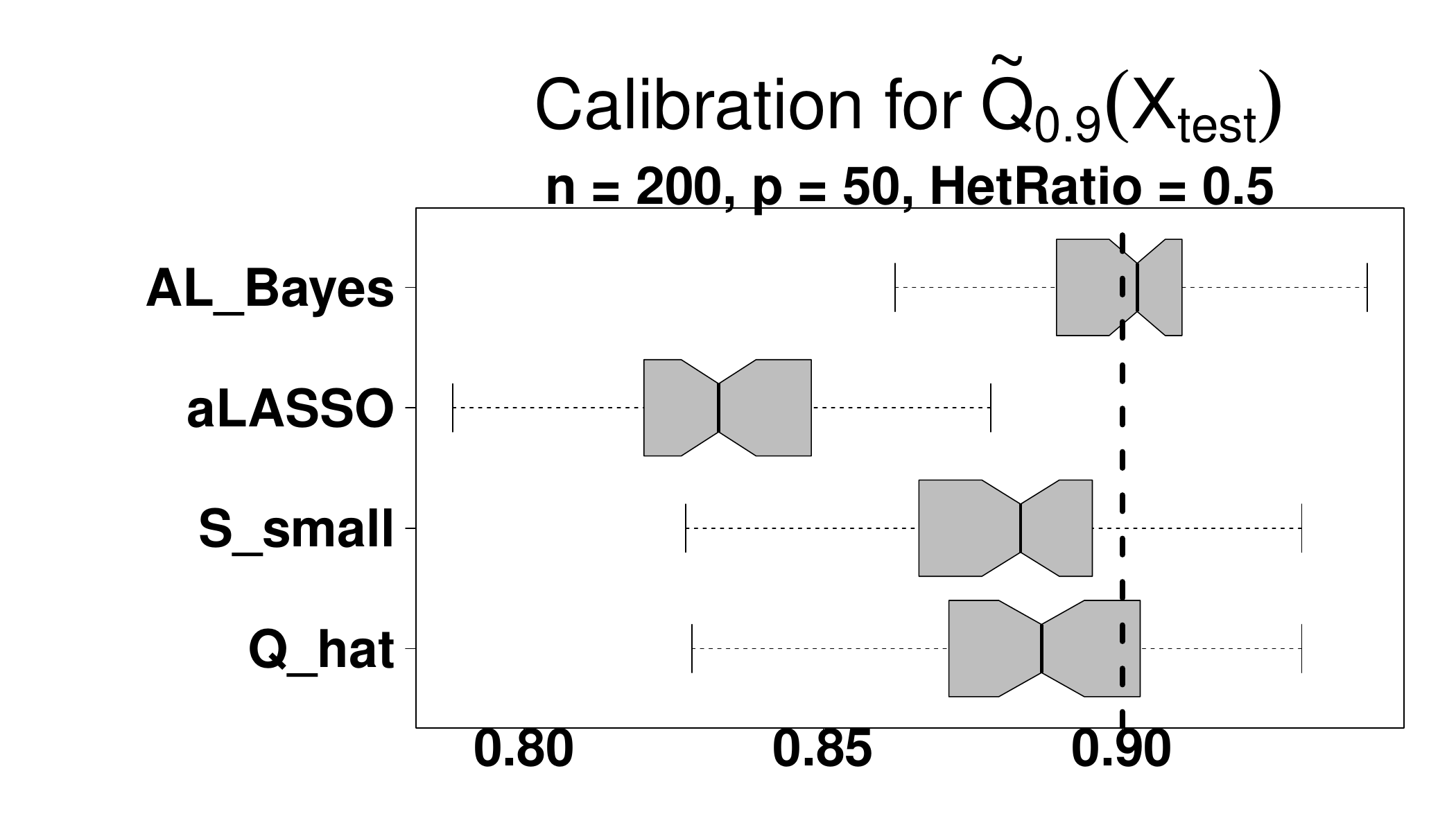}
    \includegraphics[width = .32\textwidth,keepaspectratio]{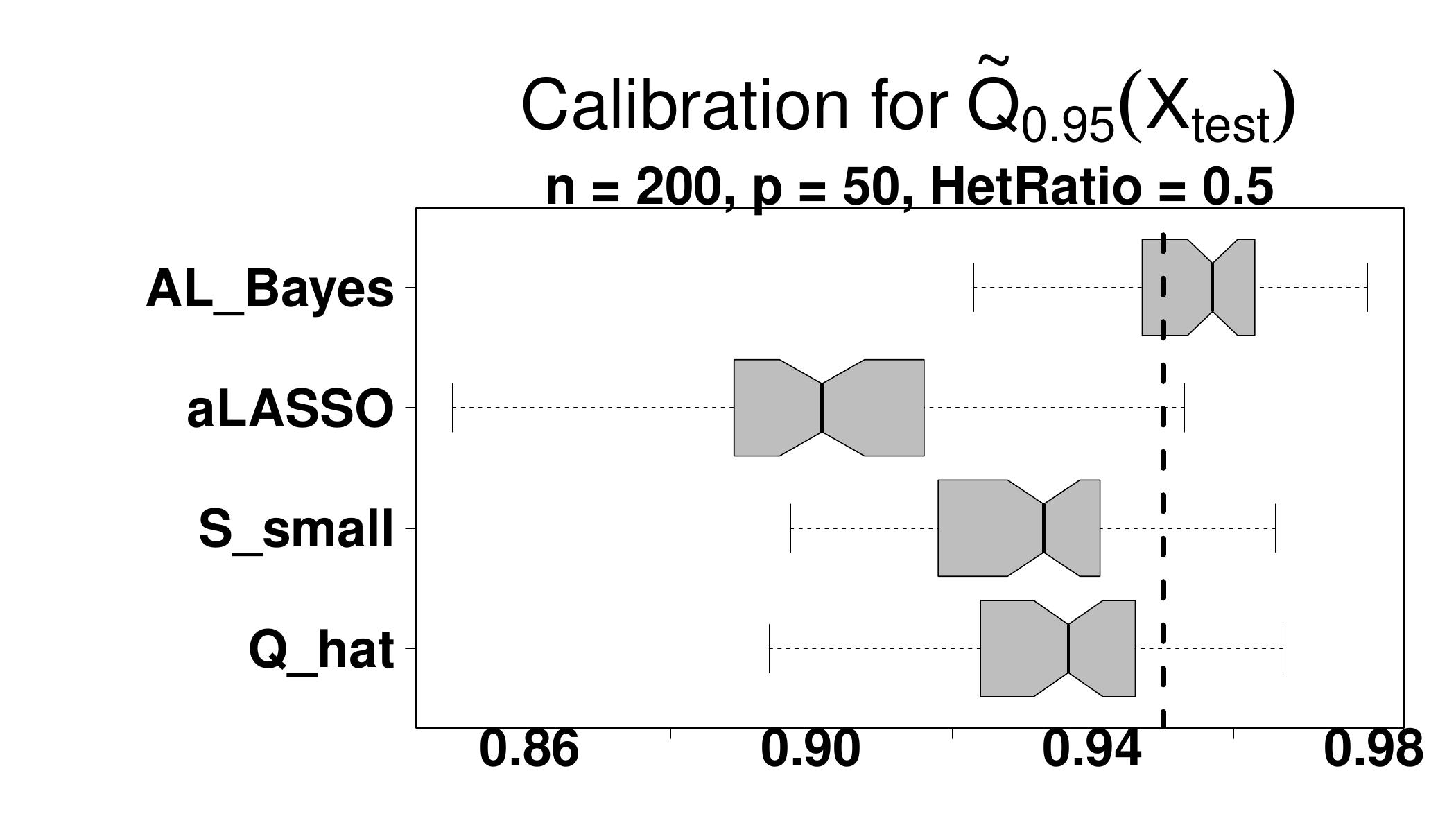}
   \includegraphics[width = .32\textwidth,keepaspectratio]{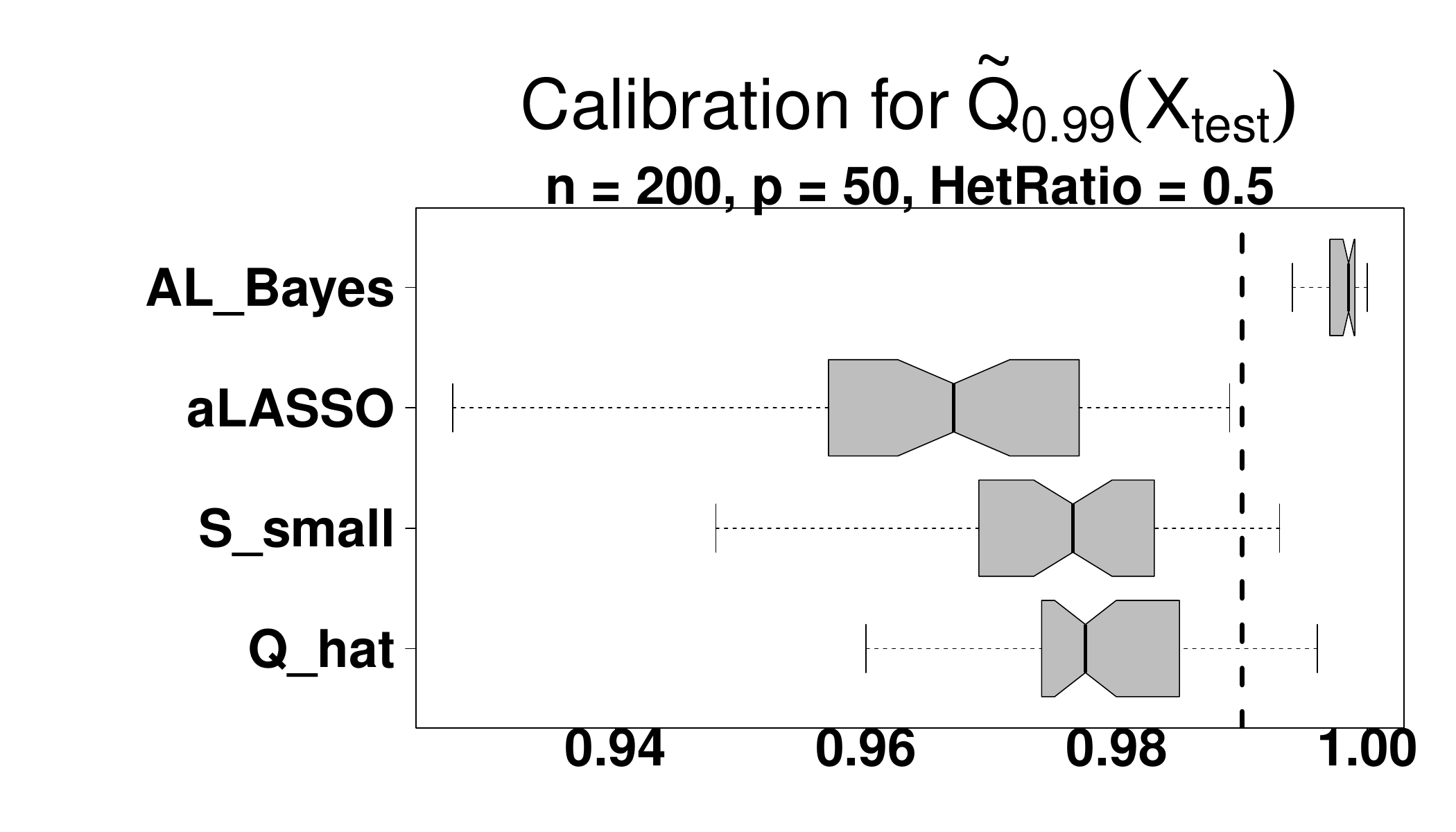}

\end{figure}
\begin{figure}[H]
    \centering
        \caption{\textbf{Calibration}: $\boldsymbol{n = 200, p= 50, \mbox{\textbf{HetRatio} }= 1}$}
  
    \label{Fig11_append}
    \includegraphics[width = .32\textwidth,keepaspectratio]{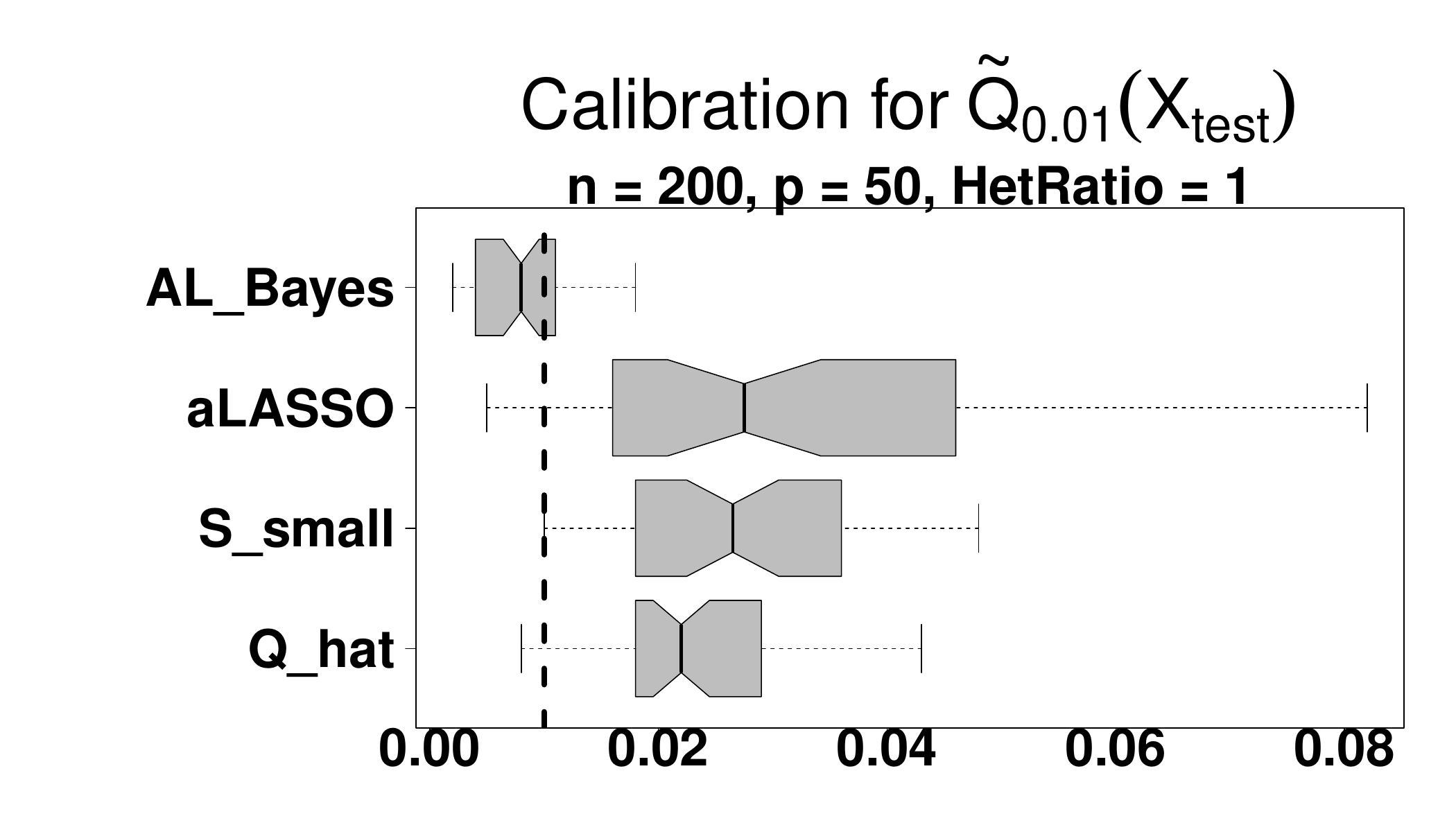}
    \includegraphics[width = .32\textwidth,keepaspectratio]{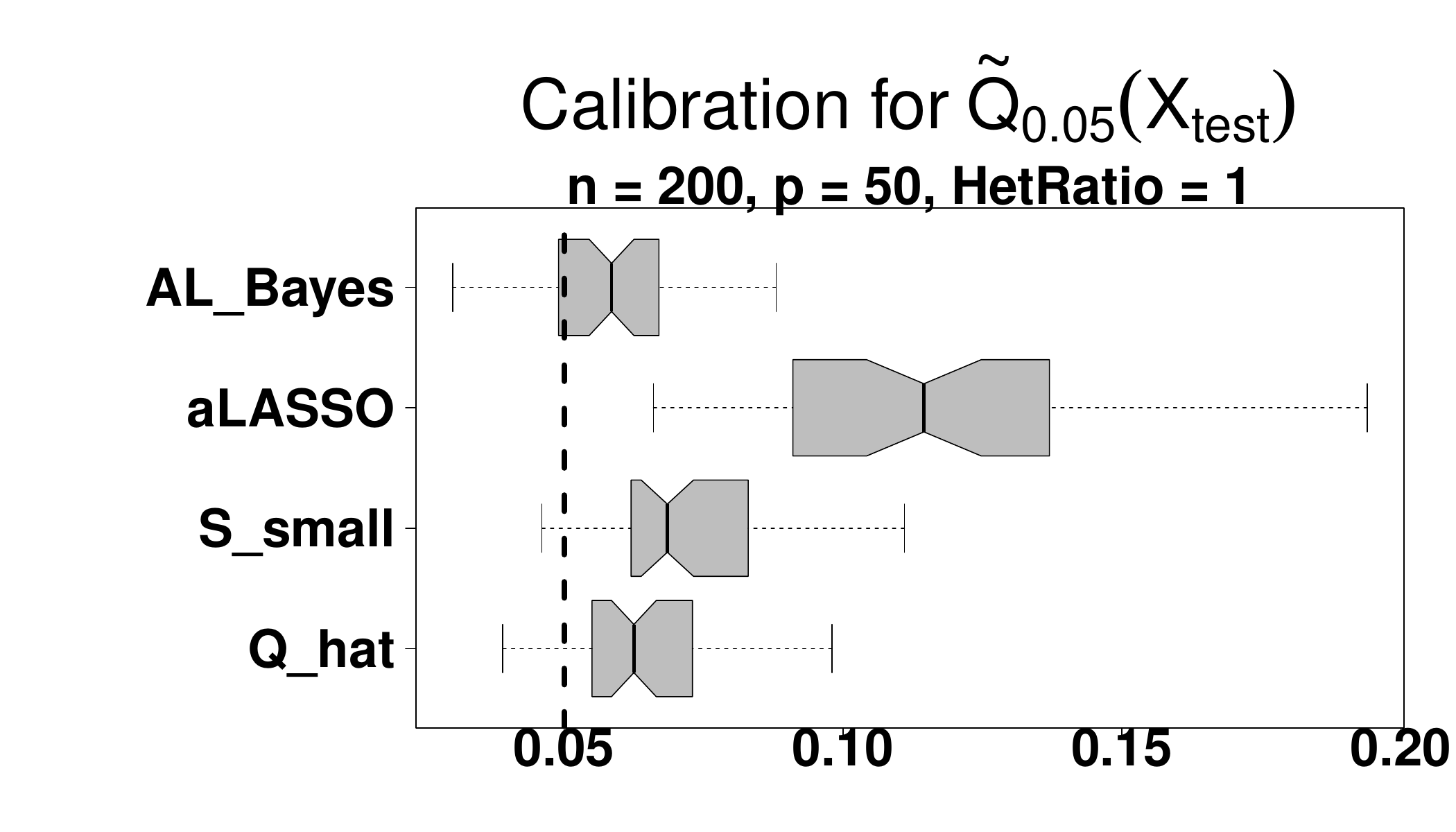}
    \includegraphics[width = .32\textwidth,keepaspectratio]{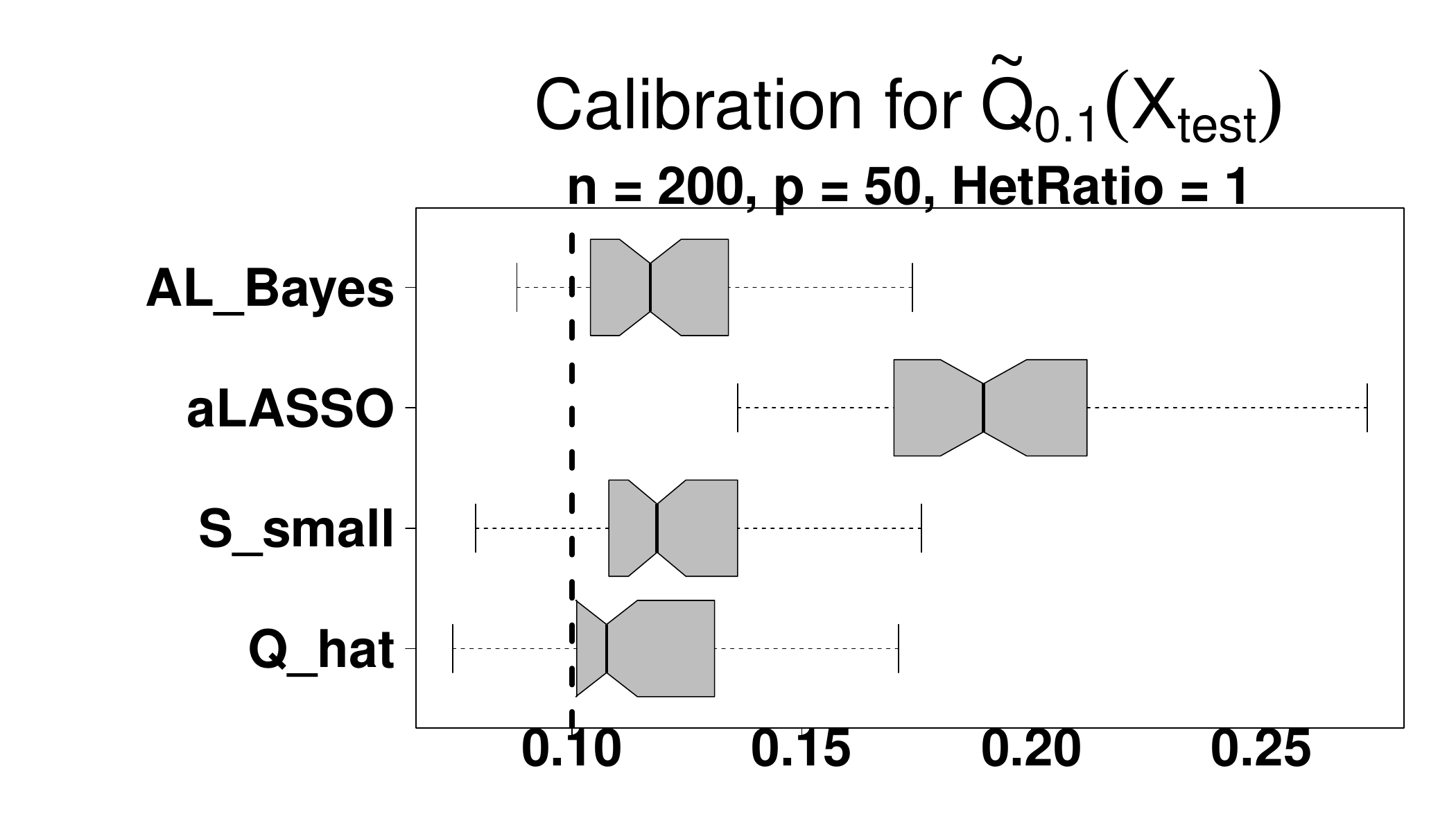}
    \includegraphics[width = .32\textwidth,keepaspectratio]{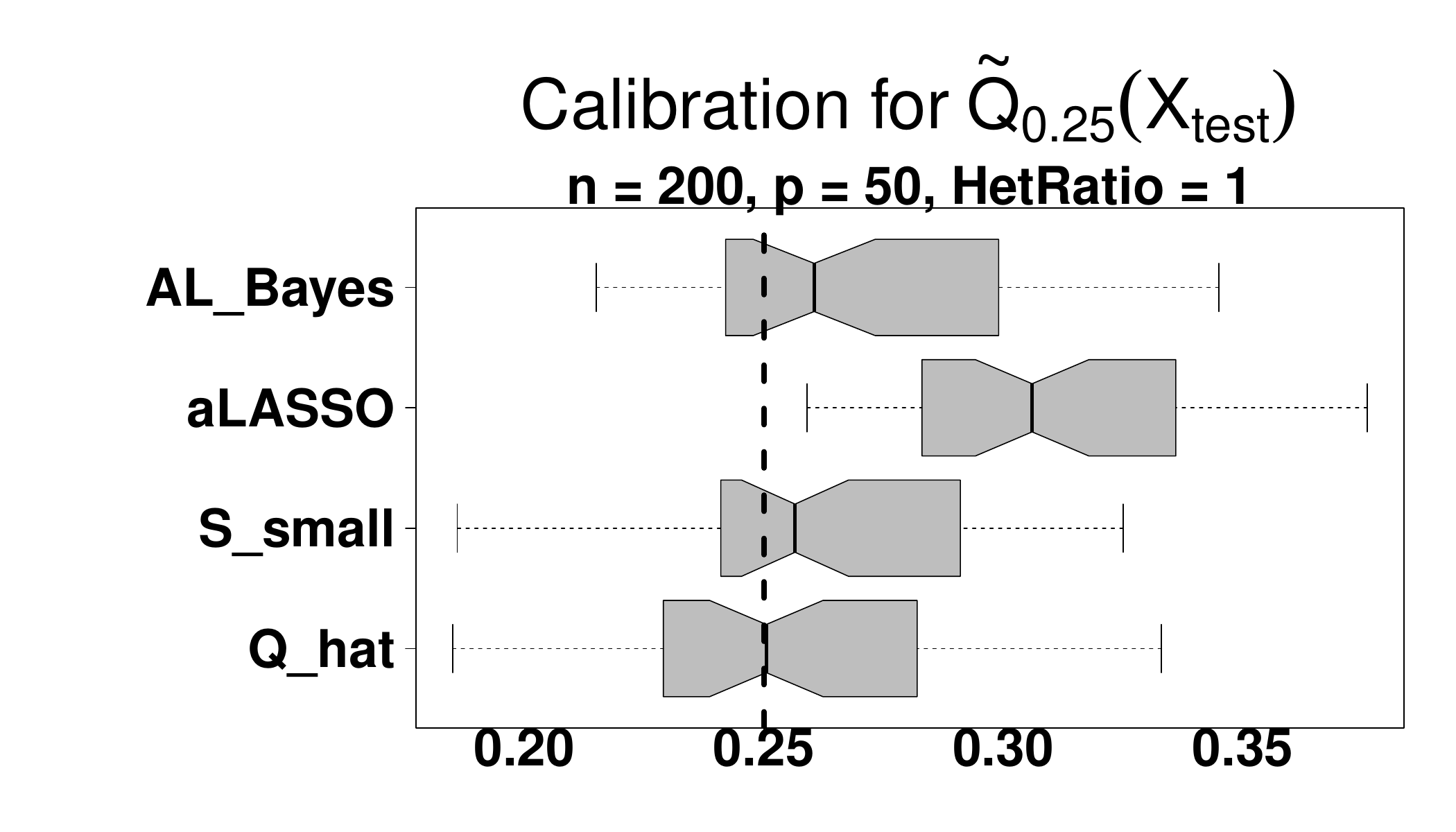}
    \includegraphics[width = .32\textwidth,keepaspectratio]{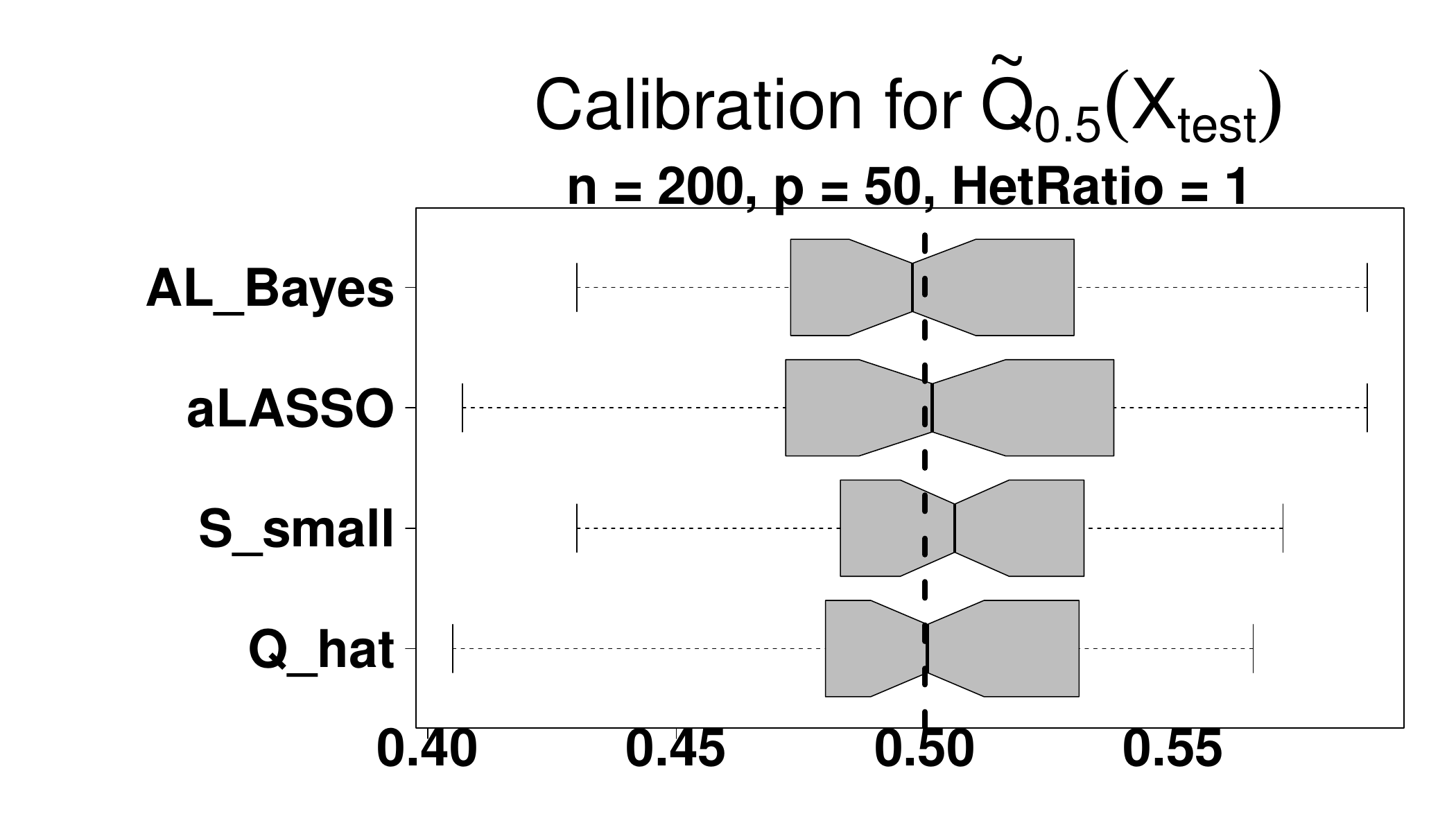}
    \includegraphics[width = .32\textwidth,keepaspectratio]{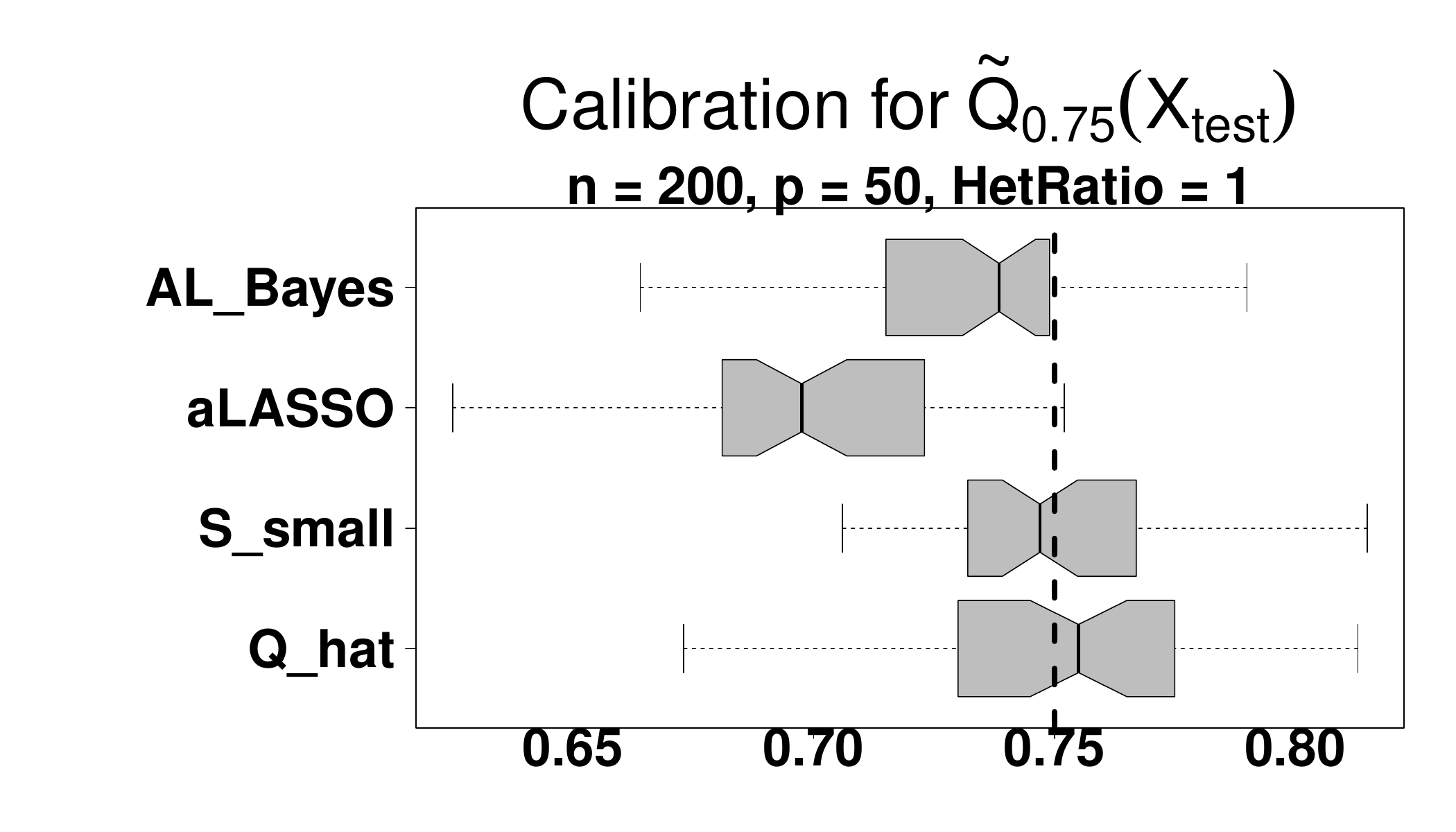}
   \includegraphics[width = .32\textwidth,keepaspectratio]{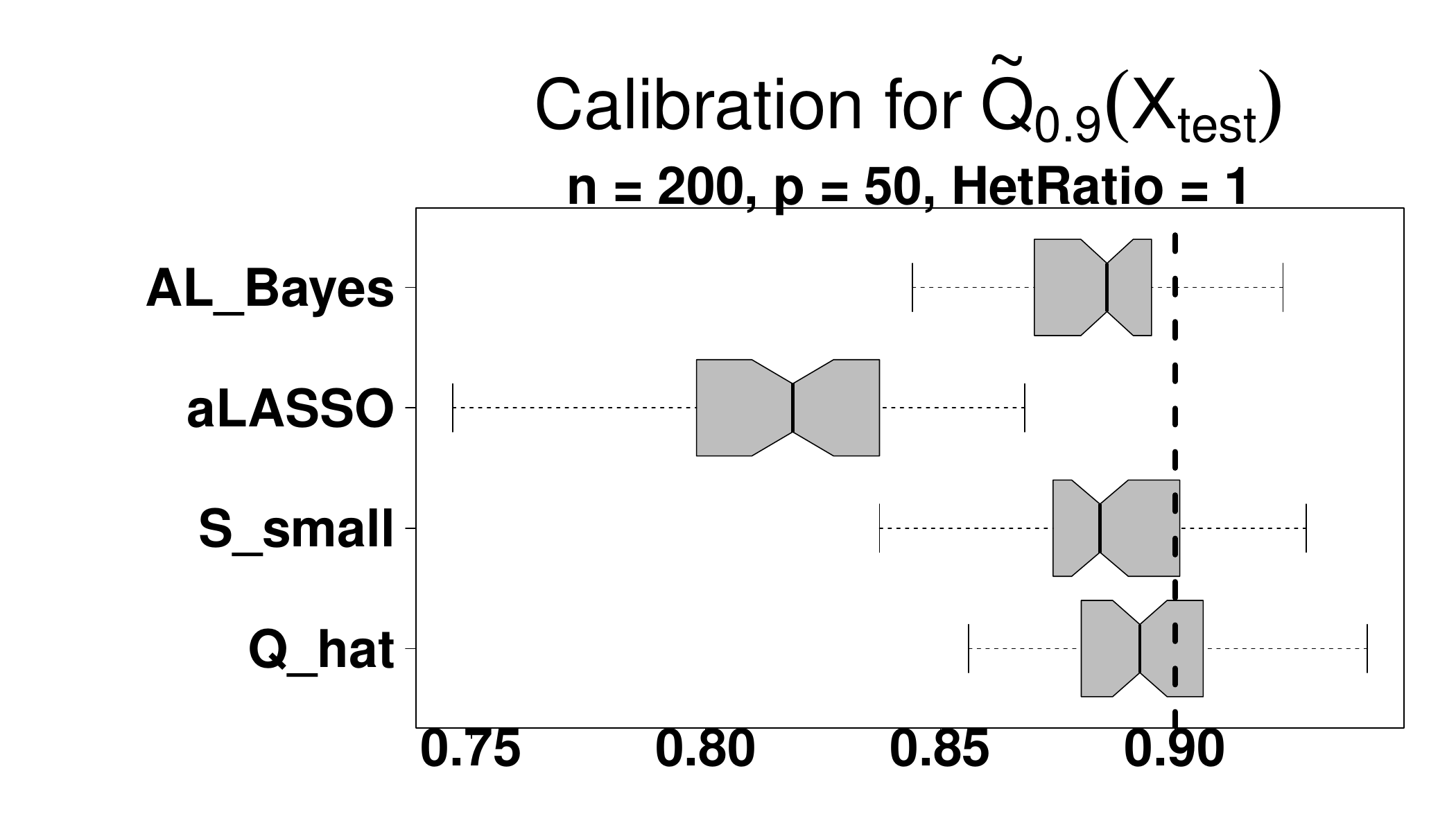}
    \includegraphics[width = .32\textwidth,keepaspectratio]{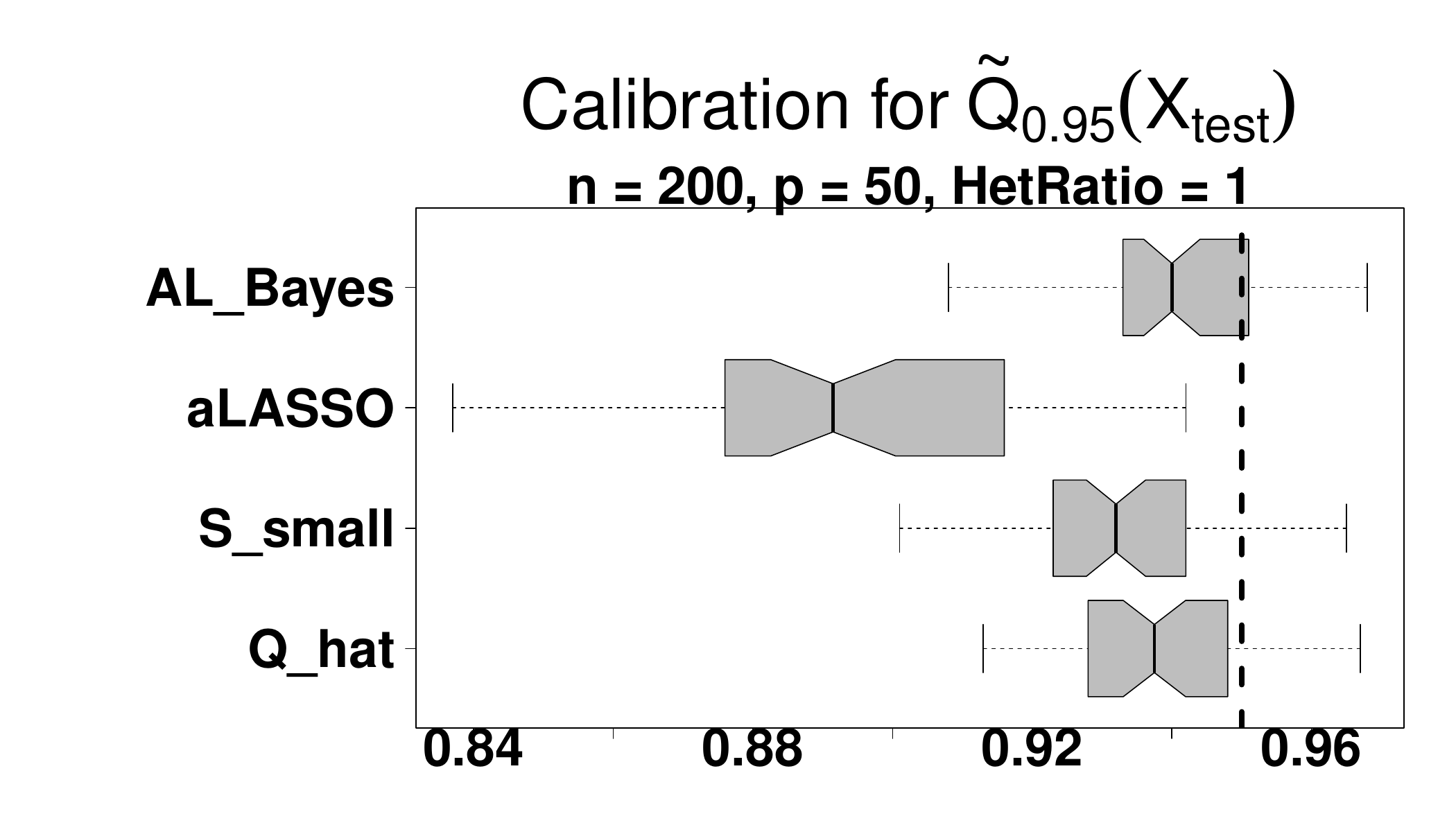}
   \includegraphics[width = .32\textwidth,keepaspectratio]{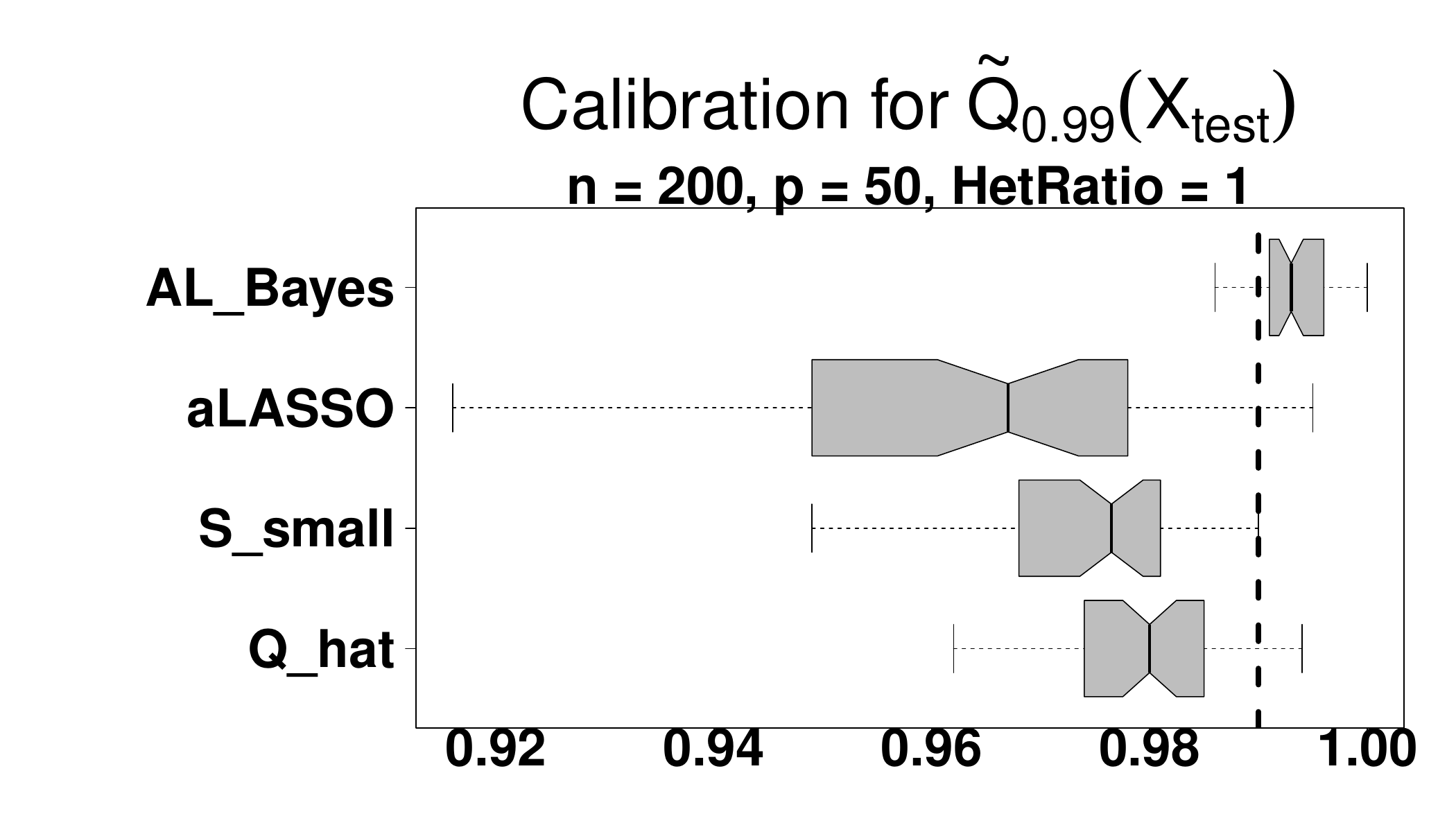}

\end{figure}

\subsection{Variable Importance}
In the main paper, we include results on the variable selection capabilities of the proposed approach based on the single subset $\mbox{S}_{small}(\tau)$ which is a member of each quantile-specific acceptable family $\mathbb{A}_{0.05}(\tau)$. However, a primary advantage of curating acceptable families for any quantile is that it removes reliance on a singular subset for capturing variable importance. As such, we evaluate the variable importance metric $\mbox{VI}_{j}(\tau)$ \eqref{VI} across simulation repititions. This quantity is valuable: it provides an informative summary of the acceptable family  \eqref{accept} that is more comprehensive than any single subset, including $\mbox{S}_{small}(\tau)$. 

In each simulation repetition and for each enumerated quantile, we average the variable importance for the homogeneous predictors and the predictors with zero coefficients ($zero$), and compute the variable importance for the single heterogeneous predictor. We average these metrics across simulation repetitions. The results for $n = 200, p = 50, \mbox{HetRatio = 1}$ are presented in Table \ref{tab6}, with similar results for the other settings, including those with independent covariates.
\setcounter{table}{5}
\begin{table*}[ht]
    \centering

    \caption{Average $\mbox{VI}_{j}(\tau)$: $n = 200, p = 50, \mbox{HetRatio = 1}$}     \label{tab6}
\begin{tabular}{rr|ccccccccc}
   & $\tau$ & 0.01 & 0.05 &0.25 & 0.5 & 0.75 &  0.95 & 0.99\\
\hline
\multirow{3}{*}{Indices ($j$)}& $het$ & 0.97 & 0.96 &  0.84 & 0.44 & 0.84 & 0.99 & 0.99\\
& $hom$ &0.70 & 0.77 &  0.87 & 0.91 & 0.87 &  0.79 & 0.72\\
& $zero$ & 0.34 & 0.35 &  0.36 & 0.35 & 0.36 & 0.35 & 0.34\\\hline
\end{tabular}
    \captionsetup{labelformat=empty}
    \caption*{Average  marginal variable importance for the covariate with heterogeneous effects ($het$), the covariates with homogeneous effects ($hom$) and the variables with no effect ($zero$) on the response distribution (top three rows). The quantile-specific acceptable families demonstrate broad agreement about the importance of the heterogeneous covariate for each quantile. This includes correctly identifying low importance for the median, in which case $\boldsymbol{\beta}_{het}(0.5) = 0$. Furthermore, the importance of the homogeneous covariates increase as the maginitude of $\boldsymbol{\beta}_{het}(\tau)$ decreases. Finally, the zero coefficients are correctly deemed as having  marginal importance for each quantile. }

\end{table*}

By examining $\mbox{VI}_{j}(\tau)$, we decide that the heterogeneous predictor is nearly vital for prediction of each quantile besides the median, when $\boldsymbol \beta_{het}(0.5) = 0$. This is informative, especially in conjunction with Table \ref{Table1} in the main paper. In the $n = 200, p = 50, \mbox{HetRatio} = 1$ setting, $\mbox{S}_{small}(\tau)$ does not achieve near 100\% true positive rates for any quantile, but variable importance maintains that it is a vital component of the quantile model. In addition, the importance of the homogeneous predictors increases as the magnitude of the heterogeneous predictor decreases. Finally, the zero predictors are deemed non-essential, as seen by low average variable importance.  Variable importance computed using the quantile-specific acceptable family expands the analysis beyond a single subset, providing useful and accurate information on informative covariates across the response distribution.

\subsection{Quantile Crossing}\label{cross}
We also investigate the quantile crossing properties of the competing approaches. Under a coherent probability model for $Y \mid \boldsymbol{x}$, quantiles cannot cross: 
$\tilde{Q}_{\tau}(\boldsymbol{x})< \tilde{Q}_{\tau'}(\boldsymbol{x})$ for any  $\tau < \tau'$. However, only the model-based  quantiles  ($\mbox{Q}_{hat}$) enforce this property; the competing methods, including the proposed approach, do not explicitly enforce quantile non-crossing. Thus, we seek to quantify the abundance of quantile non-crossing for each method.

For any $\tau < \tau'$, we compute the out-of-sample non-crossing rate (NCR) between neighboring quantile predictions at the testing points:
\begin{equation}\label{NCR}
\mbox{NCR}(\tau, \tau') = n_{test}^{-1} \sum_{i=1}^{n_{test}} \mathbbm{1}\{\tilde{Q}_{\tau}(\boldsymbol x_{test_i})< \tilde{Q}_{\tau'}(\boldsymbol x_{test_i})\}.
\end{equation}
When $\mbox{NCR}(\tau, \tau') =1$, there is no quantile crossing between the $\tau$th and $\tau'$th quantiles. 

We compute $\mbox{NCR}(0.01,0.05)$ and $\mbox{NCR}(0.95,0.99)$ averaged across simulations for each method (Table \ref{Tab6}). Remarkably, the proposed approach renders quantile crossing negligible \emph{without} any explicit constraints in the decision analysis for estimation or selection. These results showcase a key advantage of the Bayesian decision analysis \eqref{post-dec}: by fitting to the model-based fitted quantiles via \eqref{OLS_sol}, the optimal linear actions benefit from the implicit quantile non-crossing of $\hat Q_\tau(\boldsymbol x)$ under $\mathcal{M}$. Thus, we (nearly) acquire the primary advantage of simultaneous quantile regression methods (Section~\ref{simult}), but without the need for unwieldy constraints.  By comparison, the competing frequentis is subject to abundant quantile crossing, especially with larger $p$ and stronger heterogeneity. The Bayesian competitor preserves non-crossing for smaller $p$ settings, but this property is badly violated for $p = 100$. For both methods, this limits the interpretability of the estimated linear coefficients and suggests that the estimated quantiles may be unreliable for prediction or inference. 



\begin{table}[h]
\centering
    \caption{$\boldsymbol{n = 200, p = 50, \textbf{\mbox{HetRatio}} =1}$}\label{Tab6}
  \begin{tabular}{c | c | c }

     & $\mbox{NCR}(0.01,0.05)$ & $\mbox{NCR}(0.95,0.99)$ \\\hline
     $\mbox{S}_{small}(\tau)$ &0.99&0.99\\
     $\mbox{AL}_{Bayes}$  &0.99& 0.99\\
     $\mbox{aLASSO}$ &0.88 &0.86\\\hline

    \end{tabular}
\captionsetup{labelformat = empty}
        \caption*{$\boldsymbol{n = 200, p = 50, \textbf{\mbox{HetRatio}} =0.5}$}
             \begin{tabular}{c | c | c }

     & $\mbox{NCR}(0.01,0.05)$ & $\mbox{NCR}(0.95,0.99)$ \\\hline
     $\mbox{S}_{small}(\tau)$ &0.99&0.98\\
     $\mbox{AL}_{Bayes}$  &0.99& 0.99\\
     $\mbox{aLASSO}$ &0.86 &0.85\\\hline
    \end{tabular}
\captionsetup{labelformat = empty}
        \caption*{$\boldsymbol{n = 500, p = 20, \textbf{\mbox{HetRatio}} =1}$}
         \begin{tabular}{c | c | c }

     & $\mbox{NCR}(0.01,0.05)$ & $\mbox{NCR}(0.95,0.99)$ \\\hline
     $\mbox{S}_{small}(\tau)$ &0.997&1.00\\
     $\mbox{AL}_{Bayes}$  &0.99& 0.99\\
          $\mbox{aLASSO}$ &0.95 &0.94\\\hline

    \end{tabular}

    \captionsetup{labelformat = empty}
    \caption*{$\boldsymbol{n = 500, p = 20, \textbf{\mbox{HetRatio}} =0.5}$}\label{Tab9}
    \begin{tabular}{c | c | c }

     & $\mbox{NCR}(0.01,0.05)$ & $\mbox{NCR}(0.95,0.99)$ \\\hline
     $\mbox{S}_{small}(\tau)$ &1.00 &1.00\\
          $\mbox{AL}_{Bayes}$  &0.99& 1.00\\
     $\mbox{aLASSO}$ &0.93 &0.94\\\hline

    \end{tabular}
\captionsetup{labelformat = empty}
        \caption*{$\boldsymbol{n = 100, p = 100, \textbf{\mbox{HetRatio}} =1}$}
         \begin{tabular}{c | c | c }

     & $\mbox{NCR}(0.01,0.05)$ & $\mbox{NCR}(0.95,0.99)$ \\\hline
     $\mbox{S}_{small}(\tau)$ &0.97&97\\
     $\mbox{AL}_{Bayes}$  &0.46& 0.64\\
          $\mbox{aLASSO}$ &0.93 &0.89\\\hline

    \end{tabular}

    \captionsetup{labelformat = empty}
    \caption*{$\boldsymbol{n = 100, p = 100, \textbf{\mbox{HetRatio}} =0.5}$}\label{Tab9}
    \begin{tabular}{c | c | c }

     & $\mbox{NCR}(0.01,0.05)$ & $\mbox{NCR}(0.95,0.99)$ \\\hline
     $\mbox{S}_{small}(\tau)$ &1.00 &1.00\\
          $\mbox{AL}_{Bayes}$  &0.35& 0.71\\
     $\mbox{aLASSO}$ &0.95 &0.94\\\hline

    \end{tabular}

    \captionsetup{labelformat=empty}
    \caption*{Quantile non-crossing rates between the 1st and 5th and 95th and 99th quantiles.  
    Unlike aLASSO and the competing Bayesian approach, the proposed approach $\mbox{S}_{small}(\tau)$ renders quantile crossing a non-issue.}

\end{table}

\section{North Carolina Data Analysis}
\subsection{Correlation Among Covariates}
\begin{figure}
    \centering
    \includegraphics[width =1.2\textwidth,keepaspectratio]{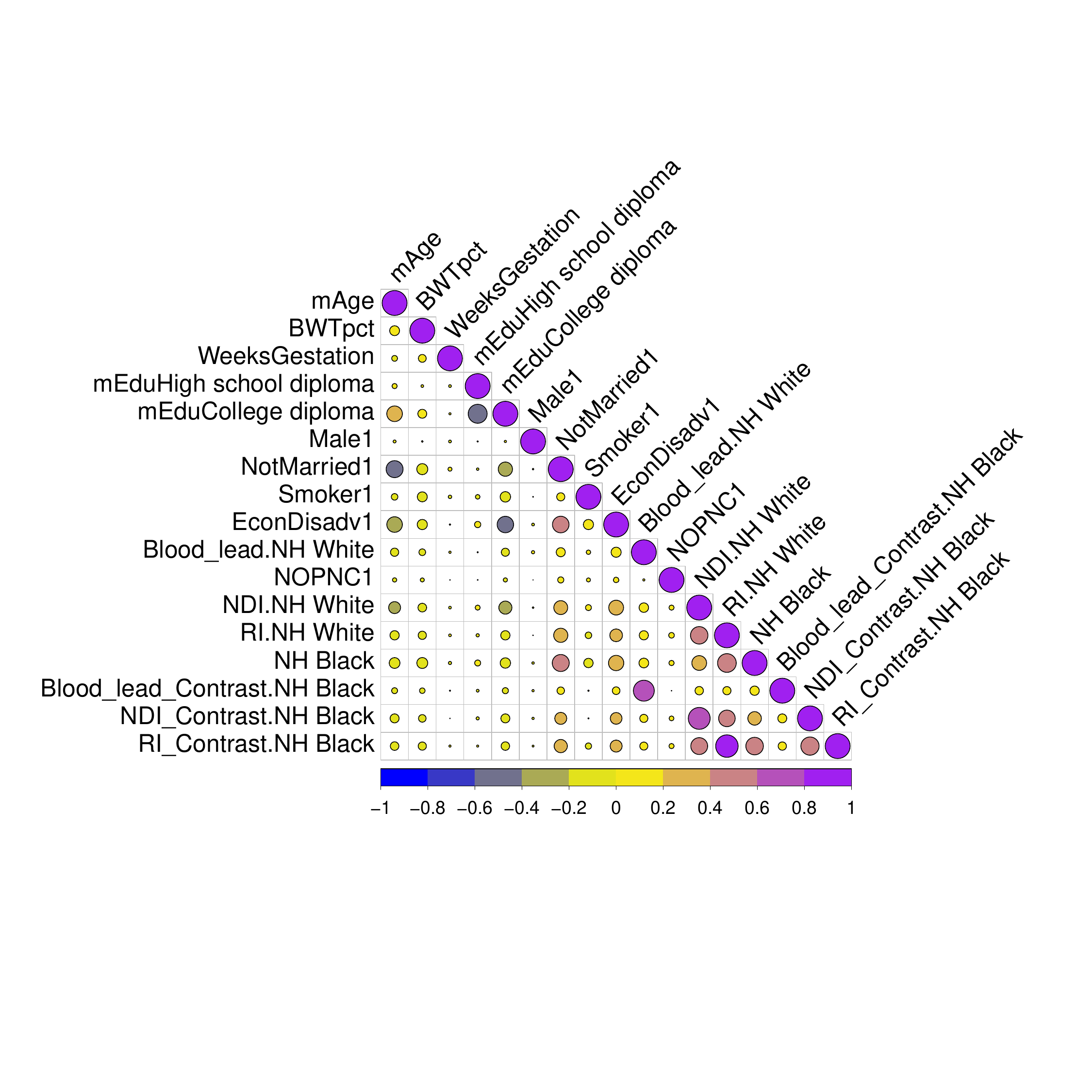}
    \caption{Pairwise pearson's correlation between covariates in the North Carolina data. Varying degrees of association can be observed. Thus, interchanging highly correlated predictors in linear quantile regression will likely sacrifice little predictive power. This supports the curation of acceptable families for each quantile. }
    \label{corrplot}
\end{figure}

Many of the covariates in the augmented North Carolina data set are highly correlated, which is evident in Figure \ref{corrplot}. This promotes the collection of many, near-optimal and  competing explanations for the same quantile function.

\subsection{Evaluating the LL-LS Model Fit on the North Carolina Data}

\begin{figure}
    \centering
    \includegraphics[width = 0.5\textwidth,keepaspectratio]{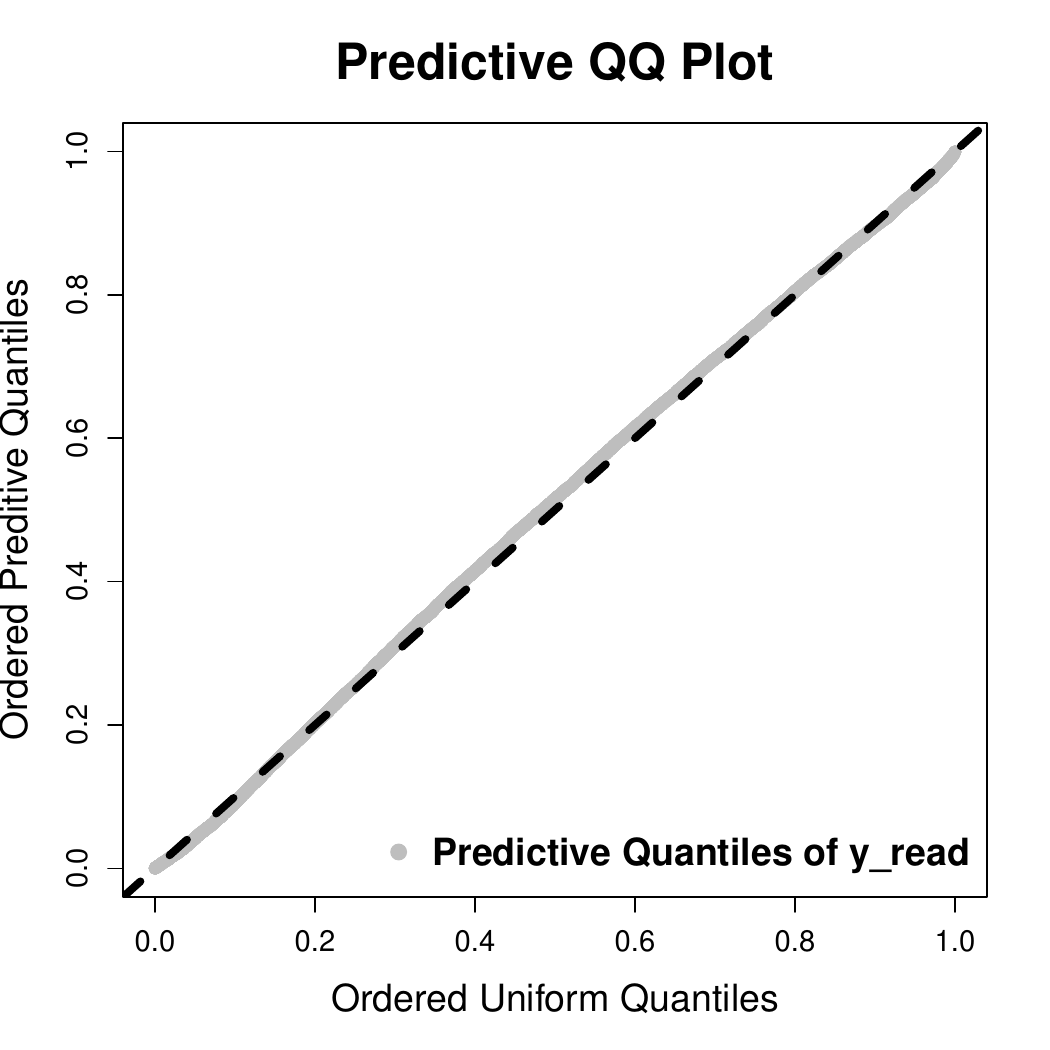}
    \caption{Predictive QQ plot for reading scores under the LL-LS model. Because the predictive quantiles of the observed reading scores are approximately uniform, as evidenced by little deviation from the 45-degree line, the model is approximately calibrated to the data.}
    \label{predQQ}
\end{figure}

Prior to posterior summarization, it is vital to first ensure that the Bayesian model $\mathcal{M}$, which is specified as the LL-LS model \eqref{M}, is calibrated to the North Carolina data. To do so, we utilize posterior predictive QQ-plots \citep{pratola2020heteroscedastic}. Under a well-calibrated model, the quantiles of the observed response variables at each covariate value under the posterior predictive distribution should be approximately uniform. For each covariate value we generate 2500 draws from the corresponding posterior predictive distribution of reading scores using posterior samples from the LL-LS model. We then calculate the empirical quantiles of the observed response at each covariate value based on the posterior predictive draws. The ordered sample quantiles of the observed reading scores are plotted against uniform quantiles in Figure \ref{predQQ}. Overlaid onto the plot is 45-degree line. We observe that the sample quantiles are approximately uniform, providing evidence that the LL-LS model is calibrated to the data.

We next determine the extent to which the North Carolina data displays predictor-dependent heteroscedasticity, which motivates a quantile regression analysis. We base our comparisons to a Bayesian homoscedastic linear regression (BHLR) fit to the North Carolina data using the same predictors and response, which assumes constant variance for any covariate value. We specify independent horseshoe priors on the BHLR regression coefficients \citep{carvalho2009handling}, and an inverse gamma prior on the error variance.

For these comparisons, traditional metrics like root mean squared error (RMSE) are unsatisfying since they are geared toward estimating predictive power, rather than the quality of a model's higher order properties. Thus, we rely on graphical displays of the variance process to determine whether the LL-LS model better detects predictor-dependent heteroscedasticity over BHLR. If the variance in end-of-grade reading scores can be explained by the covariates in our data set, we hypothesize that the model-based conditional quantiles will have heterogeneous covariates effects.

\begin{figure}
    \centering
    \includegraphics[width = 0.5\textwidth,keepaspectratio]{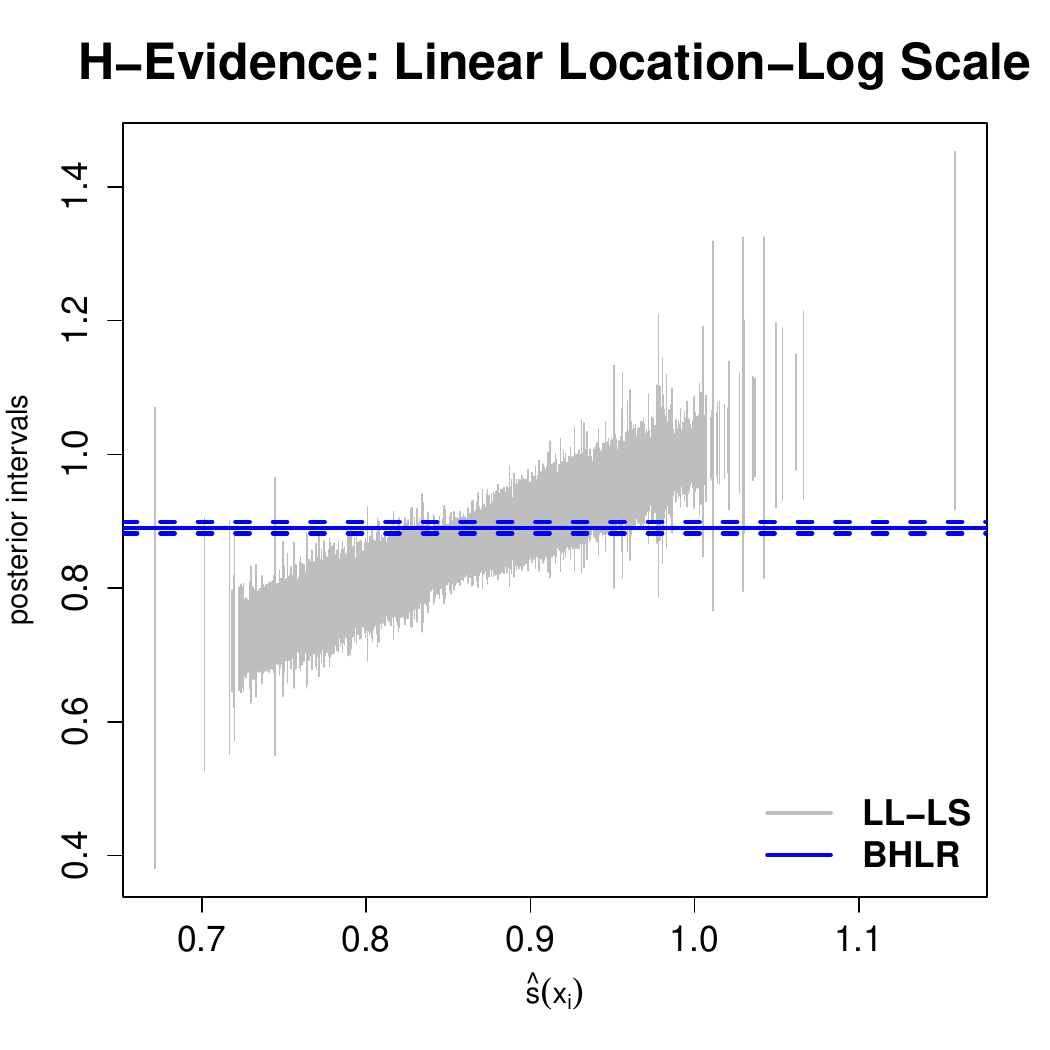}
    \caption{H-evidence for the linear location-log scale model $\mathcal{M}$. Relative to homoscedastic linear regression, $\mathcal{M}$ provides evidence that the error variance is predictor dependent, as evidenced by the significant number of posterior intervals for $\{s(\boldsymbol x_i)\}$ that do not overlap with the homoscedastic estimate.}
    \label{h-ev}
\end{figure}

We summarize the variance process under the LL-LS model using an \textit{H-evidence} plot \citep{pratola2020heteroscedastic} in Figure \ref{h-ev}, which plots $95\%$ posterior intervals  for $\{s(\boldsymbol x_i)\} = \{\sigma\exp(\boldsymbol x_i^{\intercal}\boldsymbol\gamma)\}$. We further sort these intervals by their posterior means $\hat{s}(\boldsymbol x_i)$ which aids in visualization.  By comparing the estimates of  $\{s(\boldsymbol x)\}$ under the LL-LS to the constant error variance estimate obtained from BHLR, we detect whether there is sufficient evidence in the data to determine that $s(\boldsymbol x)$ is non-constant. Thus, we overlay onto the intervals the posterior mean of the error variance under the BHLR, with accompanying $95\%$ credible bands.

 Figure \ref{h-ev} provides evidence that the variance is heteroscedastic:  many of the posterior intervals for $\{s(\boldsymbol x_i)\}$ do not overlap with the constant variance estimator under BHLR.  As such, we anticipate that linear summaries of the model-based conditional quantiles will uncover covariates for which estimated quantile regression coefficients $\boldsymbol{\beta}_{j}(\tau)$ vary in $\tau$.

\subsection{Posterior Summarization of the Bayesian Quantile Regression}
We also conducted the proposed quantile regression with subset selection using posterior samples from Bayesian quantile regression under the asymmetric Laplace likelihood with adaptive LASSO priors on the regression coefficients. We evaluate the sensitivity of the variable importance  and quantile-specific smallest acceptable subsets to the underlying Bayesian model.

Given the wide posterior uncertainty under this model for the quantile regression coefficients under the Bayesian quantile regression, particularly for extreme quantiles (Figures \ref{main} - \ref{intrctt}), we curate quantile-specific acceptable families usign $\varepsilon = .25$ for $\tau = \{0.01,0.99\}$ and $\varepsilon = .10$ for the other quantiles. This enforces that there be stronger evidence under the model that each subset fits better than the anchor. In addition, we extract 80\% credible intervals under the posterior action to quantify uncertainty among the covariates included in $S_{small}(\tau)$.

\begin{figure*}[h]
    \centering
        \includegraphics[width = .85\textwidth, keepaspectratio]{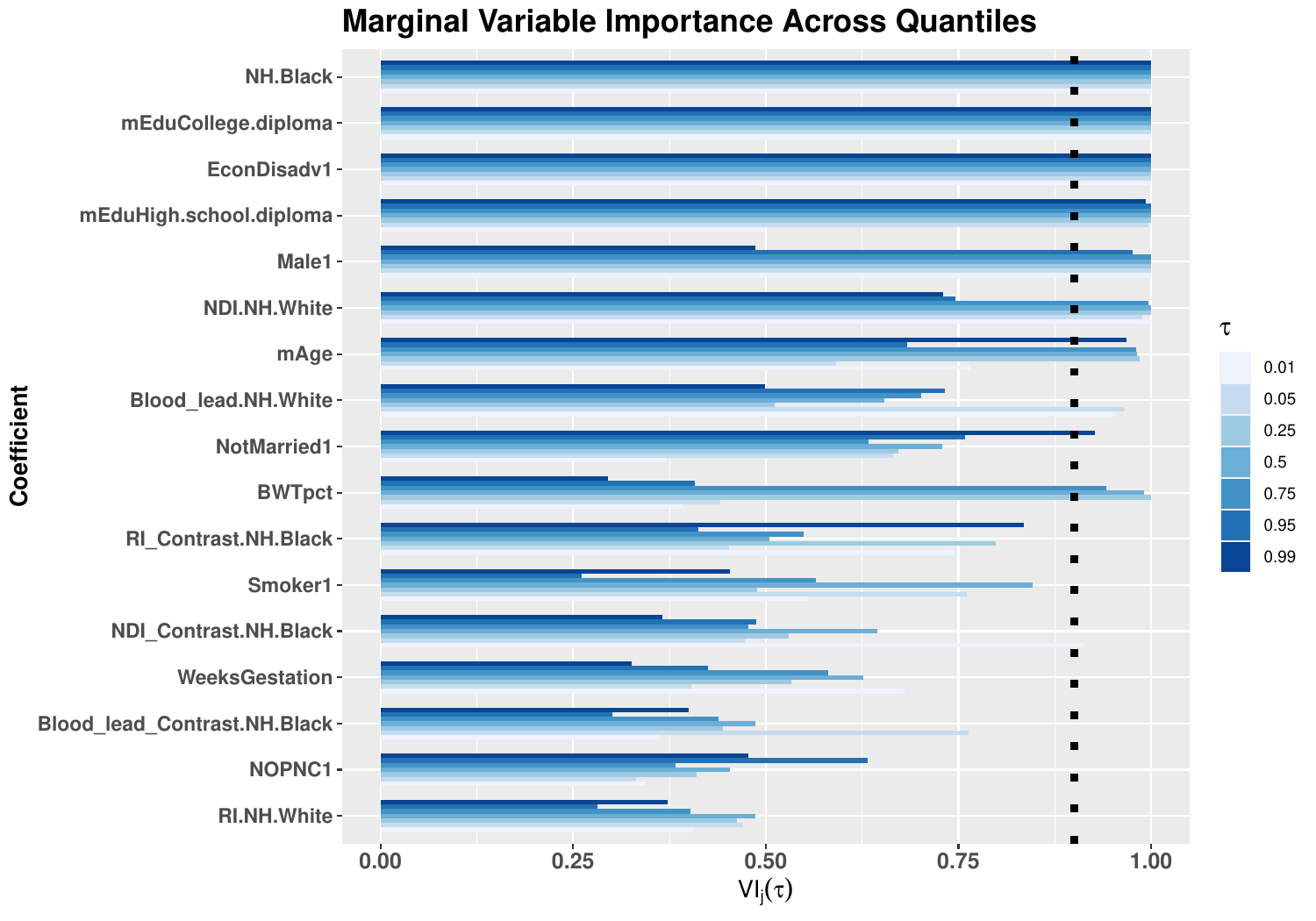}
    \includegraphics[width = .85\textwidth, keepaspectratio]
    {images/varimp_cntrst.pdf}
    \caption{Variable importance $\mbox{VI}_{j}(\tau)$ from \eqref{VI}, colored by quantile; the dashed line indicates 0.90. Large values indicate that the covariate appears in many of the acceptable subsets. The top plot provides variable importance under the Bayesian quantile regression, while the bottom row is what is presented in Figure \ref{vimps} for the LL-LS model. We conclude that overall, the covariates included in the  quantile-specific acceptable families are quite similar between the two models.}
    \label{vimpscomp}
\end{figure*}

We first present the marginal variable importance for each covariate from the quantile-specific acceptable families, and compare these to Figure \ref{vimps} in the main paper in Figure \ref{vimpscomp}. We observe between the models that the metrics are similar for each covariate.

Furthermore, the variables included in at least one $S_{small}(\tau)$ across the variables were identical between the LL-LS model and the Bayesian quantile regression. We compare the inference under the posterior action between the two models for the main effect covariates in Figure \ref{ssmall comp main}. The effect sizes and directionality across $\tau$ are quite similar. Notably, the non-monotone pattern across quantiles observed under the posterior inference for \vtt{AL_Bayes} in Figure \ref{main} is now partially corrected for \vtt{EconDisadv} and \vtt{mRace}.

\begin{figure}[h]
    \centering
    \includegraphics[width = \textwidth,keepaspectratio]{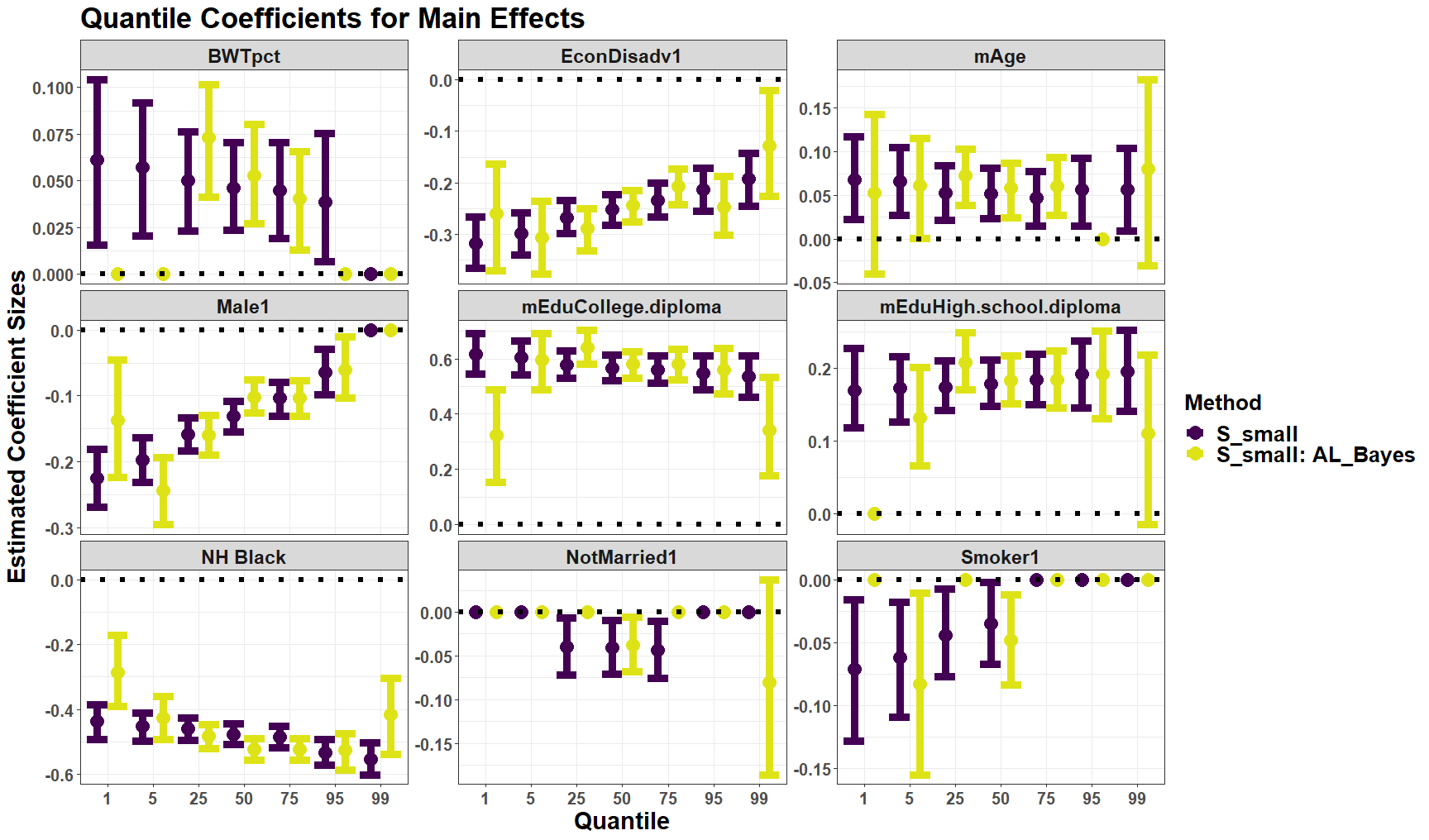}
    \caption{Inference for the main effects included in at least one quantile-specific smallest acceptable subset obtained through summarization of the LL-LS model and Bayesian quantile regression. The directionality of the coefficients is similar between the two models, while the uncertainty is greater under the Bayesian quantile regression. This is due to the asymmetric laplace likelihood, which does not capture the data generating process as well as the LL-LS model.}
    \label{ssmall comp main}
\end{figure}

To complete the analysis, we include comparisons of the inference for the interaction terms between smallest acceptable subsets, as was done in Figure \ref{intrctt} in the main paper.

\begin{figure}[h]
    \centering
    \includegraphics[width = \textwidth,keepaspectratio]{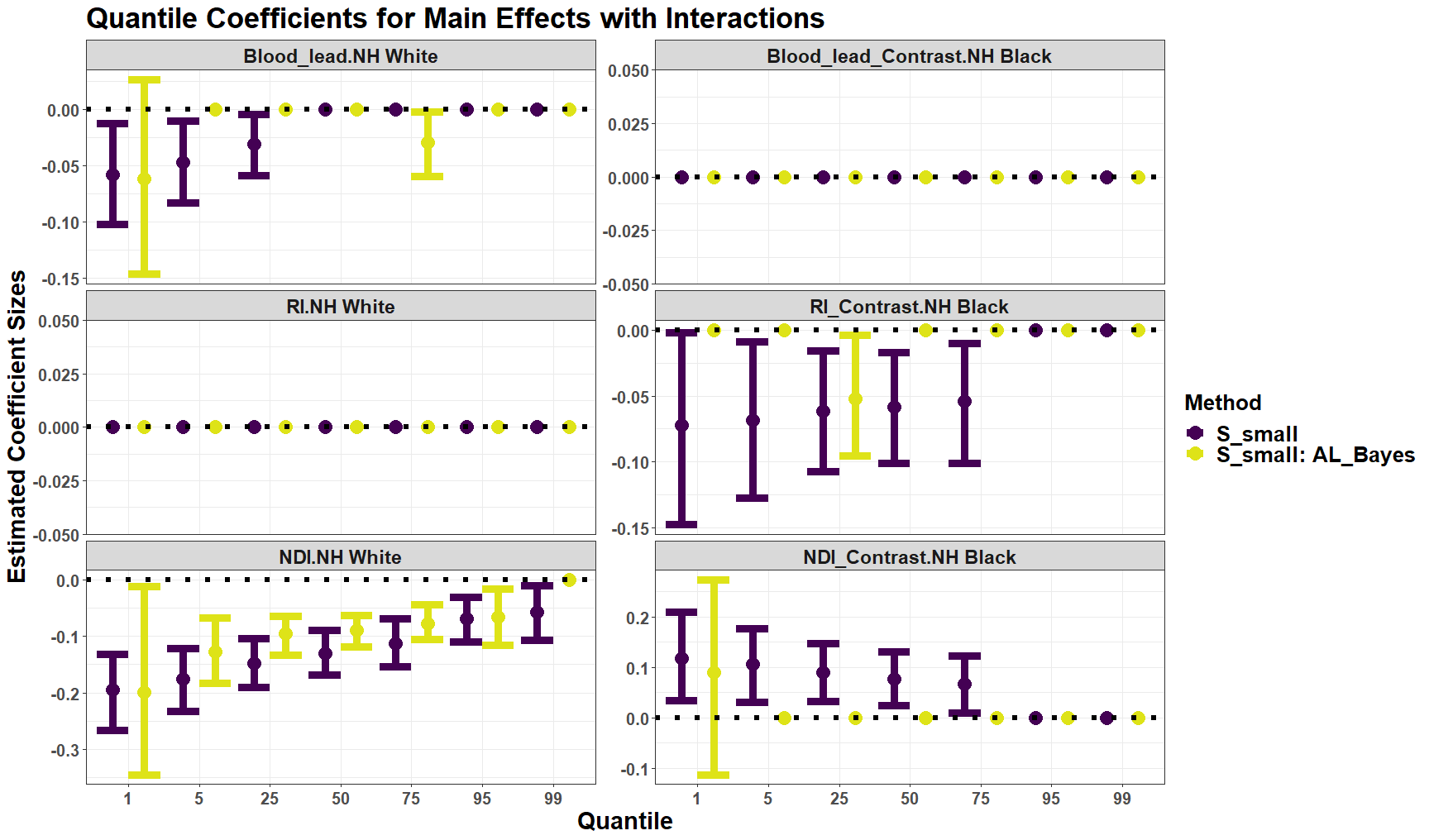}
    \caption{Inference for quantile regression coefficients obtained through posterior summarization under the LL-LS model and Bayesian quantile regression. The smallest acceptable subset from the Bayesian quantile regression detects heterogeneous effects for \vtt{Blood_lead} and \vtt{NDI}, as well as evidence of interactive effects between both \vtt{RI} and  \vtt{NDI} with \vtt{mRace}. }
    \label{ssmall comp main}
\end{figure}

For these coefficients, $S_{small}$ obtained via the Bayesian quantile regression detects heterogeneity across quantiles for \vtt{Blood_lead} and \vtt{NDI}, and provides evidence of interactions between both \vtt{NDI} and \vtt{RI} with \vtt{mRace}. This is consistent with $S_{small}$ obtained under the LL-LS model. However, $S_{small}$ under the Bayesian quantile regression does not include as many of these interactive effects across quantiles, which is due to the wide posterior uncertainty under the Bayesian quantile regression model which causes acceptable summaries to be generally more sparse. 


  \bibliographystyle{apalike}
\bibliography{reference.bib}